\documentclass[twoside,11pt,openany]{uwuedphy}%openany avoids blank pages after chapters, openright inserts them

\newcommand{\smed}{\textit{Smed}}
\DeclareMathOperator{\starup}{^{\text{*}}}

\newcommand{\revi}[1] {{\color{black} #1}}%red
\newcommand{\dt}{\Delta t}
\newcommand{\dx}{\Delta x}
\newcommand{\dy}{\Delta y}
\newcommand{\dr}{\Delta r}
\newcommand{\dxs}{\Delta x^2}

\newdimen\mycfsdim
\newcommand{\mycfs}[1]{\mycfsdim=#1pt \mycfsdim=1.2\mycfsdim
											\fontsize{#1pt}{\the\mycfsdim}\selectfont} 

\ifpdf  
    \pdfcatalog { /PageMode (/UseOutlines)
                  /OpenAction (fitbh)  }
\fi

\title{Growth and Scaling during\\ Development and Regeneration}

\ifpdf
  \author{\href{mailto:stwerner@pks.mpg.de}{Steffen Werner}}
  \cityofbirth{3. September 1984 in Bad Mergentheim}
   \researchgroup{\href{http://www.pks.mpg.de/}{Max-Planck-Institut f\"ur}\\ Physik komplexer Systeme}
	\faculty{\href{https://tu-dresden.de/die_tu_dresden/fakultaeten/fakultaet_mathematik_und_naturwissenschaften/}{Fakult\"at f\"ur Mathematik und Naturwissenschaften}}
  \collegeordept{\href{http://tu-dresden.de/die_tu_dresden/fakultaeten/fakultaet_mathematik_und_naturwissenschaften/fachrichtung_physik/fachrichtung/institute/fachrichtung_physik/itp}{Institut f\"ur Theoretische Physik}}
  \university{\href{https://tu-dresden.de}{Technische Universit\"at Dresden}}

  \crest{\includegraphics[width=12cm]{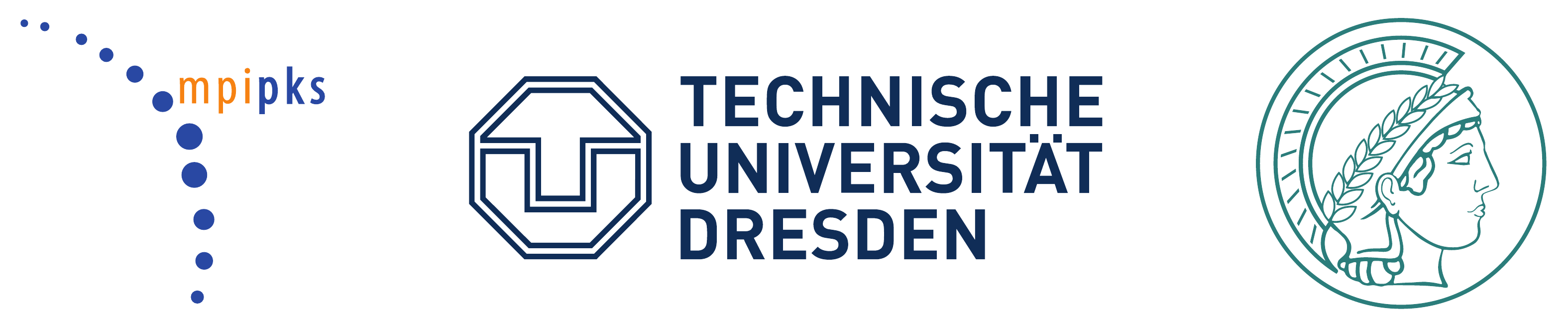}}
  
\else
  \author{Steffen Werner}
  \cityofbirth{3. September 1984 in Bad Mergentheim}
	\researchgroup{Max-Planck-Institut f\"r\\ Physik komplexer Systeme}
	\faculty{Fakult\"at f\"ur Mathematik und Naturwissenschaften}
  \collegeordept{Institut f\"ur Theoretische Physik}
  \university{Technische Universit\"at Dresden}
  \crest{\includegraphics[width=6cm]{logowue02}}
\fi

\degreedate{Dresden, 2016}

\begin{document}

%\language{english}

% sets line spacing
\renewcommand\baselinestretch{1.2}
\baselineskip=18pt plus1pt

%: ----------------------- generate cover page ------------------------

\setcounter{page}{-1}	%so there is only one warning for the missing page numbers of the title page
\maketitle  % command to print the title page with above variables
%\newcounter{olpanu} % keep track of the roman page numbering
%\setcounter{olpanu}{\value{page}}

%: ----------------------- abstract ------------------------

% Your institution may have specific regulations if you need an abstract and where it is to be placed in the document. The default here is just after title.

\frontmatter %separate the frontmatter from the \mainmatter
%\setcounter{page}{\value{olpanu}}

% Thesis Abstract -----------------------------------------------------

\begin{abstractsEnglish}        %this creates the heading for the abstract page

{\small

\noindent {\bf Life presents fascinating examples of self-organization and emergent
phenomena.} In multi-cellular organisms, a multitude of cells interact to form and maintain highly complex body plans.
This requires reliable communication between cells on
various length scales. First, there has to be the right number of cells to
preserve the integrity of the body and its size. Second, there have to be
the right types of cells at the right positions to result in a functional
body layout. In this thesis, we investigate theoretical feedback mechanisms for both
self-organized body plan patterning and size control.
\vskip0.2cm
%, based on long range communication together with simple rules for each cell.

\noindent {\bf The thesis is inspired by the astonishing scaling and regeneration
abilities of flatworms.} These worms can perfectly regrow their entire body plan even from tiny amputation fragments like the tip of the tail.
Moreover, they can grow and actively de-grow by more than a factor of 40 in length depending on feeding conditions, scaling up and down all body parts while maintaining their functionality. These capabilities prompt 
for remarkable physical mechanisms of pattern formation.\vskip0.2cm

\noindent {\bf First, we explore pattern scaling in mechanisms previously proposed to describe bio\-logical pattern formation.} We systematically extract requirements for scaling and highlight the limitations of these previous models in their ability to account for growth and regene\-ra\-tion in flatworms. In particular, we discuss a prominent model for the spontaneous formation of biological patterns
introduced by Alan Turing. We characterize the hierarchy of steady states of such a Turing mechanism and demonstrate that Turing patterns do not naturally scale.\vskip0.2cm

\noindent {\bf Second, we present a novel class of patterning mechanisms yielding entirely
self-organized and self-scaling patterns.}
 Our framework combines a Turing system with our
derived principles of pattern scaling and thus captures essential features
of body plan regeneration and scaling in flatworms.
We deduce general signatures of pattern scaling using dynamical systems
theory. These signatures are discussed in the context of experimental data.\vskip0.2cm

\noindent {\bf Next, we analyze shape and motility of flatworms.}
By monitoring worm motility, we can identify movement phenotypes upon gene
knockout, reporting on patterning defects in the locomotory system.
Furthermore, we adapt shape mode analysis to study 2D body deformations of wildtype worms,
which enables us to
characterize two main motility modes: a smooth gliding mode
due to the beating of
their cilia and an inchworming behavior based on muscle contractions.
Additionally, we apply this technique to investigate shape variations between different flatworm
species. With this approach, we aim at relating form and function in flatworms.\vskip0.2cm

\noindent {\bf Finally, we investigate the metabolic control of cell turnover and growth.}
We establish a protocol for accurate measurements of growth dynamics in
flatworms. We discern three mechanisms of metabolic energy storage; theoretical
descriptions thereof can explain the observed organism growth by rules on
the cellular scale. From this, we derive specific predictions to be tested
in future experiments. \vskip0.2cm

\noindent {\bf In a close collaboration with experimental biologists, we combine minimal theoretical descriptions with state-of-the-art experiments and data analysis.} This allows us to identify generic principles of scalable body plan patterning and growth control in flatworms. 

}

\end{abstractsEnglish}

% ---------------------------------------------------------------------- 

% Thesis Abstract -----------------------------------------------------

\begin{abstractsGerman}        %this creates the heading for the abstract page

{\small

\noindent {\bf Die belebte Natur bietet uns zahlreiche faszinierende Beispiele f\"ur die Ph\"anomene von Selbstorganisation und Emergenz.} In Vielzellern interagieren Millionen von Zellen miteinander und sind dadurch in der Lage komplexe K\"orperformen auszubilden und zu unterhalten. Dies verlangt noch einer zuverl\"assigen Kommunikation zwischen den Zellen auf verschiedenen L\"angenskalen. Einerseits ist stets eine bestimmte Zellanzahl erforderlich, sodass der K\"orper intakt bleibt und seine Gr\"osse erh\"alt. Anderseits muss f\"ur einen funktions\-t\"uchtigen K\"orper aber auch der richtige Zelltyp an der richtigen Stelle zu finden sein. In der vorliegenden Dissertation untersuchen wir beide Aspekte, die Kontrolle von Wachstum sowie die selbstorganisierte Ausbildung des K\"orperbaus.\vskip0.2cm

\noindent {\bf Die Dissertation ist inspiriert von den erstaunlichen Skalierungs- und Regene\-rations\-f\"ahigkeiten von Plattw\"urmern.} Diese W\"urmer k\"onnen ihren K\"orper selbst aus winzigen abgetrennten Fragmenten -- wie etwa der Schwanzspitze -- komplett regenerieren. Dar\"uber\-hinaus k\"onnen sie auch, je nach F\"utterungsbedingung, um mehr als das 40fache in der L\"ange wachsen oder schrumpfen und passen dabei alle K\"orperteile entsprechend an, wobei deren Funktionalit\"at erhalten bleibt. Diese F\"ahigkeiten verlangen nach bemerkenswerten physikali\-schen Musterbildungsmechanismen.\vskip0.2cm

\noindent {\bf Zun\"achst untersuchen wir das Skalierungsverhalten von fr\"uheren Ans\"atzen zur Beschreibung biologischer Musterbildung.} Wir leiten daraus Voraussetzung f\"ur das Skalieren ab und zeigen auf, dass die bekannten Modelle nur begrenzt auf Wachstum und Regene\-ration von Plattw\"urmern angewendet werden k\"onnen. Insbesondere diskutieren wir ein wichtiges Modell f\"ur die spontane Entstehung von biologi\-schen Strukturen, das von Alan Turing vorgeschlagen wurde. Wir charakterisieren die Hierarchie von station\"aren Zust\"anden solcher Turing Mecha\-nismen und veranschaulichen, dass diese Turingmuster nicht ohne \mbox{weiteres} skalieren. \vskip0.2cm

\noindent {\bf Daraufhin pr\"asentieren wir eine neuartige Klasse von Musterbildungsmechanismen, die vollst\"andig selbstorgansierte und selbstskalierende Muster erzeugen.} Unser Ansatz vereint ein Turing System mit den zuvor hergeleiteten Prinzipien f\"ur das Skalieren von Mustern und beschreibt dadurch wesentliche Aspekte der Regeneration und Skalierung von Platt\-w\"urmern. Mit Hilfe der Theorie dynamischer Systeme leiten wir allgemeine Merkmale von skalierenden Mustern ab, die wir im Hinblick auf experimentelle Daten diskutieren. \vskip0.2cm

\noindent {\bf Als n\"achstes analysieren wir Form und Fortbewegung der W\"urmer.}
Die Auswertung des Bewegungsverhaltens, nachdem einzelne Gene ausgeschaltet wurden, erm\"oglicht R\"uck\-schl\"ussse auf die Bedeutung dieser Gene f\"ur den Bewegungsapparat.
Dar\"uber hinaus wenden wir eine Hauptkomponenten\-analyse auf die Verformungen des zweidimensionalen Wurmk\"orpers w\"ahrend der nat\"urlichen Fortbewegung an.~Damit sind wir in der Lage, zwei wichtige Fortbewegungsstrategien der W\"urmer zu charakterisieren: eine durch den Zilienschlag angetriebene gleichm\"assige Gleitbewegung und eine raupenartige Bewegung, die auf Muskelkontraktionen beruht. Zus\"atz\-lich wenden wir diese Analysetechnik auch an, um Unterschiede in der Gestalt von verschiedenen Plattwurm\-arten zu untersuchen. Grunds\"atzlich zielen alle diese Ans\"atze darauf ab, das Aussehen der Plattw\"urmer mit den damit verbundenen Funktionen verschiedener K\"orperteile in Beziehung zu setzen.\vskip0.2cm

\noindent {\bf Schlussendlich erforschen wir den Einfluss des Stoffwechsels auf den Zell\-austausch und das Wachstum.} Dazu etablieren wir Messungen der Wachstumsdynamik in Plattw\"urmern. Wir unterscheiden drei Mechanismen f\"ur das Speichern von Stoffwechselenergie, deren theoreti\-sche Beschreibung es uns erm\"oglicht, das beobachtete makroskopische Wachstum des Organismus mit dem Verhalten der einzelnen Zellen zu erkl\"aren. Basierend darauf leiten wir Vorhersagen ab, die nun experimentell getestet werden. \vskip0.2cm

\noindent {\bf In enger Zusammenarbeit mit Kollegen aus der experimentellen Biologie f\"uhren wir minimale theoretische Beschreibungen mit modernsten Experimenten und Analyse\-techniken zusammen.} Dadurch sind wir in der Lage, Grundlagen sowohl der skalierbaren Ausbildung des K\"orperbaus als auch der Wachstumskontrolle bei Plattw\"urmern herauszuarbeiten.

}

\end{abstractsGerman}

% ---------------------------------------------------------------------- 

% Thesis Acknowledgements ------------------------------------------------

%\begin{acknowledgementslong} %uncommenting this line, gives a different acknowledgements heading
\begin{acknowledgements}      %this creates the heading for the acknowlegments

I would like to acknowledge all the people who supported me during my PhD. My special thanks go to:\vskip0.5cm

\noindent My supervisors \textbf{Frank J\"ulicher} und \textbf{Benjamin Friedrich} for their great support and assistance and for teaching me their inspiring way of doing science. Thanks to Frank J\"ulicher for his enthusiasm about my projects, for always making sure to be available for insightful discussions and advice and generally for all the stimulating input. Thanks to Benjamin Friedrich for his valuable  and highly appreciated guidance, for encouraging me to pursue my own ideas and for always being there to help structuring my work, to make sure I stay on track and to provide new food for thoughts.

\noindent \textbf{Jochen Rink} for being the driving force in this exciting and fruitful colla\-boration, for his open-mindedness towards our theoretical ideas and for sharing his visionary thoughts on the flatworm project with us. Thanks also for creating such an enjoyable and inspiring atmosphere in the lab.

\noindent \textbf{Tom St\"uckemann} and \textbf{Albert Thommen} not only for working hard to produce some amazing data but also for sharing this data with me, which triggered major parts of this thesis. Thanks for all the motivating discussions and the productive team spirit.

\noindent \textbf{Nicole Alt}, \textbf{Johanna Richter} and \textbf{many further student helpers} for putting a lot of effort and dedication in taming the beasts under the macroscope, generating beautiful movies for me to analyze.

\noindent \textbf{Sarah Mansour}, \textbf{Olga Frank}, \textbf{James Cleland} and \textbf{Shang-Yun Liu} for the exciting times we are having discussing science and beyond.

\noindent \textbf{All the other members of the Rink lab} for creating such a fun atmos\-phere to work in.

\noindent \textbf{Lutz Brusch} and \textbf{Michael K\"ucken} for many fruitful discussions during the theory gatherings.

\noindent \textbf{Daniel Aguilar-Hidalgo} for being a motivating sparring partner with respect to the scaling project and for the great time we had developing theories.

\noindent The summer interns \textbf{Manuel Beir\'an-Amigo} and \textbf{Yihui Quek} for the intruiging questions they asked and for their enthusiasm about the flatworm project creating an extra thrust pushing it forward.

\noindent \textbf{Ulrike Burkert and many other members of the administration and the IT department} for making everything run so smoothly and for the quick and non-bureaucratic help with no matter what question.

\noindent\textbf{Vivien Scherr and her colleagues from the library} for the great efforts to track down any exotic piece of literature I was asking for.

\noindent\textbf{My colleagues and friends} for contributing in many ways to provide a great working environment and to make my time at the PKS really enjoyable. Thanks for all the discussions and the advice about science and life in general, all the coffee breaks, kicker sessions, whisky tastings, tango lessons, movie nights, sciency pub outings ...

\noindent\textbf{My parents and my brother Jochen} for their enduring support during my whole life and for the great feeling that I always can rely on them.

\noindent\textbf{Ruth} for all her patience and support, for her understanding of my late night working sessions and for giving me great backup during the final stretch as well as for the proofreading of my thesis.\vskip0.5cm

I also gratefully acknowledge the funding by
the \textbf{Max Planck Society} and
the \textbf{German Federal Ministry of Education and Research (BMBF)}.

\end{acknowledgements}
%\end{acknowledgmentslong}

% ------------------------------------------------------------------------

%: ----------------------- contents ------------------------

\setcounter{secnumdepth}{3} % organisational level that receives a numbers
\setcounter{tocdepth}{2}    % print table of contents for level 3
\tableofcontents            % print the table of contents
% levels are: 0 - chapter, 1 - section, 2 - subsection, 3 - subsubsection (not in toc)

%: ----------------------- glossary ------------------------

% Tie in external source file for definitions: /0_frontmatter/glossary.tex
% Glossary entries can also be defined in the main text. See glossary.tex
%\include{0_frontmatter/glossary} 

%\begin{multicols}{2} % \begin{multicols}{#columns}[header text][space]
%\begin{footnotesize} % scriptsize(7) < footnotesize(8) < small (9) < normal (10)

%\printnomenclature[1.5cm] % [] = distance between entry and description
%\label{nom} % target name for links to glossary

%\end{footnotesize}
%\end{multicols}

% --------------------------------------------------------------
%:                  MAIN DOCUMENT SECTION
% --------------------------------------------------------------

% the main text starts here with the introduction, 1st chapter,...
\mainmatter
\pagestyle{fancy}
%\renewcommand{\chaptername}{} % uncomment to print only "1" not "Chapter 1" (only neccessary for book class)

%: ----------------------- subdocuments ------------------------

% Parts of the thesis are included below. Rename the files as required.
% But take care that the paths match. You can also change the order of appearance by moving the include commands.

% this file is called up by thesis.tex
% content in this file will be fed into the main document

%: ----------------------- introduction file header -----------------------
\chapter{Introduction}

% the code below specifies where the figures are stored
%\ifpdf
%    \graphicspath{{1_introduction/figures/PNG/}{1_introduction/figures/PDF/}{1_introduction/figures/}}
%\else
%    \graphicspath{{1_introduction/figures/EPS/}{1_introduction/figures/}}
%\fi

% ----------------------------------------------------------------------
%: ----------------------- introduction content ----------------------- 
% ----------------------------------------------------------------------

%potentially also use \par\noindent and not the empty line but no \\

\section{Development, growth and regeneration}%essentially characterizes important aspects of life in general
\begin{figure}[tbp]
  \centering
  \includegraphics[width=1\textwidth]{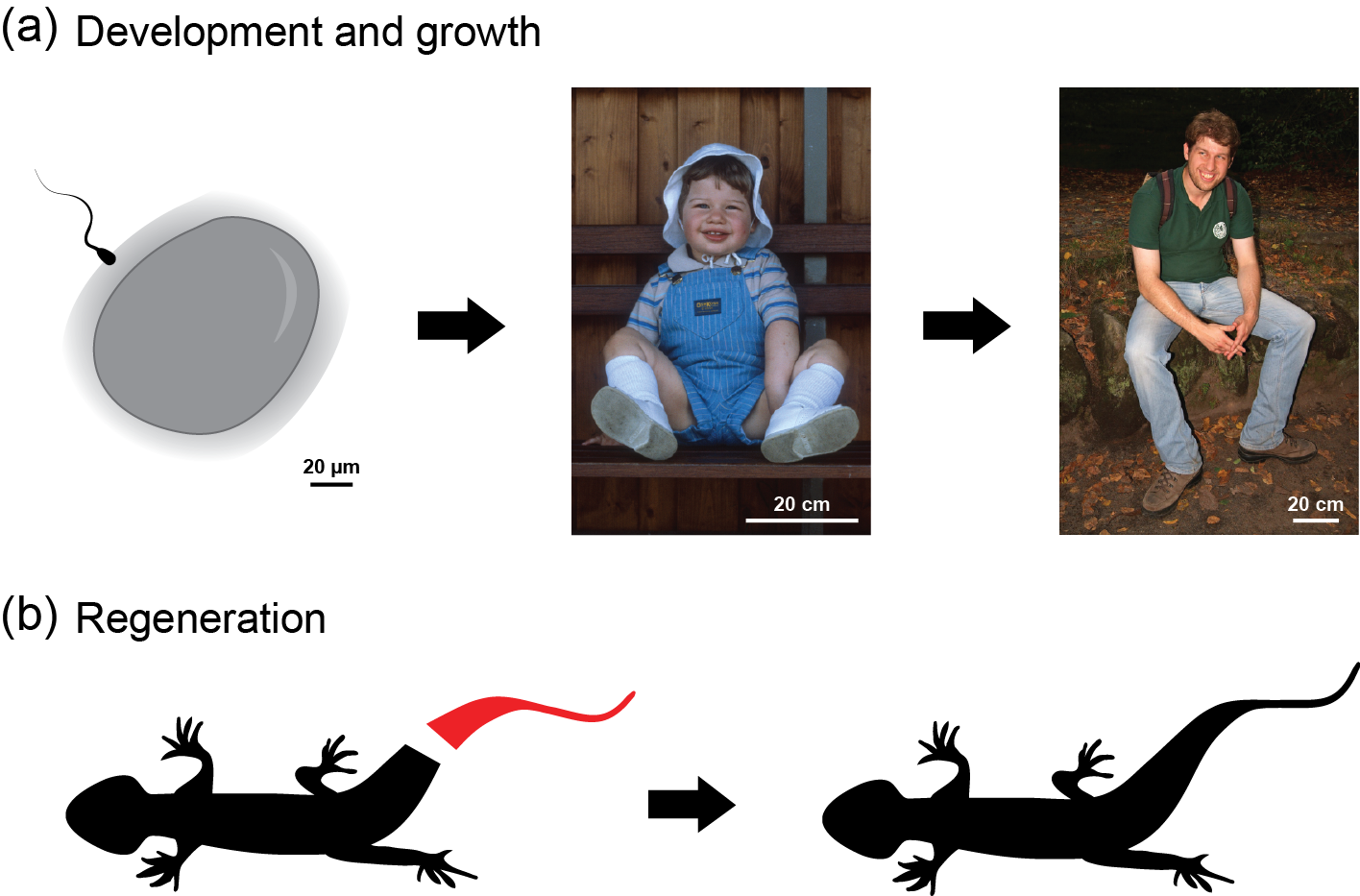}
  \caption[Development and Regeneration]
   {(a) Human development starts from a single fertilized egg cell, from which the body plan emerges and grows to its final size. (b) Many multi-cellular organisms like lizards and salamanders can regenerate major parts of their body.}
   \label{fig:DevelopmentRegeneration}
\end{figure}

The world around us is populated by a great variety of organisms of very different shapes, sizes and levels of complexity.  Many of the most complex organisms, including
\sidenote{0.45\textwidth}{``Cell and tissue, shell and bone, leaf and flower, ... Their problems of
form are in the first instance mathematical problems, their problems of growth are essentially physical problems, and the morphologist is, ipso facto, a student of physical \mbox{science}.'' \;---\; D'Arcy W. Thompson, On Growth and Form, 1945 \cite{thompson1945growth}}
\noindent humans, develop from a single fertilized egg cell, see Fig.~\ref{fig:DevelopmentRegeneration}(a). 
The egg divides multiple times to give rise to the many cells that form the different tissues of the adult organism \cite{gilbert2014developmental, wolpert2011principles}. This embryonic \textbf{development} results in a well-defined body plan of the organism, which eventually can reproduce again.
One important aspect of develop\-ment is \textbf{growth}, i.e.~the increase in organism size. The growth at the scale of the organism follows from processes at the cellular scale: (i) an increase in cell number by cell division, (ii) an increase in cell size by cellular growth and (iii) an increase in the extra-cellular material \cite{wolpert2011principles}.

\noindent During most of its lifetime, an organism maintains shape and function of its body, despite the fact that cells continuously become damaged and get lost \cite{pellettieri2007cell, gilbert2014developmental, wolpert2011principles}. This \textbf{homeostasis} requires the sustained addition of new cells by cell division as well as mechanisms of controlled cell death such as apoptosis. Importantly, the turnover processes have to be well orchestrated at the cell, tissue and organism level. Imperfect homeostasis results in aging of the organism \cite{pellettieri2007cell}.

\noindent Furthermore, many organisms can regenerate after injury to some extent, see Fig.~\ref{fig:DevelopmentRegeneration}(b) \cite{sanchezalvarado2006bridging, gilbert2014developmental, wolpert2011principles}. 
\textbf{Regeneration} refers to de-novo formation of large parts of tissues and organs that have been damaged or lost. In contrast to embryonic development, which comprises a fixed sequence of morphogenetic events starting from the ferti\-lized egg as a well-defined initial condition, the starting point of regeneration strongly depends on the injury and is thus variable.
 It is a major question to what extent both processes are guided by the same principles and depend on the same mechanisms \cite{sanchezalvarado2006bridging}.

\noindent

\noindent The ability to reproduce and the permanent struggle against decay are important cha\-rac\-teristics of life in general \cite{schroedinger1944what}. We are only beginning to understand the respective processes of development, growth and regeneration in simple model systems with the help of mo\-dern molecular biology. In this thesis, we combine minimal theoretical descriptions with the analysis of biological data in flatworms as model systems in order to extract fundamental physical principles for body plan patterning and growth control in multicellular organisms.

\subsection{From cells to tissues to organisms}
\noindent The question of how an organism forms has puzzled natural philosophers and researchers for more than two millennia. It was Aristotle who performed one of the earliest do\-cumented experiments in developmental biology in 350 BC \cite{aristotle350BChistory,gilbert2014developmental, wolpert2011principles}. He opened chicken eggs at various time intervals after fertilization and observed that the embryo gradually resembles a chicken. The gradual formation of an organism, called \textbf{epigenesis}, was confronted with the alternative hypothesis of preformation \cite{gilbert2014developmental, wolpert2011principles}. The latter assumes a completely pre-formed miniature body, which then only grows. Despite the work of Aristotle, the theory of preformation was still prevalent in Europe until the 18th century and is embodied in the idea of the ``homunculus'', the tiny version of a person encapsulated in the sperm cell \cite{gilbert2014developmental, wolpert2011principles, vonGoethe1933faustII}. Later a related discussion arose about the concepts of pre-encoded and emergent complexity, as will be mentioned below \cite{sander1997landmarks}.
%We discuss these alternatives again with respect to the experiments by Roux and Driesch below.%This illustrates the difficulties to grasp the concept of de-novo formation of a complex body plan.

\noindent The basic building blocks of higher order organisms are the cells, which come at very different shapes and sizes \cite{gilbert2014developmental, wolpert2011principles}.
For example, a muscle cell and a blood cell have a completely different appearance and very different properties but they both originate from a single egg cell, which has divided many times to give rise to all the cells of the body \cite{gilbert2014developmental, wolpert2011principles}. The cells become committed to fulfill specific tasks and change their properties during the process of \textbf{differentiation}. Uncommitted cells that can differentiate into other cell types are called \textbf{stem cells}. Often, there are also tissue-specific stem cells that are already partially committed and can only turn into a subset of cell types. In some organisms, differentiated cells are able to de-differentiate into less committed cells \cite{baguna2012planarian,tata2013dedifferentiation}.

\begin{figure}[tbp]
  \centering
  \includegraphics[width=1\textwidth]{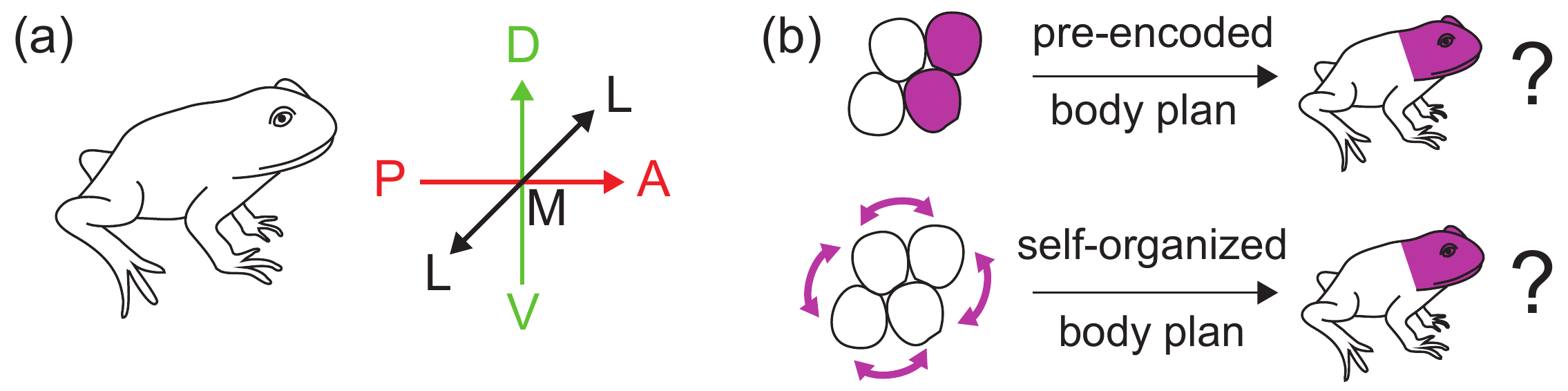}
  \caption[Body axes and morphogenesis]
   {(a) The bilaterian body plan, for example of a frog, is characterized by three perpendicular axes in anterior-posterior (AP), dorso-ventral (DV) and medio-lateral (ML) direction. (b) Two concepts of morphogenesis: mosaic development based on pre-encoded structures (exemplified by purple cells forming the head), self-organized formation of a body plan as an emergent phenomenon based on mutual interactions between cells.}
   \label{fig:AxesDevelopment}
\end{figure}

\noindent During embryonic development and regeneration, cells differentiate and organize in a position dependent manner to form a well-defined body plan. Most modern animals are \textit{Bilaterians}, characterized by three main body axis, see Fig.~\ref{fig:AxesDevelopment}(a): the anterior-posterior axis from tail to head, the dorso-ventral axis from the front to the backside and a mirror symmetric medio-lateral axis \cite{niehrs2010growth}. Yet, how is the cellular behavior choreographed with respect to this internal coordinate system?

\noindent With the advancements in light microscopy
%, especially attributed to van Leeuwenhoek, Hooke and later Abbe and Zeiss,
and the ability to observe microscopic structures, biologists could address this question and perform experi\-ments, in which they selectively perturbed a specific part of an organism in order to reveal its functions in body plan patterning \cite{kriss1998history,mocek1971wilhelm,morgan1901regeneration, driesch1908science, gilbert2014developmental, wolpert2011principles}. 
As a result, biological research changed from a descriptive to an experimental science. Here, we will highlight three early experiments by Chabry, Driesch and Morgan to discuss important concepts of morphogenesis.

\noindent Chabry selectively killed individual cells in the early embryo of the marine invertebrate \textit{Tunicate} after the first or second cell division with a needle. In consequence, only certain parts of the organism developed, depending on which cells he had destroyed \cite{gilbert2014developmental,sander1997landmarks}. The results were later confirmed by completely removing the two muscle precursor cells of the 8-cell embryo \cite{whittaker1982muscle,gilbert2014developmental}.
These separated cells became muscle cells by themselves, while the remaining embryo was lacking the muscles. 
%Roux killed one of the two cells after the first cell division of the frog with a hot needle and reported that exactly half of the embryo develops as if cut in two with a razor blade \cite{driesch1908science,wolpert2011principles,sander1997landmarks}.
Driesch performed a similar experiment in sea-urchins but obtained completely opposite results \cite{driesch1908science}. He separated the two cells after the first round of cell division and observed that
a single cell can develop into a perfectly patterned organism, just of a smaller size.
At the same time, Morgan was performing various regeneration experiments especially with the freshwater polyp \textit{Hydra} and flatworms, and he reported that these animals could re-grow perfectly patterned heads, tails and other body parts after amputation \cite{morgan1901regeneration}.

\noindent

%\noindent Hence, in some cases the developmental program is not flexible enough to compensate for the loss of cells but there are also many examples in which cells seem to reorganize and form missing pieces at the right place and with correct proportions.
\noindent These observations can be discussed in the light of two fundamental concepts of morphogenesis, see Fig.~\ref{fig:AxesDevelopment}(b) \cite{sander1997landmarks,gilbert2014developmental}: First, the theory of self-differentiation (or mosaic development) builds on the idea of a pre-encoded (hidden) complexity in the early partitioning of the tissue that then only enfolds. Second, the converse theory (sometimes called conditional specification) considers complexity of the body plan as an emergent phenomenon by the interaction of different parts. Today, we begin to appreciate that embryonic development combines both paradigms. The second one has the appeal to account for regeneration in a natural fashion and will be studied in this thesis in the context of self-organized body plan patterning.

\subsection{Cellular communication and chemical signals}
\noindent At the end of the 19th and the beginning of the 20th century, the existence of ``formative substances'' was postulated to control cell fate during development and regeneration \cite{rogers2011morphogen,morgan1905polarity, morgan1901regeneration, sander1997landmarks}. It was proposed that these substances are found in graded abundance originating from the animal poles as the organizing centers and that they determine polarity and growth of an organism by controlling the cellular behavior \cite{rogers2011morphogen,morgan1905polarity, morgan1901regeneration}. Related concepts were built on a ``physiological gradient'', inspired for example by the fact that the regenerative capability in some flatworm species varies along the body axis from head to tail \cite{child1911studies}.

\noindent These considerations lead to the notion of \textbf{morphogens} as specific signaling molecules that are secreted in distinct source regions and spread in the tissue. The morphogen concentrations provide chemical cues that control division and differentiation of cells \cite{gurdon2001morphogen,neumann1997morphogens, gilbert2014developmental, wolpert2011principles}. 
The term was originally coined by Turing, who proposed a purely theoretical mechanism for the spontaneous emergence of chemical patterns as a template for the body plan layout \cite{turing1952chemical}.
As a complementary theoretical approach, it was discussed how a pre-existing organizing region can instruct tissue patterning by secretion of morphogens \cite{wolpert1969positional, rogers2011morphogen}.
It was proposed that graded concentration profiles provide cells with the information about their position within the tissue. We will discuss a well-known illustration of this idea, the French flag model, in Section \ref{Intropatterningtheory}. Next, we provide biological examples for organizing regions and concentration gradients of signaling molecules. 

\noindent Early experiments by Mangold and Spemann found evidence for an organizing region that instructs body plan patterning in the embryo of the frog \textit{Xenopus laevis} \cite{spemann1924induktion}. When transplanting the now so-called Spemann organizer into another frog embryo, the latter developed a second perfectly patterned head. Furthermore, experimental evidence for ``organizing substances'' at the animal poles was found in leaf hoppers \cite{sander1975patterning,sander1960analyse,kalthoff1976analysis}. After splitting the embryo in a head  and a tail fragment, in most cases neither part developed normally. Yet, if substances from the tail tip were moved to the head fragment, the head fragment developed into a complete embryo. Interestingly, also the tail fragment developed further, indicating a concentration-dependent effect of these tail substances.

%A substance at the posterior pole of the embryo has been reported, containing symbiotic bacteria, which can be used as a marker. After splitting the embryo in a head (anterior) and tail (posterior) fragment, the anterior fragment either does not develop further or forms only head structures. However, after transplanting the mentioned posterior substance to the posterior side of the head fragment, a complete embryo can form. Interestingly, also the tail fragment develops further after part of the posterior substance was removed, indicating a concentration-dependent effect of this posterior substance.
%- leaf hopper
%- transplanting egg material
%- posterior pole symbiotic bacteria->marker
%- shifting to head part before splitting (abschnueren)
%  experiments by Sander, von Ubisch and Stumpf found evidence that substances can influence cell fate in a concentration-dependent manner \cite{rogers2011morphogen,sander1997landmarks}.

\begin{figure}[tbp]
  \centering
  \includegraphics[width=1\textwidth]{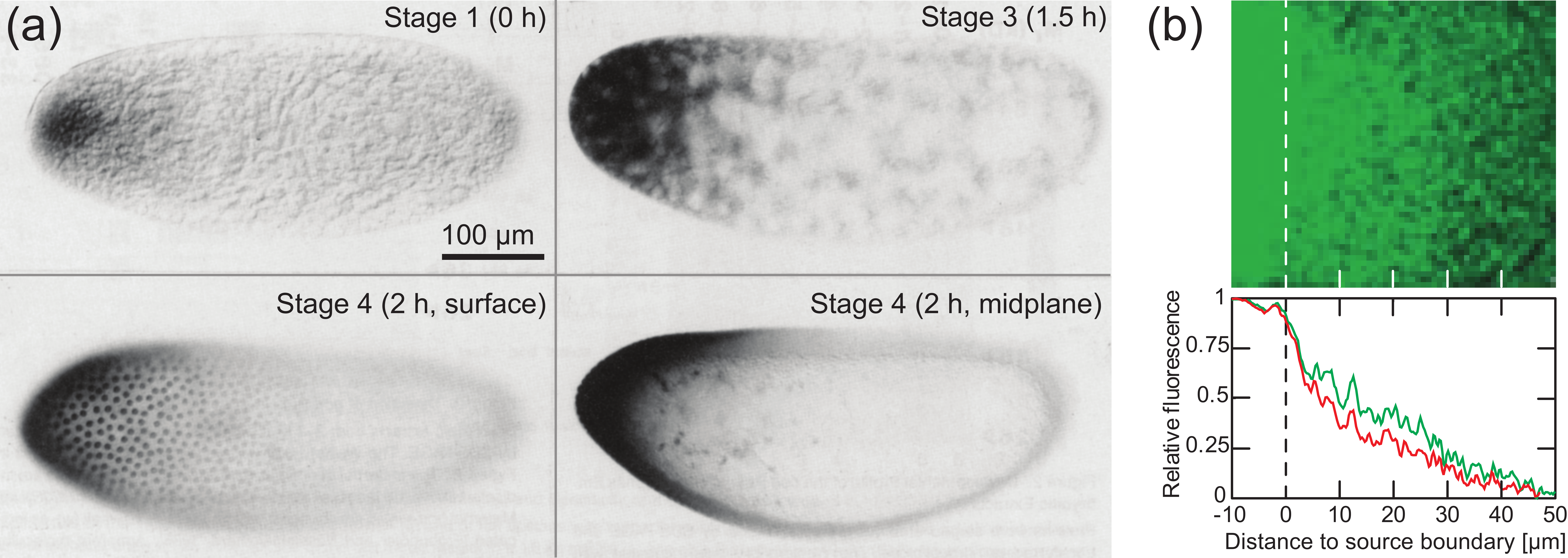}%png
  \caption[Morphogen gradients]
   {(a) Bicoid protein gradient (black) in the embryo of the fruit fly \textit{Drosophila melanogaster} at different stages of development (anterior side at the left, modified with permission from \cite{driever1988gradient}, scale bar and approximate timing added by the author) (b) Decapentaplegic protein in the imaginal wing disc of the fruit fly labeled by GFP and quantification by GFP fluorescence (green) and GFP immunostaining (red) (modified with permission from \cite{kruse2004dpp}).}
   \label{fig:MorphGrad}
\end{figure}

\noindent  Pioneering work by N\"usslein-Volhard and colleagues could identify the protein Bicoid as a signaling molecule in the embryo of the fruit fly \textit{Drosophila melanogaster} and demonstrated its instructive role in tissue patterning, see Fig.~\ref{fig:MorphGrad}(a) \cite{frohnhoefer1986organization,driever1990autonomous,ephrussi2004seeing}. They also visualized its graded concentration profile decrea\-sing from the anterior tip, which could be fitted by an exponential function \cite{driever1988gradient,driever1988bicoid}. Additionally, they showed that Bicoid influences cellular behavior in a concentration-dependent manner \cite{driever1988bicoid}.

\noindent There are several key signaling systems for patterning and growth control that are widespread across the animal kingdom. Prominent examples belong to the Transforming growth factor $\beta$ (TGF-$\beta$) superfamily and to the Wnt family.
TGF-$\beta$ proteins can be found in a wide range of organisms from simple worms to mammals. They control growth, patterning, tissue homeostasis and even the immune system  \cite{herpin2004transforming}.
In this thesis,  we will encounter four examples of these proteins: Activin in the clawed frog \textit{Xenopus} \cite{gurdon1995direct,green1992responses,green1990graded} and Decapentaplegic (Dpp) in the fruit fly \cite{spencer1982decapentaplegic,wartlick2011dynamics,affolter2007decapentaplegic,kruse2004dpp}, see Fig.~\ref{fig:MorphGrad}(b), as well as Bone morphogenetic protein (Bmp) and Anti-dorsalizing morphogenic protein (Admp) in flatworms \cite{gavino2011bmp,adell2010gradients}.
%concerning capital letters: bmp - gene, Bmp - protein, BMP - family, pathway
Also Wnt family members are found in orga\-nisms from invertebrates to humans \cite{almuedocastillo2012wnt,oberhofer2014wnt,raspopovic2014digit,Davidson2012Wnt,komiya2008wnt,niehrs2010growth}. The name is a portmanteau of Wingless (the corresponding protein in the fruit fly), and Integration 1 (the homolog originally identified in mammal cancer research). These proteins control the division, differentiation and migration of cells as well as the specification of the main body axes. In this thesis, we especially consider the head-tail polarity in flatworms associated with Wnts.

\noindent

\subsection{From signals to genes and back}
\noindent The function of a cell is largely determined by the proteins inside \cite{wolpert2011principles}. There are several classes of proteins. Housekeeping proteins for the maintenance of the basic cellular functions such as protein synthesis, structural support and cell metabolism are present in all cells under physiological conditions. Other proteins are only found in certain types of cells and are involved in specific tasks such as division, developmental signaling, force generation, sensory perception or immune responses.

\begin{figure}[tbp]
  \centering
  \includegraphics[width=1\textwidth]{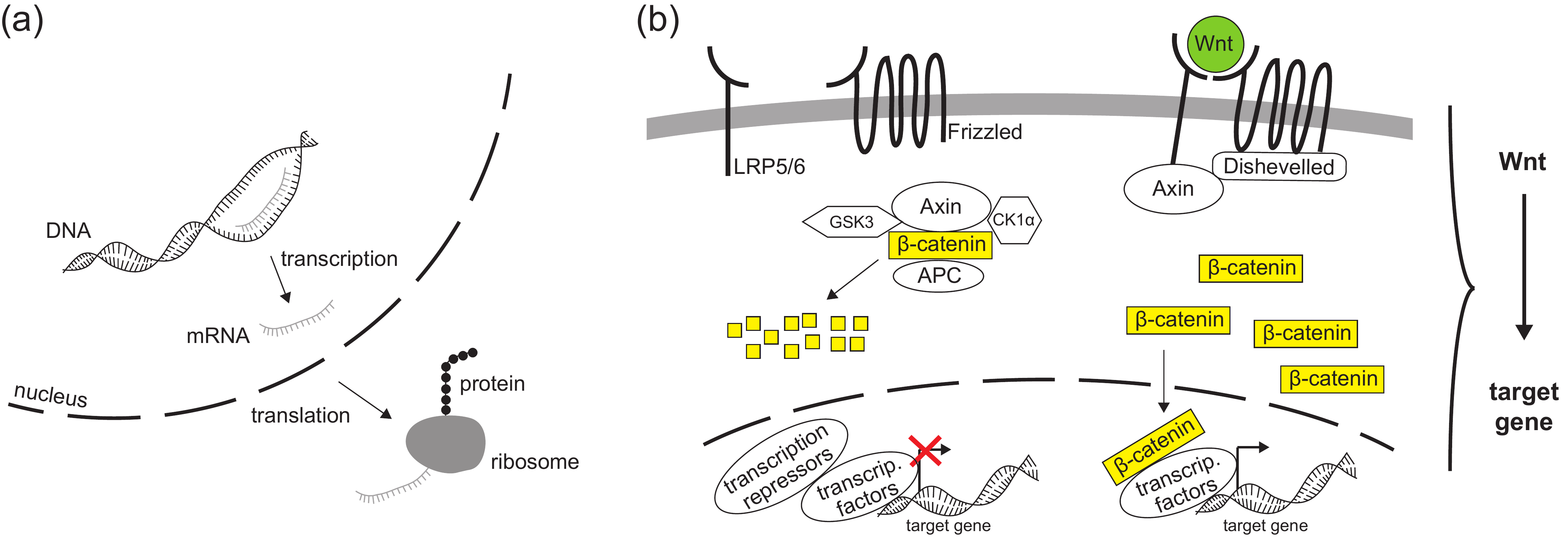}
  \caption[Gene expression]
   {(a) Gene expression: proteins are synthesized by transcribing the genetic information saved in the DNA to mRNA molecules, which then act as a blueprint for the protein.
   (b) Gene expression can be activated or de-activated by signaling molecules (here exemplified for canonical Wnt signaling). In the absence of Wnt molecules, $\beta$-catenin is tagged for degradation by a destruction complex, which includes Axin, Adenomatosis polyposis coli (APC), Glycogen synthase kinase 3 (GSK3) and Casein kinase $1\alpha$ (CK$1\alpha$). If Wnt is bound to the Frizzled receptors and the Low density lipoprotein receptor-related protein 5 or 6 (LRP5/6), the formation of the destruction complex is suppressed and $\beta$-catenin can translocate to the nucleus to act as a co-activator of various genes \cite{gilbert2014developmental, wolpert2011principles,nusse2013wnt}.
%(b) Genes can set up a positive or negative feedback loop if the produced protein acts (directly or indirectly) on the gene that encodes it.
   }
   \label{fig:Genes}
\end{figure}

\noindent The blueprint for all proteins is chemi\-cally encoded in the \textbf{genome}. The genome consists of Deoxyribonucleic acid (DNA) macromolecules, in which characteristic base pair sequences (genes) represent the proteins.
This genetic information can be read out from the DNA strands by a gene expression pathway as depicted in Fig.~\ref{fig:Genes}(a). First, the DNA unfolds at the respective gene site and the gene sequence is successively copied to a messenger ribonucleic acid strand (mRNA) during a process called \textbf{transcription}. Again a sequence of nitrogenous bases encodes the specific protein. Second, during \textbf{translation}, the mRNA acts as a template for protein synthesis with the help of ribosomes (i.e.~large complexes consisting of proteins and RNA strands).

\noindent All somatic cells (with a few exceptions) are genetically equivalent because they all stem from the same initial egg cell. During cell division, the DNA becomes duplicated and one copy remains in each daughter cell \cite{gilbert2014developmental, wolpert2011principles}. The cells acquire different fates if different genes are activated and thus different proteins are present in the cells. This activation of genes is in turn also controlled by signaling molecules. Fig.~\ref{fig:Genes}(b) illustrates the signaling cascade of canonical Wnts as an example. In the absence of Wnt molecules, $\beta$-catenin is tagged for degradation by a destruction complex involving several molecules such as Adenomatous poluposis coli (APC) and Axin. Upon binding of Wnt molecules to the Frizzled receptors and co-receptors (LRP5/6), Axin is recruited to the membrane and the formation of the destruction complex is suppressed. Thus, $\beta$-catenin can accumulated and reach the nucleus, where it acts as a transcription co-activator for specific target genes. In effect, Wnt has an activating effect on the expression of these target genes.

\noindent The resulting proteins can fulfill certain tasks for the cell in which they have been synthesized, yet they can also be released and act as morphogens to activate or de-activate parts of the genome in other cells. This realizes positive and negative feedback loops, from which complex cellular signaling networks are built.

\subsection{Gene expression can be modified in experiments}
Experimentalists can interfere with the synthesis of specific proteins by exploiting the control and error correction machinery of the cells, which ensures robustness of the important processes of DNA duplication, transcription and translation and modifies their outcome \cite{gilbert2014developmental}. One such mechanism is RNA interference (RNAi), for which small interfering RNA (siRNA) pieces are used by the cell to target specific mRNA strands, mainly for destruction \cite{agrawal2003rna}. For example, this can be an immune response against exogenous RNA introduced by viruses.
Similarly, experimentalists can artificially suppress a protein of choice by introducing a RNA sequence for this protein in the cell. Such RNAi techniques are applied to obtain some of the data presented in this thesis.\\

%\noindent Experimentalists can interfere with the synthesis of proteins, targeting specific genes. For example, suppression of a protein of choice can be achieved by introducing a corres\-ponding RNA sequence.
%This technique exploits the control and error correction machinery of the cells, which ensures robustness of the important processes of DNA duplication, transcription and translation and modifies their outcome \cite{gilbert2014developmental}. One such cellular mechanism is RNA interference (RNAi), for which small interfering RNA (siRNA) pieces are used by the cell to identify specific mRNA strands mainly for destruction \cite{agrawal2003rna}. For example, this can be an immune response against exogenous RNA introduced by viruses. It also enables the experimentalist to knockout specific genes, as has been 
%done for data presented in this thesis.

%The cells are equipped with a large control and error correction machinery to avoid failures of the important processes of DNA duplication, transcription and translation and to modify their outcome \cite{gilbert2014developmental}. One such mechanism is RNA interference (RNAi), for which small interfering RNA (siRNA) pieces are used by the cell to target specific mRNA strands mainly for destruction \cite{agrawal2003rna}. This can for example be an immune response against exogenous RNA introduced by viruses. Similarly, by introducing a RNA sequence for a protein of choice, experimentalists can artificially suppress this protein. Such RNAi techniques are applied to obtain experimental data presented in this thesis.

\section{Flatworms as a model organism for scaling, growth and regeneration}
%\noindent \sidenote{0.45\textwidth}{``Imo semel diu adservatam portionem anticam tandem post plurium dierum intervallum inspiciens, defectum in integrum restituisse vidi, renataque pars multo tenerior, pellucidiorque aliquamdiu mansit reliquo corpore.'' \;---\; Peter S. Pallas, Spicilegia zoologica quibus novae imprimis et obscurae animalium species iconibus, descriptionibus atque commentariis illustrantur, 1774 \cite{pallas1774spicilegia}}
%\noindent Flatworms have been already described in 863 AD in China and their astonishing ability to regenerate is known in Europe at least since the 18th century \cite{pallas1774spicilegia,elliott2013history,duan863you}.
\noindent Classic experiments on flatworm regeneration already inspired the idea of morphogenetic gradients \cite{morgan1905polarity, morgan1901regeneration,child1911studies}.
In recent years, the flatworm \textit{Schmidtea mediterranea} (\textit{Smed}) has become increasingly popular as a model organism to study regeneration and growth, aging, and even behavior \cite{newmark2002not,oviedo2008planarians,gentile2011planarian, talbot2011quantitative, salo2009planarian, prados2012cue, shomrat2013automated, pellettieri2007cell, rompolas2009schmidtea}.
There are many different flatworm species populating very diverse habitats all around the world. They are found in saltwater, in freshwater and in the soil; some live more than \mbox{1000 m} under the sea and some parasitic species (like flukes and tapeworms) in the body of other organisms \cite{oviedo2008planarians,rink2013stem,rompolas2009schmidtea,sluys1998new}. \smed{} is a non-parasitic flatworm living in freshwater. Such free-living species are sometimes also referred to as ``planarians'' \cite{egger2006freeliving}.

\begin{figure}[tbp]
  \centering
  \includegraphics[width=1\textwidth]{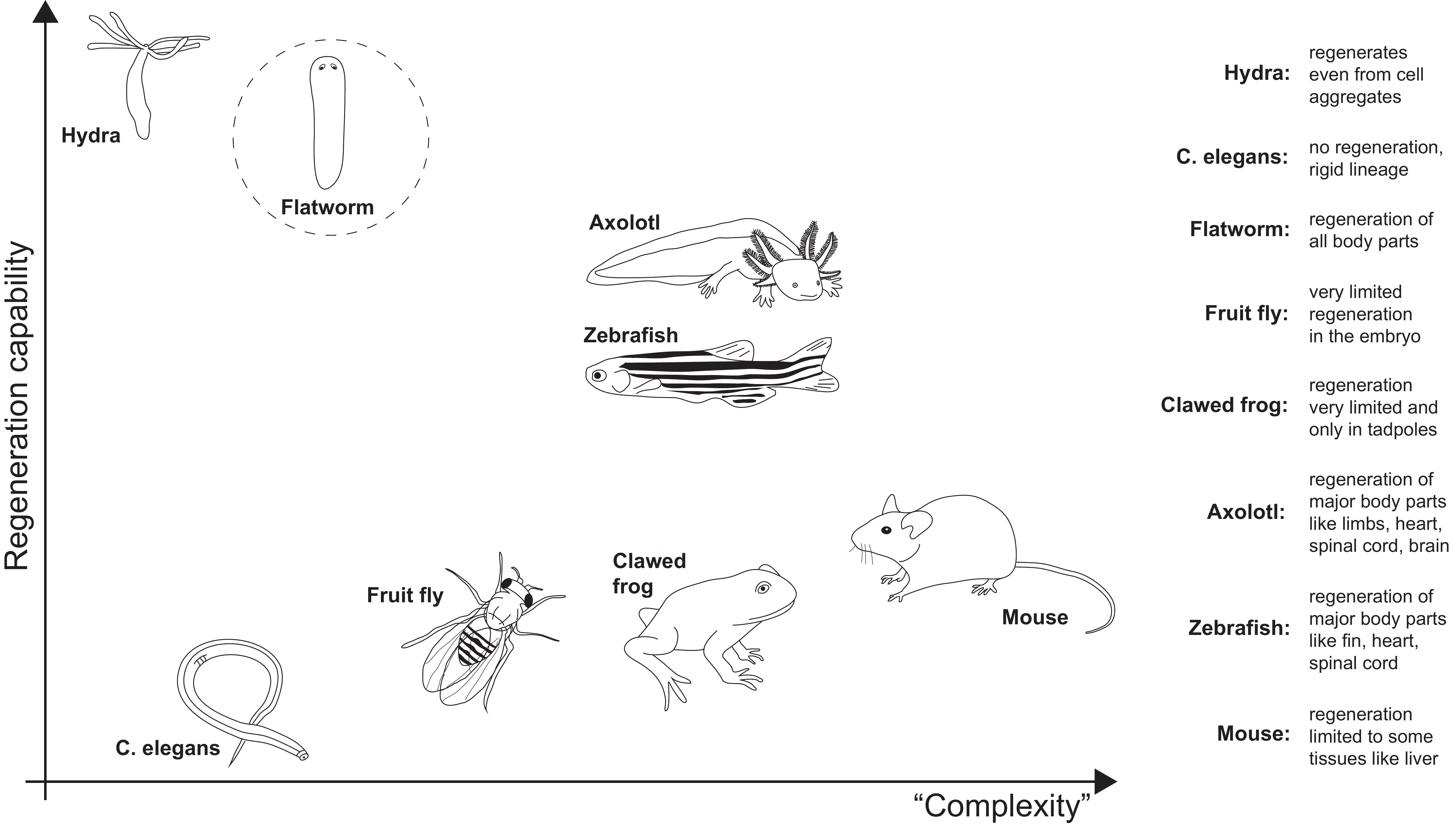}
  \caption[Complexity and regneration capabilities]
   {Flatworms are the most complex model organisms that can still regene\-rate every tissue. This regeneration capability is shared with simpler organisms such as \textit{Hydra}. At the same time, flatworms possess organ systems like a centralized nervous system and two distinct brain lobes, which are characteristic for the most complex organisms \cite{sanchezalvarado2006bridging,newmark2002not,morgan1901regeneration,egger2006freeliving,gilbert2014developmental, wolpert2011principles,nusse2013wnt}.}
   \label{fig:ComplexityRegeneration}
\end{figure}

\noindent Flatworms (Greek: \textit{Platyhelminthes}) bridge the gap between other model organisms of lower complexity such as the freshwater polyp {\it Hydra} %\cite{bode2003head,gamba2012critical}
and the {\it C. elegans} worm 
%\cite{sulston1977postembryonic,riddle1997celegans}
and
those of higher complexity, such as fruit fly, clawed frog,
%\cite{smithbolton2009regenerative}
 axolotl, zebrafish and mouse \cite{wolpert2011principles, sanchezalvarado2006bridging}, see Fig.~\ref{fig:ComplexityRegeneration}. Many members of the flatworm phylum seem to represent the most evolved organisms that are still able to regenerate any part of their body \cite{sanchezalvarado2006bridging,newmark2002not,morgan1901regeneration,egger2006freeliving}. For example, \smed{} can restore its complete body plan from amputation fragments as tiny as the very tip of the tail with only about $10^4$ cells \cite{montgomery1974minimal}.
By re-growing missing body parts and re-modeling oversized organs, they recover their normal shape scaled to the size of the amputation fragment within 1-2 weeks \cite{adell2010gradients, rink2013stem}. This astonishing regene\-ration capability is comparable to the much simpler \textit{Hydra} and distinguishes flatworms from other model organisms with much less body plan plasticity \cite{pellettieri2007cell,sanchezalvarado2006bridging}.

\noindent

\begin{figure}[t]
  \centering
  \includegraphics[width=0.85\textwidth]{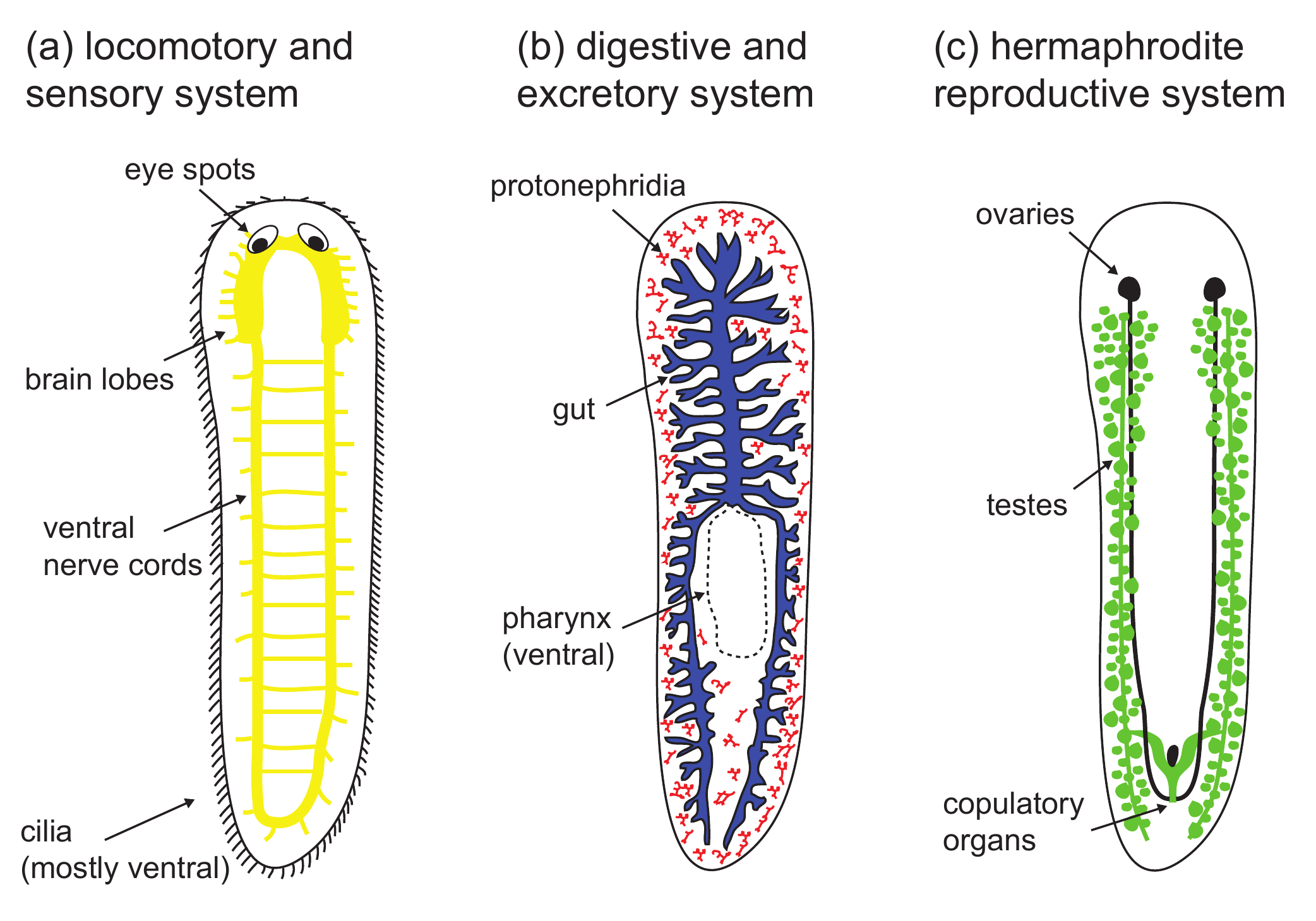}
  \caption[Organs of \smed{}]
   {The body plan of the flatworm \textit{Schmidtea mediterranea} (\smed{}) shows already key characteristics that are usually associated with higher order organisms \cite{robertsgalbraith2015organ,Rink2011maintenance, chong2011molecular,wang2010functional,yazawa2009planarian,elliott2013history}.}
   \label{fig:PlanarianOrgans}
\end{figure}

\noindent  Thereby, the body plan of \smed{} shows already key characteristics that are usually associated with higher order organisms \cite{newmark2002not}, see Fig.~\ref{fig:PlanarianOrgans}. The bilaterally symmetric \smed{} possess a central nervous system with a distinct bilobed brain and two ventral nerve cords connected by commissural neurons \cite{reddien2004fundamentals,Rink2011maintenance, inoue2004morphological, rink2013stem, salo2009planarian,gonzalezestevez2009autophagy}. The sensory system processes information from chemo-, rheo- and photorecetors leading to a complex behavioral repertoire \cite{Rink2011maintenance, inoue2004morphological,gonzalezestevez2009autophagy,paskin2014planarian}.
%For example the worms show a variety of phototactic responses depending on the wavelength of light \cite{paskin2014planarian, inoue2004morphological}. 
Their usual mode of motility is a gliding motion on a secreted layer of mucus being propelled by the beating of numerous short flagella  (or cilia) that project from their multi-ciliated ventral epithelium \cite{Rink2009planarian, rompolas2010outer, rompolas2009schmidtea, almuedocastillo2011deshevelled}. However the worms also use the numerous muscles situated along their inner body wall for (i) steering, (ii) quick escape responses, (ii) exploratory head motion and (iv) a back-up movement strategy \cite{Rink2009planarian, oviedo2010longrange, travis1981replicating, paskin2014planarian, rompolas2010outer}.

\noindent

\noindent \smed{} belong to the taxon \textit{Tricladida}, which is reflected by the fact that their highly ramified gut splits in three main branches \cite{reddien2004fundamentals, newmark2002not, rink2013stem, gonzalezestevez2009autophagy}, see Fig.~\ref{fig:PlanarianOrgans}(b). During feeding, the carnivorous worms suck in food through their extensible pharynx opening. After digestion, the pharynx also functions as an anus for excretion. Protonephridia constitute a further part of the excretory system, which performs similar functions as the kidneys in humans \cite{Rink2011maintenance}.
%The protonephridia are found throughout the mesenchyme, the tissue that fills up the space between the organs and also contains the stem cells of the worm \cite{reddien2004fundamentals, baguna2012planarian}.

\noindent Turnover, growth and regneration  completely relies on a pool of stem cells called neoblasts. The fraction of neoblasts among all cells had been estimated to be as large as $25-35\%$ \cite{baguna1989regenerationIII,baguna2012planarian, baguna1981quantitative, baguna1990growth}.
If the worms are depleted of all neoblasts by $\gamma$-irradiation, they will show a regression of the body starting from the head and finally they fall apart \cite{pellettieri2010cell, rink2013stem, dubois1949contribution, baguna2012planarian}. The time scale until the decay sets in varies between species in the range from several days to a few weeks \cite{pellettieri2010cell,dubois1949contribution}. %dubois, Contribution a l'tude de la migration des cellules de rgnration chez les Planaires Dulcicoles.
This indicates that there is no de-differentiation of cells to restore the stem cell pool. A subpopulation of neoblasts is pluripotent (and maybe even totipotent) and can develop into every other cell type \cite{handbergthorsager2007planarian,rink2013stem,gentile2011planarian,newmark2002not, pellettieri2007cell, baguna2012planarian}. Wagner \textit{et al.}~have shown that irradiated worms can be rescued by transplanting a single pluripotent neoblast from an intact worm \cite{wagner2011clonogenetic}. 
%Some neoblasts are very likely even totipotent and can potentially form a germ line as suggested by: (i) the transformation of asexual into sexual strains by either transplantation of neoblasts or feeding of minced worms \cite{baguna1989regeneration, nodono2012stem}, and (ii) the expression of germ line markers in stem cells of asexual worms and the formation of (yet not functional) reproductive organs in asexual worms at low temperature conditions \cite{handbergthorsager2007planarian}.
Even though neoblasts are well defined by their progression through the cell cycle resul\-ting in cell division, there is increasing evidence that the neoblast population is not homogeneous \cite{vanwolfswinkel2014singlecell, scimone2014neoblast, moritz2012heterogeneity, handbergthorsager2008stemcells}. Most likely, they also comprise lineage-restricted subpopulations and transiently amplifying cells that go through a few more rounds of cell division during differentiation.

\noindent \smed{} show both sexual and asexual reproduction \cite{newmark2002not,rink2013stem, dunkel2011memory,salo2009planarian}. Some strains are herma\-phrodites, which develop testes and ovaries symmetrically along their body axis as
well as copulatory organs for cross-fertilization, see Fig.~\ref{fig:PlanarianOrgans}(c).
In contrast, asexual
\sidenote{0.45\textwidth}{``Wenn einem beim Duell ein Ohr oder sonst ein Glied abgeschlagen wurde, so wuchs innerhalb von acht Tagen erstens ein neues Ohr an den Menschen und zweitens ein neuer Mensch an das Ohr. ... Wer sich vermehren wollte, schnitt sich zum Beispiel einen oder zwei oder zehn Finger ab und wartete acht Tage lang.'' \;---\; Joachim Ringelnatz, Abseits der Geographie, 1924 \cite{ringelnatz1924abseits}}
strains do not possess a reproductive system and reproduce by fissioning: the worm attaches its tail to the substrate and glides on until the body is ripped into two or more pieces, which develop into new worms after regeneration \cite{newmark2002not}. Fissioning depends strongly
on the environmental conditions, in particular on temperature, feeding, light
and worm density \cite{morita1984effects,dunkel2011memory,pigon1974cephalic, newmark2002not}. For example, fissioning frequency is reduced
in crowded environments.
Like regeneration abilities, reproduction strategies also vary widely among flatworm species \cite{egger2006freeliving}. In particular, one observes various approaches to asexual reproduction. Some species like \smed{} first split and regenerate afterwards, others already grow the respective organs of the new body plan before fissioning. The se\-cond stra\-tegy comes in two forms: as paratomy with the new body aligned to the old axis and budding with a non-aligned outgrowth at the side or pointing backward.

%\noindent In recent years, the flatworm \textit{Schmidtea mediterranea} (\textit{Smed}) has become a popular model organism for studies on regeneration and growth \cite{Newmark:2002not,oviedo2008planarians,gentile2011planarian}. 
%Flatworms (greek: \textit{Platyhelminthes}) bridge the gap in complexity between other model organisms as {\it Hydra} and {\it C. elegans} on the one hand and {\it D. melanogaster}, {\it Axolotl}, zebrafish and mouse on the other hand. They represent some of the simplest organisms with bilateral body plan, yet possess a distinct brain with two lobes, setting them apart from simpler worms like \textit{C. elegans} \cite{gentile2011planarian}. \\
%Flatworms are found in virtually all parts of the world, living in both salt- and freshwater, 
%and include parasitic species like the cause of bilharzia.
%\textit{Schmidtea mediterranea} belong to a subset of non-parasitic species, commonly referred to as `planarians'.  Their flattened and elongated body morphology is kept in shape by a deformable extracellular matrix material and the contraction status of their muscular plexus. Planarians glide over the substratum, being propelled by the beating of numerous short flagella 
%(or cilia) that project from their multi-ciliated ventral epithelium. They can steer their path in response to light, chemical stimuli, and temperature \cite{paperDec14}. Even a limited ability for learning has been proposed, including habituation and Pavlovian conditioning \cite{Shomrat:2013}.\\

\begin{figure}[t]
  \centering
  \includegraphics[width=1\textwidth]{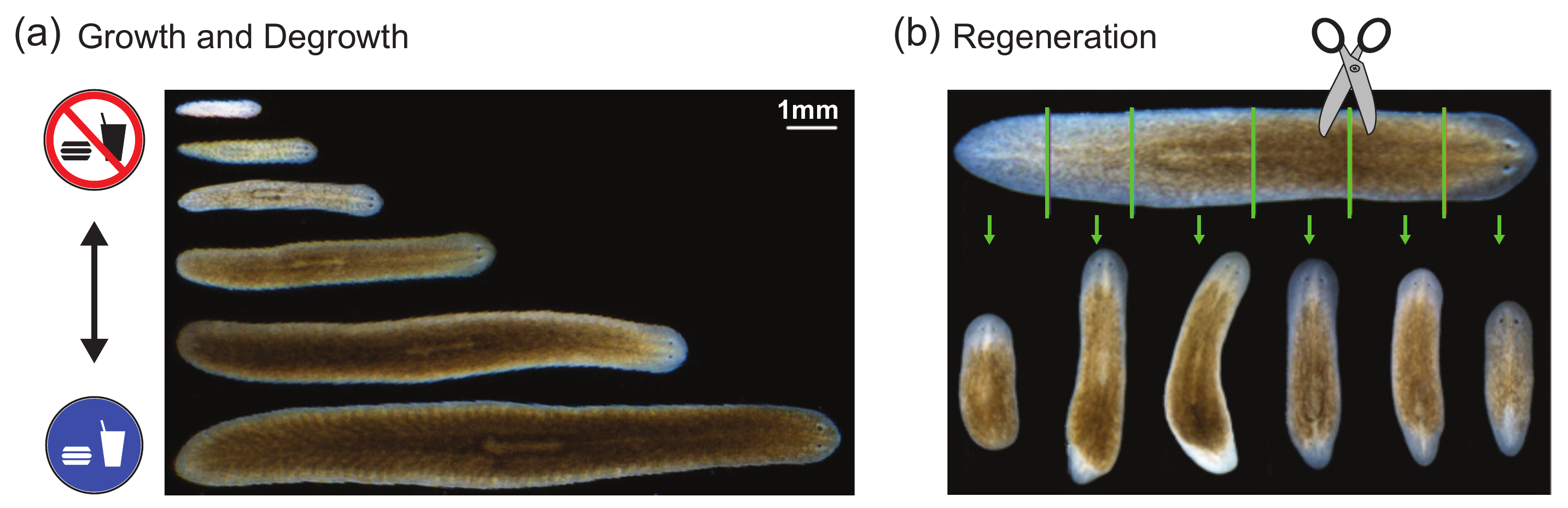}
  \caption[Growth and regeneration of \smed{}]
   {Growth and regeneration in flatworms: (a) \textit{Schmidtea mediterranea} (\smed{}) can reversibly grow over a 40fold range depending on feeding conditions (Images taken by Nicole Alt under the supervision of the author). (b) Amputation fragments regenerate to form a perfectly shaped worm within 2 weeks (green lines mark the cuts, white tissue parts indicate the regeneration site before all pigments have been reestablished, modified with permission from \cite{liu2013reactivating}).}
   \label{fig:PlanarianGrowthReg}
\end{figure}

\noindent \smed{} are well-suited model organisms to study growth and cell turnover. First, all somatic cells are continuously replaced by cell turnover at a time scale of weeks \cite{rink2013stem, pellettieri2007cell}.
Second, growth and cell turnover rely completely on the division of a large population of neoblasts, some of which are pluripotent \cite{handbergthorsager2008stemcells,rink2013stem,gentile2011planarian,newmark2002not, pellettieri2007cell, baguna2012planarian}. Third, depending on feeding conditions, the worms can grow and actively shrink (usually referred to as ``degrow''), while scaling their body plan over more than one order of magnitude in length in the range of 0.5 mm to 2 cm \cite{newmark2002not,oviedo2008planarians,rink2013stem}, see Fig.~\ref{fig:PlanarianGrowthReg}(a). Active degrowth enables the worms to survive long starvation periods of several months. It has been suggested that worms recycle apoptotic material during degrowth to fuel the metabolism of the remaining cells \cite{gonzalezestevez2009autophagy,gonzalezestevez2007gtdap1}.
%It has been discussed that shrinking worms, in fact, rejuvenate and regain the characteristics of younger worms \cite{lange1968possible, baguna2012planarian, mouton2011lack, baguna1990growth, gonzalezestevez2012decreased}.

\noindent

\noindent\smed{} are also well-suited to study scalable patterning during growth and regeneration. Being able to reversibly grow by a factor of 40 in length and perfectly regenerate even from scrambled body fragments prompts for patterning mechanisms that are not only highly robust and self-organizing but also functional across a wide range in sizes, see Fig.~\ref{fig:PlanarianGrowthReg}.
%Although the impressive regeneration capabilities of flatworms have already been described in the 18th century \cite{pallas1774spicilegia}, only state-of-the-art techniques of modern biology allow us to unravel the molecular details of scalable pattern formation.
For example, regeneration comprises a tightly controlled sequence of responses \cite{almuedocastillo2014jnk,adell2010gradients, lobo2012modeling, reddien2004fundamentals, gonzalezestevez2009autophagy,baguna1976mitosisI, pellettieri2007cell, morgan1901regeneration, rink2013stem,gurley2010expression, baguna1976mitosisII, wenemoser2010planarian}. After an immediate muscle contraction to close the wound, a peak in cell division together with the migration of stem cells within the first 12 h generates an outgrowth of undifferentiated tissue at the wound site, the blastema. In a second step, the now oversized, remaining body parts are re-patterned to match the size of the amputation fragment. The corresponding process is called morphollaxis and comprises a more sustained proliferation and cell death response during the following days. All this pattern formation is guided by an internal coordinate system that is re-established as one of the earliest cues in the remaining worm fragment independently of cell differentiation \cite{niehrs2010growth, gurley2010expression}.

\noindent Our work is especially focussed on the anterior-posterior (AP) axis specifying head and tail position. The Wnt/$\beta$-catenin system is a conserved pathway for such AP patterning in many animals and it is also key for the AP polarity in flatworms \cite{almuedocastillo2012wnt,adell2010gradients, forsthoefel2009emerging,gurley2008betacatenin,petersen2008smed,gurley2010expression, niehrs2010growth}. There are 9 Wnt genes in flatworms. In particular the canonical Wnts (activating $\beta$-catenin signaling) are highly expressed in the tail. %\cite{adell2010gradients, forsthoefel2009emerging,gurley2010epression} 
Inhibition of the Wnt/$\beta$-catenin system leads to formation of additional heads, while overexpression induces tail identity everywhere in the worm \cite{adell2009smedevi, gurley2008betacatenin, petersen2008smed, gurley2008betacatenin}. Remarkably, reducing the level of $\beta$-catenin is sufficient to induce full regenerative power to some flatworm species that are usually deficient in head regeneration \cite{liu2013reactivating, umesono2013molecular, sikes2013restoration}. Taken together, the experiments suggest that $\beta$-catenin is instructive for positional information along the AP axis in a concentration-dependent manner.

\noindent Perpendicular to the AP axis, the dorsal-ventral (DV) axis and the medial-lateral (ML) axis are formed \cite{forsthoefel2009emerging, lobo2012modeling}. For example, the DV axis is specified by the interplay \mbox{between} the TGF-$\beta$ family members Bmp and Admp. Together, all three body axes are charac\-terized by gradients in abundance and expression of specific molecules, in neoblast mitotic activity as well as in membrane potential \cite{adell2010gradients, beane2013bioelectric, simanov2012flatworm, oviedo2007smedinx11, gavino2013tissue, almuedocastillo2012wnt, gurley2010expression} and it is a major open question how these gradients robustly guide body plan patterning irrespective of the size of the organism.\\

\section{Theories of body plan patterning by morphogens}\label{Intropatterningtheory}
\begin{figure}[tbp]
  \centering
  \includegraphics[width=1\textwidth]{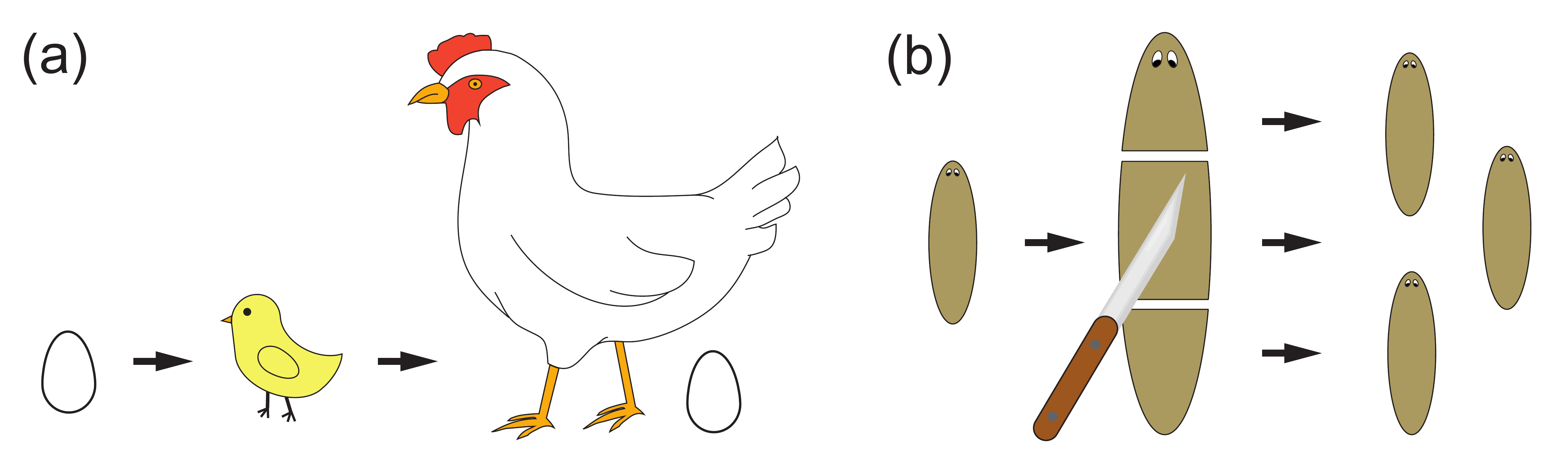}
  \caption[Sequential development vs. regeneration from arbitrary initial conditions]
   {(a) Idea of sequential development: The emerging tissues in the fertilized egg determine the future body plan of the juvenile chicken and later of the adult hen. 
% The layout in the egg determines the body plan of the juvenile chicken, which determines the body plan of the adult hen.
  Finally maternal signals break again the spatial symmetry in the new egg and guide development of another chicken. (b) Regeneration in flatworms: Diverse initial fragments are repatterned to form the same body plan scaled to fragment size.}
   \label{fig:chickeggplanarian}
\end{figure}

Here, we will discuss two ideas of pattern formation based on cell-cell communication via morphogens. The first type of mechanisms describes how a pre-existing structure can act as an organizing center for the development of further patterns. Such theories are built on the idea of a well-defined, sequential
developmental program where existing patterns determine the formation of new patterns, see Fig.~\ref{fig:chickeggplanarian}(a). Maternal patterning cues provide a pre-pattern for the embryonic tissues forming from an egg cell and the layout in the embryo determines the future body plan of the animal \cite{riverapomar1996gradients,abrams2009early,fu2012asymmetrically}. For example the localization of bicoid mRNA to the anterior pole of the fly embryo is a maternal effect.
However, the impressive capabilities of flatworms to regenerate from almost arbitrary initial fragments challenge these concepts, see Fig.~\ref{fig:chickeggplanarian}(b). In contrast to the sequential patterning from predefined cues, Alan Turing introduced a second class of mechanisms for self-organized pattern formation \cite{turing1952chemical}. We will discuss both approaches and assess them from a biological perspective.

\noindent Body plan patterning requires that cell fates are assigned depending on the relative position of the cells in the tissue or the organism \cite{wolpert2011principles,gilbert2014developmental}. In order to obtain the information about their spatial position, cells sense their environment and communicate with each other. For that, cells respond to mechanical cues as well as chemical signals such as morphogens \cite{howard2011turing,oster1988lateral,urdy2012evolution,rogers2011morphogen,lawrence2001morphogens}.

\begin{figure}[t]
  \centering
  \includegraphics[width=1\textwidth]{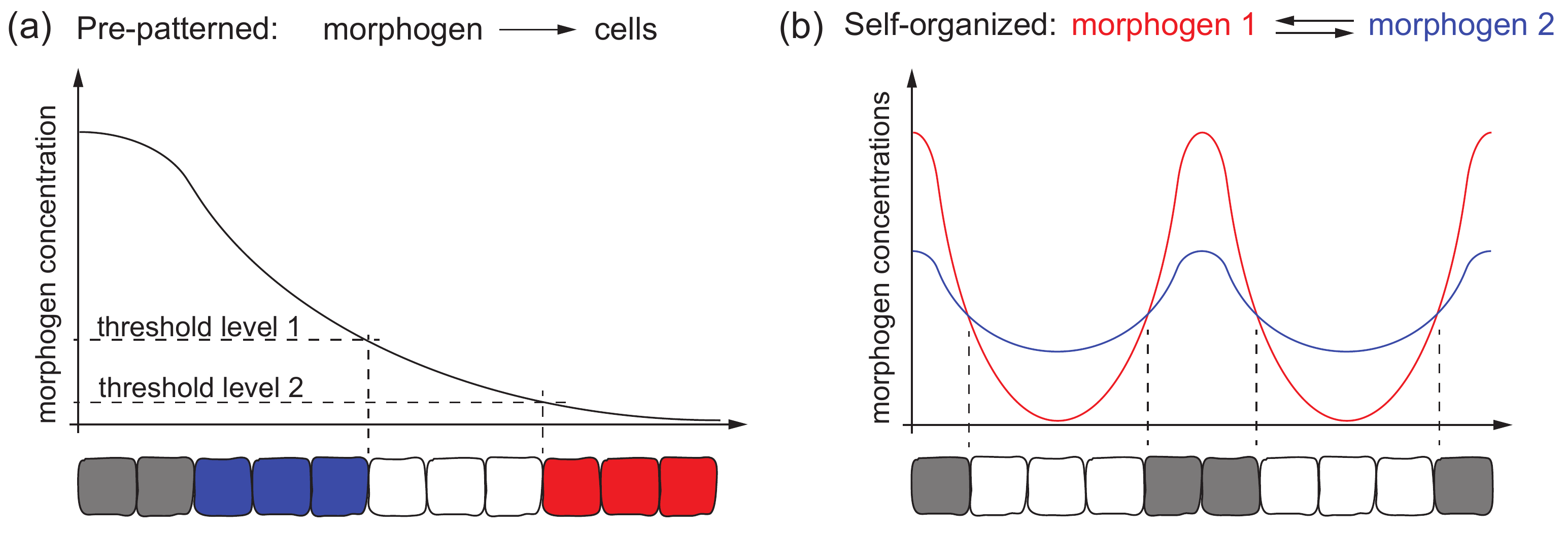}
  \caption[French flag model and Turing mechanism]
   {(a) The French flag model substantiates the idea that a graded morphogen profile provides positional information for cells in a tissue. Specifically, cells adopt distinct cell fates (sketched blue, white, red) by responding to the local morphogen concentration depending on whether a certain threshold is met. The model assumes a pre-defined source region (grey), which secretes the morphogens. (b) In a Turing system, at least two chemical species (morphogen 1 and 2, red and blue) interact. Source regions (gray) of the morphogens are established in a self-organized way (here: a source forms where the concentration of the first morphogen is larger than the concentration of the second morphogen).}
   \label{fig:FrenchFlagTuring}
\end{figure}

\noindent Signaling molecules often establish long-range, graded concentration profiles, which can be accounted for by the interplay of transport and degradation \cite{umulis2013mechanisms, gregor2005diffusion,wartlick2009morphogen}. Cells respond to the concentration of these molecules in their local environment. Thus, a graded morphogen concentration can provide cells with the information about its
spatial distance from the morphogen source \cite{wolpert1969positional}, see Fig.~\ref{fig:FrenchFlagTuring}(a). This idea forms the basis for the French flag model, which draws a simplified picture of how body plan patterning might be guided by graded morphogen profiles. 
For example, a stripe pattern can be generated if cellular differentiation depends on discrete, genetically encoded threshold levels. In Fig.~\ref{fig:FrenchFlagTuring}(a), cells turn blue if the morphogen concentration is above the first threshold level and red if it is below the second threshold level. Several intersecting gradients can generate more complex patterns.

\noindent Dose-dependent responses have indeed been observed in experiments, yet in a more complex way than given by the simplified French flag model \cite{schier2009nodal,green2015positional,gurdon1995direct,green1992responses,green1990graded,wilsom1994mesodermal,kieker2001morphogen}. For example, cells of the frog \textit{Xenopus} show a distinct differentiation response to at least five concentration thresholds of Activin \cite{green1992responses,green1990graded}. However, the early response of the individual cells is less specific and very inhomogeneous \cite{wilsom1994mesodermal,green1994slow}. It appears that only the interaction between cells leads to the well-defined behavior of the cell aggregate. Furthermore, the cells in the tissue are reported to respond in a ratchet-like manner to the highest level of Activin to which they were exposed \cite{gurdon1995direct}. Similar dose-dependent responses have also been recently reported for Wnts in the frog embryo with respect to the AP axis patterning \cite{kieker2001morphogen}.

\noindent Thus, although the simple concept of a direct threshold-comparison has been questioned in some organisms, the French flag model continues to provide a pictorial representation of the case where a feedback on the concentration gradient by the responding cells can be neglected.
As an extension, it also has been proposed that the cells might compare spatial and temporal differences in morphogen concentration and correspondingly adjust cell division, differentiation and motility \cite{wartlick2014growth,wartlick2011dynamics, mumcu2011selforganized, wartlick2011understanding,richards2015spatiotemporal,romanczuk2015optimal}.

\noindent The described mechanisms can explain patterning, given a pre-defined source region where morphogens are produced locally. Yet, how is this source region established in the first place? One possibility is that another graded morphogen profile specified this source by a threshold rule and a sequential developmental program establishes one pattern from an already existing one, see Fig.~\ref{fig:chickeggplanarian}(a). An alternative explanation dates back to Alan Turing. In a seminal paper from 1952, he proposed a general framework for the spontaneous formation of biological patterns, independent of pre-patterning cues \cite{turing1952chemical}. Turing's framework was later further explored by Meinhardt and Gierer \cite{meinhardt1982models, gierer1988biological, Koch1994Biological, gierer1972theory, gierer1981generation}.

\noindent Turing demonstrated how the interaction of at least two diffusing molecular species can result in chemical patterns in a completely self-organized manner, see Fig.~\ref{fig:FrenchFlagTuring}(b). Thereby, these (often periodic) patterns specify their own production regions. It is an appealing idea that the chemical patterns layout the body plan of an animal. Yet for the following 50 years, there was only little experimental evidence for Turing's ideas and the focus of developmental biology was shifted to other patterning mechanisms like the French flag model \cite{green2015positional}. Certainly this was partly due to the fact that it is generally difficult to demonstrate experimentally that a specific pattern is generated by a Turing mechanism. While a unidirectional relationship like in the French flag model (where a concentration of one molecule has a particular effect on other molecules or the cells) is straight forward to analyze, this is less so for Turing patterning which includes cross-reaction terms. The behaviour of a Turing system is often counterintuitive and its analysis might easily yield misleading results. For example, the concentration of some Turing molecules peak at the maximum concentration of their repressors, compare to Section \ref{TuringGeneralFrame}. Furthermore, if there are more than two chemical species involved, the concentration of one of them might decrease upon knockout of its direct activator because of additional indirect effects, see Appendix \ref{appreactdiff:knockout} for an example. Despite the challenging task of revealing the existence of Turing patterns and identifying the involved molecules, recently more and more evidence has been accumulated that Turing's ideas might be the guiding principles for the formation of a wide range of biological patterns, ranging from the formation of digits in vertebrate limbs to the emergence of left-right asymmetry \cite{nakamura2006generation, mueller2012differential,sheth2012hox, raspopovic2014digit,schier2009nodal,economou2012periodic,marcon2012turing}. Some of these examples combine Turing patterning with a French flag model.

\noindent In fact, in many biological systems, pattern formation might result from a  combination of both concepts \cite{green2015positional}. The feedback loop of a Turing mechanism ensures high robustness and possesses the ability to generate chemical patterns from random fluctuations. In particular, Turing systems can spontaneously generate graded concentration profiles like required for the French flag model. In turn, biological patterns hardly emerge in a completely homogeneous environment without any pre-patterning cues, as Turing already remarked \cite{turing1952chemical}. During development as well as during regeneration, pre-existing morphogen profiles and tissue structures can guide the formation of chemical patterns. Thus, Turing mechanisms might often be found downstream of morphogen profiles that modify the respective patterns \cite{green2015positional, raspopovic2014digit}.\\

\section{Turing mechanism yields self-organized patterns}\label{TuringGeneralFrame}
In 1952, Alan Turing introduced a generic framework for the self-organized formation of chemical patterns in biology \cite{turing1952chemical}. He asked the very general question whether there are conditions, under which a system of two or more diffusing and reacting chemical species possesses a homogeneous steady state, which is linearly unstable with respect to inhomogeneous perturbations, such that inhomogeneous patterns form spontaneously. Very importantly, in consequence, any model for self-organized patterning based on diffusing and cross-reacting molecules can be understood within the Turing framework \cite{turing1952chemical,gierer1988biological,segel1972dissipative,gierer1981generation, meinhardt1982models, oster1988lateral}. Recently, there are attempts to further generalize the Turing mechanism to generic interactions between two players without a diffusion term \cite{Watanabe2015,nakamasu2009interactions}.

\noindent In the following we show that the conditions for a Turing instability can be derived from the linear stability analysis of the homogeneous steady state. For this, we briefly recall the basic Turing model comprising two chemical species with concentrations $A$ and $B$. Details on the derivations are provided in Appendix \ref{appreactdiff:linstabTuring} or in the literature \cite{turing1952chemical, segel1972dissipative, gierer1988biological, gierer1981generation, okubo2002diffusion}. The general reaction-diffusion system for two molecular species in one space dimension is
\begin{eqnarray}
\label{eq:GenReactDiff}
\partial_t A&=&R_A(A,B)+D_A\,\partial_x^2\,A\nonumber\\
\partial_t B&=&R_B(A,B)+D_B\,\partial_x^2\,B\,,
\end{eqnarray}
with diffusion coefficients $D_A$ and $D_B$ and two generic functions $R_A$ and $R_B$ describing the reactions between the different molecules and the effects of each molecular species on itself. In order to spontaneously form stable patterns from random fluctuations, these reaction functions have to fulfill two requirements: the homogenous steady state $(A^*_h,B^*_h)$, defined by $R_A(A^*_h,B^*_h)=0$ and $R_B(A^*_h,B^*_h)=0$, has to be (i) stable with respect to homogeneous perturbations (to avoid a diverging behavior), yet (ii) unstable with respect to inhomogeneous perturbations.

\noindent It is possible to derive a set of necessary conditions for spontaneous pattern formation from these  two conditions above by applying linear stability analysis. For this, we consider a small perturbation $(a, b)$ about the homogeneous steady state: $A=A^*_h+a$ and $B=B^*_h+b$. The linearized form of Eq.~\ref{eq:GenReactDiff} can be written as
\begin{equation}\label{eq:linTuringsys}
\large\partial_t\begin{pmatrix} a_s\\[0.2cm] b_s\end{pmatrix}=\text{\huge M}_{\text{\large s}}\begin{pmatrix} a_s\\[0.2cm] b_s\end{pmatrix}\,.
\end{equation}
Here, the perturbation modes $a_s(t)$ and $b_s(t)$ with wavenumber $s$
represent the spatial Fourier transform of $a(t,x)$ and $b(t,x)$, respectively:
\begin{equation} a_s(t)=\int a(x,t)\,e^{-2\pi s x/L}\,dx\;,\quad b_s(t)=\int b(x,t)\,e^{-2\pi s x/L}\,dx\end{equation}
The matrix $M_s$ is given by
\begin{equation}\label{eq:linTuringmat}
\text{\huge M}_{\text{\large s}}=\begin{pmatrix} \partial_A R_A-D_A(2\pi s/L)^2 & \partial_B R_A\\[0.2cm] \partial_A R_B & \partial_B R_B-D_B(2\pi s/L)^2\end{pmatrix}\,,
\end{equation}
where derivatives are evaluated at $A=A^*_h$, $B=B^*_h$.

\noindent In order to fulfill the two conditions on the stability above, (i) the real parts of both eigenvalues $q^{I}_s$ and $q^{II}_s$ of the matrix $M_s$ have to be negative for $s=0$ and (ii) at least one has to be positive for $s\neq0$. As shown in Appendix \ref{appreactdiff:linstabTuring}, this results in the following conditions for the trace Tr and the determinant Det of the matrix $M_s$:
\begin{equation} \text{Tr}[M_0]<0\;,\quad \text{Det}[M_0]>0\quad\text{and}\quad \exists\, s>0\;\text{with}\; \text{Det}[M_s]<0\,.
\end{equation}

\begin{figure}[tbp]
  \centering
  \includegraphics[width=0.95\textwidth]{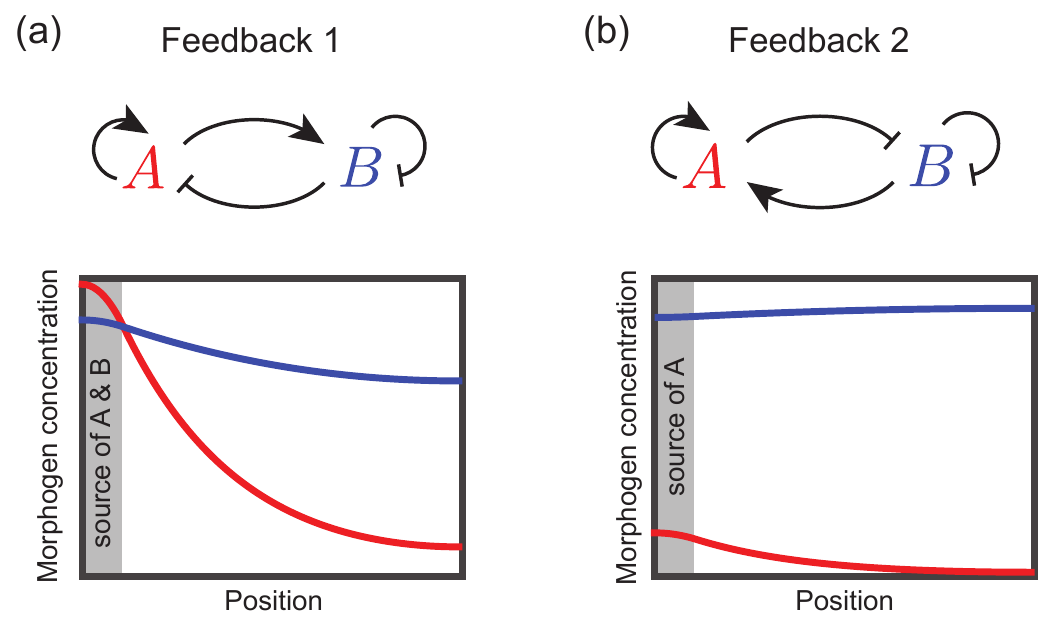}
  \caption[Two possible topologies with two Turing molecules]
   {There are two possible feedback topologies that lead to pattern formation in a Turing system with two molecular species ($A$ and $B$): (a) an activator-inhibitor scheme with activator $A$ and inhibitor $B$, (b) a second topology where both chemical species have activating and inhibiting effects. We show two examples of typical concentration profiles corresponding to the two feedback topologies. The self-organized source regions are indicated in gray.
%   given by Eq.~\ref{EqReactDiffChoice} with Eq.~\ref{eq:prodtheta} and by Eq.~\ref{EqReactDiffChoiceAlternative}, respectively. Parameters: $D_B/D_A = 30$, $\alpha_B/\alpha_A = 4$, $\beta_B/\beta_A = 2$, $B_c/(\alpha_B/\beta_B)=1$.
   }
   \label{fig:turingtopologies}
\end{figure}

\noindent In consequence, we obtain several constraints on the reaction design, which have been summarized by the principle of local activation and lateral inhibition \cite{gierer1981generation, meinhardt1982models,okubo2002diffusion, gierer1988biological, oster1988lateral}, see Appendix \ref{appreactdiff:linstabTuring} for details and Fig.~\ref{fig:turingtopologies} for illustration of the allowed feedback topologies. The necessary conditions for spontaneous pattern formation are:\\
1. One molecule has to be self-activating and the other self-inhibiting. In the following, we choose $A$ to be the self-activator and $B$ the self-inhibitor:
\begin{equation}
\partial_A R_A>0\quad\text{and}\quad\partial_B R_B<0\,.
\end{equation}
2. There have to be cross-reaction terms of opposing sign:
\begin{equation}
\Big(\partial_A R_B\Big)\Big(\partial_B R_A\Big)<0
\end{equation}
\noindent 3. The diffusion coefficient of the self-activator has to be sufficiently smaller than the diffusion coefficient of the self-inhibitor:
\begin{equation}
D_A<D_B\,.
\end{equation}
\noindent As a result for a Turing system with two chemical species, there are two possible network topologies as depicted in Fig.~\ref{fig:turingtopologies}. In Turing feedback 1, the concentration $A$ has only activating effects and $B$ has only inhibiting effects both on itself and the respective other player. In this case, we can shortly refer to $A$ and $B$ as the concentrations of activator and inhibitor instead of self-activator and self-inhibitor. In Turing \mbox{feedback 2}, both molecular species have activating and inhibiting effects. This includes depletion models, where, for example, binding of both molecular species enhances the production of $A$ but consumes $B$ \cite{gierer1972theory}. Note that there is a formal correspondence between the two Turing topologies by the replacement $B\rightarrow B_{c}-B$ with a constant parameter $B_c$.

\noindent How the principle of local activation and lateral inhibition leads to spontaneous pattern formation can be nicely demonstrated for the Turing feedback 1 of Fig.~\ref{fig:turingtopologies}(a). At the source, the activator dominates because the inhibitor is diffusing away more quickly.  Consequently, the source region stabilizes itself and initially tends to expand. In contrast, some distance away from the source, the inhibitor dominates due to its fast diffusion. This creates at least two distinct regions, which already comprises a simple pattern. In contrast to other systems where diffusion homogenizes a pattern, here diffusive spreading in combination with specific reactions enhances small inhomogeneities in the concentrations. Therefore, it is sometimes also referred to as a diffusion-driven instability \cite{ruan1998diffusion, okubo1980diffusion, okubo2002diffusion}.\\

%\noindent \begin{wrapfigure}{o}{0.45\textwidth}
%\vspace{-5pt}
%%\fbox{\minipage[t]{\dimexpr\linewidth-2\fboxsep-2\fboxrule\relax}
%       {\color{myrulecolor} \fontfamily{pzc}\selectfont
%       %\calligra
%``... living matter, while
%not eluding the 'laws of physics' as established
%up to date, is likely to involve 'other laws of
%physics' hitherto unknown, which, however, once
%they have been revealed, will form just as
%integral a part of this science as the former.''\\
%--- E. Schr\"{o}dinger, What is life?}
%%    \endminipage}
%\end{wrapfigure}

\section[Open questions in growth control and scalable body patterning]{Open questions in the study of growth control and scalable body patterning}
\noindent  During development, growth and regeneration, cells communicate with each other and mutually influence their behavior, especially by changing the expression status of their genes. This information exchange orchestrates cell division, cell death and differentiation. It can also guide cell migration and can elicit the release of further signals, e.g.~the secretion of  morphogens.
As a common theme of this thesis, we devise theoretical descriptions of how cell fate decisions based on local rules 
on the microscopic scale result in the formation and maintenance of a macroscopic body plan, drawing inspiration from flatworms as a specific model organism. In particular, we address the following questions:
%\vskip0.3cm

\noindent \textbf{How is the turnover of cells regulated during growth and homeostasis and how are fluxes in cell number related to fluxes of metabolic energy?}
\mbox{Organisms} have to control the number of cells to ensure homeostasis and growth in a well-defined manner. As changes in cell number depend on the balance between cell proliferation and cell loss, there needs to be a communication between dividing and differentiated cells \cite{pellettieri2007cell}.
%For example, cell loss likely induces cell division to maintain an intact tissue. Conversely, cell division might also affect cell death in differentiated cells to enable both tissue renewal and growth.
Furthermore, the environment and in particular the availability of food provides an external stimulus to influence cell behavior, such as proliferation and cell death \cite{wolpert2011principles,layalle2008tor}. 
%\vskip0.3cm

\noindent \textbf{What are minimal requirements for self-organized patterns that scale with organism size?}
Organisms also have to control the type and position of cells for a reliable body plan patterning. Again, this requires a communication between cells, ranging from direct neighbor-neighbor interactions to long range signaling via mobile molecules such as morphogens \cite{wolpert2011principles,gilbert2014developmental,forsthoefel2009emerging}.
Importantly, pattern formation can be observed on all length scales from the development of a fertilized egg to growth and regeneration of large scale tissues in mature organisms. Yet, patterning mechanisms often possess fixed characteristic length scales defined by the intrinsic physical properties of the system \cite{umulis2013mechanisms,gregor2005diffusion,wartlick2009morphogen}, which is challenged by cases of biological pattern scaling like regeneration in flatworms.
%\vskip0.3cm

\noindent We aim to understand scalable body plan patterning and cell turnover dynamics by building on the framework of dynamical systems theory and birth-death processes. Biological systems add a new perspective to those classical concepts. They usually operate far from equilibrium and form patterns with unconventional properties \cite{meinhardt1982models, haynie2001biological, cross1993pattern}. Development and regeneration are subject to a high level of noise ranging from external perturbation like a variable environment to the intrinsically stochastic nature of gene expression \cite{tkacik2011information, tkacik2014information,gregor2007probing}. Thus, robust growth control and body plan patterning require reliable sensing, transmitting and processing of noisy information \cite{dubuis2013positional, tkacik2014information}. We address the question of robustness with respect to initial conditions and physical parameters as well as the structural robustness of the models.
%This thesis is mainly inspired by the astonish scaling and regeneration capabilities of flatworms, yet the results obtained here and the application to nonlinear dynamics and pattern formation in general are likely to be also relevant for other biological systems.\\

\noindent This thesis is mainly inspired by the astonishing scaling and regeneration capabilities of flatworms, yet the results obtained here are likely also to be relevant for other biological organisms as well as related questions in biological physics, nonlinear dynamics and pattern formation.\\
\clearpage

\section{Organization of the thesis}
In this thesis, we investigate biological shape and size control on various levels, for which we combine theoretical descriptions and state-of-the-art analyses of experimental data in flatworms.

\noindent First, we explore mechanisms for self-organized and self-scaling pattern formation (Chapter \ref{patterningA}). We discuss to what extent previously proposed theories can account for the scaling of morphogen profiles. For this, we adhere to a strict mathematical definition of scaling and illustrate the difference between perfect scaling and approximate scaling with system size. Furthermore, we demonstrate the absence of scaling in classical Turing pattering, going beyond linear stability analysis.

\noindent Second, flatworms like \smed{} challenge existing theories on body plan patterning, which typically can at most only explain one of both: either scaling or self-organization. Here, we introduce and characterize a novel class of mechanisms that combine both features (Chapter \ref{patterningB}). The developed theory can act as a framework to understand robust body plan scaling during growth and regeneration in flatworms.

\noindent Next, worm shape variations cannot only be observed between different stages of development, growth and regeneration. Individual worms also show large body deformation during movement, driven by muscle contractions. We apply a method based on ``Principal component analysis'' to characterize shape dynamics and analyze movement patterns. A similar shape mode analysis also enables us to compare shapes of different related species. Analyzing motility patterns and comparing shape variations between worm species is the first step towards relating form and function (Chapter \ref{pca}).

\noindent Finally, we analyze growth and degrowth dynamics and their dependence on feeding conditions. We investigate mechanisms for size control by metabolic energy balances to explain the macroscopic growth and degrowth behavior in terms of microscopic cell turnover dynamics. Our theory makes testable predictions for the ongoing experiments. Beyond, we develop the theoretical basis for additional measurements on various scales to reveal further details about the control logic of cell turnover (Chapter \ref{growth}).

% ----------------------------------------------------------------------	% background information
% this file is called up by thesis.tex
% content in this file will be fed into the main document

%\chapter[Scaling and non-scaling in morphogen systems]{Scaling and non-scaling\\ in morphogen systems} 
\chapter[Scaling in morphogen systems]{Scaling in morphogen systems} 
\label{patterningA}% top level followed by section, subsection

% the code below specifies where the figures are stored
%\ifpdf
%    \graphicspath{{2_patterning/figures/PNG/}{2_patterning/figures/PDF/}{2_patterning/figures/}}
%\else
%    \graphicspath{{2_patterning/figures/EPS/}{2_patterning/figures/}}
%\fi

% ----------------------- contents from here ------------------------

\section{Scaling of biological patterns}
The development of multi-cellular organisms with a well-defined, complex body plan is one of the most fascinating processes in nature. A series of patterning events take place
 \sidenote{0.45\textwidth}{``The continuous change in form that takes place from hour to hour puzzles us by its very simpli\-city. The geometric patterns that present themselves at every turn invite mathematical analysis.'' \;---\; Experimental embryology, Thomas H. Morgan, 1927 \cite{morgan1927experimental}}
across various length scales leading to a distinct layout, which is scaled to match the size of the organism. The scaling of body plan patterns becomes especially apparent if a juvenile  organism already resembles its adult counterpart or if individuals of different, yet related species look very much alike besides their great differences in size \cite{benzvi2011scaling,gregor2005diffusion,umulis2013mechanisms}. A third, more subtle example is the robust formation of proportionate patterns in the same organism during development despite size variations that arise from varying environmental conditions or stochastic fluctuations of growth rates \cite{benzvi2010scaling, gregor2005diffusion}.

\noindent In this chapter, we analyze to what extent previously proposed theories can account for pattern scaling. First, we introduce a \revi{mathematical} definition of gradient \revi{scaling}. Based on these considerations, we revisit and assess scaling mechanisms proposed for morphogen gradients in pre-patterned systems and extract the main principles underlying scaling. This will later allow us to point out the important differences between scaling mechanisms for pre-patterned and self-organized systems in Chapter \ref{patterningB}. Furthermore, we discuss the absence of scaling in self-organized Turing systems beyond the classical approach of linear stability analysis.
The latter has been published in Werner \textit{et al.} \cite{werner2015scaling}.\\

\pagebreak

\section{Morphogen dynamics and concentration profiles}\label{MorphDyn}
One of the most simple morphogen systems that result in graded concentration profiles draws on diffusion from a localized source with an effective diffusion coefficient $D$ and a linear degradation with rate $\beta$ \cite{wartlick2009morphogen}. It has been frequently applied to describe morphogen gradients in the fruit fly \cite{driever1988bicoid,wartlick2011dynamics,benzvi2010scaling,gregor2007stability,gregor2005diffusion}. Note that effective diffusion might result from a wide range of underlying undirected processes, such as active transport or even signaling between neighboring cells without secretion of motile molecules.

\noindent For simplicity, we consider a one-dimensional system of size $L$ with reflecting boundary conditions. The corresponding time evolution of the morphogen concentration $C=C(t,x)$ is given by
\begin{eqnarray}\label{eq:mdiff}
&& \partial_t C=D\,\partial_x^2\,C-\beta\,C+\nu\\
&& \partial_x C|_{x=0}=\partial_x C|_{x=L}=0\,.\label{eq:mdiffbound}
\end{eqnarray}
Here, $\nu(x)=\alpha\,\Theta(w-x)$ describes localized morphogen production with rate $\alpha$ in a source of width $w$. $\Theta$ denotes the Heaviside step function.

\noindent Eq.~\ref{eq:mdiff} also holds in two and three dimensions, if the system is symmetric with respect to the other dimensions and, thus, the morphogen concentration still only depends on $x$, see Appendix~\ref{appreactdiff:generaldyn}. If the tissue grows at a time scale comparable to the time scale of the morphogen dynamics, one has to consider additional terms for advection and dilution.

\noindent The steady state solution to Eq.~\ref{eq:mdiff}-\ref{eq:mdiffbound} is computed in Appendix~\ref{appreactdiff:steadystatenogrowth}. Here, we are mainly concerned with the steady state concentration outside the source region, which is given by
\begin{equation}\label{eq:mst}
C^*(x)=C_0\,\frac{\cosh(L/\lambda-x/\lambda)}{\cosh(L/\lambda)}\quad\text{for }w\leq x\leq L
\end{equation}
with amplitude
\begin{equation}\label{eq:mampl}
C_0=\frac{\alpha\,\sinh(w/\lambda)}{\beta\,\tanh(L/\lambda)}
\end{equation}
and a characteristic length scale
\begin{equation}\label{eq:lambda}
\lambda=\sqrt{D/\beta}\,.
\end{equation}
Within the course of a characteristic time scale (which is of the order of $1/\beta$ in our example, see Appendix~\ref{appreactdiff:relaxtosteadystate}), the morphogen will relax towards this steady state profile.
Analyzing the steady state is well justified in a case of separated time scales, where other dynamics (e.g.~growth and differentiation) are slow in comparison to the relaxation time of the morphogen profile. In the following,  we will often assume such a quasi steady state. One should keep in mind that this is a simplifying assumption which is not always fulfilled.
In fact, it has been argued that many gradients might be already read out before the steady state is reached as a means to increase robustness \cite{barkai2009robust,bergmann2007presteadystate,richards2015spatiotemporal,fried2014dynamic}.
Yet, even in these cases, the steady state is the reference state and provides a first order approximation to the concentration the cells actually respond to.\\

\section{\revi{Scaling of concentration profiles with system size}}\label{scalingprofiles}
Profiles of signaling molecules have been measured for example in the fruit fly \textit{Droso\-phila melanogaster}, which is an important model organism to study biological pattern formation \cite{bate1993development,stjohnston1992origin}. Quantifications of morphogen concentrations in the developing wing and eye of the fly at different stages of development
revealed that the concentration profiles maintain an approximately constant shape relative to the changing size of the growing tissue \cite{wartlick2011dynamics,benzvi2011expansion, hamaratoglu2011dpp, wartlick2014growth}.
%This means that the morphogens spreads further as the system grows.
\revi{This biological observation of profiles that scale with system size inspired the development of a number of theoretical mechanisms that can account for pattern scaling \cite{benzvi2011expansion,benzvi2010scaling, benzvi2011scaling, wartlick2011understanding, averbukh2014scaling, wartlick2011dynamics,othmer1980scale, ishihara2006turing,pate1984applications, mumcu2011selforganized,umulis2013mechanisms}. In this section, we discuss examples, which represent two major classes of these mechanisms. We define a mathematical property of \textit{perfect shape scaling} and analyze approximations of this property.}

\begin{figure}[tbp]
  \centering
  \includegraphics[width=0.85\textwidth]{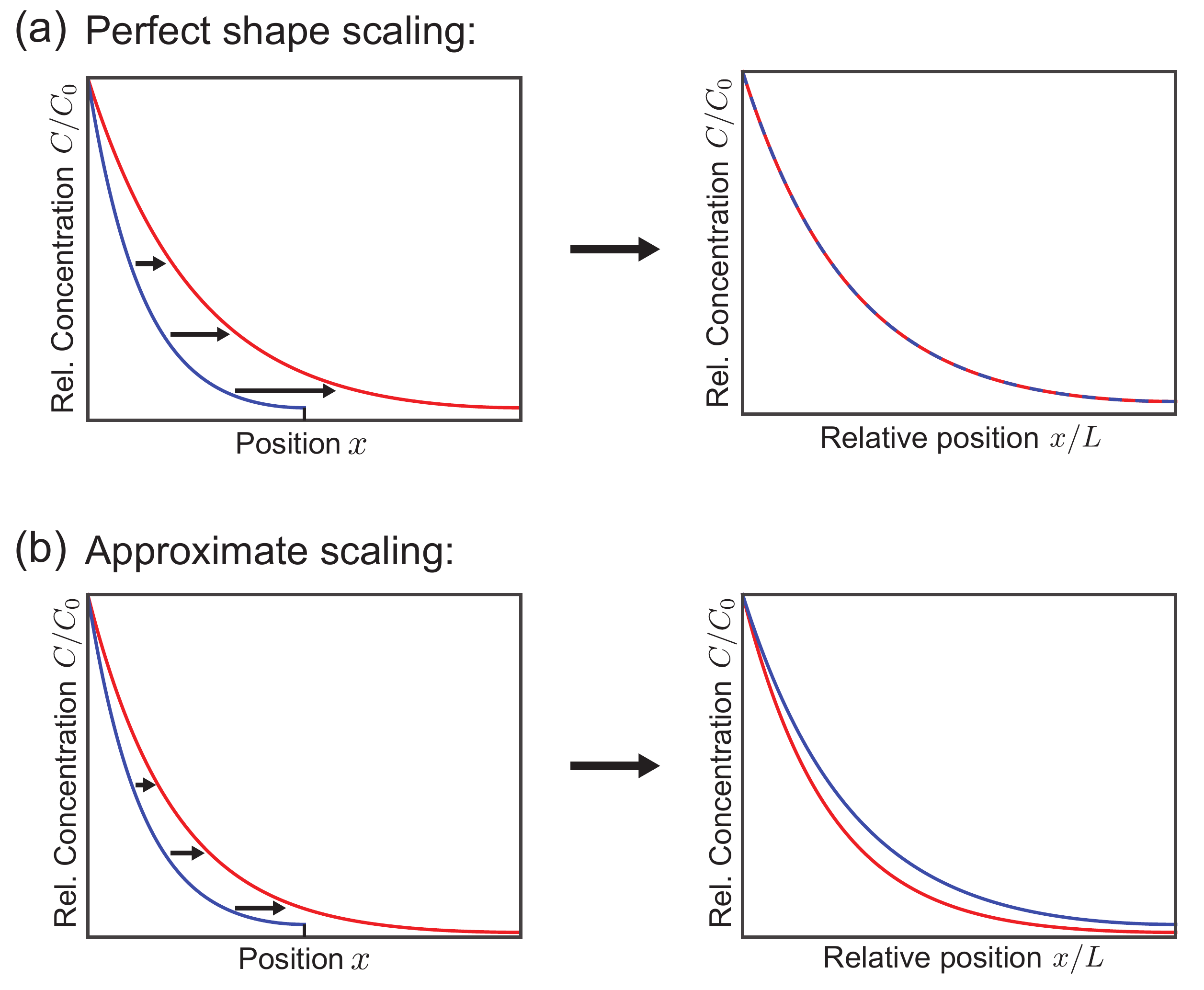}
  \caption[Perfect versus approximate shape scaling of morphogen gradients]
   {\revi{Illustration of the mathematical concept of \textit{perfect shape scaling} and \textit{approximate scaling}: (a) Perfect shape scaling: normalized concentration profiles collapse onto a master curve when plotted as a function of the relative position. (b) In a more general case, concentration profiles might expand with system size, though the collapse onto a single master curve is only approximate. We classify this case as approximate scaling}.}
   \label{fig:GradientScalingExpansion}
\end{figure}

\noindent We define \revi{\textit{perfect shape scaling}} as the ideal case of strictly proportional \revi{scaling} of a concentration profile with system size. In this case, the concentration at steady state can be written as 
\begin{equation}\label{eq:gradscal}
C(x; L)=C_0(L)\cdot \mathcal{Z}(x/L)\,,
\end{equation}
where $C_0(L)$ denotes a \revi{possibly size-dependent} amplitude and $\mathcal{Z}(x/L)$ defines the shape of the profile, which only depends on the relative spatial coordinate $x/L$. \revi{As a result, scaling profiles from systems of different sizes perfectly collapse onto a single master curve} if plotted as a function of relative coordinates and normalized by their amplitude, see Fig.~\ref{fig:GradientScalingExpansion}(a).
For the steady state solution in Eq.~\ref{eq:mst}, \revi{perfect shape scaling} arises if the characteristic gradient range $\lambda$ is not constant but \revi{strictly} proportional to the length of the system $L$:
\begin{equation}\label{eq:lambdaL}
\lambda\propto L\,.
\end{equation}
\revi{In a more general case,} if the length scale $\lambda$ increases with $L$ in a \revi{monotonic, yet nonlinear fashion, we will classify this as \textit{approximate scaling}}, see Fig.~\ref{fig:GradientScalingExpansion}(b):
\begin{equation}
\partial_L \lambda>0\,.
\end{equation}

%\noindent \revi{We emphasize that approximate scaling can be virtually indistinguishable from the ideal case of perfect shape scaling, both in measurements and simulations.} 
%\revi{In fact,} measured morphogen profiles in the fruit fly show a collapse similar to Fig.~\ref{fig:GradientScalingExpansion}(a) indicating scaling, yet there is certainly always some variability in the data \cite{wartlick2011dynamics,benzvi2011expansion, hamaratoglu2011dpp, wartlick2014growth}.
%This can be due to noise in the system and inaccuracies in the measurement or because the sepa\-ration of time scales between system growth and morphogen dynamics does not hold. \revi{Importantly, it might also be the case 
%that the underlying biological mechanism inherently cannot lead to perfect shape scaling in the sense of Eq.~\ref{eq:gradscal} because a mechanism for approximate scaling is sufficient to fulfill the biological function.}
%%Below, we will discuss such an example. 
%%Still, throughout this thesis, we carefully distinguish
%\revi{Nevertheless, in the following, we will strictly} distinguish between the two notions of \revi{\textit{approximate scaling}} and \revi{\textit{perfect shape scaling} as introduced above. Although the experimental discrimination between them might be next to impossible,
%%\cite{benzvi2011expansion,benzvi2010scaling,fried2014dynamic,benzvi2011scaling}.
%this strict distinction will be useful in the subsequent mathematical analysis.}

\noindent \revi{In the following, we will strictly distinguish between the two notions of \textit{approximate scaling} and \textit{perfect shape scaling}.  This slightly academic distinction will prove useful in the subsequent mathematical analysis.
Nonetheless, we want to emphasize that approximate scaling can be virtually indistinguishable from the ideal case of perfect shape scaling, both in measurements and simulations.
%In fact, measured morphogen profiles in the fruit fly show a collapse similar to Fig.~\ref{fig:GradientScalingExpansion}(a) indicating scaling, yet there is certainly always some variability in the data \cite{wartlick2011dynamics,benzvi2011expansion, hamaratoglu2011dpp, wartlick2014growth}.
In contrast to our theoretical description, real systems are generally prone to intrinsic fluctuations and measurement inaccuracies. Furthermore, the sepa\-ration of time scales between system growth and morphogen dynamics assumed here will always be an approximation, such that the morphogen concentration never exactly corresponds to the theoretically computed steady state profile. Therefore, even if the underlying mechanism is in principle able to generate profiles that perfectly scale like in Fig.~\ref{fig:GradientScalingExpansion}, we will never observe a perfect collapse of the measured concentration profiles.
On the contrary, mechanisms of approximate scaling that cannot yield perfect shape scaling in the sense of Eq.~\ref{eq:gradscal}, may still fully account for gradient scaling as observed in biological systems.
%On the contrary, mechanisms which inherently cannot lead to perfect shape scaling in the sense of Eq.~\ref{eq:gradscal} might  also be suitable descriptions for the approximate collapse of profiles observed in biological systems.
%From an evolutionary point of view, it might be more advantageous that a mechanism is robust with respect to perturbation than that it theoretically allows for perfect shape scaling.
%There is no general reason why a mathematically perfect scaling mechanism should be selected by evolution to govern a noisy biological system.
As an example for a mechanism that yields approximate scaling,
we discuss a simple realization of a prominent scaling mechanism, the
expansion-repression model.
We emphasize that this mechanism has been successfully applied to analyse profile scaling in fly wing development \cite{benzvi2011expansion,benzvi2010scaling,benzvi2011scaling}.}
%robustness of parameter

\noindent According to the definition of $\lambda$ in Eq,~\ref{eq:lambda}, \revi{scaling} can be achieved if either the diffusion coefficient or the degradation rate or both depend on system size. Two simple choices are: (i) $D\propto L^2$, $\beta=$ const and (ii) $\beta\propto L^{-2}$, $D=$ const.

\begin{figure}[t]
  \centering
  \includegraphics[width=1\textwidth]{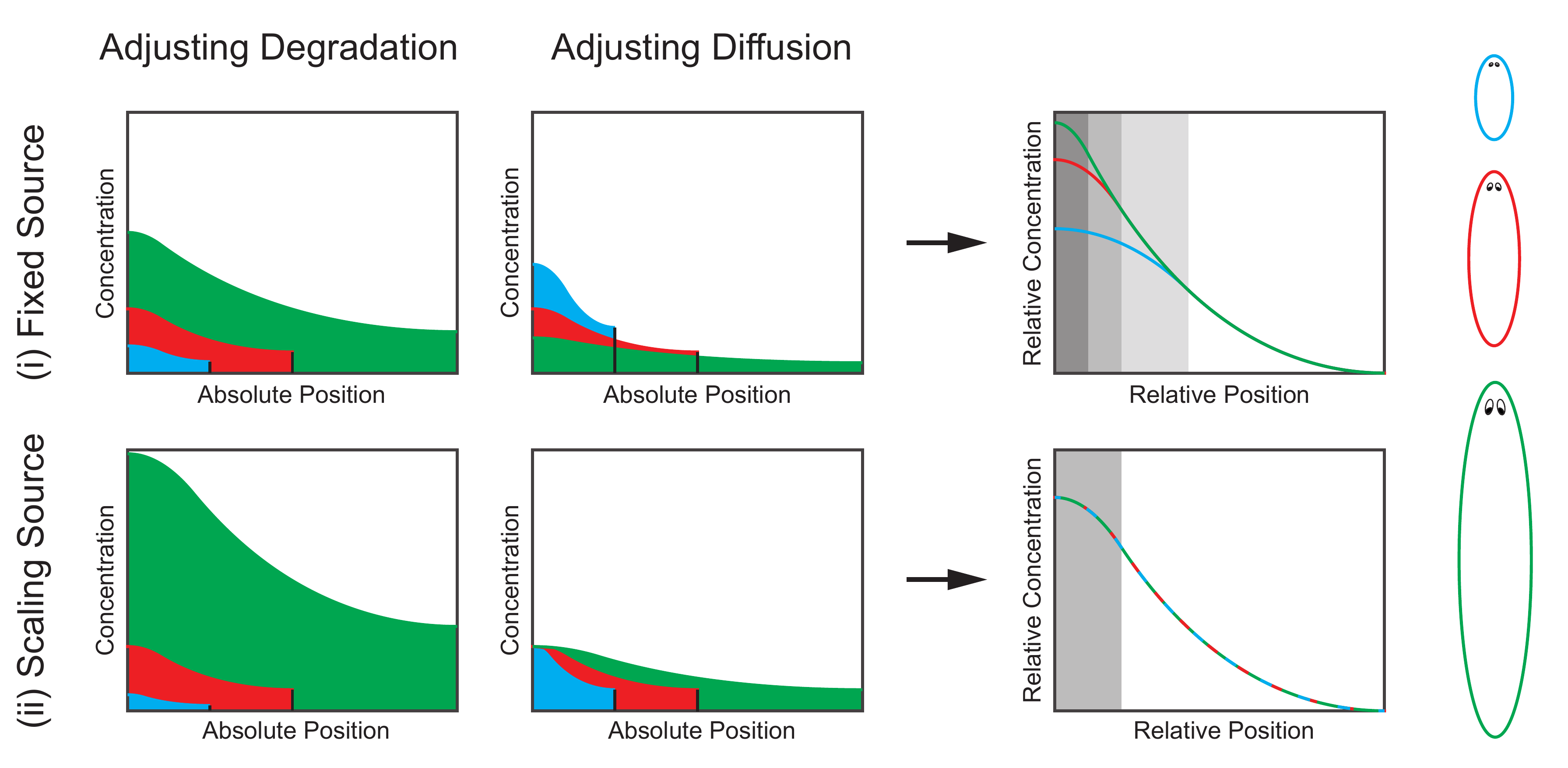}
  \caption[Gradient scaling and implications for the amplitude]
   {Size-dependent amplitudes of scaling profiles: Models for the scaling of morphogen profiles often assume the adjustment of degradation rate $\beta$ or diffusion coefficient $D$ with system size $L$. As a result, the amplitude of the morphogen concentration might change depending on whether the morphogen source size (gray) is fixed (upper row) or scales with $L$ (lower row). The reference profile (red) is the same in all plots (Eq.~\ref{eq:mst}, $\alpha/\beta=10$, $L/\lambda=2$, $L/w=5$). The other profiles correspond to either half (blue) or twice the system size (green) with constant ratio $L/\lambda$ and either scaling or non-scaling source size $w$.}
   %Reference (solid line): nu=0.1,\beta=0.01,L=20,D=1,w=4
   %small (dotted line): L=10, \beta=0.04 or D=4, w=2 if scaling
   %large (dashed line): L=40, \beta=0.0025 or D=0.25, w=8 if scaling
   \label{fig:GradientScaling}
\end{figure}

\noindent \revi{In Eq.~\ref{eq:gradscal}, we defined scaling of morphogen profiles as a shape property, irrespective of the amplitude $C_0(L)$. This amplitude} might be a function of system size $L$, compare also to \cite{umulis2013mechanisms}. The variation of the amplitude with $L$ can reveal details of the underlying scaling mechanism, see Fig.~\ref{fig:GradientScaling}. We discuss this for the two limiting cases of (i) a source of fixed width $w$ and (ii) a scaling source $w\propto L$.\\
(i) Fixed source: If we increase $\lambda$ by enhancing diffusion, the amplitude will decrease because the same amount of morphogen is distributed more homogeneously in the system, compare to Eq.~\ref{eq:mampl} or Eq.~\ref{eq:mstexp}. In contrast, if we increase $\lambda$ by reducing degradation, the amplitude will increase due to a longer lifetime of the molecules. If $D$ and $\beta$ both depend on size, the amplitude can be kept constant.\\
(ii) Scaling source: Eq.~\ref{eq:mampl} shows that for this case the amplitude does not explicitely depend on $D$ but depends inversely on $\beta$. Scaling by changing the diffusion properties does not affect the amplitude but scaling by adjustment of degradation leads to a strong size dependence $C_0\propto L^2$.

\noindent Measurements of the signaling protein Dpp in the fruit fly found a decreasing degradation rate with size as well as the characteristic increase in the amplitude \cite{gregor2007stability,gregor2005diffusion,wartlick2011dynamics,wartlick2011understanding,mumcu2011selforganized}. 
In the clawed frog, both cases have been reported, adjustment of diffusion properties for Bmp as well as adjustment of degradation for the Bmp antagonist Chordin \cite{benzvi2008scaling,inomata2013scaling}.

\noindent Next, we discuss two classes of simple models for the \revi{scaling} of morphogen profiles, \revi{which represent minimal versions of scaling mechanisms proposed in the literature \cite{benzvi2011expansion,benzvi2010scaling, benzvi2011scaling, wartlick2011understanding, averbukh2014scaling, wartlick2011dynamics,othmer1980scale, ishihara2006turing,pate1984applications, mumcu2011selforganized,umulis2013mechanisms}. We will illustrate that models of the second class rely on a changing amplitude during scaling. Thus, the amplitude of the morphogen profile encodes the system size.}\\

\section{Scaling of morphogen profiles in pre-patterned systems}\label{GradientScaling}
\subsection{The concept of an expander as a chemical size reporter}
In order to achieve scaling of a morphogen profile in a simple set-up as introduced in Section~\ref{MorphDyn}, the length scale of the concentration profile has to couple to the size of the system. Several mechanisms have been proposed for the expansion of a morphogen profile with system size $L$ that assume an additional molecular species, often called ``expander" \cite{benzvi2010scaling, benzvi2011scaling, wartlick2011understanding, averbukh2014scaling, wartlick2011dynamics,othmer1980scale, ishihara2006turing,pate1984applications, mumcu2011selforganized}.  %old: barkai2009robust, wartlick2014growth
If the concentration $E$ of such an expander is a function of $L$, the interaction of these expander molecules with the morphogen can result in profile expansion with system size. Depending on whether the expander concentration increases or decreases with system size, the expander might enhance or reduce either diffusion or degradation of the morphogen, respectively.

%\noindent Numerous mechanisms can effectively lead to an enhanced degradation rate: binding of the expander could disintegrate or alter the morphogen such that it cannot fulfill its signaling task. Yet, the expander could also act as a co-receptor that facilitates internalization and thus removes the morphogen from the system.
%Analogously, reduced degradation can emerge if the expander prevents disintegration, modifications %conformational changes?
%or internalization by binding to the morphogen or its suppressors. Similarly, the expander may reduce diffusion by increasing the size of the morphogen molecule when being bound to it. Alternatively, the expander might also enhance binding affinity to a substrate or reduce solubility or hinder active transport processes and thus, reduce the spreading by effective diffusion. Again, the opposite effects of the expander would enhance effective diffusion.\\

%How to do that: - degradation means either that a molecule is altered or disintegrated such that it cannot fulfil its signaling task or it is removed from the system; diffusion might be changed by changing the size of the molecule making it larger by binding or smaller by compaction. Or fascillitating or inhibiting binding what slows it down.

\noindent Yet, how does the expander obtain its size-dependent concentration and become a che\-mical size reporter?
In the following, we discuss two main classes of expander me\-chanisms that have been proposed for the scaling of morphogen profiles. The first class is characterized by the fact that the expander couples to the system size independently of the morphogen. The second class subsumes mechanisms that allow for self-scaling of the morphogen profile by a feedback loop between the expander and the morphogen. We illustrate the scope and the limits of the mechanisms at hand. This allows us to systematically extract the main principles that lead to robust scaling as applied later to self-organized systems.

\subsection{A simple mechanism of gradient scaling:\\
The expander-dilution model}\label{Sec:ExpDil}
\begin{figure}[tbp]
  \centering
  \includegraphics[width=0.81\textwidth]{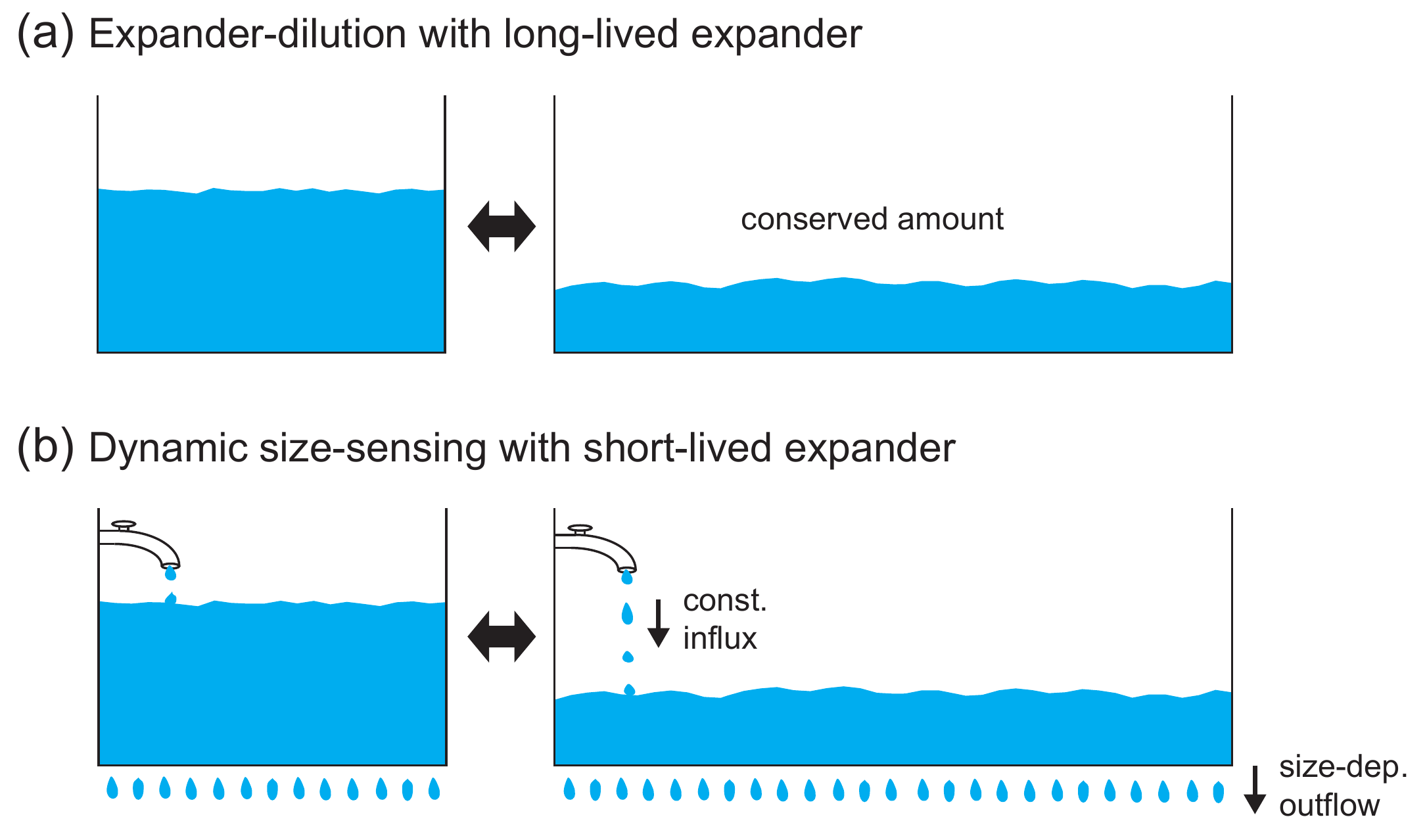}
  \caption[Scaling with an autonomously controlled expander]
   {Schematic illustration of how the concentration of an expander becomes size-depend. (a) Expander-dilution model: a conserved amount of expander gets diluted in a larger system. (b) Dynamic size-sensing in a system with constant influx and size-dependent outflow of the expander.}
   \label{fig:ExpDilScheme}
\end{figure}

The most simple idea for how an expander concentration can report on system size is given by the expander-dilution mechanism \cite{wartlick2011understanding, wartlick2011dynamics, mumcu2011selforganized}, see Fig.~\ref{fig:ExpDilScheme}(a). It requires a constant amount of a long-lived expander, which is neither produced nor degraded. As the system size increases, the expander gets diluted and, thus, the expander concentration depends on system size. Hence, $E\propto 1/L$ in a one-dimensional system. However, such a mechanism is highly vulnerable to the loss of expander and in particular could not easily cope with amputations of parts of the system.

\noindent A similar mechanism can overcome this issue by reading out system size from geometrical features like area-volume ratios. As a specific example, we assume that the expander of concentration $E$ is degraded everywhere in the one-dimensional system and produced in a source of constant width $w_E$ at the boundary:
\begin{equation}\label{eq:ediffED}
\partial_t E(t,x)=D_E\,\partial_x^2\,E(t,x)-\beta_E\,E(t,x)+\alpha_E\,\Theta(w_E-L+x)\,,
\end{equation}
Again, $D_E$ is an effective diffusion coefficient, $\beta_E$ the degradation rate and $\alpha_E$ the production rate of the expander.
In the limit of fast expander diffusion ($\lambda_E=\sqrt{D_E/\beta_E}\gg L$), the steady state concentration is given by
\begin{equation}\label{eq:est_eg}
E^*=\frac{\alpha_E\,w_E}{\beta_E}\,\frac{1}{L}\,.
\end{equation}
Thus, the expander level $E^*$ encodes the system size $L$, see Fig.~\ref{fig:ExpDilScheme}(b). In order for the morphogen to scale with $L$, the morphogen degradation could for example be coupled to the expander as
\begin{equation}\label{eq:edcoupling}
\beta\propto E^2\,.
\end{equation}
The size read-out can also be established differently from the specific choice of Eq.~\ref{eq:ediffED}, see Appendix~\ref{appreactdiff:autonexp} for a general derivation. For example, the expander could be produced everywhere in the system and only be degraded at the boundary, as discussed by other authors \cite{hunding1988size, ishihara2006turing, othmer1980scale, pate1984applications, gregor2007stability}.

\noindent The mechanism is robust against loss of expander, but it requires $\lambda_E\gg L$ irrespective of system size. Note that a small degradation rate results in a large $\lambda_E$, yet also in a slow relaxation to the steady state, see Appendix~\ref{appreactdiff:relaxtosteadystate}. Furthermore, the mechanism relies on a tightly controlled size of the source and the degradation zone. In the example above, the expander source has to keep its width although the system size changes. This shifts the problem from scaling the morphogen profile to establishing a source pattern for the expander. Finally, regeneration of arbitrarily shaped fragments seems unlikely to agree with the notion of a fixed geometry of source and degradation zone. Therefore, we next discuss the second class of \revi{scaling} mechanisms, in which the morphogen itself controls the expander expression leading to a \revi{self-regulating} feedback loop.

\subsection[Perfect shape scaling vs.~approximate scaling:\\ The expansion-repression model]{\revi{Perfect shape scaling vs.~approximate scaling}:\\ The expansion-repression model}\label{Sec:ExpRep}
The discussion on morphogen scaling has been enriched by the proposal of a class of models that \revi{adjust gradient shape to system size} in a self-organized manner \cite{benzvi2010scaling,benzvi2011expansion}. These me\-cha\-nisms build on a feedback loop between morphogen and expander. Interestingly, this relates to the engineering framework of an integral feedback control, where the gradient is adjusted by the expander as long as it does not meet its target range \cite{barkai2009big,benzvi2010scaling}. One prominent example is the expansion-repression model.  In order to illustrate the main principles, we discuss one specific simple realization of this model, in which the expander production is only turned on where the morphogen level falls below a threshold $C_{th}$ like considered in \cite{mumcu2011selforganized}, see Fig.~\ref{fig:exprep}(a). Alternative feedback topologies are analyzed in Appendix~\ref{appreactdiff:ercontprod} and lead qualitatively to the same conclusions.

\noindent For our choice, Eq.~\ref{eq:ediffED} of the expander dynamics is modified to
\begin{equation}\label{eq:ediffER}
\partial_t E(t,x)=D_E\,\partial_x^2\,E(t,x)-\beta_E\,E(t,x)+\alpha_E\,\Theta\big(C_{th}-C(x)\big)\,.
\end{equation}
In order to close the feedback loop and adjust the morphogen length scales, the expander has to affect either diffusion or degradation of the morphogen, see Section~\ref{scalingprofiles}.

\vskip0.2cm
\noindent {\it A simple feedback scheme \revi{that yields approximate scaling, though not perfect shape scaling}. --- }
\revi{Let us start with a simple realization of the expansion-repression mechanism, which has been first proposed for gradient scaling in the developing fly wing \cite{benzvi2010scaling}. In this realization, it is considered that the expander suppresses the degradation of the morphogen \cite{benzvi2010scaling,benzvi2011expansion,mumcu2011selforganized}.} Thus, $\beta$ in Eq.~\ref{eq:mdiff} is not constant anymore but obeys
\begin{equation}\label{eq:BarkaiExpRep}
\beta=\frac{\beta_{0}}{1+E/E_{th}}
\end{equation}\vskip0.2cm
\noindent with two constant parameters $\beta_{0}$ and $E_{th}$. In consequence, increasing expander levels enlarge the range of the morphogen. Yet, in turn, this suppresses the expander production. Eventually, a steady state is reached, in which the expander production is just cancelled by the total expander degradation in the system. The source size of the expander in this steady state is determined by
\begin{equation}\label{eq:exprepstM}
C^*(L-w^*_E)=C_{th}\,,
\end{equation}
where $C^*$ is given by Eq.~\ref{eq:mst} and $w_E^*$ denotes the size of the self-organized source region of the expander, see Fig.~\ref{fig:exprep}(a).
In the limit of a quickly spreading expander ($\lambda_E\gg L$), its steady state concentration is
\begin{equation}\label{eq:exprepstE}
E^*=\frac{\alpha_E\,w^*_E}{\beta_E\,L}\,.
\end{equation}
From this, we can note the following observations: First, the expander source (established by the feedback with the morphogen) must not scale ($w_E^*\not\propto L$), because this would result in a constant expander level irrespective of system size. Second, the morphogen amplitude must change for perfect scaling. Otherwise, if the amplitude were constant with a scaling gradient range $\lambda\propto L$, we would obtain $w_E^*\propto L$ from Eq.~\ref{eq:exprepstM} and the expander would not encode the system size. \revi{This generally arises from the fact that in such a mechanism for scaling via a feedback loop both players, morphogen and expander, have to encode the system size in the steady state. Third, the particular feedback scheme including Eq.~\ref{eq:BarkaiExpRep} yields approximate scaling of the concentration profiles, though not a perfectly linear relationship of $\lambda\propto L$ as required for perfect shape scaling, see Fig.~\ref{fig:exprep}(b)-(c).}

\begin{figure}[tbp]
  \centering
  \includegraphics[width=1\textwidth]{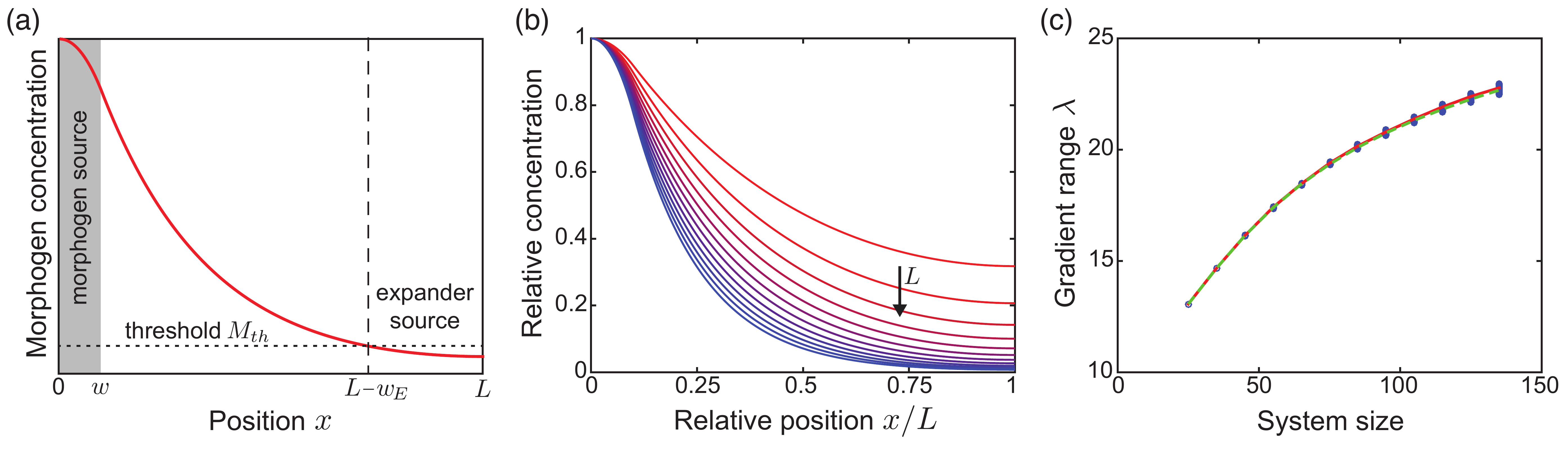}
  \caption[Expander-repression model and deviations from scaling]
   {\revi{A simple realization of the expansion-repression models that yields approximate scaling, yet no perfect shape scaling}. (a) In this specific model the morphogen confines the expander source to the region where the morphogen concentration is below $C_{th}$. (b) Rescaled steady state profiles do not perfectly collapse. (c) The gradient range $\lambda$ increases with system size in a sub-linear fashion. We determine $\lambda$ by (i) computing $\lambda=\sqrt{D/\beta}$ at various positions in the system (blue dots), (ii) fitting the solution of Eq.~\ref{eq:mst} (solid red line) and (iii) calculating it directly in the limit of $\lambda_E\ll L$ (dashed green line). We use typical parameters for the fly wing: $D=1\,\mu$m$^2$/s, $D_E=10\,\mu$m$^2$/s, $\alpha=0.01\,\mu$M/s, $\alpha_E=0.001\,\mu$M/s, $w/L=0.1$, $C_{th}=0.1\,\mu$M, $\beta_E=10^{-4}$/s, $\beta_{0}=10^{-2}$/s, $E_{th}=1\,\mu$M, $L=25...135\,\mu$m \cite{benzvi2011expansion,benzvi2010scaling}. \revi{This choice of parameters allows us to illustrate deviations from perfect shape scaling. Other parameter sets might approximate scaling better and might be practically indistinguishable from the mathematical limit of perfect shape scaling.}}
   \label{fig:exprep}
\end{figure}

\noindent The latter can be understood most clearly if we ask what relationship between $\beta$ and $E$ would be needed for the steady-state morphogen range $\lambda^*=\chi_{\raisemath{-2pt}{\lambda}} L$ to scale with a constant factor $\chi_{\raisemath{-2pt}{\lambda}}$. In Appendix~\ref{appreactdiff:erstepprod}, we provide a detailed derivation for different scenarios. Here, we only discuss the case of a scaling morphogen source ($w \propto L$) as a representative
 example. If we combine Eq.~\ref{eq:mst}, \ref{eq:exprepstM} and \ref{eq:exprepstE}, we obtain 
 %that $\beta^*$ has to obey
\begin{equation}\label{eq:ExpRepScaling}
\lambda^*=\chi_{\raisemath{-2pt}{\lambda}} L\quad\Rightarrow\quad\beta^*\propto \cosh\left(\frac{\beta_E}{\alpha_E \chi_{\raisemath{-2pt}{\lambda}}}\, E^*\right)
\end{equation}
%for $\lambda^*$ to \revi{perfectly} scale at steady state.
\revi{Thus, this functional relationship between $\beta^*$ and $E$, which would be
required to achieve perfect shape scaling, is different from the relationship, we had originally assumed in
Eq.~\ref{eq:BarkaiExpRep}. Note the qualitative difference between the two relationships: whereas Eq.~\ref{eq:ExpRepScaling} corresponds to a positive feedback loop, Eq.~\ref{eq:BarkaiExpRep} describes a negative feedback loop.
%Yet, this relationship is not in agreement with Eq.~\ref{eq:BarkaiExpRep}.  \revi{
%Note the qualitative difference since Eq.~\ref{eq:ExpRepScaling} corresponds to a positive feedback loop, whereas Eq.~\ref{eq:BarkaiExpRep} describes a negative feedback loop.
%In fact, it is not even qualitatively similar as it shows that the degradation rate has to increase with expander concentration to result in a scaling steady state.
We conclude that the considered case of Eq.~\ref{eq:ediffER} and Eq.~\ref{eq:BarkaiExpRep} does not yield perfect shape scaling, though approximate scaling is nonetheless possible. Such approximate scaling can be practically indistinguishable from perfect shape scaling as has been demonstrated by numerical simulations of the discussed feedback and related versions \cite{benzvi2010scaling,benzvi2011expansion,mumcu2011selforganized}.}\\

\revi{\subsection{Other feedback schemes for self-organized gradient scaling}\label{Sec:OtherExpRepSchemes}
The classical expansion-repression mechanism relies on a feedback in which the morphogen suppresses the expander while the expander increases the range of the morphogen. In consequence, this forms an overall negative feedback loop. In Sec.~\ref{Sec:ExpRep} we discussed an example which does robustly lead to approximate scaling, though not perfect shape scaling. Would perfect shape scaling be possible if we deviate from the scheme of a negative feedback loop?

\noindent Eq.~\ref{eq:ExpRepScaling} provides the functional relationship between $\beta$ and $E$ that would be required for \revi{perfect shape} scaling in the steady state.} Thus, we could replace Eq.~\ref{eq:BarkaiExpRep} by 
\begin{equation}\label{eq:ExpRepScalingUnstable}
\beta\propto \cosh\left(E/E_{th}\right)\,.
\end{equation}
Then, the system possesses a steady state pattern that scales perfectly by construction. However, this steady state is unstable in numerical simulations \revi{as it forms an overall positive feedback loop.} An excess in expander would reduce the level of morphogen by enhanced degradation and thus the expander production and consequently the expander level would increase even further. Depending on the initial conditions, the system shows two distinct dynamics. For small expander levels below the unstable steady state, the expander source will eventually be suppressed completely and the morphogen concentration will diverge. For large expander levels, the system will converge to a second, non-scaling steady state.

\vskip0.2cm
\noindent {\it Variations of the expander feedback show qualitatively similar results. --- }
\revi{In Appendix~\ref{appreactdiff:erstepprod}-\ref{appreactdiff:ercontprod}, we discuss further variations of the feedback scheme of Eqs.~\ref{eq:ediffER}-\ref{eq:BarkaiExpRep}. For these variations, we likewise find approximate scaling, though no perfect shape scaling in a strict mathematical sense over a large size range.
%, compare to \cite{benzvi2010scaling,benzvi2011expansion,mumcu2011selforganized}.
As a general rule, the ideal case of perfect shape scaling and stability are not easily teamed up. We strongly emphasize that this does not imply that the mechanisms for approximate scaling cannot provide suitable theoretical descriptions for gradient scaling in biological systems. Furthermore, more complex feedback schemes might be able to achieve perfect shape scaling or at least closely mimic perfect scaling over a considerable size range \cite{benzvi2010scaling}.}

\subsection{Conclusions from our analysis of expander models}\label{Sec:ScalingLesson}
\begin{figure}[tb]
  \centering
  \includegraphics[width=0.9\textwidth]{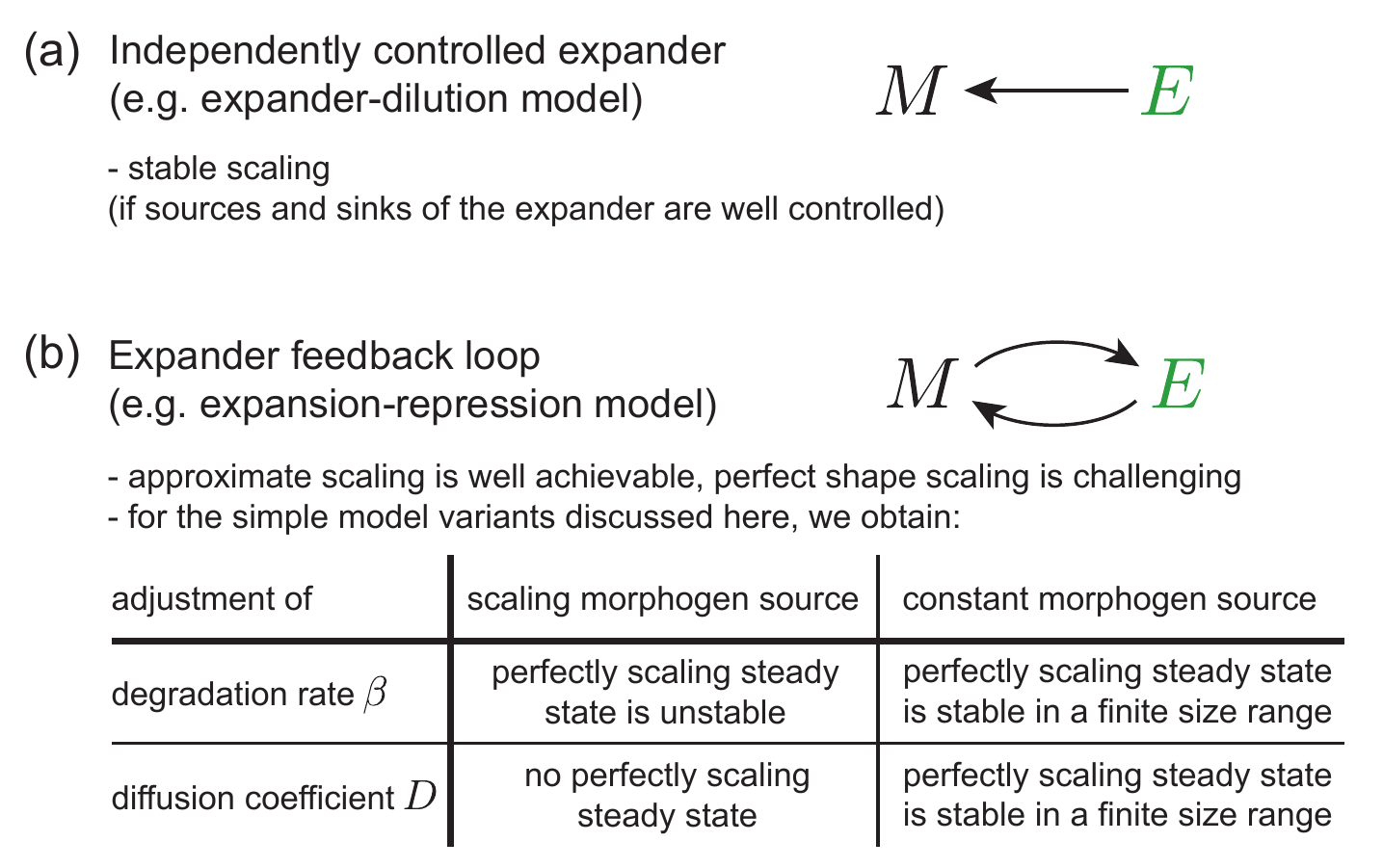}
  \caption[Two main classes of expander models]
   {Various mechanisms have been proposed for the scaling of morphogen profiles using an expander as a chemical size-reporter: (a) the first class of models considers an expander which reads out system size independently of the morphogen, see \cite{wartlick2011understanding, wartlick2011dynamics, mumcu2011selforganized,hunding1988size, ishihara2006turing, othmer1980scale, pate1984applications, gregor2007stability} for examples, (b) the second class of models assumes a feedback loop between the morphogen and the expander such that the morphogen profile scales itself \cite{benzvi2010scaling,benzvi2011expansion}. We demonstrated that in this case it is challenging to obtain a \revi{perfectly scaling steady state.} The table summarizes the results for a morphogen that acts on the expander production like in Eq.~\ref{eq:ediffER}, see Appendix~\ref{appreactdiff:erstepprod}-\ref{appreactdiff:ercontprod} for details. For all cases, it is possible to construct a robust feedback yielding \revi{good approximations to the idealized case perfect shape scaling}.}
   \label{fig:ExpTab}
\end{figure}

Above, we discussed \revi{simple realizations of scaling mechanisms for morphogen profiles, representing two main classes of such mechanisms}, see Fig.~\ref{fig:ExpTab}. If the expander measures system size independently of the morphogen, this can result in a robust size read-out and potentially lead to scaling \cite{wartlick2011understanding, wartlick2011dynamics, mumcu2011selforganized,hunding1988size, ishihara2006turing, othmer1980scale, pate1984applications, gregor2007stability}. Yet, it requires a tightly controlled layout of the system, e.g.~pre-patterned sources and sinks for the expander. Thus, it does not easily comply with the ability to regenerate and might just shift the problem from the scaling of the morphogen gradient to the establishment of a source and sink pattern for the expander.

\noindent The second class of models aims to establish scaling of the morphogen profile in a self-organized manner by a feedback loop between the morphogen and the expander  \cite{benzvi2010scaling,benzvi2011expansion}. It turns out that it is challenging to find mechanisms with steady state patterns that are both: perfectly scaling \revi{in the sense of Eq.~\ref{eq:gradscal}} and stable. \revi{If a feedback is chosen, such that a perfectly scaling steady state exists like in Eq.~\ref{eq:ExpRepScalingUnstable}, this fixed point might not to be stable. Still, many models like the one defined by Eqs.~\ref{eq:ediffER}-\ref{eq:BarkaiExpRep} yield stable steady states, which can well approximate perfect scaling, depending on the choice of parameters \cite{benzvi2011expansion, benzvi2010scaling}.}

\noindent The intuitive picture is that an overall negative feedback loop between morphogen and expander ensures a stable steady state. In contrast, two mutually suppressing molecular species, which might yield \revi{perfect shape scaling}, form an overall positive feedback loop. Thus, they will not result in a stable steady state \revi{unless there are additional feedback loops.
%further auto- or cross-regulatory effects.
In Chapter~\ref{patterningB}, we discuss a novel self-organizing scaling mechanism that builds on such feedback regulations.}

\noindent An important signature of the second class of models is that the morphogen amplitude varies with size $L$. This is a direct consequence of the feedback with a quickly spreading expander, which ensures a uniform expansion of the gradient range across the system. It can be understood intuitively because the morphogen not only couples to the expander but also the expander couples to the morphogen, hence, both have to encode the system size. This can result in constraints on the system design. For example, it led to the observation that the combination of a scaling morphogen source and an adjustment of morphogen diffusion by the expander could not yield scaling of the morphogen profile. We will have to take this into account when discussing scaling in self-organized systems in Chapter~\ref{patterningB}.

%\noindent The analysis suggests the following general construction principle for scaling feedbacks in the second class of models, which will be applied later to self-organized systems: (i) choose an interaction between morphogen and expander as one half of the feedback loop, (ii) derive the remaining interaction such that the morphogen profile scales in the steady state, (iii) analyze the stability of the combined system with morphogen and expander.\\

\section{Revisiting the absence of scaling in Turing patterns}\label{TuringNoScaling}

\noindent In this thesis, we consider one specific choice of a Turing system, which is particularly suitable for analytical treatment.
We restrict our analysis to a one-dimensional system with reflecting boundary conditions. The dynamics of the two molecules are described by
\begin{eqnarray}
\label{EqReactDiffChoice}
\partial_t A&=&\alpha_A\,P(A,B)-\beta_A\,A+D_A\,\partial_x^2\,A\nonumber\\
\partial_t B&=&\alpha_B\,P(A,B)-\beta_B\,B+D_B\,\partial_x^2\,B\,.
\end{eqnarray}
Thus, we specifically consider linear degradation with rates $\beta_A$, $\beta_B$ and production with rates $\alpha_A$, $\alpha_B$. The function $P(A,B)$ implements a switch-like response to the concentrations $A$ and $B$. Production is switched on if the activator concentration exceeds the inhibitor concentration, $A\gg B$. We choose a Hill function, which arises naturally from cooperative and competitive chemical reactions in biological systems, see Appendix~\ref{appreactdiff:HillFunc}:
\begin{equation}
P(A,B)=\frac{A^h}{A^h+B^h}\,.
\label{EqHill}
\end{equation}
In the limit $h\rightarrow \infty$, the Hill function can be replaced by the Heaviside theta function
\begin{equation}
P(A,B)=\Theta(A-B)\,.
\label{eq:prodtheta}
\end{equation}
As a technical point, we have to define $\Theta(0)\stackrel{!}{=}0$, in order for the homogeneous steady state to always exist, see Appendix~\ref{appreactdiff:HomSteady}.
A Fermi function could be another choice to account for switch-like production, yielding similar results.

\noindent In the case of the production switch in Eq.~\ref{eq:prodtheta}, the relative source size is defined by
\begin{equation}  \ell/L=\langle P\rangle = \frac{1}{L}\int_0^L P\,dx\,,\label{eq:relsource}\end{equation}
where brackets denote a spatial average over the system.
We use the same quantity also to define the source size for the Hill-type production function. This is especially well justified if the Hill exponent $h$ in Eq.~\ref{EqHill} is large.

%\noindent Our choice of a Turing system corresponds to the Turing feedback 1 in Fig.~\ref{fig:turingtopologies}(a). By replacing $B\rightarrow B_{c}-B$, one obtains the second type of feedback topology. Let us illustrate this correspondence between the two topologies for the case of the source switch given by Eq.~\ref{eq:prodtheta}:
%\begin{eqnarray}
%\label{EqReactDiffChoiceAlternative}
%\partial_t A&=&\alpha_A\,\Theta(A+B-B_c)-\beta_A\,A+D_A\,\partial_x^2\,A\nonumber\\
%\partial_t B&=&\beta_B\,B_c-\alpha_B\,\Theta(A+B-B_c)-\beta_B\,B+D_B\,\partial_x^2\,B\,.
%\end{eqnarray}
%Now, $B_c$ acts as a threshold to activate or inhibit production, respectively, while $\beta_B\,B_c$ describes a default production of $B$ everywhere in the system. For the complete production term to be positive, we have to choose $\beta_B\,B_c\geq\alpha_B$. Fig.~\ref{fig:turingtopologies} compares the concentration profiles in the steady state between both types of reaction topologies. Note that in the first topology, the concentration $A$ peaks at the maximum of the concentration $B$ of its inhibitor.

\noindent Next, we analyze our example of a Turing system given by Eq.~\ref{EqReactDiffChoice} and 
demonstrate the absence of scaling. We derive the full hierarchy of steady state patterns and discuss their existence and stability, thereby going significantly beyond linear stability analysis of the homogeneous steady state. This will put us in the position to understand the mechanisms for scaling of a Turing system.

\subsection{A hierarchy of steady state solutions}\label{Sec:TuringSteadyStates}
Eq.~\ref{EqReactDiffChoice} together with the production function in Eq.~\ref{EqHill} possesses a unique homogeneous steady state:
\begin{equation} A^*_h=\frac{\alpha_A/\beta_A}{1+(\beta_A\alpha_B/(\alpha_A\beta_B))^h}\quad,\quad B^*_h=\frac{\alpha_B/\beta_B}{1+(\beta_A\alpha_B/(\alpha_A\beta_B))^h}\,.\label{Eq:TuringHomFixedPoint} \end{equation}
For $h\rightarrow\infty$, the steady state concentrations approach zero, $A^*_h=B^*_h=0$, see Appendix~\ref{appreactdiff:HomSteady}.

\begin{figure}[tbp]
\centering
\includegraphics[width=1\textwidth]{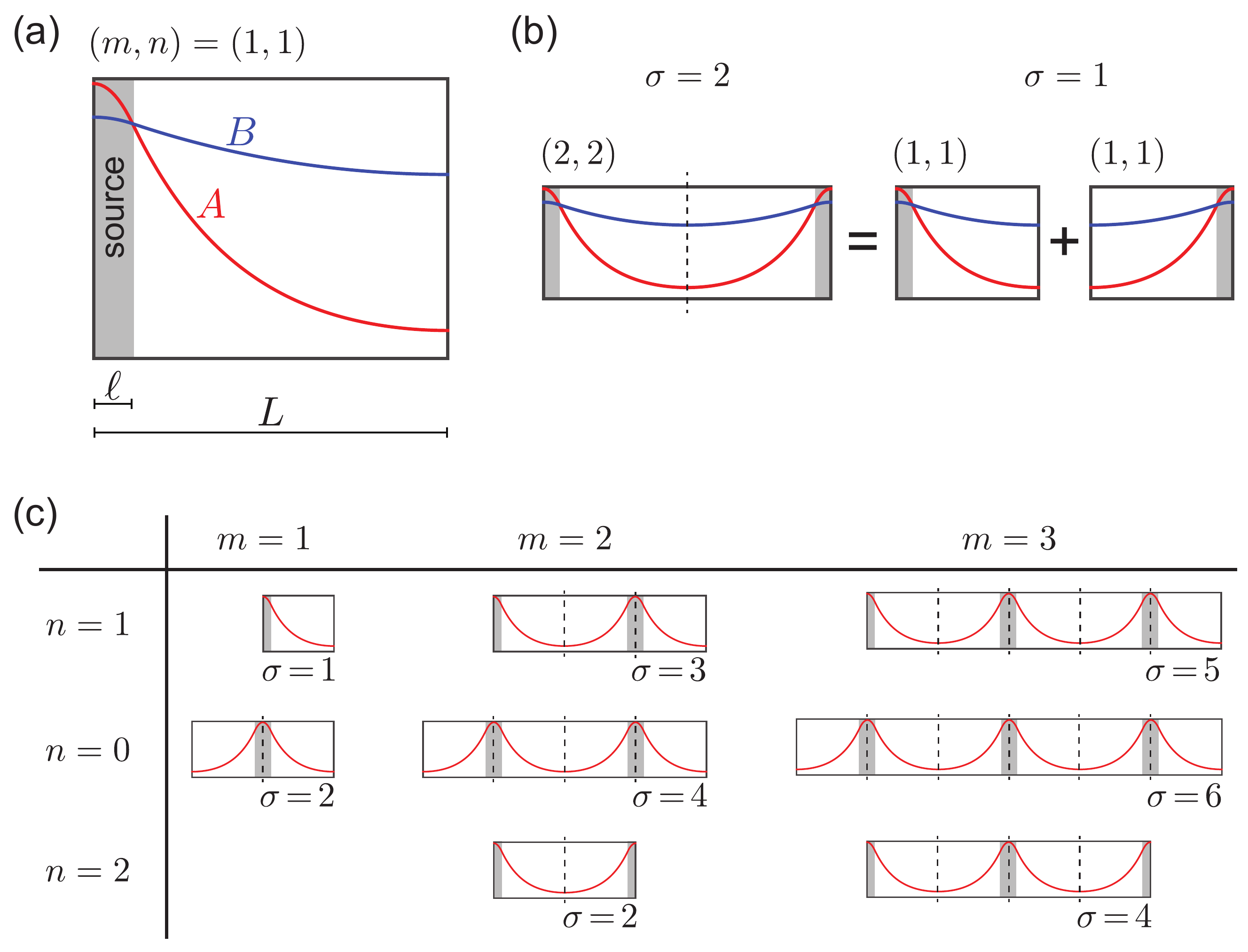}
\caption[Hierarchy of Turing patterns]{\small 
(a) The steady state solutions of Eq.~\ref{EqReactDiffChoice} with Eq.~\ref{eq:prodtheta} can be characterized by a pair of pattern numbers $(m,n)$, where $m$ is the number of contiguous source regions (gray) with $A>B$ and $n$ is the number of source regions touching the system boundaries.
(b) Any $(m,n)$-pattern can be constructed as a concatenation of $\sigma=2m-n$ copies of the $(1,1)$-pattern as a basic building block, here shown for $(m,n)=(2,2)$.
(c) Thus, we obtain a complete hierarchy of steady state solutions, exemplified for the concentration $A$ up to $m=3$ sources.
Parameters: $D_B/D_A = 30$, $\alpha_B/\alpha_A = 4$, $\beta_B/\beta_A = 2$, $\lambda_A/L=\sqrt{0.1}\approx 0.3$, $\lambda_B/L=\sqrt{1.5}\approx 1.2$.
}
  \label{fig:TuringSteadyStates}
\end{figure}

\noindent The inhomogeneous steady states can be computed analytically in the limit of $h\rightarrow\infty$, corresponding to the source switch of Eq.~\ref{eq:prodtheta}, see Appendix~\ref{appreactdiff:InhomSteady}. We find a hierarchy of steady state patterns, which can be characterized by the number $m$ of contiguous source regions with $A>B$ and the number $n$ of source regions touching the system boundaries, see Fig.~\ref{fig:TuringSteadyStates}.  Any $(m,n)$-pattern can be constructed as a concatenation of $\sigma=2m-n$ copies of the $(1,1)$-pattern as a basic building block, see Appendix~\ref{appreactdiff:InhomSteady}. Note that $s=\sigma/2$ is the corresponding leading order Fourier mode of the spatial patterns.

\begin{figure}[tbp]
  \centering
  \includegraphics[width=1\textwidth]{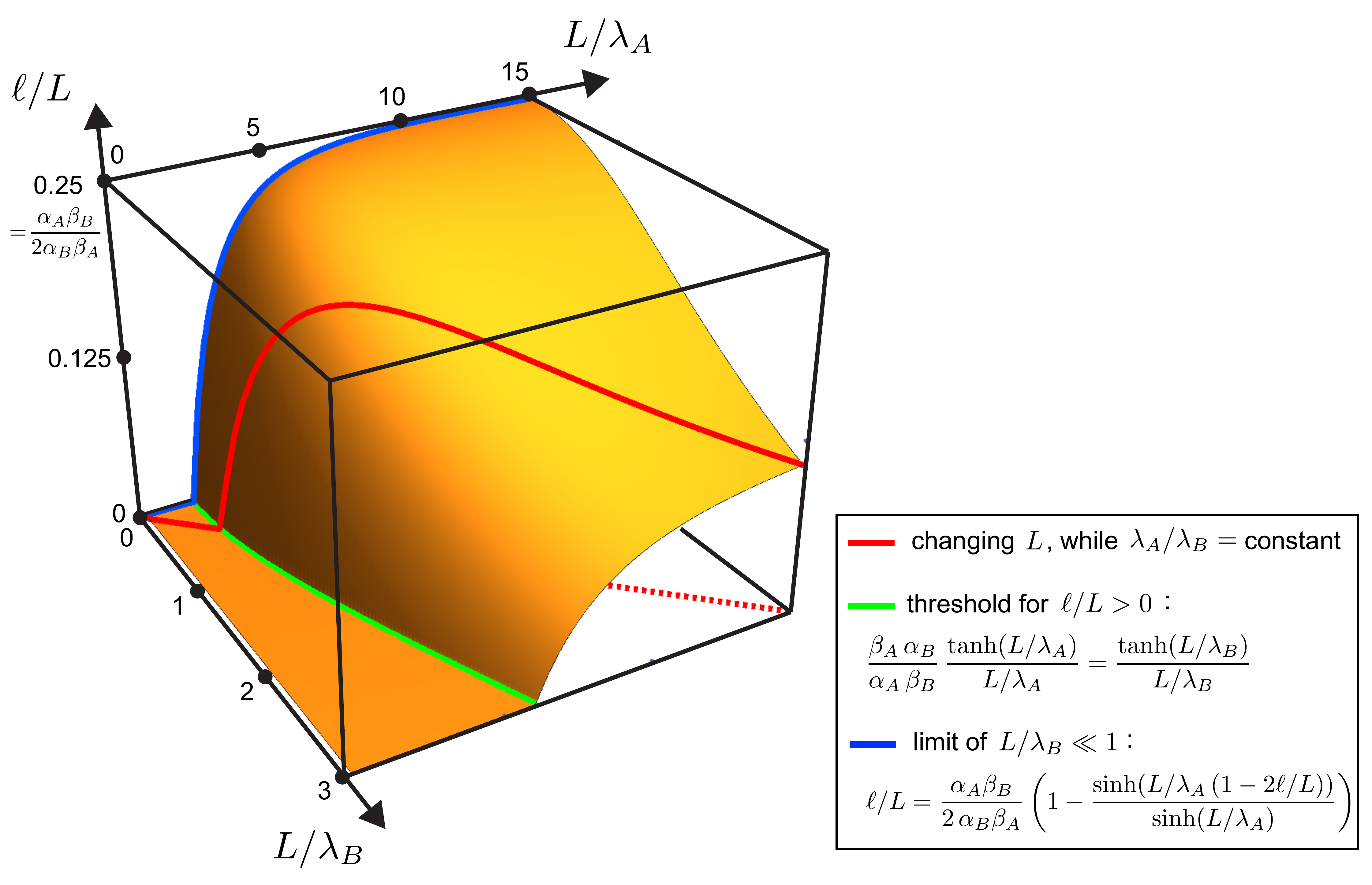}
  \caption[Relative source size of a Turing system]
   {Pattern length scales set size of source: The relative source size $\ell/L$ of the $(1,1)$-pattern at steady state is uniquely determined by the pattern length scales $L/\lambda_A$ and $L/\lambda_B$.
   If only the system size is varied, while the other parameters are kept constant, the relative source size changes along the red curve (here: \mbox{$\lambda_B/\lambda_A=5$}). Following the red curve, the source size approaches $\ell/L\propto \lambda_B/L$ for $\lambda_B\ll L$. Furthermore, we obtain a threshold for the existence of the $(1,1)$-pattern (green) and a scaling regime in the limit of $\lambda_B\gg L\gg \lambda_A$ (saturation of blue curve).
Parameter: $\alpha_A\beta_B/(\beta_A\alpha_B)=0.5$.}
   \label{fig:RelSource}
\end{figure}

\noindent The $(1,1)$-pattern with a single source, located in the interval $0\leq x<\ell$, is given by
\begin{eqnarray}
A^*_{(1,1)}&=&\frac{\alpha_A}{\beta_A}\,\begin{cases}1-\frac{\sinh(L/\lambda_A-\ell/\lambda_A)}{\sinh(L/\lambda_A)}\,\cosh\left(\frac{x}{\lambda_A}\right)& ,0\leq x\leq \ell\\[0.2cm]
\frac{\sinh(\ell/\lambda_A)}{\sinh(L/\lambda_A)}\,\cosh\left(\frac{x-L}{\lambda_A}\right)& ,\ell<x\leq L
\end{cases}\label{Asol}\\[0.2cm]
B_{(1,1)}^*&=&\frac{\alpha_B}{\beta_B}\,\begin{cases}1-\frac{\sinh(L/\lambda_B-\ell/\lambda_B)}{\sinh(L/\lambda_B)}\,\cosh\left(\frac{x}{\lambda_B}\right)& ,0\leq x\leq \ell\\[0.2cm]
\frac{\sinh(\ell/\lambda_B)}{\sinh(L/\lambda_B)}\,\cosh\left(\frac{x-L}{\lambda_B}\right)& ,\ell<x\leq L\,.
\end{cases}\label{Bsol}
\end{eqnarray}
Patterns are characterized by two pattern length scales for $A$ and $B$, respectively, analogous to Eq.~\ref{eq:lambda}.:
\begin{equation} \lambda_A=\sqrt{D_A/\beta_A}\quad\text{and}\quad\lambda_B=\sqrt{D_B/\beta_B}\,. \label{eq:lambdaAB}\end{equation}
The source size $\ell$ has to satisfy the implicit equation
\begin{equation}
\frac{\alpha_A\,\beta_B}{\beta_A\,\alpha_B}\,\frac{\sinh(\ell/\lambda_A)}{\sinh(L/\lambda_A)}\,\cosh\left(\frac{\ell-L}{\lambda_A}\right)=
\frac{\sinh(\ell/\lambda_B)}{\sinh(L/\lambda_B)}\,\cosh\left(\frac{\ell-L}{\lambda_B}\right)\label{EqSource}
\end{equation}[0.1cm]
Fig.~\ref{fig:RelSource} depicts the relative source size $\ell/L$, which is a function of only three dimensionless ratios: $L/\lambda_A$, $L/\lambda_B$ and $\alpha_A\beta_B/(\beta_A\alpha_B)$, see also Appendix~\ref{appreactdiff:InhomSteady}. A dimensional analysis of Eq.~\ref{EqReactDiffChoice} shows that this is true also for finite $h$ and other production functions. The red line illustrates the variation of the relative source size $\ell/L$ as a function of $L$, when all other parameters are kept constant. Note that an identical curve could be obtained if instead of $L$, the length scales $\lambda_A$ and $\lambda_B$ would be varied via a common control parameter. This will become important in Chapter~\ref{patterningB} when we combine the Turing system with an expander feedback.

\noindent

\noindent The source size is zero in systems that are very small relative to the characteristic length scales, see Fig.~\ref{fig:RelSource} (green curve). For these small systems, the homogeneous steady state is the only existing solution. The $(1,1)$-pattern begins to appear only above a critical size $L_1$, where $L_1$ obeys
 \begin{equation} \frac{\beta_A\,\alpha_B}{\alpha_A\,\beta_B}\,\frac{\tanh(L_1/\lambda_A)}{L_1/\lambda_A}=\frac{\tanh(L_1/\lambda_B)}{L_1/\lambda_B}\,.\end{equation}
 This equation implies in particular that $\lambda_A$ has to be sufficiently small in comparison to $L$ and $\lambda_B$ for the existence of inhomogeneous patterns, see Appendix~\ref{appreactdiff:InhomSteady}.
 
 \noindent The relative source size $\ell/L$ strongly increases with $L$ beyond $L_1$. In the limit of $\lambda_B\gg L$, it eventually saturates, see Fig.~\ref{fig:RelSource} (blue curve). This observation of source scaling for an existing $(1,1)$-pattern has been reported before \cite{gierer1974biological, gierer1972theory, meinhardt1982models}. In contrast, following the red curve for a finite inhibitor range $\lambda_B$, the source size $\ell$ approaches a constant value independent of $L$. In the limit $\lambda_B\ll L$, we obtain $\ell/L\propto \lambda_B/L$. The source does not sense the system boundaries but instead is restricted by $\lambda_B$.
 
\begin{figure}[tbp]
\centering
\includegraphics[width=0.9\textwidth]{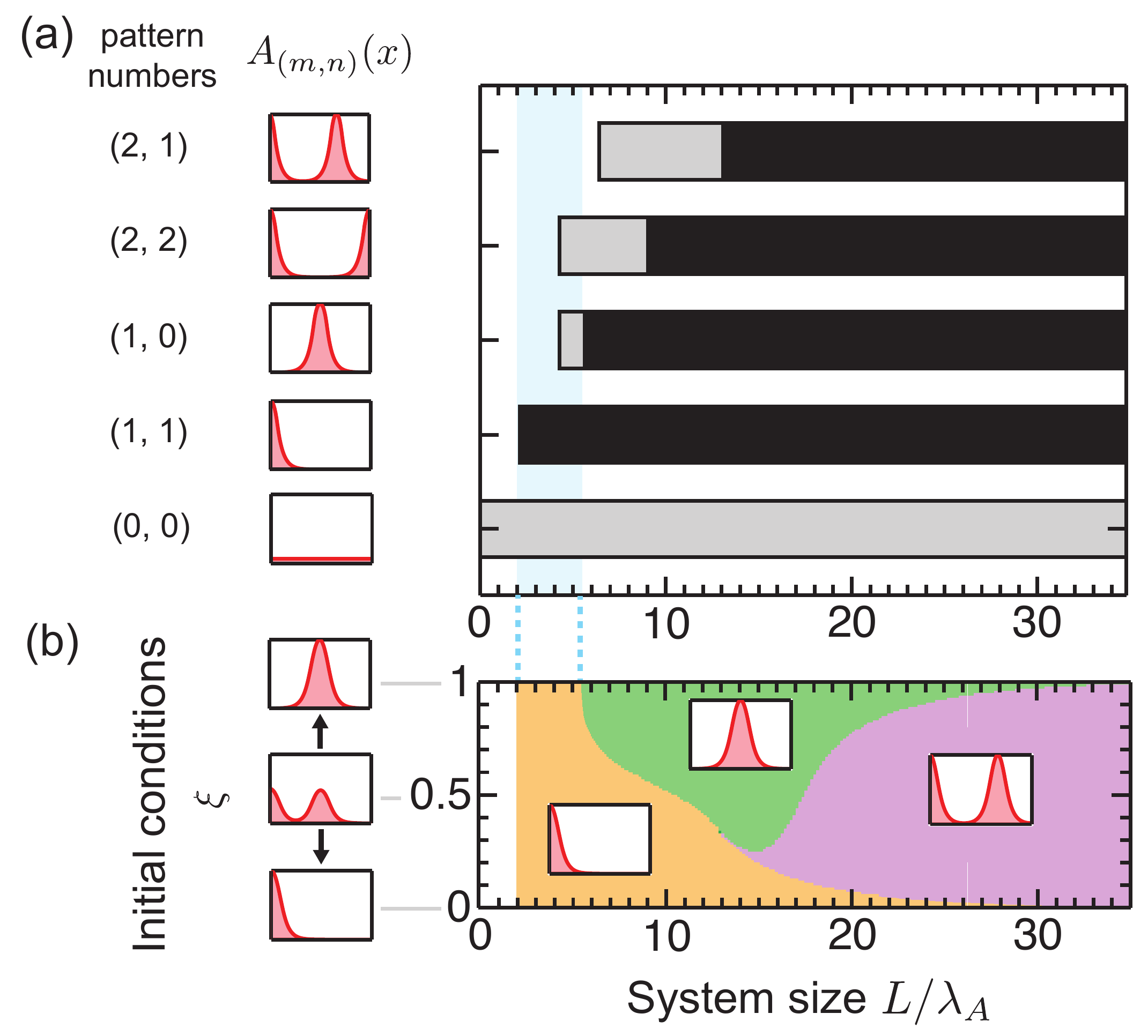}
\caption[Absence of scaling in classcial Turing patterns]{\small 
Classical Turing patterning implies that 
in larger systems higher-order patterns form.
(a) Typical profiles of the activator concentration $A_{(m,n)}(x)$ for the $(m,n)$-pattern are shown in red as steady state solutions to Eq.~\ref{EqReactDiffChoice}.
Size ranges are shown, where the $(m,n)$-pattern is linearly stable (black),
or exists, but is not stable (gray).
In the blue region, the $(1,1)$-pattern is the only stable pattern.
(b)
Basins of attraction: 
final pattern type at steady state as a function of system size on the horizontal axis and initial conditions on the vertical axis.
Initial conditions linearly interpolate 
between the (1,1)- and (1,0)-pattern, i.e.,
$A(x,t{=}0)=(1-\xi)\,A_{(1,1)}(x)+\xi \,A_{(1,0)}(x)$, 
and analogously for $B(x,t{=}0)$.
Parameters: 
$D_B/D_A = 30$, $\alpha_B/\alpha_A = 4$, $\beta_B/\beta_A = 2$,
$h{\rightarrow}\infty$ (a), $h = 5$ (b).
}
  \label{ModeStabPlot}
\end{figure}

\noindent Analogous to the $(1,1)$-pattern, any higher order $(m,n)$-pattern only exists if the system size $L$ exceeds a critical size $L_{\sigma}=\sigma\,L_1$. Fig.~\ref{ModeStabPlot}(a) shows the region of existence (gray) for patterns of different pattern numbers.

 \noindent

\subsection{Higher order patterns form in larger systems}\label{Sec:TuringNotScale}
We have shown that the number of coexisting patterns increases with increasing size. Next, we investigate which of these patterns are stable and can in fact be observed in systems of different sizes.

\vskip0.2cm
\noindent {\it Higher order modes become linearly unstable in larger systems. --- }
The homogeneous steady state is linearly unstable with respect to inhomogeneous perturbations for parameter values that fulfill the Turing conditions discussed in Section~\ref{TuringGeneralFrame}. A linear stability analysis as detailed in  Appendix~\ref{appreactdiff:HomSteady} provides the corresponding parameter range for our specific choice of Eq.~\ref{EqReactDiffChoice} and Eq.~\ref{EqHill}. For a certain choice of parameters and a specific system size, there is a range of perturbation modes $s\in [s_{min},s_{max}]$, which grow exponentially and might result in inhomogeneous patterns. Importantly, the mode number $s$ scales with system size $L$ as can already be seen from the structure of the general linear stability matrix in Eq.~\ref{eq:linTuringmat}. Thus, both interval bounds $s_{min}$ and $s_{max}$ linearly depend on L as does the mode that becomes unstable first, when changing one control parameter. Therefore, in a larger system, more and higher order perturbation modes become unstable \cite{gierer1981generation,ishihara2006turing,othmer1980scale}.

\noindent Nevertheless, the linear instability of the homogeneous steady state is only a necessary condition for the formation of inhomogeneous patterns. The mode number $s$ of the perturbation can provide a first clue about the type of pattern that might form, yet there is no general relationship between the growing perturbation modes and final patterns. In fact, it is not even clear that an inhomogeneous steady state exists in a system of finite size, even if the homogeneous steady state is linearly unstable. We encounter such a situation in our system.\\

%\vskip0.2cm
\noindent {\it  The Turing system possesses a rich dynamics including oscillations due to a finite system size and the multistability of several steady states. --- }
Fig.~\ref{ModeStabPlot}(a) summarizes size ranges of existence (gray) and stability (black) of steady state patterns of our model. %, revealing a rich dynamics.
For very small systems with $L<L_1$, the homogeneous steady state is the only steady state pattern that exists in agreement with reflecting (and also periodic) boundary conditions. Still, the homogeneous steady state can be linearly unstable. As there are no diverging terms in Eq.~\ref{EqReactDiffChoice}, we observe homoclinic orbits and oscillations. These oscillations induced by boundary effects take place far away from the homogeneous steady state. Thus, they appear even if all eigenvalues of the linear perturbation matrix $M_s$ are real.

\noindent For larger systems with $L>L_2$, several inhomogeneous steady states coexist and are linearly stable.
We numerically determined the region of linear stability (black) for several inhomogeneous steady states, see Appendix~\ref{appreactdiff:StabInhom}. We did not find any upper size limit for the linear stability for a particular pattern. However,  
we observed increasingly smaller basins-of-attraction in systems of increasing size. Thus, lower order patterns become unstable with respect to finite-amplitude perturbations
in favor of higher wavenumber patterns, as exemplified in Fig.~\ref{ModeStabPlot}(b).

\vskip0.2cm
\noindent {\it In conclusion, steady state patterns of increasing wavenumber emerge with increasing system size. --- }
In particular, the interplay between the characteristic length scales $\lambda_A$ and $\lambda_B$ and the system size $L$ determines whether a certain pattern fits into the system and whether it will preferentially form. For example, the $(1,1)$-pattern is globally stable only in a limited size range, see Fig.~\ref{ModeStabPlot}(a) (blue shading). Therefore, only if the system is pinned to this blue region by changing the characteristic length scales with system size, the $(1,1)$-pattern emerges reliably from arbitrary initial conditions. In a hypothetical case, in which $\lambda_A$ and $\lambda_B$ scale with $L$, while $\alpha_A\beta_B/(\beta_A\alpha_B)$ stays constant,
we can hope to find a range of parameters, for which 
a particular $(1,1)$-pattern with a scaling source %can be expected to be 
is the only stable steady state, see Eq.~\ref{EqSource} and Fig.~\ref{fig:RelSource}.
In the next Chapter, we introduce a corresponding feedback that couples the Turing system to the dynamics of an expander molecule.\\
\pagebreak

\section{Summary}
\noindent In this chapter we analyzed scaling of morphogen profiles in pre-patterned and self-organized systems. For this, we especially paid attention to the difference between (perfect) scaling of the profiles and approximate scaling.

\noindent For the case of pre-patterned systems, several  scaling mechanisms have been proposed, drawing on the idea of additional expander molecules, which encode system size \cite{wartlick2011understanding, wartlick2011dynamics, mumcu2011selforganized,benzvi2010scaling,benzvi2011expansion}. We distinguish two classes of scaling mechanisms. In the first class, the expander reads out system size independently of the morphogen like in the expander-dilution model. The second class comprises mechanisms of self-scaling by a feedback loop between expander and morphogen like in the expansion-repression model. These latter models typically yield robust approximate scaling but often do not perfectly scale. We highlight several requirements of such expander feedbacks, in particular the fact that the amplitude of the morphogen profile has to vary with system size.

\noindent As a prototype example for self-organized patterning, we discussed a classical Turing system with two players. Our specific choice of equations is especially suitable for analytical treatment. We illustrated the absence of scaling in a comprehensive way: Besides the linear stability analysis of the homogeneous steady state, we also assessed existence and stability of the inhomogeneous patterns. We found a hierarchy of inhomogeneous steady state patterns, for which higher order patterns are favored with increasing system size. Thereby, the behavior is governed by the ratio of system size $L$ to the characteristic length scales $\lambda_A$ and $\lambda_B$ of the two considered molecules, indicating that scaling might be achieved by adjusting these length scales with $L$. In fact similar ideas have already been explored in earlier works \cite{hunding1988size, ishihara2006turing, othmer1980scale, pate1984applications}. It has been shown that by coupling $\lambda_A$ and $\lambda_B$ to an autonomous chemical size reporter like in the expander-dilution model, the Turing patterns scale. In Chapter~\ref{patterningB}, we demonstrate that we can also devise an expander feedback loop in the spirit of the expansion-repression model to generate self-organized scaling of a Turing system.

% ---------------------------------------------------------------------------
% ----------------------- end of thesis sub-document ------------------------
% ---------------------------------------------------------------------------						
% this file is called up by thesis.tex
% content in this file will be fed into the main document

\chapter[Scaling and regeneration of self-organized patterns]{Scaling and regeneration\\ of self-organized patterns} \label{patterningB}% top level followed by section, subsection
\chaptermark{Scaling \& regeneration of self-organized patterns} %only change chapter heading

% the code below specifies where the figures are stored
%\ifpdf
%    \graphicspath{{2_patterning/figures/PNG/}{2_patterning/figures/PDF/}{2_patterning/figures/}}
%\else
%    \graphicspath{{2_patterning/figures/EPS/}{2_patterning/figures/}}
%\fi

% ----------------------- contents from here ------------------------

\section{Self-scaling and self-organization}
\begin{figure}[tp]
\centering
\includegraphics[width=1\textwidth]{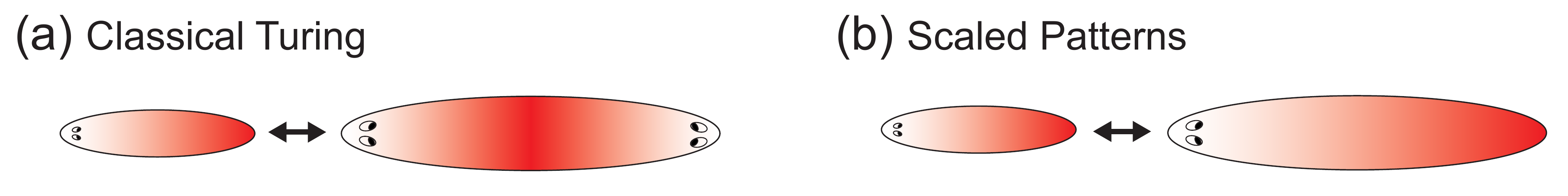}
\caption[Turing patterns vs.~scaling patterns]{\small 
Classical Turing patterns show more
periodic repeats in larger systems as a result of fixed intrinsic
length scales (a), instead of being a scaled-up version of the
patterns in small systems (b).}
\label{fig:PatternScalingWorm}
\end{figure}

Flatworms challenge previous theories of pattern scaling and self-organization. Regene\-ration from minute amputation fragments of strongly varying size prompts for  patter\-ning mechanisms that are both: scalable and self-organizing. Turing systems, as prominent frameworks for self-organized patterning, do not naturally scale \cite{umulis2013mechanisms,othmer1980scale,green2015positional}. They typically generate the same patterns repeated multiple times in a growing system as depicted in Fig.~\ref{fig:PatternScalingWorm}. In Chapter~\ref{patterningA}, we have illustrated that this is due to fixed characteristic length scales, which define the wave length of the patterns.

\noindent However, it has been demonstrated that Turing patterns can in fact scale if the charac\-teristic length scales are coupled to a chemical size-reporter, which we called expander
 \sidenote{0.45\textwidth}{``Everywhere nature works true to scale, and every\-thing has its proper size accordingly.'' \;---\; D'Arcy W. Thompson, On Growth and Form, 1945 \cite{thompson1945growth}}
 \cite{hunding1988size, ishihara2006turing, othmer1980scale, pate1984applications}. These previous works assumed an expander that is independently controlled by a pre-patterned system like in Section~\ref{Sec:ExpDil}. Thus, the proposed mechanisms account for scaling, but not in a fully self-organized way.
Conversely, self-scaling mechanisms have been discussed previously (e.g.~for fly development), yet for systems that are not fully self-organized \cite{benzvi2010scaling,benzvi2011scaling,umulis2013mechanisms,wartlick2011dynamics, averbukh2014scaling}, see Section~\ref{Sec:ExpRep}.

 \noindent

\noindent Now, in this chapter, we combine the two features of scaling and self-organization and present a generic mechanism that yields fully self-organized and self-scaling patterns. Importantly, the expander feedback in a Turing system with a self-organized source follows a very different control logic than in systems with a pre-existing source. In consequence, perfectly scaling steady states are stable across a wide range of system sizes.
Our novel class of self-scaling Turing patterning provides a conceptual framework to address scaling and regeneration in flatworms and guides the design and interpretation of experiments.
This result has been published in Werner \textit{et al.} \cite{werner2015scaling}.

\noindent First, we present a specific example which combines ideas from Turing patterning, discussed in Section~\ref{TuringNoScaling}, with theoretical concepts for the scaling of morphogen profiles, analyzed in Section~\ref{GradientScaling}. We demonstrate that this system is capable of self-organized scaling and explain the underlying mechanism. In a second step, we discuss generalizations of the mechanism and predictions for experiments in flatworms.

%\label{Sec:TuringScaling}
\section{A minimal model for self-organized pattern scaling}
\noindent Analogous to Section~\ref{GradientScaling}, we add another molecular species, termed expander, whose concentration $E$ will provide a read-out of system size $L$. As a specific case, we \mbox{assume} the expander is produced homogeneously, spreads by diffusion and is subject to degradation
\begin{equation} \partial_t E=\alpha_E-\beta_{E}\,E+D_E\,\partial_x^2\,E\,.\label{EqEdot}
\end{equation}
Again, $\alpha_E$ denotes the respective production rate, $\beta_E$ the degradation rate and $D_E$ an effective diffusion coefficient.

\noindent In order to achieve self-organized scaling, the dynamics of the three molecular species have to be mutually coupled in a similar manner as for the expander feedback in Section~\ref{Sec:ExpRep}. As an illustrative example, we choose a feedback loop, in which the inhibitor of the Turing system controls the degradation rate of the expander via
\begin{equation}\label{eq:betaE}
 \beta_E=\kappa_E\,B
 \end{equation}
 with a positive constant $\kappa_E$.
 This choice ensures that the expander concentration is approximately homogeneous even for small $D_E$ because the inhibitor concentration itself is typically rather homogeneous in a Turing system.
 
\noindent In turn, the expander also feeds back on the Turing system and changes the length scales of the morphogen profiles, see Fig.~\ref{SourceGradientScaling1}(a). Analogously to Eq.~\ref{eq:betaE}, we choose a linear regulation of the degradation rates by the expander (with $\kappa_{A}$, $\kappa_{B}>0$)
\begin{equation} 
\beta_A=\kappa_{A}\, E\;,\quad  \beta_B=\kappa_{B}\, E\,.
\label{EqbeE}
\end{equation}
Note that Eqs.~\ref{EqReactDiffChoice} and \ref{EqEdot} can be considered as a classical three component Turing system, yet with specifically chosen coupling terms given by Eqs.~\ref{EqHill}, \ref{eq:betaE} and \ref{EqbeE}.

\section{Numeric solution shows scaling and pattern regeneration}
\begin{figure}[tbp]
\centering
\includegraphics[width=1\textwidth]{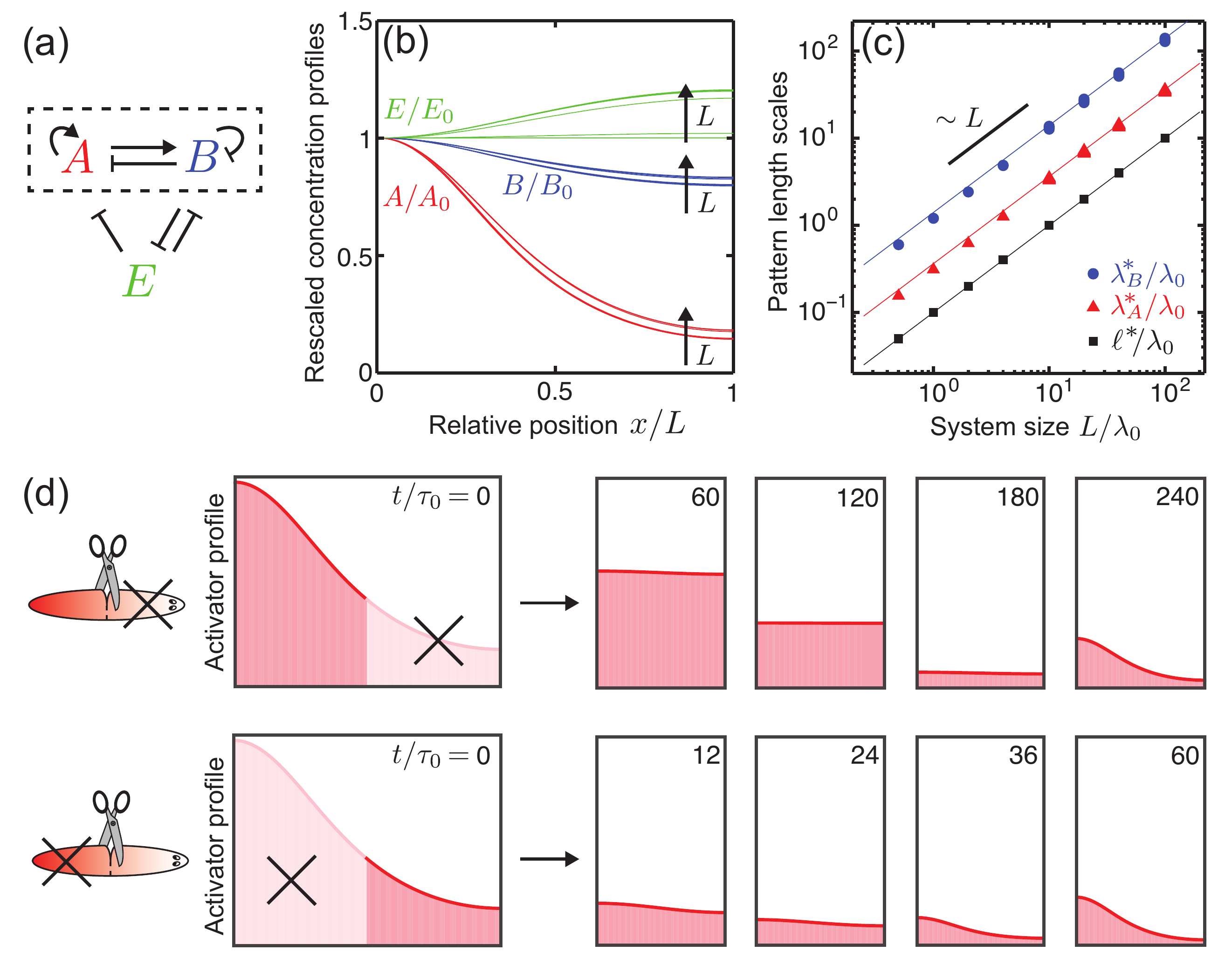}
\caption[Scaling in a Turing system with expander feedback]{\small 
Scalable pattern formation in a Turing system with expander feedback. 
(a) 
The Turing system and the expander mutually control their degradation rates, resulting in a stable feedback loop. 
(b) 
Scaling corresponds to morphogen profiles that collapse as a function of relative position $x/L$ 
(normalized by respective concentrations $A_0$, $B_0$, $E_0$ at $x=0$).
(c) 
The feedback self-consistently adjusts the length scales $\lambda_A$ and $\lambda_B$ of the morphogen profiles 
and thus the source size $\ell$ with system size
(symbols: numerical results; lines: analytical solution of Eqs.~\ref{EqReactDiffChoice} and \ref{EqEdot} at steady state for homogeneous expander concentration and $h{\rightarrow}\infty$).
Here,
$\lambda_0=[D_A/\sqrt{\alpha_A \kappa_A}]^{1/2}$ denotes the characteristic length scale of the system.
(d) Regeneration of concentration profiles: after cutting the system in two (initial size $L/\lambda_0=20$), the expander feedback adjusts the activator concentration to fit the smaller system size both in a tail fragment (upper row) and head fragment (lower row).  Repatterning of the tail fragment takes significantly longer due to the initially higher morphogen concentration (time given in units of $\tau_0=1/\sqrt{\alpha_A \kappa_A}$).
Parameters: 
$D_B/D_A=30$, $D_E/D_A=10$, $\alpha_B/\alpha_A=4$, $\alpha_E/\alpha_A=0.4$, $\kappa_B/\kappa_A=2$, $\kappa_E/\kappa_A=2$, $h=5$.
}
\label{SourceGradientScaling1}
\end{figure}

Before analyzing this system of equations analytically, we numerically demonstrate that the steady state pattern scales with system size over several orders of magnitude, see Fig.~\ref{SourceGradientScaling1}(b)-(c).
Small deviations from the perfect collapse of the concentration profiles arise due to a finite diffusion coefficient of the expander and a finite $h$ of the production function. Still, the source size $\ell^*$ at steady state and the characteristic length scales $\lambda^*_A$ and $\lambda^*_B$ show almost perfect scaling with system size. In the limit of an homogeneous expander and a switch-like source with $h\rightarrow\infty$, we analytically show in Section~\ref{Sec:InsightsTuringScaling} that the scaling becomes exact.

 \noindent

\noindent Pattern scaling does not come at the cost of robustness for the Turing system. We can challenge patterning by perturbations that mimic amputation experiments, see Fig.~\ref{SourceGradientScaling1}(d). Here, the activator concentration $A$ reliably adjusts to fit the smaller system size. Note that the system part with the initially oversized source and thus higher morphogen concentrations takes longer to re-adjust.

\section{Dynamical systems analysis of scaling and regeneration}\label{Sec:InsightsTuringScaling}
\begin{figure}[tbp]
\centering
\includegraphics[width=1\textwidth]{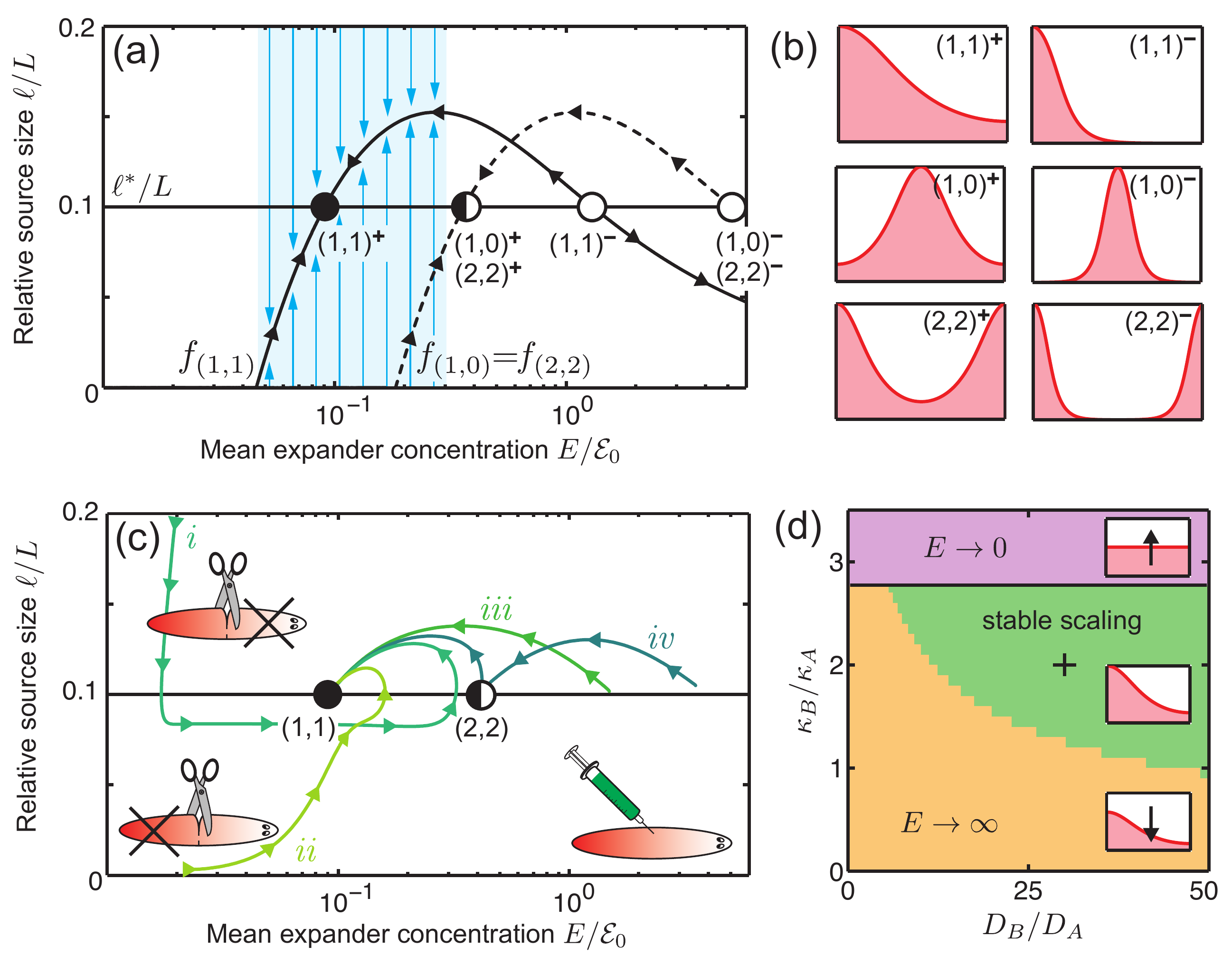}
\caption[Fixed points and dynamics in a Turing system with expander feedback]{\small 
Fixed points and dynamics in a Turing system with expander feedback. 
(a)-(b) Two-dimensional projection of the phase space: all fixed points (circles) are characterized by the same relative source size.
For adiabatically slow expander dynamics, the system relaxes along the nullclines of the Turing system $f_{(m,n)}$ 
(shown for $h\rightarrow\infty$, $\lambda_E\gg L$, corresponding to Eq.~\ref{EqSource}). 
As each nullcline intersects the steady-state condition of Eq.~\ref{Eqrelsourceconst} twice, the system possesses two fixed points $(n,m)^+$ and $(n,m)^-$ for each pair $(n,m)$. In the blue region, the (1,1)-pattern is the only stable steady state of the Turing system, compare to Fig.~\ref{ModeStabPlot}, implying that all trajectories starting there must converge to this fixed point. Here, $\mathcal{E}_0=(\alpha_A/\kappa_A)^{1/2}$ denotes a characteristic concentration of the system.
(c) 
Example trajectories, mimicking amputation experiments (labeled \textit{i},\textit{ii}), 
and uniform, one-time injection of the expander (\mbox{labeled} \textit{iii},\textit{iv}); 
all converge to the same stable fixed point, an appropriately scaled (1,1)-pattern.
(d)
Parameter regions for stable, self-scaling pattern formation (green),
and regions of expander divergence (orange, purple). Parameters of panel (c) and Fig.~\ref{SourceGradientScaling1} indicated by cross.
Parameters: 
$D_B/D_A=30,\quad D_E/D_A=10,\quad \alpha_B/\alpha_A=4,\quad\alpha_E/\alpha_A=0.4,\quad\kappa_B/\kappa_A=2,\quad\kappa_E/\kappa_A=2,\quad 
h=5$, $L/\lambda_0=10$, unless indicated otherwise.
}
\label{SourceGradientScaling2}
\end{figure}

Next, we provide insight into how and why scaling works.
First, we identify steady states, each of which scales with system size. 
For the simple case of adiabatically slow expander dynamics, 
we then show that the (1,1)-pattern is a stable steady state.
A phase space description allows us to characterize fixed points and understand the non-linear dynamics.
We illustrate the behavior by considering a two-dimensional projection of the phase space of our dynamical system, spanned by the relative source size given by Eq.~\ref{eq:relsource} and the mean expander level, see Fig.~\ref{SourceGradientScaling2}(a).

\vskip0.2cm
\noindent {\it An implicit scaling relation holds at steady state. --- }
We first show that all steady states of the extended Turing system are characterized by the same relative source size $\ell^*/L= \langle P^*\rangle$ independent of system size $L$. Thus, in Fig.~\ref{SourceGradientScaling2}(a) all steady states (circles) are found on a horizontal line.
By spatial averaging of Eq.~\ref{EqReactDiffChoice} and \ref{EqEdot}, we obtain for steady state concentrations $B^*$ and $E^*$
\begin{eqnarray}
0&=&\alpha_B \langle P^*\rangle - \kappa_B \langle B^* E^*\rangle\\
0&=&\alpha_E - \kappa_E\langle B^*E^*\rangle\,.
\end{eqnarray}
Hence, the relative source size in the steady state is constant
\begin{equation}
\frac{\ell\starup}{L}=\frac{\alpha_E\,\kappa_B}{\alpha_B\,\kappa_E}\,.
\label{Eqrelsourceconst}
\end{equation}
In contrast, the length scales defined in Eq.~\ref{eq:lambdaAB} are not constant anymore but change with expander level:
\begin{equation} \lambda_A=\sqrt{\frac{D_A}{\kappa_A E}}\quad\text{and}\quad\lambda_B=\sqrt{\frac{D_B}{\kappa_B E}}\,.\end{equation}\vskip0.2cm
\noindent We have seen in Fig.~\ref{SourceGradientScaling1}(c) that $\lambda_A^*$ and $\lambda_B^*$ at steady state scale with high precision with system size. In order to explain this scaling behavior, we consider the limit of a spatially homogeneous expander concentration $E$, for which we can show that the scaling becomes exact. This limit corresponds to either a large expander range $\lambda_E=\sqrt{D_E/(\kappa_E B)}\gg L$ or a large inhibitor range $\lambda_B=\sqrt{D_B/(\kappa_B E)}\gg L$. The latter condition is in compliance with the requirements for a typical Turing system and is promoted by the scaling feedback.

\noindent If the expander level was imposed as constant $E=E_0$, we would obtain a Turing system without expander feedback, as discussed in Section~\ref{TuringNoScaling}. Thus, the concentrations would converge towards one of the $(m,n)$-patterns discussed in Section~\ref{Sec:TuringSteadyStates} with pattern length scales $\lambda_A(E_0)$ and $\lambda_B(E_0)$. The relative source size $f_{(m,n)}=l/L$ of such a pattern depends on $E_0$ only via the dimensionless ratios $\lambda_A(E_0)/L$ and $\lambda_B(E_0)/L$, see Eq.~\ref{EqSource}. This fact is illustrated by Fig.~\ref{fig:RelSource}, where the red curve displays $f_{(1,1)}$ in the limit of $h\rightarrow \infty$. Importantly, $f_{(m,n)}=f_{(m,n)}(L^2 E_0)$ is only a function of $L^2E_0$ and changing $E_0$ has analogous effects on the relative source size as changing $L^2$.

\noindent The same argument also implies that a $(m,n)$-pattern can only exist above a critical value of $E_0$, 
corresponding to the minimum system size for the existence of patterns in Fig.~\ref{ModeStabPlot}(a). 
Below this critical value, $f_{(m,n)}$ is zero. 
Above this value, $f_{(m,n)}$ displays a nonmonotonic dependence on $E_0$, 
which results from opposing effects of the 
pattern length scales of the activator and the inhibitor on the source size $\ell$,
see Fig.~\ref{SourceGradientScaling2}(a).

\noindent The intersections of the curves $f_{(m,n)}$ (obtained from the steady states of the Turing system without expander) with the
constant value $\ell^*/L$ given by Eq.~\ref{Eqrelsourceconst} define the steady states of the full system with
expander feedback.
For each pattern type $(m,n)$, 
we find two steady-state patterns, denoted $(m,n)^+$ and $(m,n)^-$, with respective expander levels $E^{+}_{(m,n)}<E^{-}_{(m,n)}$, 
see the black and white circles in Fig.~\ref{SourceGradientScaling2}(a).

\noindent The fact that $f_{(m,n)}(L^2\,E^*)=\ell^*/L$ is independent of system size $L$ by Eq.~\ref{Eqrelsourceconst}, implies that also $L^2E^*$ is independent of $L$ for each steady state. 
We conclude $E^*\propto L^{-2}$ and thus 
$\lambda_A^*\propto L$, $\lambda_B^*\propto L$ at each fixed point, 
consistent with our numerical results in Fig.~\ref{SourceGradientScaling1}(c).

\vskip0.2cm
\noindent {\it The stability in the vicinity of the scaling fixed points can be assessed in the
simple limit of adiabatically slow expander feedback.  --- }%
In this limit, the source size first relaxes to $\ell/L=f_{(m,n)}(L^2\,E)$ for some
$(m,n)$, corresponding to the fast time scale of the Turing system. 
Then, by Eq.~\ref{EqEdot}, the system moves slowly along this nullcline according to
\begin{equation}\label{eq:NullclineRelax}
\partial_t E=\alpha_E-\frac{\kappa_{E}\,\alpha_B}{\kappa_B}\,f_{(m,n)}(L^2\,E)\,.
\end{equation}
Stability of steady-state patterns requires $\partial_E f_{(m,n)}>0$, which
is observed to hold only for $E_{(m,n)}^{+}$, see
Fig.~\ref{SourceGradientScaling2}(a).

\noindent Which branch $f_{(m,n)}$ is selected for arbitrary initial conditions by the fast Turing dynamics? 
As we consider the limit of slow expander feedback, the expander concentration is approximately constant within the time scale of the Turing relaxation. Thus, the question of which pattern $(m,n)$ is selected is formally equivalent to the stability of $(m,n)$-patterns in the Turing system without expander feedback as a function of system size $L$. 
The blue region in Fig.~\ref{ModeStabPlot}(a), in which the $(1,1)$-pattern is the only stable pattern, can be associated with the blue range of expander concentration in Fig.~\ref{SourceGradientScaling2}(a). Therefore, we can deduce that this represents a basin-of-attraction for the $(1,1)^{+}$-pattern.

\vskip0.2cm
\noindent {\it A numerical analysis reveals a large basin-of-attraction for the first order pattern. --- }%
Fig.~\ref{SourceGradientScaling2}(c) shows that the $(1,1)^{+}$-pattern is an attractive fixed point also for trajectories starting outside this blue region and for nonadiabatic expander dynamics.  Two example trajectories, labelled (\textit{i}) and (\textit{ii}),
corresponding to head and tail fragments, respectively, 
converge to the $(1,1)$-pattern, after a transient under- and over-shoot of the source size.
Two additional trajectories, labeled (\textit{iii}) and (\textit{iv}), simulating uniform injection of the expander, likewise converge to this fixed point. We can understand these observations by drawing on the results for the Turing system without expander in Section~\ref{Sec:TuringSteadyStates} and \ref{Sec:TuringNotScale}.

\vskip0.2cm
\noindent {\it The dynamics for large expander concentrations are characterized by a successive movement through higher order patterns.  --- } %
The $(1,1)^+$-state is also attractive for trajectories like ($iv$) with initially high values of $E$. Here, a $(2,2)$-pattern with an additional source is  transiently emerging. Yet, according to the expander dynamics given by Eq.~\ref{eq:NullclineRelax}, the system moves towards the fixed point $(2,2)^+$ as long as the $(2,2)$-pattern comprises a stable fixed point of the Turing system without expander. However, the fixed point $(2,2)^+$ of the Turing system with expander lies in a region of the phase space where the $(2,2)$-pattern of the Turing system alone is unstable for our choice of parameters. Thus, the fixed point  $(2,2)^+$ is in fact a saddle point, attractive with respect to the expander dynamics along the nullcline $f_{(2,2)}$ but unstable with respect to the Turing system. Again, this can be understood from Fig.~\ref{ModeStabPlot}(a) by noting the formal correspondence between changes in $E$ for a fixed system size $L$ and changes in $L^2$ for the classical Turing system without expander. Fig.~\ref{ModeStabPlot}(a) illustrates that the $(2,2)$-pattern is unstable outside the blue region. The result would be different for the $(1,0)$-pattern. Note that the nullclines of the $(1,0)$-pattern and the $(2,2)$-pattern are congruent in our projection of the phase space. In consequence, there exists a trajectory starting with a $(1,0)$-pattern that looks identical to ($iv$), besides the fact that it ends in the fixed point $(1,0)^+$, which is stable for our choice of parameters, as illustrated in Fig.~\ref{ModeStabPlot}(a).

\vskip0.2cm
\noindent {\it Dynamics of homogeneous states allows to understand regeneration of patterns.  --- } %
For small expander values, there are no inhomogeneous steady state patterns of the Turing system, yet homogeneous patterns exist. To understand the relaxation of the full system, we will discuss two limiting cases, assuming again the separation of time scales between Turing dynamics and expander system:

\noindent a) In the limit of fast Turing dynamics, the Turing system will first approach the homogeneous steady state, given by Eq.~\ref{Eq:TuringHomFixedPoint} for a certain value of $E$. On the larger time scale of the expander relaxation, the system moves along the corresponding nullcline $\ell/L=f_{(0,0)}$ according to 
\begin{equation} \partial_t E=\alpha_E-\frac{\alpha_B \kappa_E}{\kappa_B}\,f_{(0,0)}\quad\text{with}\quad f_{(0,0)}= P(A_h^*,B_h^*)\,.
\end{equation}
If $f_{(0,0)}<\ell^*/L$, the expander level increases and consequently the Turing length scale decreases until the homogeneous pattern becomes unstable and inhomogeneous patterns can form.

\noindent b) As a second case, we consider the opposite limit, for which the Turing dynamics are adiabatically slow and the inhibitor concentration $B$ does hardly change within the time scale of relaxation of  the expander towards $\alpha_E/(\kappa_E\,B)$. We find a qualitatively similar behavior. It exists another homogeneous, yet dynamic solution of the Turing system, in which the ratio $\chi=B/A$ and thus $P(A,B)=g(\chi)$ is approximately constant. The ratio $\chi$ is determined by the following implicit relation, see Appendix~\ref{appreactdiff:LowExpander}:
\begin{equation}  \frac{1}{1+\chi^h}=\frac{\alpha_E}{\kappa_E}\,\frac{\kappa_A-\kappa_B}{\alpha_A\,\chi-\alpha_B}\,.\label{Eq:xi}
\end{equation}
In order for $E$ to increase such that the system enters the regime where inhomogeneous patterns exist, $B$ has to decrease. According to
\begin{equation} \partial_t B=\alpha_B\, g(\chi) - \frac{\kappa_B\,\alpha_E}{\kappa_E}\,,\label{eq:BdynHomS}\end{equation}
this requires that $g(\chi)<\ell^*/L$.

\noindent Depending on the choice of parameters, both transient homogeneous states (corresponding to the two limiting cases of a separation of time scales) can be observed when mimicking regeneration experiments in the Turing system with expander. In Fig.~\ref{SourceGradientScaling2}(c), the horizontal stretch of trajectory ($i$) corresponds to $g(\chi)$. Both production terms $ f_{(0,0)}$ and $g(\chi)$ of the transient homogeneous states have to be smaller than $\ell^*/L$, explaining the transient undershoot during regeneration. Note that from $g(\chi)=\ell^*/L$ follows that also $ f_{(0,0)}=\ell^*/L$. The set of parameters corresponding to this joint condition mark the breakdown of robust regeneration, see horizontal line in Fig.~\ref{SourceGradientScaling2}(d). Beyond this thres\-hold, the homogeneous morphogen concentrations diverge while the expander vani\-shes (purple region). If the parameters obey this constraint, the basin-of-attraction of the $(1,1)$-state extends towards small values of $E$ outside the blue region in Fig.~\ref{SourceGradientScaling2}(a).

\section{As single source pattern is attractive for a large region of the phase space} %
\begin{figure}[tbp]
\centering
\includegraphics[width=1\textwidth]{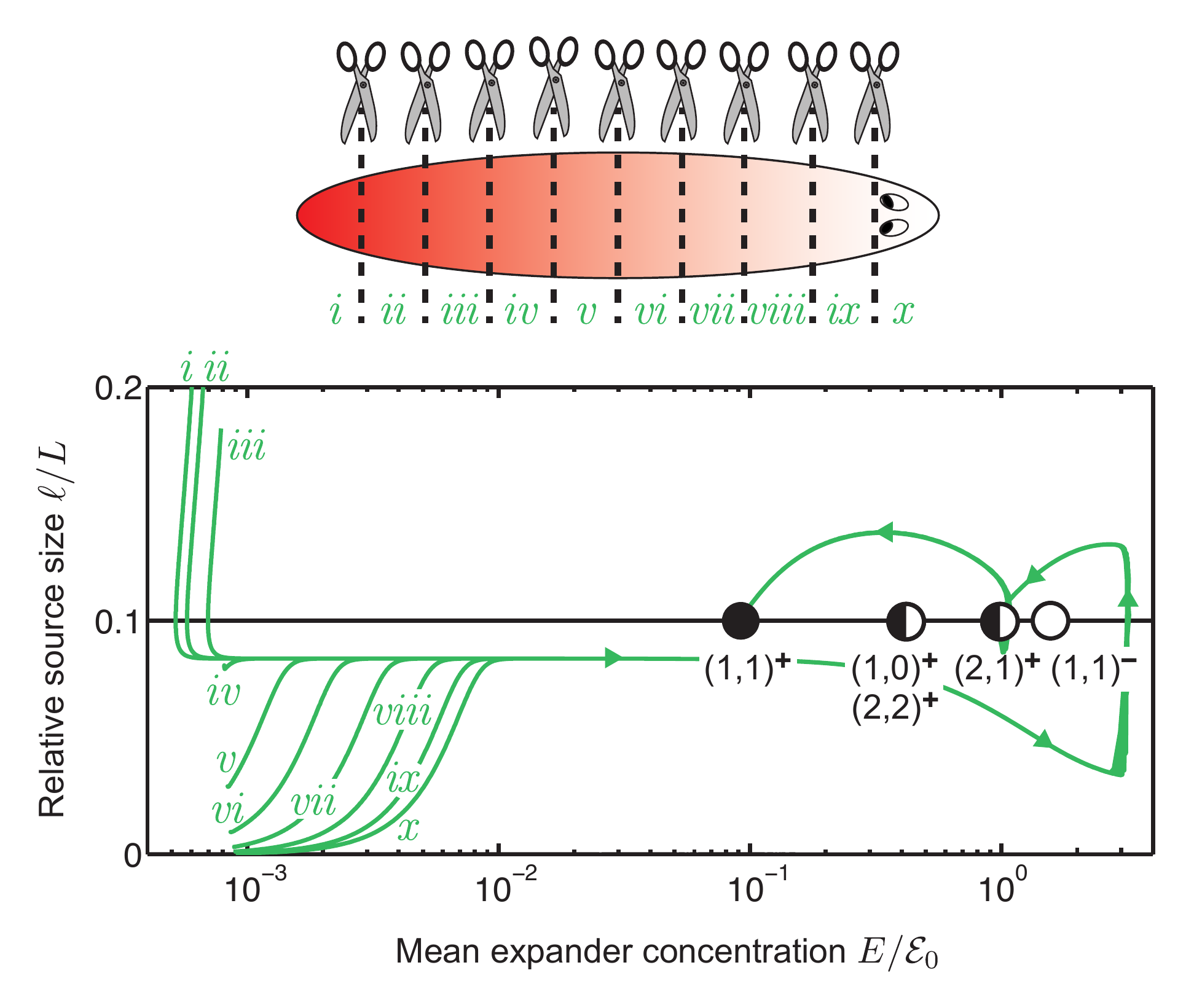}
\caption[Amputation experiment with ten cuts]{\small 
Numerical solution of the Turing system with expander mimicking an amputation experiment with 10 cuts. All trajectories converge to the fixed point $(1,1)^+$ after a transient formation of the $(2,1)$-pattern. Parameters: 
$D_B/D_A=30, D_E/D_A=10, \alpha_B/\alpha_A=4,\alpha_E/\alpha_A=0.4,\kappa_B/\kappa_A=2$, $\kappa_E/\kappa_A=2$, 
$h=5$, $L/\lambda_0=10$.
}\vskip0.5cm%%%
\label{Fig:TenCuts}
\end{figure}

Based on the discussion above, we expect the $(1,1)$-pattern to be re-established even when simulating more drastic amputation experiments, as shown in Fig.~\ref{Fig:TenCuts}. Again, we find the source $\ell/L$ of each smaller system first approaching the horizontal line corresponding to $g(\chi)$. Eventually, the expander value is sufficiently large such that the homogeneous state becomes unstable and inhomogeneous patterns can form. Transiently, the $(2,1)$-pattern arises but as the $(2,1)^+$-solution is unstable with respect to the Turing dynamics, the additional source vanishes and the system returns to the $(1,1)$-pattern.

\noindent Details of the relaxation dynamics and whether $(1,1)^+$ is the only stable fixed point or whether a second stable fixed point $(1,0)^+$ with a smaller basin-of-attraction exists like in our example, depends on the choice of parameters. In general, we observe robust pattern scaling for a vast parameter range,
provided (i) inhibitor diffusion is sufficiently fast (a necessary
condition for pattern formation in any Turing system)
and (ii) the feedback strength is similar for both Turing molecules,
see Fig.~\ref{SourceGradientScaling2}(d). The threshold above which the expander concentration vanishes, resulting in a divergence of $A$ and $B$ (purple), can be analytically determined from the condition $f_{(0,0)}=g(\chi)=\ell^*/L$. This corresponds to the fact that the expander should increase for homogeneous morphogen concentrations.

\section{Structural robustness for pattern scaling}
Several generalizations of the proposed minimal mechanism are conceivable. 
First, the feedback of the Turing system on the expander level could be likewise implemented via the production rate, 
e.g.~$\alpha_E\propto B$, instead via the degradation rate $\beta_E=\kappa_E B$. 
Then, scaling would require $\beta_A\propto1/E\,,\;\beta_B\propto1/E$ as well as a fast spreading expander, yielding analogous steady states.

\noindent As a second possibility for pattern scaling, the feedback in Eq.~\ref{EqbeE} could also be mediated by $A$ (instead of $B$ or additional to $B$), provided the expander diffuses sufficiently fast. For example, the following system can be interpreted in the sense that pairs of $A$ and $E$ as well as $B$ and $E$ are only degraded together when bound to each other
\begin{eqnarray}
\partial_t A&=&\alpha_A\,P(A,B)-\kappa_A\,A E+D_A\,\partial_x^2\,A\nonumber\\
\partial_t B&=&\alpha_B\,P(A,B)-\kappa_B\, EB+D_B\,\partial_x^2\,B\nonumber\\
\partial_t E&=&\alpha_E-\kappa_A\,A E-\kappa_B \,BE+D_E\,\partial_x^2\,E\,.
\end{eqnarray}
More generally, the degradation term does not even have to be linear in $A$, $B$ and $E$. Two increasing functions of $A$ and $E$, and $B$ and $E$, respectively, that fulfill the Turing conditions together with $P(A,B)$ are sufficient to generate robust scaling by the same arguments as above, see Appendix~\ref{appreactdiff:GenScalTuring}.

\noindent Remarkably, 
the feedback loop considered here, featuring two mutually suppressing concentration profiles ($B$ and $E$),
would be unstable for a pre-patterned morphogen source of fixed size
as discussed for the expander feedback in Section~\ref{Sec:ExpRep}. The feedback mechanism becomes stable, only due to the self-stabilizing effect of the Turing system itself.

\noindent Our mechanism yields pattern scaling by scaling of the morphogen profiles via a feedback on the degradation rates of $A$ and $B$. As discussed in Section~\ref{scalingprofiles}, scaling of exponential profiles can in principle also be achieved by mechanisms affecting the diffusion of morphogens. However, controlling only diffusion is not compatible with self-organized pattern scaling as presented here. Our mechanism relies on a size-dependent amplitude of the morphogen profiles, which is lacking for pure diffusion control, as discussed in Section~\ref{Sec:ScalingLesson}.

\noindent The morphogen amplitude in our feedback scheme changes quadratically with system size as can be seen from Eq.~\ref{Asol}-\ref{Bsol}. It is interesting to note that the flux $\beta_A A$ has a size-independent amplitude. The spatial profile of this flux could provide a read-out of the scaling morphogen profiles independent of their amplitudes, see Appendix~\ref{appreactdiff:ScalConstAmpl}. For example, if degradation only happens by internalization upon binding to a receptor, the concentration of down-stream targets of the morphogen $A$ would be proportional to the flux $\beta_A A$.\\ 
\clearpage%%%

\section{Signatures of self-scaling patterns}\label{Sec:TuringPredictions}
Our minimal model for the scaling of self-organized patterns can act as a framework to understand scaling and regeneration in flatworms. We find a number of signatures characterizing this class of patterning mechanism, which can be identified in experiments.

\noindent (i) First, the over-all morphogen levels depend on system size. As already discussed in Section~\ref{Sec:ScalingLesson} for self-scaling feedback loops, this is necessary for the morphogens to influence the expander in a size-dependent manner. However, downstream targets might scale with a constant amplitude. 

\noindent (ii) Second, for the simple scalable Turing systems discussed here, a size-dependent morphogen amplitude requires the adjustment of degradation rates. For the experi\-ments, we have to take into account that many processes can effectively lead to a change in degradation by expander molecules. For example, binding of the expander could disintegrate or alter the morphogen such that it cannot fulfill its signaling task. Alternatively, the expander could act as a co-receptor that facilitates internalization and thus removes the morphogen from the system.
Conversely, the expander might reduce degradation by preventing disintegration, modifications %conformational changes?
or internalization by binding to the morphogen or its suppressors.

\noindent (iii) Third, the expander by definition shows a size-dependent concentration. If the expander level is proportionally to $L^2$, it can couple linearly to the degradation rate to achieve scaling. Note that the expander is likely not a single molecule but an entire signaling module, for which some components might already be expressed in a size-dependent manner. Thus, analyzing changes in expression levels between animals of different size can help to identify the expander mechanism.

\noindent (iv) Fourth, a perturbation of the expander feedback by over-expression or gene knockout or also by hindered spreading is expected to lead to a change in the wave number of the pattern. Depending on the coupling of the expander to the morphogen dynamics, one might observe the most severe effects in either small or large animals.

\noindent (v) Fifth, the source is expected to exhibit a non-monotonic dynamics after amputation. In particular, an initially oversized source might show an undershoot during re\-generation.
%This corresponds to a fast response of the Turing system in comparison to the expander dynamics.

\noindent (vi) Sixth, there might be additional fixed points of the system with small basins-of-attraction. In our example, the $(1,0)^+$-pattern was also stable and locally attractive. If the system is initialized close to this pattern, the pattern will be maintained, even in the presence of small perturbations.

\section{Comparison to experiments in flatworms}\label{Sec:TuringPredictionsFlatworms}
\begin{figure}[tbp]
\centering
\includegraphics[width=0.7\textwidth]{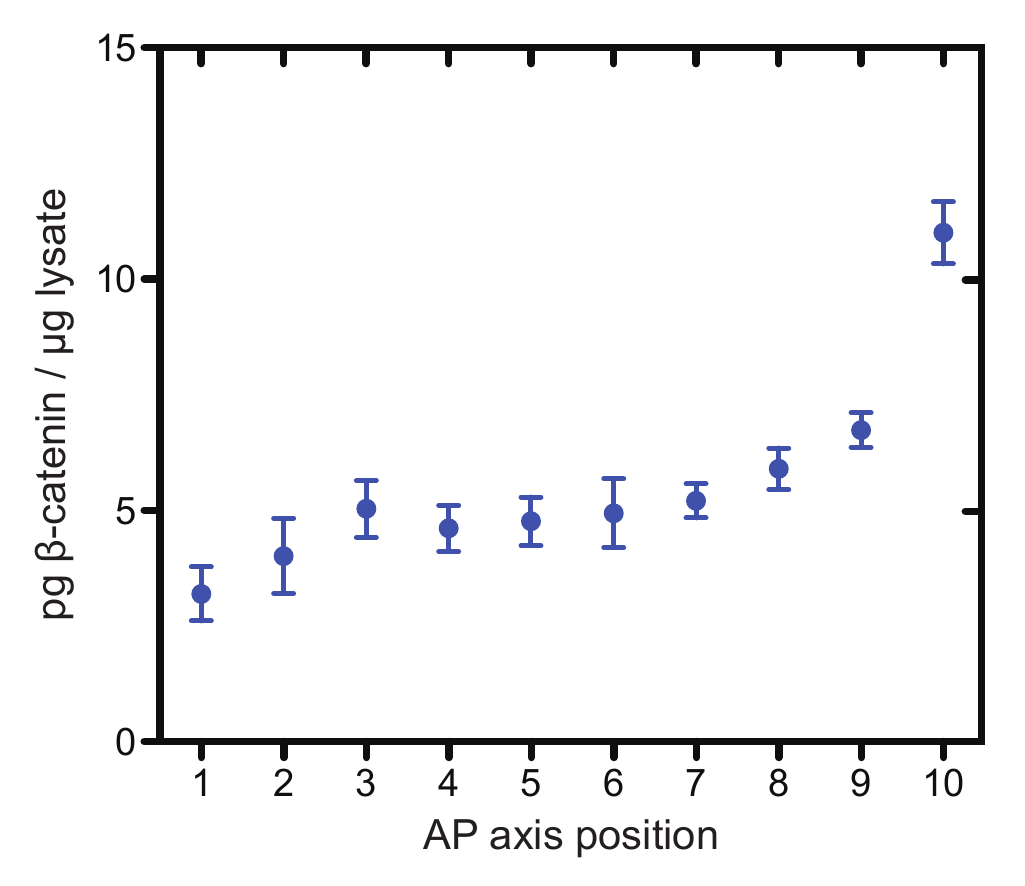}
\caption[$\beta$-catanin gradient in flatworms]{\small 
The $\beta$-catenin concentration forms a gradient from tail to head defining the AP axis in flatworms. Quantitative Western Blot by Tom St\"uckemann in the group of Jochen Rink (lysate refers to dissolved cell material). 
}
\label{Fig:betacat}
\end{figure}

The signatures discussed in Section~\ref{Sec:TuringPredictions} can to some extent be found in flatworms, in particular for the Wnt/$\beta$-catenin pathway associated with head-tail polarity  \cite{almuedocastillo2012wnt,adell2010gradients, forsthoefel2009emerging,gurley2008betacatenin,petersen2008smed,gurley2010expression, niehrs2010growth, elliott2013history}. The Wnt molecules are expressed in the posterior part of the animal and shape the graded $\beta$-catenin profile with a maximum at the tail, see Fig.~\ref{Fig:betacat}.

\noindent The expression regions of components of the Wnt pathway as well as independent anterior and posterior markers fit to the size of the worm \cite{gurley2010expression}. Furthermore, first preliminary data by our experimental collaborators suggests that also the $\beta$-catenin profile matches worm size with a constant amplitude, see (i).

\noindent After amputation, this graded cue has to be quickly scaled to fit the remaining body fragment. Indeed, it has been observed that the position-dependent expression pattern along the AP axis changes within the existing cells to adjust to the new fragment size in the course of a few days \cite{yazawa2009planarian,gurley2010expression, gurley2008betacatenin}. Thus, the robust re-establishment of a scaled chemical pattern seems to predate the remodelling of the body plan.
Thereby, the expression region of one of the Wnt molecules is reported to show similar dynamics as the source size in our model, see (v). After cutting the worm, the expression of Wnt11.5 is strongly reduced and undershoots its target range before it eventually scales up again to fit the fragment size \cite{gurley2010expression}.

\noindent Knockdown of $\beta$-catenin leads to the formation of multiple heads \cite{adell2009smedevi, gurley2008betacatenin, petersen2008smed, gurley2008betacatenin, elliott2013history}. Interestingly, after a transient drug treatment, the double-headed phenotype persists even after consecutive amputations \cite{lobo2012modeling}. This agrees with our discussed example for which one second order pattern (and even more specifically the $(1,0)$-pattern) seems to be a locally attractive fixed point, see (vi).

\noindent Additionally, very narrow slices of the worm can also sometimes result in a double-headed phenotype \cite{adell2010gradients, lobo2012modeling, elliott2013history,morgan1898experimental}. Again, this compares to our model, for which the morphogen profile transiently flattens after amputation. In the model, the profile becomes the more homogeneous the smaller the fragment. Starting from a homogeneous profile can occasionally result in the emergence of a higher order pattern with a small basin-of-attraction, see (vi).

\noindent Candidates for the expander molecules still have to be identified. However, a first sequencing screen by our experimental collaborators found several genes which are expressed in a size-dependent manner. Some of them are even approximately homogeneously expressed and show a strong (possibly quadratic) dependence on worm length, see (iii).

\noindent For a homogeneous effect of the expander on the morphogen, the expander might be a rather small and fast diffusion molecule. Such small molecules could for example quickly travel between cells via gap junctions. Interesting, it has been reported that the blocking of gap junctions in flatworms leads to body patterns with a larger wavelength \cite{oviedo2010longrange}, see (iv).

\noindent In summary, we could identify several signatures of our simple model, which help us to lay out the route for further experiments such as targeted gene knockouts. Thereby, it is important not to confuse patterns with processes. The same steady state pattern can arise from very different underlying mechanisms. This is particularly true for the case of Turing models \cite{oster1988lateral}. Already for a simple three component model as we discussed in this chapter, the knockout of activator or inhibitor might easily provoke misleading interpretations of the data, see Appendix~\ref{appreactdiff:knockout}. Considering not an endpoint assay but a time course of the system after amputations or gene knockouts might to some extent circumvent this problem. The dynamics typically reveal additional, valuable details that might enable us to distinguish between different mechanisms.\\

\section{Summary and discussion}
In this chapter, we introduced a new class of pattern forming mechanisms, which gene\-rate patterns that perfectly scale with system size in a completely self-organized manner.
The patterning scheme relies on expander molecules that dynamically adjust the degradation rates of morphogens in a Turing system.
Thereby, the expander controls the pattern length scales and the source size of the resulting Turing patterns.
Conversely, the expander concentration is dynamic itself and is regulated by the concentrations of the Turing morphogens.
For the feedback introduced here, the relative source size at steady state is always independent of the system size, see Eq.~\ref{Eqrelsourceconst}.
Furthermore, we showed that a head-tail polarity pattern with a single source region 
scales as a function of the system size, is stable with respect to perturbations, 
and regenerates in amputation fragments.

\noindent Previous works have already discussed scaling of morphogen profiles and patterns in pre-patterned systems. Either the morphogen source was specified by other pre-existing cues \cite{benzvi2010scaling,benzvi2011expansion} or the morphogen profile was self-organized according to a Turing mecha\-nism but scaling relied on a predefined geometry \cite{hunding1988size, ishihara2006turing, othmer1980scale, pate1984applications}. Such frameworks are challenged by the regeneration capabilities of flatworms. In addition, we will show in Chapter~\ref{growth} that the length-width ratio of the worms does not stay constant during growth, which is also not compatible with mechanisms relying on a fixed geometry. Now, our self-organized model for pattern scaling is able to read out system size in the absence of any pre-patterns and without strong dependence on specific geometrical features.

\noindent Certainly, body plan patterns in flatworms do not emerge in the total absence of potential pre-patterning cues. For example, it has been shown that the contact of dorsal and ventral tissue at the wound site causes a regeneration response \cite{kato1999role}. Still, our self-organized patterning system captures main characteristics of body plan scaling and regeneration in flatworms and allows to discuss generic concepts. Thereby, it provides the basis for further investigations and more sophisticated modeling approaches.

\noindent In the minimal theory formulated here, we neglected spontaneous expander degradation.  Such spontaneous degradation would cap the expander concentration and set a lower size limit for pattern scaling.
Additionally, in size-monitoring systems as considered here, a key challenge relates to the simple fact that these obviously require long-range communication across the scale of the system. This implies a trade-off between an upper size limit for scaling, and the time-scale of pattern formation. 
Here, this time scale is set by morphogen diffusion and system size. 
For example, assuming a maximum diffusion coefficient of $100\, \mu$m$^2$/s and a maximum organism size of $20$ mm, relevant for the flatworms considered, we infer a patterning time scale of $3{-}30$ days, roughly consistent with the experimental range of $1{-}2$ weeks 
for the restoration of body plan proportions after amputation \cite{newmark2002not,adell2010gradients}. Note that active, non-directional transport in addition to passive diffusion could accelerate morphogen dispersal and thus allow for faster pattern formation \cite{gregor2005diffusion}. 

\noindent In fact, it has been suggested for flatworms that fast long-range communication might be implemented via gap junctions or the central nervous system \cite{peiris2013gap, almuedocastillo2012wnt,oviedo2010longrange}. Besides the transmission of actual nervous signals, a molecular transport along the nerve cords was proposed \cite{Rink2009planarian,almuedocastillo2012wnt, yazawa2009planarian}. Moreover, a gradient of membrane voltage has been discussed to mediate long range patterning and scaling of body parts \cite{beane2012chemical, beane2013bioelectric, levin2012molecular}. Yet, even when considering faster transport processes than normal diffusion, eventually we can expect a physical limit to scaling, which would be revealed by the emergence of higher order body plan patterns.
Interestingly, it has been reported that a second head forms in very large animals of some species \cite{jenkins1963bipolar, egger2006freeliving}.

\noindent So far, we were mainly concerned with the Wnt/$\beta$-catenin system and the gradient formation along the anterior-posterior (AP) axis. Yet, there are also two additional axes, which are specified by other signaling molecules
 \cite{niehrs2010growth}. For example, the Admp/Bmp system appears to be responsible for the establishment of a dorsal and a ventral side  \cite{gavino2011bmp}. In flatworms, these molecules seem to cross-react analogous to a Turing system. Previously, they have also been proposed to be a Turing pair in the frog \textit{Xenopus laevis} \cite{meinhardt2009models} and a shuttling mechanism has been discussed as a means to adjust the Bmp gradient to the size of the frog embryo \cite{benzvi2008scaling}.

\noindent The theory developed in this chapter provides a framework to understand robust axis formation and scalable body plan patterning.
In the future, it will be important to test the generic concepts presented here
in regeneration experiments and to quantify spatial profiles of signaling molecules and genetic activity. With this, we aim to identify the key modules leading to scaling and regeneration.

% ---------------------------------------------------------------------------
% ----------------------- end of thesis sub-document ------------------------
% ---------------------------------------------------------------------------						
% this file is called up by thesis.tex
% content in this file will be fed into the main document

\chapter{Flatworm shape dynamics and motility} \label{pca}% top level followed by section, subsection

% the code below specifies where the figures are stored
%\ifpdf
%    \graphicspath{{4_pca/figures/PNG/}{4_pca/figures/PDF/}{4_pca/figures/}}
%\else
%    \graphicspath{{4_pca/figures/EPS/}{4_pca/figures/}}
%\fi

\renewcommand{\r}{\mathbf{r}}
\newcommand{\C}{\mathrm{C}}
\renewcommand{\v}{\mathbf{v}}

% ----------------------- contents from here ------------------------

\section{Modes, movement and morphology}
While the previous chapters were concerned with the patterning of a specific body layout, this chapter aims to analyze motility-induced and inter-species variations of the body shapes as a first step to relate form and function. Ultimately, we would like to understand why worms look and behave the way they do. Most of the  results of this chapter have been published in Werner \textit{et al.} \cite{werner2014shape}.

\noindent  Evolutionary adaption to a specific environment manifests itself in the emergence of characteristic body morphologies, which among others allow the organism to effi\-ciently sense its environment and steer its path. Thus, motility phenotypes reveal abnormal body plan patterns. Flatworms usually display a smooth gliding motility,
\sidenote{0.45\textwidth}{``Whether it be the sweeping eagle in his flight, or the open apple-blossom, the toiling work-horse, the blithe swan, the branching oak, the win\-ding stream at its base, the drifting clouds, over all the coursing sun, form ever follows function, and this is the law.'' \;---\; Louis H. Sullivan, The Tall Office Building Artistically Considered, 1896 \cite{sullivan1896tall}}
resulting from a coordinated beating of the cilia in their densely ciliated ventral epithelium \cite{Rink2009planarian, rompolas2010outer, rompolas2009schmidtea, almuedocastillo2011deshevelled}. For the cilia to act in synchrony, they have to be oriented in parallel to each other, which is coordinated by a structural polarity of the epi\-thelial cells in the plane of the epithelial tissue. This planar cell polarity system (PCP) comprises several molecules that asymmetrically localize to the anterior or posterior side of each cell, respectively, and bind through the membrane to the corresponding counterpart in the neighbouring cells \cite{vladar2009planar,seifert2007frizzled,elliott2013history}. It has been described that the PCP system, which is based on local cell-cell interactions, couples to global polarity cues provided by organism-scale / tissue-scale concentration gradients of signaling molecules as discussed in the previous chapter \cite{vladar2009planar,seifert2007frizzled,elliott2013history}. Together with our experimental collaborators, we aim to explore this coupling by quantifying the gliding speed after knocking out genes of the Wnt/$\beta$-catenin pathway as a long range patterning system. 

\noindent In a second approach, we analyze and characterize the movement patterns of flatworms in detail. How do the worms steer their path? We find that a characteristic bending posture correlates with active turning during gliding, presumably caused by unilateral muscle contractions. Thus, during gliding motility, speed is generated by beating cilia whereas steering is controlled by muscles. 
Occasionally, flatworms also show a second type of motility behavior, called ``inch-worming'', sometimes also referred to as ``peristaltic movement'' \cite{Rink2009planarian, rompolas2010outer,rompolas2009schmidtea,almuedocastillo2011deshevelled}. Switching to this motility mode is associated with escape responses and impaired cilia functionality. We  characterize inch-worming by stereotypic width changes of these worms. Our method reveals regular lateral contraction waves with a period of about $4\,\mathrm{s}$ in inch-worming worms. Since then, body contractions have been further analyzed and a recent publication reports on distinct differences between ``scrunching'' as a quick escape response and ``peristalsis'' as a more persistent gait if cilia functionality is impaired \cite{cochetescartin2015scrunching}. We will compare these results to our previous findings.

\noindent Finally, we employ shape analysis methods to characterize the morphologies of various flatworm species, which come at very different shapes and sizes. The taxonomic identification is usually challenging, relying largely on the time-consuming mapping of internal characters. Here, we illustrate how to systematically and quantitatively distinguish diffe\-rent flatworm species by head shape. The head is one of the most characteristic external hallmarks of the flatworm body plan. Many sensory organs for various stimuli (e.g. light, temperature, touch, chemical signals) are located in the head \cite{inoue2004morphological,barnard2004animal}. %Barnard,p.102 
This includes two eyespots and the auricles with a high density of nerve cells. Yet, also other, less prominent shape characteristics might be crucial for the worm to survive, for example by allowing for more efficient movement and navigation. The systematic characterization and quantitative comparison of body plan morphologies provides the basis for relating worm shape to fitness of a species in its specific environmental niche \cite{kavanagh2013developmental,shoval2012evolutionary}.

\noindent We analyzed the motility patterns of \smed{} and head shapes of different flatworm species using shape mode analysis based on principal component analysis (PCA) \cite{Pearson:1901,Jackson2005,Jolliffe:2005}. The technique has been previously applied to biological data to e.g.~analyze the dynamics of human arm postures, crawling {\it C. elegans} worms or swimming sperm cells \cite{Sanger:2000, stephens2011emergence, stephens2010modes, Ma:2014,Gallagher:2013,Stephens:2008}. In contrast to those examples, which are well described by their center line, the flatworm body is constantly deforming during movement depending on the contraction of their muscular plexus due to the absence of skeletal elements or a rigid body wall. Thus, we adapted PCA to analyze and quantify motility patterns of a closed non-convex boundary outline.

\noindent The starting point of PCA is typically a large data set comprising $N_a$ measurements of $N_b$ features. An example would be the curvature at $N_b$ discrete positions along a winding worm at $N_a$ time points. In biological data sets the individual features are often correlated, e.g.~the curvature at adjacent positions is similar due to a finite stiffness of the body. Therefore, the data might be well described by a much smaller set of characteristic shape modes. For example a sine wave might be a good first approximation to the wiggling of a worm. The most descriptive shape modes are found in a systematic way by determining the largest variations in the data set. At best, a small number of empirical modes characterizes most of the data and can be interpreted in a meaningful way. As a side-effect, this method reduces measurement noise by averaging over several, partially redundant features. PCA can also be seen as a rotation of the phase-space coordinate system such that it is more suitable for the data at hand. Afterwards, the data can be projected  with only negligible information loss onto a subset of the phase-space spanned by a small number of axes. These axes correspond to the shape modes.

\section{Worm motility reports on patterning defects upon gene knockout}
\begin{figure}[tbp]
  \centering
  \includegraphics[width=0.85\textwidth]{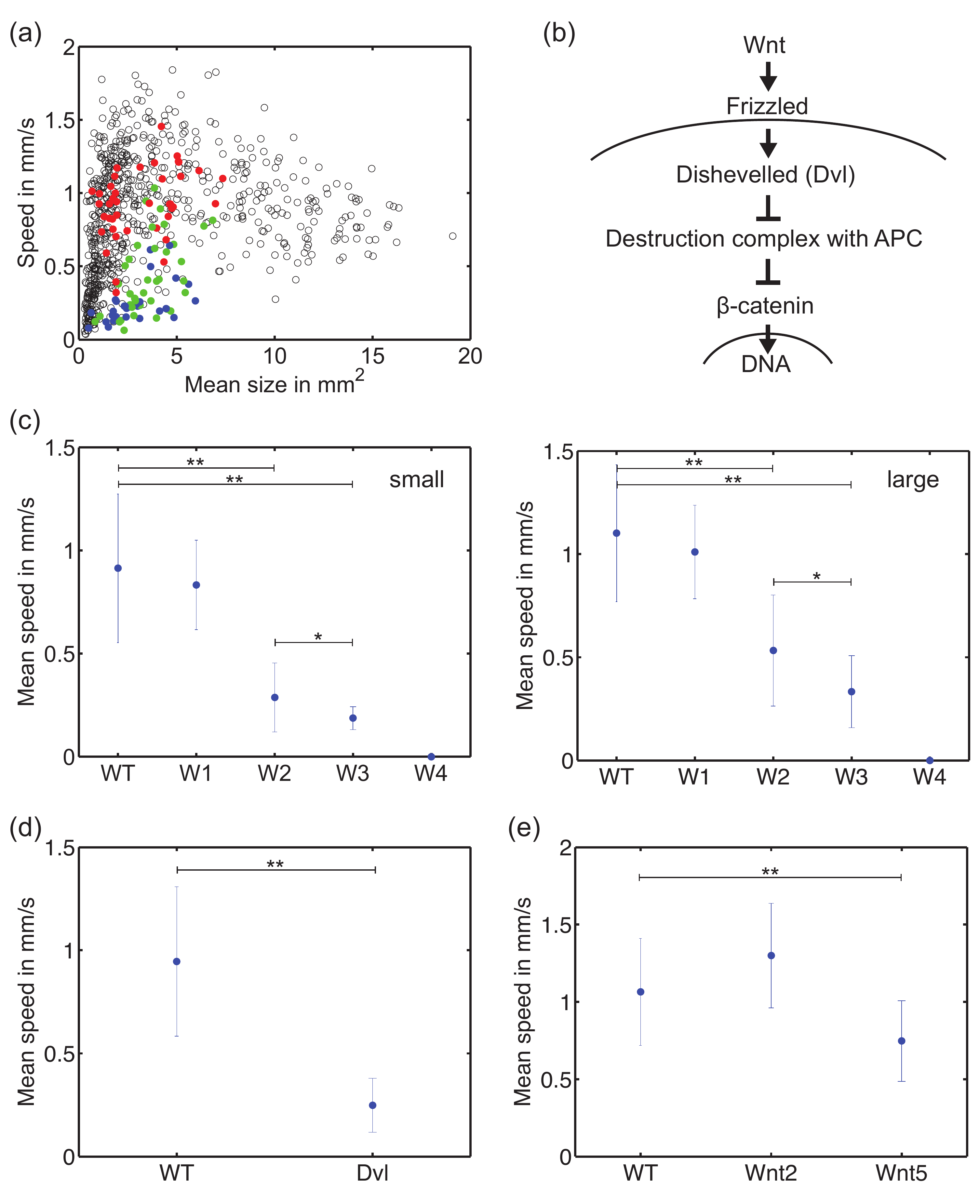}
  \caption[Flatworm speed and phenotypes]
   {(a) Speed of wild type worms (black circles) and worm speed after APC RNAi feeding for one week (red dots),  two weeks (green dots) and three weeks (blue dots), respectively (705 wild type worms, 41/38/39 RNAi worms). (b) Canonical Wnt pathway: binding of Wnt ligands leads to $\beta$-catenin translocation to the nucleus (via Dvl activation and subsequent APC inhibition), adapted with permission from \cite{gurley2008betacatenin}. (c) Quantification of the APC RNAi experiment in panel (a) for small worms (1-3 mm$^2$) and large worms (3-7 mm$^2$). At week 4 worms still show body contractions but no net displacement anymore. (small: 271 wild type worms, 20/15/14 RNAi worms; large: 130 wild type worms, 19/22/12 RNAi worms) (d) Speed measurements after 3 weeks of Dvl RNAi feeding (322 wild type worms, 12 RNAi worms of size 1-4 mm$^2$). (e) Speed measurements after 3 weeks of RNAi of the non-canonical Wnt2 and Wnt5 (154 wild type worms, 17 Wnt2 RNAi worms and 25 Wnt5 RNAi worms of size 2.2-5 mm$^2$). (Imaging of wild type worms by Nicole Alt under the supervision of the author. RNAi experiments by Sarah Mansour. All analyses performed by the author. Asterisks denote the significance level of the \mbox{p-value} test: * $5\%$ and ** $1\%$. Error bars correspond to standard deviations.)} 
   \label{fig:SpeedPheno}
\end{figure}

We analyze movies of gliding worms as described in Appendix~\ref{appmeasurement} and extract the center line of the worm body. The midpoint of the center line is tracked to determine the median worm speed.
We observe that wild type worms actively regulate their gliding speed up to a maximum of $2$ mm/s with a weak dependence on size, see black circles in Fig.~\ref{fig:SpeedPheno}(a). The range in speed is in quantitative agreement with earlier works in \smed{} and \textit{Dugesia tigrina} \cite{talbot2011quantitative,pavlova2000ciliary}. Next, we compare this result to the speed of modified worms, for which pathway components of the Wnt/$\beta$-catenin patterning system have been silenced by RNA interference (RNAi).
 Imaging of wild type worms has been performed by Nicole Alt under the supervision of the author. The RNAi experiments have been performed by Sarah Mansour in the group of Jochen Rink (two RNAi feedings per week). All analysis has been done by the author.
 
 \noindent The canonical Wnt/$\beta$-catenin signaling cascade is illustrated in Fig.~\ref{fig:SpeedPheno}(b) \cite{gurley2008betacatenin, komiya2008wnt}. In the absence of canonical Wnts, $\beta$-catenin is degraded by a destruction complex, which includes APC as a key component. If Wnt ligands bind to the Frizzled receptors, the destruction complex is inhibited via Dishevelled (Dvl). Thus, $\beta$-catenin can accumulate in the cell and reach the nucleus to control transcription. Canonical Wnt/$\beta$-catenin signaling has been associated with tail formation \cite{gurley2008betacatenin}. Knockout of Dvl by RNAi feedings leads to the emergence of multiple heads, while knockout of APC leads to the formation of a second tail after head amputation.
 
 \noindent In intact worms, we observe a motility phenotype for APC RNAi, see colored points in Fig.~\ref{fig:SpeedPheno}(a). Worm speed is reduced over the course of the experiment. We can quantify this trend for two size classes and see a significant effect already after two weeks of RNAi feeding, see Fig.~\ref{fig:SpeedPheno}(c). It remains to determine whether APC acts directly on the PCP pathway or whether a graded $\beta$-catenin signal is required to maintain planar cell polarity.
 
\noindent Similarly, when knocking down both Dvl genes of \smed{}, we observe a reduced speed suggesting a disoriented cilia carpet, see Fig.~\ref{fig:SpeedPheno}(d). In fact, it is known that Dvl is not only a component of the canonical Wnt signaling cascade, leading to a translocation of $\beta$-catenin to the nucleus, Dvl also acts in non-canonical Wnt signaling, where it activates the PCP pathway \cite{tao2010dishevelled,komiya2008wnt,almuedocastillo2011deshevelled}.  It has been shown that silencing Dvl leads to a less, shorter and more disorganized cilia \cite{almuedocastillo2012wnt,almuedocastillo2011deshevelled}.

\noindent In order to determine to what extent non-canonical Wnt signaling is related to PCP in flatworms, Sarah Mansour performed RNAi of Wnt2 and Wnt5. Unlike the canonical tail Wnts, Wnt2 is expressed in the head and Wnt5 is expressed around the body margin and especially along the central nervous system \cite{gurley2010expression,almuedocastillo2012wnt}. Especially Wnt5 has been discussed to act in non-canonical signaling in flatworms and is also involved in PCP systems of other organisms \cite{almuedocastillo2012wnt,gao2011wnt}. Fig.~\ref{fig:SpeedPheno}(e) shows no movement phenotype for Wnt2 RNAi, but Wnt5 RNAi results in a significant reduction in speed. Interestingly, the Wnt5 phenotype is rather subtle. While for APC and Dvl RNAi, the worms seem to loose the ability to glide, switch to inch-worming motility and eventually cease being motile altogether, the Wnt5 RNAi worms are still capable of normal gliding motion, just slower. It is an interesting question, whether this is due to an incomplete knockdown or the particular effect of Wnt5 on the PCP system. 

\noindent In summary, analyzing worm movement upon RNAi is a simple and non-lethal \mbox{approach} to identify patterning phenotypes. The method reveals even subtle effects as for Wnt5 and inspires further more elaborate measurements of cilia orientation and gene expression profiles.

\newpage
 
\section{Shape mode analysis of 2D worm outlines}
Next, we analyze the behavioral repertoire of flatworms in greater detail. Different movement strategies are reflected in bending of the worm body.   
Due to the large body plasticity, motility-induced changes in body posture are not well described by only the center line but require an analysis of the perimeter dynamics. Thus, we face the challenge of characterizing the shape of closed, planar curves. 
For non-convex shapes, this can be non-trivial.

\noindent We extract the midline and the perimeter of the worms from movie sequences as described in Appendix~\ref{appmeasurement}, see Fig.~\ref{fig_smed}(a). The imaging has been performed by Nicole Alt under the supervision of the author.
The worm shape corresponds to a closed curve described by a position vector $\r(s)$ as a function of 
arc-length $s$ along its circumference, see Fig.~\ref{fig_smed}(b). 
We use the tip of the worm tail as a distinguished reference point $\r_1$ 
that specifies the position of $s=0$. 
We further specify a center point $\r_0$, using the midpoint of the tracked 
center line of the worms. 
The profile of radial distances $\rho(s)=|\r(s)-\r_0|$ measured with 
respect to the center point $\r_0$ characterizes outline shape, even for non-convex outlines. 
 Shapes of convex curves might also be characterized by a profile of radial 
distances $\rho(\varphi)$ as a function of a polar angle $\varphi$. 
However, this definition does not generalize to non-convex curves 
(or, more precisely, to curves that are not radially convex with respect to $\r_0$). 
To adjust for different worm sizes, we normalize the radial distance profiles 
by the mean radius 
$\overline{\rho}=\langle \rho(s)\rangle$ as 
$\hat{\rho}=\rho(s)/\overline{\rho}$ and plot it as a function of normalized 
arc-length $\hat{s}=s/L$, where $L$ is the total length of the circumference.

\begin{figure}
\centering
\includegraphics[width=1\textwidth]{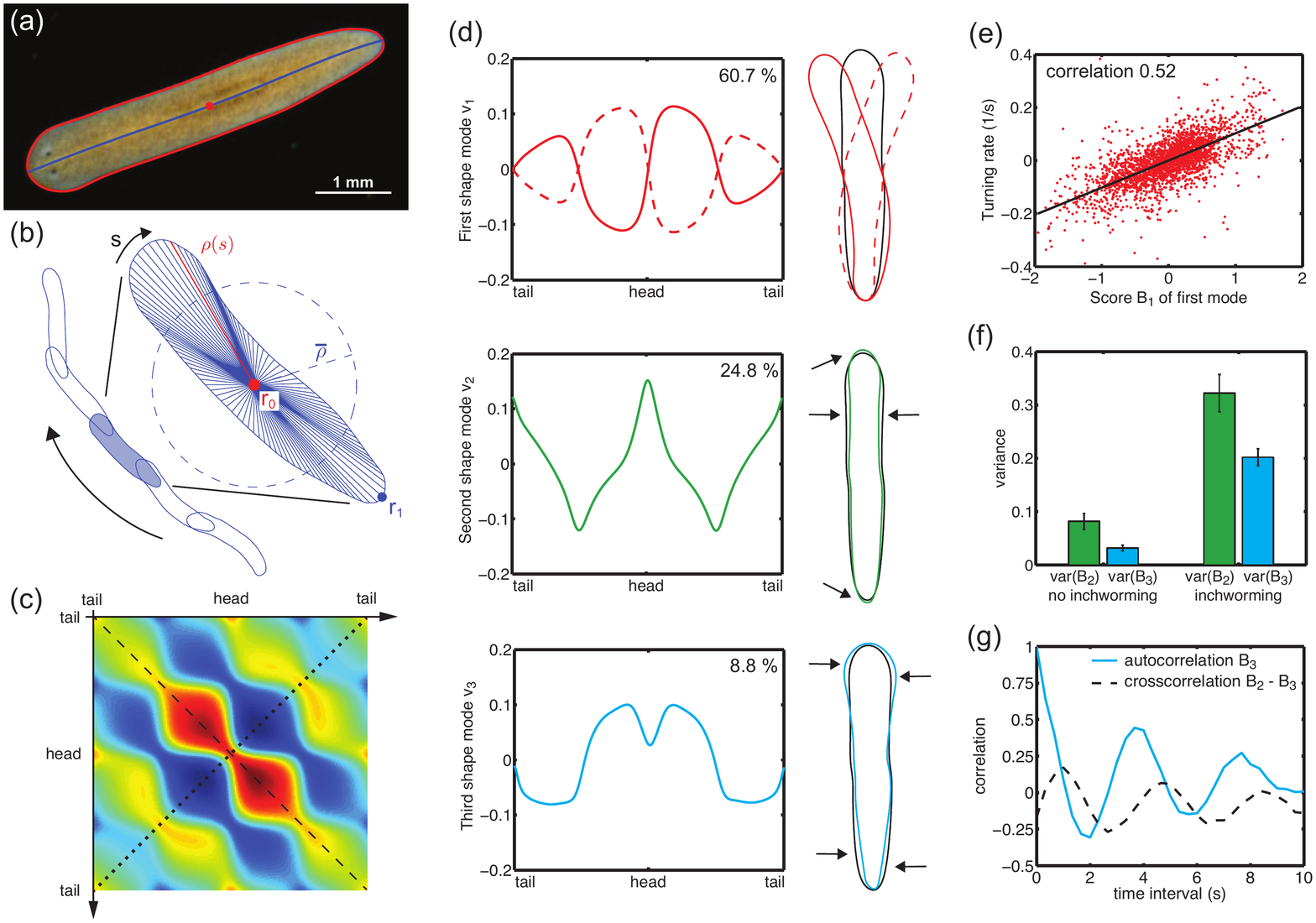}
\caption[Three shape modes characterize body shape dynamics]{
Three shape modes characterize projected flatworm body shape dyna\-mics.
(a)
Our custom-made MATLAB software tracks worms in movies and extracts worm 
boundary outline (red) and centerline (blue). 
(b)
The radial distance $\rho(s)$ between the boundary points and midpoint of the 
centerline ($\r_0$, red dot) is calculated 
as a parameterization of worm shape.
We normalize the radial distance profile of each worm by the mean radius $\overline{\rho}$.
(c) Covariance matrix of the radial distance profiles: the second symmetry axis (dotted line) corresponds to statistically symmetric behavior of the worm with respect to its midline (745 worm movies, imaging by Nicole Alt under the supervision of the author, all analysis performed by the author).
(d)
The three shape modes with the largest eigenvalues account for 94\% of the 
shape variations. 
The first shape mode characterizes bending of the worm and alone accounts for 
61\% of the observed shape variance. 
We show its normalized radial profile on the left as well as the 
boundary outline corresponding to the superposition of the mean worm shape 
and this first shape mode (solid red: $B_1=1$, dashed red: $B_1=-1$, black: 
mean shape with $B_1=0$). The second shape mode describes lateral thinning ($B_2=0.3$), 
while the third shape mode corresponds to unlike deformations of head and tail ($B_3=0.8$), 
giving the worm a wedge-shaped appearance. 
(e)
The first shape mode with score $B_1$, describing worm bending, strongly 
correlates with the instantaneous turning rate of worm midpoint trajectories. 
(f) We manually selected 30 movies where worms clearly show inch-worming and 50 movies with no inch-worming behavior. The variance of score $B_2$ and $B_3$ increases for the inch-worming worms. 
(g) The autocorrelation of mode $B_3$ and the crosscorrelation between mode $B_2$ and mode $B_3$ reveals an inch-worming frequency of approximately $1/4\,\mathrm{Hz}$, hinting at generic behavioral patterns. 
}
\label{fig_smed}
\end{figure}

\noindent As a mathematical side-note, we remark that using the signed curvature 
$\kappa(s)=(\partial_s^2\r(s))\cdot(\partial_s\r(s))$ along the circumference, instead 
of the radial distance profile $\rho(s)$, would amount to a significant 
disadvantage: The property that a certain curvature profile 
corresponds to a closed curve imposes a non-trivial constraint on the set of 
admissible curvature profiles. For the normalized radial distance profiles, 
however, there is a continuous range of distance profiles that correspond to 
closed curves, making this choice of definition more suitable for applying 
linear decomposition techniques such as shape mode analysis. In fact, given a 
particular normalized radial distance profile, the corresponding 
circumference length $L/\overline{\rho}$ is reconstructed self-consistently 
by the requirement that the associated curve must close on itself, see Appendix~\ref{apppca:reconstruction}.\\

\section{Shape dynamics during crawling and inchworming}
\subsection{A bending mode and two width-changing modes}
We extracted $N_w=29\;993$ worm outlines from a total of 745 analyzed movies.
We computed normalized radial distance profiles as described above, each 
profile being represented by $N_r=200$ radii, resulting in a large $N_w\times N_r$ data matrix $\mathcal{R}_{i,j}=\hat{\rho}_i(s_j)$.
From the average of all radial profiles, 
we define a mean worm shape that averages out shape variations $\hat{\rho}_0(s)=\sum_{i=1}^{N_w} \hat{\rho}_i(s)/N_w$, 
see Fig.~\ref{fig_smed}(d) (right inset, black). From this, we can devise a $N_w\times N_r$ matrix $\mathcal{R}^0_{i,j}=[\hat{\rho}_0; ... ; \hat{\rho}_0]$, for which all the rows are equal to the mean profile.
Next, we computed the $N_r\times N_r$ covariance matrix
\begin{equation}
\mathcal{C}=(\mathcal{R}-\mathcal{R}^0)^T\,(\mathcal{R}-\mathcal{R}^0)
\end{equation}
between the individual radial profiles, 
using the centered (mean-corrected) data matrix,
see Fig.~\ref{fig_smed}(c). The covariance matrix is by construction symmetric along the dashed line.
The approximate symmetry of the covariance matrix along the dotted diagonal 
shows that shape variations are statistically symmetric with respect to the worm midline. For example the worm bends as often to the left as to the right.

\noindent The $N_r$ eigenvectors $\v_j(s)$ or shape modes of this covariance matrix correspond to axes of a new coordinate system. In this coordinate system, the variation of the data along each axis is linearly uncorrelated and the corresponding variance is given by the respective eigenvalue.
The shape modes with the largest eigenvalues are those with maximal descriptive power. 
We can uniquely express the worm shape as
\begin{equation}
\hat{\rho}(s)=\sum_{j=1}^{N_r}B_{j}\v_j(s)\,, 
\end{equation}
where the shape scores $B_{j}$ can be computed by a linear least-square fit. Due to correlations, many data sets can be well described by a truncated sum, using only a small number of shape modes with the largest eigenvalues. 

\noindent Fig.~\ref{fig_smed}(d) shows the first three shape modes, 
which together account for $94\%$ of the observed variation in shape.
We find that the dominant shape mode $\v_1$ is anti-symmetric,
describing an overall bending of the worm. 
In contrast, the second and third mode describe symmetric width changes of the worm:
The second shape mode $\v_2$ characterizes a lateral thinning 
of the worm associated with a pointy head and tail. 
Correspondingly, a negative contribution of the second shape mode with $B_2<0$ describes lateral 
thicke\-ning of the worm (with a slightly more roundish head and tail).
The third shape mode $\v_3$, finally, is associated
with unlike deformations of head and tail, giving the worm a wedge-like appearance. 
Superpositions of these three shape modes describe in-plane bending of the worms, 
and a complex width dynamics of head and tail.

\noindent We find that shape changes control the direction of gliding motility and thus steer the worm's path:
Fig.~\ref{fig_smed}(e) displays a significant correlation between the rate of turning along the worm trajectory 
and the first shape score $B_1$, which characterizes bending of the worm.
The sign and magnitude of this ``bending score'' directly relates  to the direction and rate of turning. 
For simplicity,
we restricted the analysis to a medium size range of $8$-$10\,\mathrm{mm}$ length, 
analogous results are found for other size classes.

\subsection{The second and third modes characterize inch-worming}
In addition to cilia-driven gliding motility, 
flatworms employ a second, cilia-independent motility pattern known as inch-worming, 
which provides a back-up motility system in case of dysfunctional cilia \cite{Rink2009planarian, rompolas2010outer,rompolas2009schmidtea,almuedocastillo2011deshevelled}. 
We test whether modes two and three might relate to this second motility pattern. For this, 
we analyzed movies of small worms known to engage more frequently in this kind of behaviour.
We manually classified $80$ movies
of worms smaller than $0.9$ mm that had been starved for 10 weeks, 
yielding a number $30$ inch-worming and $50$ non-inch-worming worms
for a differentiated motility analysis (cases of ambiguity were not included).
We find that the second and third shape mode, 
which characterize dynamic variations in body width,
are indeed more pronounced in inch-worming worms, see Fig.~\ref{fig_smed}(f).
Next, we computed the temporal autocorrelation of time series of the
second shape mode $B_3$, see Fig.~\ref{fig_smed}(g) (solid blue).
We observe stereotypical shape oscillations with a characteristic frequency
of 0.26 Hz. 
From the cross-correlation between $B_3$ and $B_2$ in Fig.~\ref{fig_smed}G (dashed black), 
we find that both shape scores oscillate with a common frequency and relative phase lag of $-0.6\,\pi$ (where $B_2$ lags behind). 
Thus, both shape modes act together in an orchestrated manner to facilitate inch-worming,
hinting at coordinated muscle movements and periodic neuronal activity patterns.

\section{Discriminating flatworm species by shape}
After having developed tools to measure shape changes of the same species over time, 
we next explored the utility of shape mode analysis for comparing different species.
The model species \textit{Schmidtea mediterranea} is but one of many hundred flatworm species existing worldwide \cite{liu2013reactivating,egger2006freeliving}. We aim to introduce shape mode analysis as a technique to facilitate taxonomic classification of these species. PCA enables us to extract typi\-cal shape variations between species without ``a priori'' assumptions on characteristic features.

\begin{figure}[t]
\centering
\includegraphics[width=0.9\textwidth]{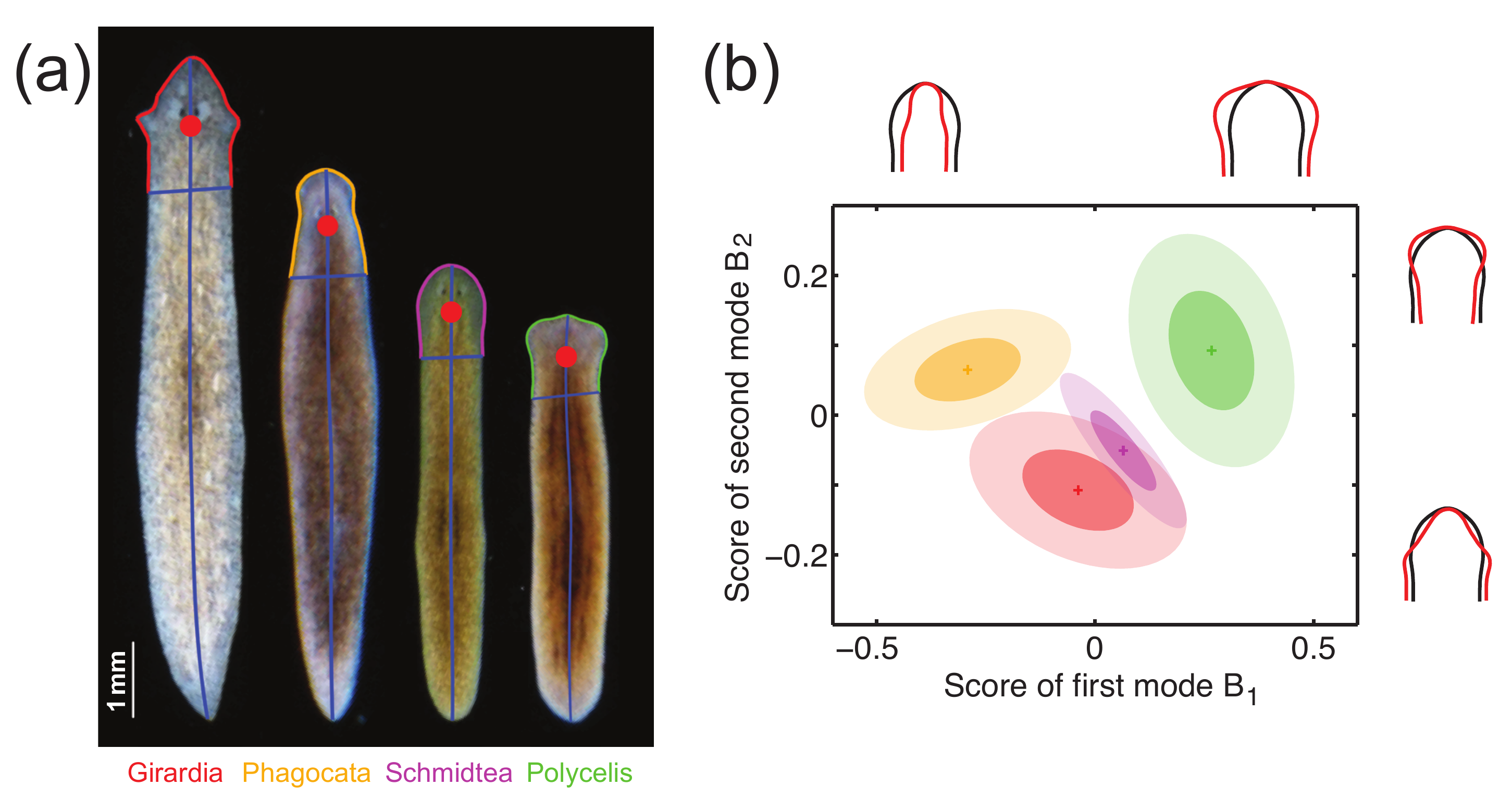}
\caption[Distinguishing head morphologies of flatworm species.]{
(a)
Application of our method to parametrize head morphology of four different flatworm species. For each species, time-lapse sequences of $4$ different worms were recorded as two independent runs of $16$ frames each. The head is defined as the most anterior $20\%$ of the worm body. Radial distances $\rho(s)$ are computed with respect to the midpoint of the head (red dot at $10\%$ of the worm length from the tip of the head).
(b)
By applying PCA to this multi-species data set, we obtain two shape modes, which together account for $88\%$ of the shape variability. Deformations of the mean shape with respect to the the two modes are shown (black: mean shape, red: superposition of mean shape and first mode with $B_1=\pm0.4$ and second mode with $B_2=\pm0.2$, respectively). We represent head morphology of the four species in a combined shape space of these two modes. Average head shapes for each species are indicated by crosses, with ellipses of variance including $68\%$ (dark color) and $95\%$ (light color) of motility-associated shape variability, respectively. (Imaging by Nicole Alt and Miquel Vila-Farr\'e, data analysis by the author)
}
\label{fig_species}
\end{figure}

\noindent Having available a large live collection of flatworm species, we choose four species 
re\-pre\-senting the genera \textit{Girardia}, \textit{Phagocata}, \textit{Schmidtea} and \textit{Polycelis}.
Besides potentially size-dependent variations in aspect ratio, 
the four species differ by their characteristic head shapes, see Fig.~\ref{fig_species}(a). 
Accordingly, we restrict the shape analysis to the head region only 
(defined as the most anterior $20\%$ of the worm body).
In analogy to the procedure described above for the full worm body, we characterize each head shape by a vector of distances from the midpoint of the head (red dot, $10\%$ of the worm length from the tip of the head) to the outline $\rho(s)$ of the head region. Next, PCA is applied to the normalized radial distance profiles like before.

\noindent We found that the first two eigenmodes captured $88\%$ of head shape variability 
within this multi-species data set. 
Fig.~\ref{fig_species}(b) shows species-specific mean shapes for each of the four species in a combined head shape space, as well as ellipses of variance covering $68\%$ (dark color) and $95\%$ (light color) of motility-associated shape variability, respectively. Shape reconstruction from individual modes is explained in Appendix~\ref{apppca:head}. The first mode seems to describe the extent of the auricles while the second mode can be associated with the auricle position.
This comparison of flatworm species repre\-sen\-ting four genera
illustrates linear dimensionality reduction as a simple means to map morphological differences across species.

\section{Discussion}
In this chapter, we have discussed several approaches to relate form and function in flatworms focussing on motility. Movement patterns are generated by various features of the body plan such as cilia and muscle layout. By analyzing worm motility, we gained first insights into their functionality and coordination. In a first approach, we measured worm speed in wild type and RNAi treated animals and could quantify subtle motility phenotypes, which hint at patterning defects of the PCP system.

\noindent In a second approach, we analyzed the shape variations during movement, adapting principal component analysis in order to apply it to 2D outlines. We characterized inchworming 
as a second motility mode, different from cilia-based gliding motility, which is driven by  well-coordinated muscle contractions with a characteristic frequency of about $0.26$ Hz. In a recent work, contractile motility gaits have been further analyzed \cite{cochetescartin2015scrunching}. Cochet-Escartin \textit{et al.}~distinguish between two behavioral traits: (i) scrunching as a transient and fast response to cutting as well as electrical, acidic and temperature shocks, (ii) peristalsis as a more persistent and slow movement strategy if cilia functionality is impaired. The authors could induce the latter in two different ways: (i) by RNAi of the iguana gene leading to defective ciliogenesis and (ii) by increasing the viscosity of the media (by adding $16\%$ ficoll). Interestingly, they extracted frequencies of the area oscillations during pe\-ristal\-sis and obtain $0.28\pm0.03$ Hz (iguana) and $0.26\pm0.02$Hz (ficoll), respectively. These values are very similar to our result. Together, the three conditions illustrate a very generic inchworming/peristalsis behavior despite their higher level of complexity compared to other model organisms such as \textit{C. elegans} \cite{sanchezalvarado2006bridging,wolpert2011principles,robertsgalbraith2015organ,stephens2010modes}. It remains to determine how this relates to the underlying structure of the muscular plexus and the nervous system.

\noindent In a third approach, we apply PCA to head shapes of different flatworm species as a tool for taxonomic classification. We captured the main characteristics of  head shapes of four species belonging to different genera by just two shape modes. The result suggests that head morphogenesis is mainly controlled by two molecular networks: one to determine the head width and a second one for the anterior positioning of the auricles.~The availability of transcriptome sequence data for these species will now provide us with the opportunity to test this hypothesis. We also have planned to extend the analy\-sis to many more species of the collection in the laboratory of Jochen Rink and relate the characteristic body plan features to the conditions in the respective environmental niche. A similar inter-species comparison with respect to evolutionary selected traits has been performed for beaks of Darwin finches, phalanxes of vertebrates, heads of ants and wings of bats \cite{kavanagh2013developmental, shoval2012evolutionary}. Based on these analyses, we expect for flatworms that a small number of shape modes with the largest eigenvalues spans the feature space, in which most of the shape variations take place. Highly specialized species have typically rather extreme morphologies and can be found at the corners of the observed shape set. In contrast, generalists show mixtures of different traits, which simultaneously optimizes the fitness to fulfil several tasks. It will be interesting to test this hypothesis for flatworm species, some of which inhabit extreme environments.

% ---------------------------------------------------------------------------
% ----------------------- end of thesis sub-document ------------------------
% ---------------------------------------------------------------------------	
% this file is called up by thesis.tex
% content in this file will be fed into the main document

\chapter[Quantitative study of flatworm growth and cell turnover]{Quantitative study of\\ flatworm growth and cell turnover} \label{growth}% top level followed by section, subsection
\chaptermark{Quantifying flatworm growth and cell turnover} %only change chapter heading

% the code below specifies where the figures are stored
%\ifpdf
%    \graphicspath{{3_growth/figures/PNG/}{3_growth/figures/PDF/}{3_growth/figures/}}
%\else
%    \graphicspath{{3_growth/figures/EPS/}{3_growth/figures/}}
%\fi

% ----------------------- contents from here ------------------------

\section{Homeostasis is a dynamic steady state}
Multicellular organisms in their adult stages have a relatively constant outward appearance but the integrity of a functional body is only maintained due to a permanent replacement of damaged or lost cells \cite{pellettieri2007cell}. In humans, it has been estimated that the total mass of cells we replace every year is almost as much as our entire body weight \cite{reed1999dysregulation}. 
The time scales of the cellular turnover range from a few days for blood cells and the stomach to several years for cells in the heart and the skeleton \cite{milo2010bionumbers}. Only a small fraction of cells like some nerve cells might never be replaced.%Bionumbers Database (BNID 101940, 107076, 107077, 107078, 109908)

\begin{figure}[ht]
  \centering
  \includegraphics[width=1\textwidth]{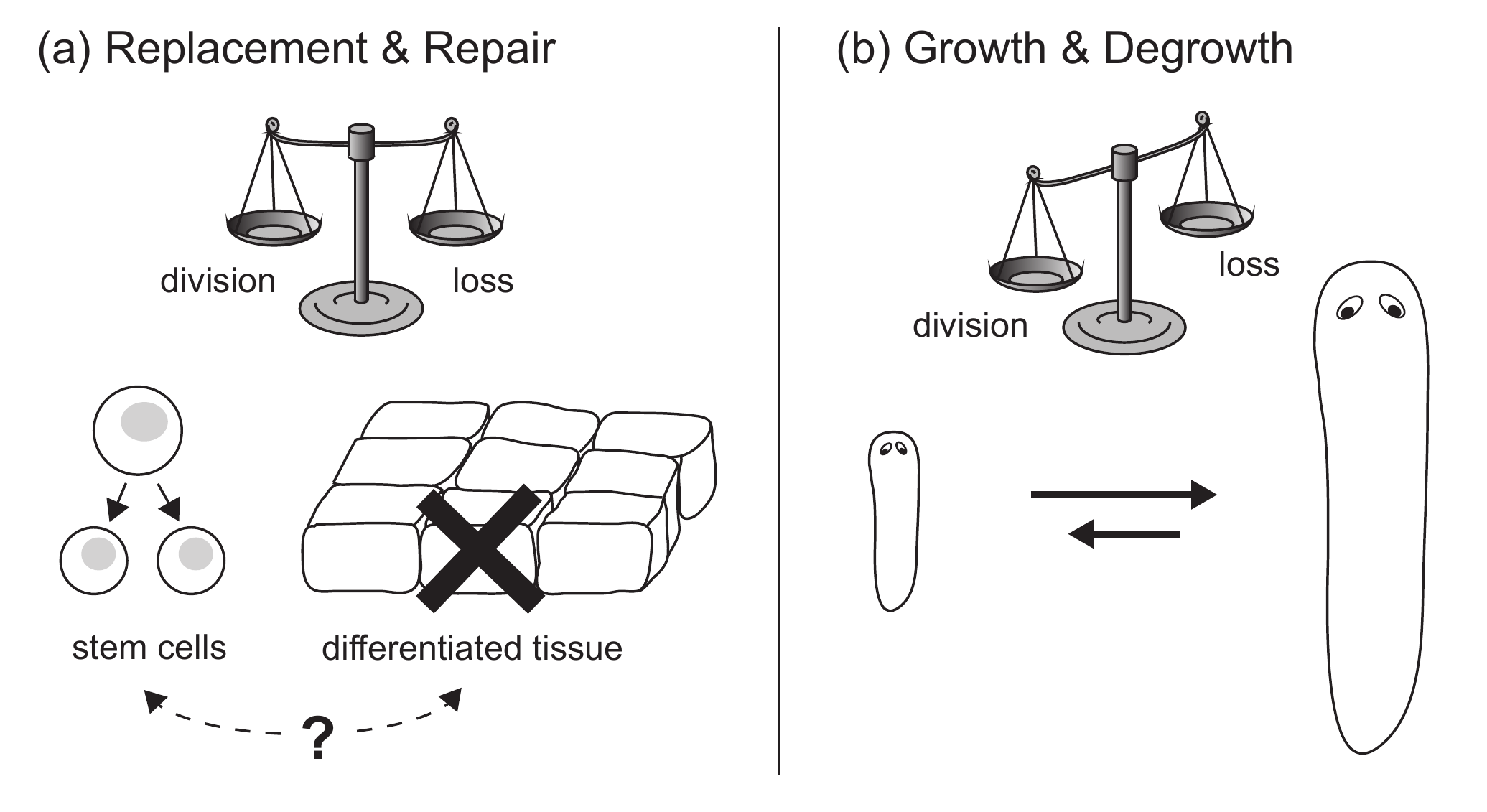}
  \caption[Cell turnover and growth]
   {(a) Stem cell division and the loss of cells have to be balanced to maintain a constant size of the tissue. (b) For growth and degrowth, the balance of these processes is slightly shifted in a well controlled manner.}
   \label{fig:ReplacementGrowth}
\end{figure}

\noindent  From the point of view of dynamical systems theory, a constant outward appearance in the face of permanent turnover can be considered as a steady state, for which inflows (i.e.~the generation of new cells) and outflows (i.e.~the loss of old cells) are balanced.
\sidenote{0.45\textwidth}{``{\selectlanguage{polutonikogreek}potamo~isi to~isin a>uto~isin >emba'inousin <'etera ka`i <'etera <'udata >epirre~i.}''
 %Trema ", Spiritus asper >, Spiritus lenis <, Akut ', Gravis `, circumflex ~
% $\pi o \tau \alpha \mu o \tilde{\iota} \sigma \iota$ $\tau o \tilde{\iota} \sigma \iota \nu$ $\alpha \upsilon \tau o \iota \sigma \iota \nu$ $\epsilon \mu \beta \alpha \iota \nu o \upsilon \sigma \iota \nu$ $\epsilon \tau \epsilon \rho \alpha$ $\kappa \alpha \iota$ $\epsilon \tau \epsilon \rho \alpha$ $\upsilon \delta \alpha \tau \alpha$ $\epsilon \pi \iota \rho \rho \epsilon \tilde{\iota}$.
 (On those stepping into rivers staying the same other and other waters flow.) \;---\; Heraclitus of Ephesus, 500 BC \cite{kirk1954heraclitus}}
The underlying processes of cell division and cell loss have to be well controlled in order to avoid a disintegration of the organism or an uncontrolled overgrowth such as cancer. For this to function robustly, there is likely a communication between the dying cells in the tissue and the dividing stem cells in their niche, see Fig.~\ref{fig:ReplacementGrowth}.  A limited capacity to replace cells is related to sickness and aging of organisms \cite{pellettieri2007cell}.
During growth, the balance between inflows and outflows is shifted to an increased cell division in comparison to cell loss. 
Importantly, growth is not only controlled by a developmental program but also influenced by the availability of food \cite{wolpert2011principles,layalle2008tor}. In many organisms, nutrition levels affect the speed of growth and the final size.

\noindent
In this chapter, we address the link between the microscopic scale of cell turnover and the macroscopic growth in our favourite multicellular organism the flatworm \smed{}.
Previous research has mainly been devoted to cell death and cell division as individual processes but little is known about how these two processes are jointly regulated and mutually influence each other to result in coordinated growth at the organism scale \cite{pellettieri2007cell}.
Flatworms are suitable model organisms to study cell turnover and growth in a comprehensive way as they permanently replace all of their cells within a few weeks and reversibly grow over a 40-fold range in size \cite{rink2013stem, baguna1981quantitative,newmark2002not,oviedo2008planarians, pellettieri2007cell}. Thereby, the cell size appears to be approximately constant and the reversible growth mainly corresponds to a change in cell number \cite{schultz1904reduktionen, newmark2002not, oviedo2003allometric, baguna1981quantitative}. Interestingly, changes in worm size has been previously related to aging and small worms have been described to be more juvenile. In Appendix~\ref{app:flatwormaging}, we discuss signatures of aging in sexual and asexual flatworms.

\begin{figure}[tbp]
  \centering
  \includegraphics[width=1\textwidth]{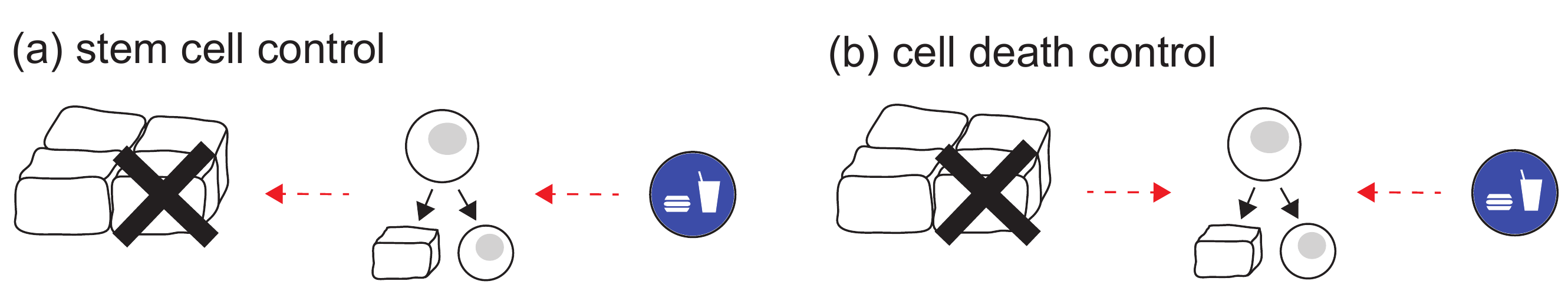}
  \caption[Stem cell control vs.~cell death control]
   {Two paradigms of turnover control: (a) stem cells and their descendants induce the death of differentiated cells, (b) dying cells induce cell division and recruit progenitor cells to the tissue. In both cases feeding enhances cell division.}
   \label{fig:StemDeathControl}
\end{figure}

\noindent Two basic models of growth and turnover control have been discussed for flatworms: (i) stem cell control and (ii) cell death control \cite{pellettieri2010cell}, see Fig.~\ref{fig:StemDeathControl}. In the first scenario, dividing stem cells or their descendants induce cell death in differentiated cells for replacement. In order to generate growth, the replacement must not be perfect. We might hypothesize that cells only respond with a certain probability (maybe depending on their age or fitness) to the death signal from the stem cell pool. In the second scenario, the dying cells send out signals to enhance cell division and recruit progenitor cells to the tissue. In both cases, an additional input from the nutritional status of the worm is needed for growth.

\noindent By studying flatworms as a model system, we aim to bridge the scales between cellular turnover processes and growth and degrowth on the organismal level. First, we quantify the size of the worms as well as growth dynamics and find a non-trivial size-dependence of the growth and degrowth rates. All experiments were performed in the laboratory of Jochen Rink. Several members have been involved in this close collaboration as we will explicitly state for each data set. In a second step, we theoretically discuss three paradigmatic models for how cell division and cell loss might be linked to the metabolic status of the worm. Each model is able to explain the observed behavior within the accuracy of the measurements. We obtain several predictions of the models that can be tested in further experiments. The results directly relate to the questions of how these worms can survive long starvation periods and what determines the limits to growth. Finally, we establish the theoretical framework for future measurements to reveal the control logic of turnover processes.

\section{Size-dependent growth and degrowth dynamics in flatworms}\label{sec:growthdegrowth}
\subsection{Allometric scaling laws}\label{sec:AllScalLaw}
\subsubsection{Measurements of area, cell number and mass of the worm}
\begin{figure}[tbp]
  \centering
\includegraphics[width=0.55\textwidth]{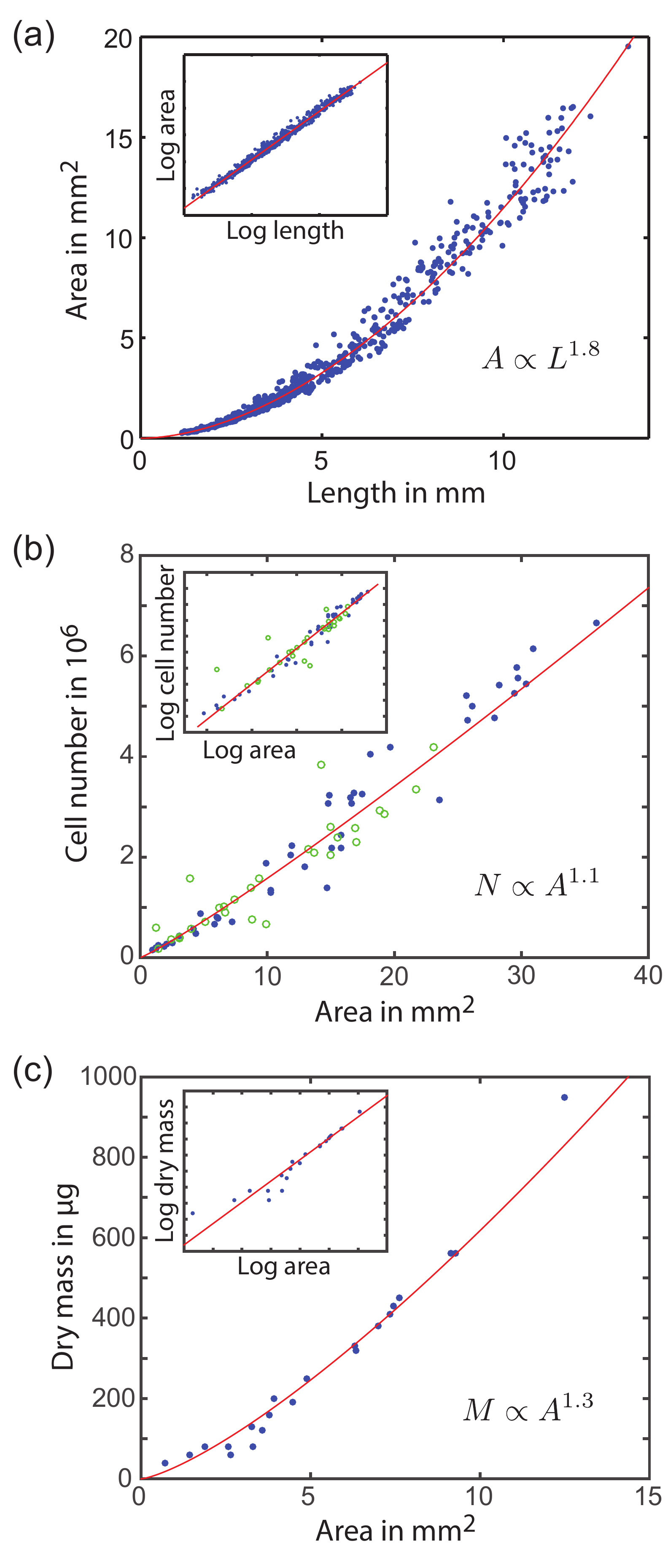}
  \caption[Allometric scaling Laws]
  {Allometric scaling laws: (a) Width and length of the worm do not change proportionately. Large worms are thinner than small worms (Imaging by Nicole Alt under the supervision of the author, analysis by the author, 722 measurements of star\-ving worms). (b) Cell numbers are measured by direct cell counting (green circles) and by histone quantifications (blue dots). The cell number changes almost propor\-tionally with worm area (Imaging and cell number quantifications by Albert Thommen, analysis by the author, 31 and 45 worms, respectively). (c) In contrast, the dry mass increases stronger than area and cell number (Imaging and mass measurements by Albert Thommen, analysis by the author, 21 worms).}
  \label{fig:ScalingLaws}
\end{figure}

The first step towards measuring growth dynamics is to accurately determine worm size. Together with our experimental collaborators we performed experiments to extract the outer dimensions like length and width as well as the cell number and the mass of the worms. From these measurements we obtained allometric scaling laws that relate the various quantities for cross-validation and to gain further information about the body plan of \smed, see Fig.~\ref{fig:ScalingLaws}.

\noindent One important quantity is the area of the worm, which refers to the  2D-projection of the worm body. Area as well as other outer dimensions are readily accessible by microscopic imaging and we have developed a protocol to measure these quantities in a precise and reproducible way, see Appendix~\ref{appmeasurement}. Several student helpers in the group of Jochen Rink have been involved in the imaging of the worms. The author designed the experiments, provided training and supervision and performed the analysis of the data.

\noindent Fig.~\ref{fig:ScalingLaws}(a) reveals an allometric scaling relation between worm area and length with an exponent of $1.81\pm 0.01$. Thus, large worms are relatively thinner than small worms. Remarkably, the scaling law holds across the entire size range with a very small transverse spread and a high reproducibility. These worms were starved for at least a week. The respective scaling law for well fed worms can be found in Appendix~\ref{app:growthallometric}. There we also discuss the details of the data analysis. The scaling exponents have been determined using a robust regression algorithm with bi-squared weights implicitly obtained from the spread of the data.

  \noindent
Ultimately, we are most interested in the changes in cell number because this quantity directly relates to the rates of cell division and cell death. However, measurements of cell numbers are fatal for the worms. As a solution, we establish a functional relationship between cell number and worm area. By using the worm area as a read-out for size, we can reduce the number of worms needed for various experiments and perform time-course experiments with well-defined initial sizes.

\noindent The cell number itself has been measured in two different ways by Albert Thommen in the group of Jochen Rink. First, he determined the amount of histones in the worm. Histones are structural units that organize the DNA in eukaryotic cells \cite{gilbert2014developmental}. The DNA strands are wrapped around the histones in the nuclei for compaction. Histones are only synthesized during cell division and the amount of histones was measured to be constant throughout the lifetime of the cells. After the amount of histones per cell has been determined, the total amount of histones in a worm can be used as a read-out for cell number. For this, the amount of histones was measured via Western blotting in a known fraction of the total protein mass. In a second experiment, cell numbers were obtained by disintegrating the worm into individual cells and automatically counting cells in microscope images of a well defined volume fraction of the worm.

\noindent The resulting cell numbers are plotted in Fig.~\ref{fig:ScalingLaws}(b) as a function of worm size. We find that the number of cells scales almost linearly with the area of the worms with a scaling exponent of $1.11\pm 0.02$. The most simple explanation would be that the height hardly changes and also the large worms remain rather flat. Alternatively, the size of cavities like the gut might increase over-proportionately, compensating for an potential increase in height for larger worms. Furthermore, it might be possible that the fraction of cell types changes, which might also explain a linear scaling despite variations in height. Irrespective of which scenario applies, the scaling exponent shows that our area measurements are a good proxy for cell number.

\noindent Finally, we also determine the dry mass of the worms. Again, the experiment has been carried out by Albert Thommen, while the analysis of the movies has been performed by the author. Although more data is needed for a more reliable result, the analysis of the first data points suggests that the scaling exponent of $1.33\pm 0.06$ for the dry mass is larger than the exponent for the cell number. Therefore, the mass per cell seems to slightly increase with size. Interestingly, some preliminary measurements suggest that in contrast the protein mass per cell is independent of worm size, see Appendix~\ref{app:growthallometric}.

\subsubsection{Comparison and interpretation of scaling laws}
\noindent Previously, contradictory results have been published on scaling laws in \smed{} and other related species, see Tab.~\ref{tab:ScalingLawsLit}. For \textit{Planaria maculata}, the scaling of worm mass with area to the power of $1.38$ agrees well with our result \cite{vonbertalanffy1940unersuchungen}. For \textit{Dugesia lugubris}, Lange measured the volume in serial sections and found a quadratic relationship with length, which is in agreement with our scaling law for the cell number \cite{lange1967quantitative, lange1968possible}. Bagu\~n\`a \textit{et al.} have measured cell number and worm volume in \smed{} \cite{baguna1981quantitative, baguna1976mitosisI}. In agreement with our data, they find similar values for the cell numbers and from them we can obtain a quadratic scaling with the length of the worm. However, a few years later, the same group published various scaling laws for four different species including sexual and asexual flatworms which do not agree with our results.

\begin{table}[t]
\begin{center}
  \begin{tabular}{ l | c | c | c }
    scaling law & species & our result (\smed{}) & reference \\[5pt] \hline
    $M\propto A^{1.38}$ & \textit{Planaria maculata}$^{?}$ & $M\propto A^{1.33}$ & \cite{vonbertalanffy1940unersuchungen} \\ \hline
    $V\propto L^{2.02}$ & \textit{Dugesia lugubris}$^{\dagger}$ & ($N\propto L^{2.01}$) & \cite{lange1967quantitative, lange1968possible} \\ \hline
        $N\propto L^{2}$ & \smed{} & $N\propto L^{2.01}$ & \cite{baguna1981quantitative, baguna1976mitosisI} \\ \hline
        $N\propto L^{1.81}$ &  & $N\propto L^{2.01}$ &  \\
        $N\propto A^{1.44}$ & \smed{} & $N\propto A^{1.11}$ & \cite{romero1991quantitative} \\
        $A\propto L^{1.26}$ &  & $A\propto L^{1.81}$ &  \\ \hline
        $M_{\text{wet?}}\propto L^{0.9}$ & \smed{} & $M\propto L^{2.40}$ & \cite{oviedo2003allometric} \\ \hline
        $M_{\text{wet}}\propto A$ & \textit{Schmidtea polychroa}$^{\dagger}$ & $M\propto A^{1.33}$ & \cite{mouton2011lack} \\ \hline
        $M_{\text{prot}}\propto A$ & \textit{Schmidtea polychroa}$^{\dagger}$  & $M_{\text{prot}}\propto A^{1.13}$ & \cite{mouton2011lack}, Appendix~\ref{app:growthallometric} \\ \hline
        $N\propto L$ & \textit{Dugesia japonica}$^{\dagger}$ & $N\propto L^{2.01}$ & \cite{takeda2009planarians} 
  \end{tabular}
\end{center}
  \caption[Scaling Laws in the Literature]
  {Scaling laws for various flatworm species found in the literature. Our experiments are done in asexual worms, $^{\dagger}$ denotes the sexual strains, $^{?}$ marks a species, which cannot be taxonomically classified.}
  \label{tab:ScalingLawsLit}
\end{table}

\noindent Further, more recent measurements of the worm mass by Oviedo \textit{et al.}~also deviate from our scaling law \cite{oviedo2003allometric}. However, a scaling with $L^{0.9}$ is a rather questionable result considering that the worm width also increases. In the same paper, the authors counted the number of cintillo cells, which are found around the margin of the head and are involved in mechanosensing. If we assume that the head can be approximated by a semicircle and that the number of cintillo cells increases proportionately to the head margin, the width of the worm at the head follows $L^{0.68}$, see Appendix~\ref{app:growthallometric}, which is similar to our results. For \textit{Schmidtea polychroa}, the wet weight and the protein content have been measured \cite{mouton2011lack}. At least  the protein content might show a similar dependence on area as in our experiment. Finally, Takeda \textit{et al.}~measured the total DNA mass in {\it Dugesia japonica} and find an approximately linear dependence on body length \cite{takeda2009planarians}. As the cell number can be assumed to be roughly proportional to the total DNA mass, this is not in agreement with our measurements in Fig.~\ref{fig:ScalingLaws}(b). In analogy to the argument about the worm mass of Oviedo \textit{et al.}, such a linear relationship of cell number with length is unlikely given the significant increase in worm width in larger worms. Note that the spectroscopic quantification of flatworm DNA is rather unreliable because it interferes with the absorption peak of the pigments. Hence, we apply different approaches.

\noindent Thus, while we could confirm several measurements of scaling laws in flatworms obtained before 1980, we find much less agreement with recently published results. This discrepancy with respect to our data might only in parts be explained by species-specific variations and differences between sexual and asexual strains and could relate to the general problem of size measurements in animals with flexible body shapes. In our experiments, we have paid particular attention to accuracy and reproducibility by analyzing movie sequences of individual worms, following a strict protocol. Extensive cross-checks as well as consistency-checks between redundant measurements were performed. 

\noindent Our results on the scaling of the cell number suggest that flatworms indeed stay rather flat. As the worms lack a blood system, this would ensure that all cells can be provided with oxygen by diffusion from the outer epithelium. While this is the most likely yet not the only explanation for the observed scaling relation in Fig.~\ref{fig:ScalingLaws}(b), further measurements of worm heights are necessary to confirm our hypothesis. More data points are also needed for the scaling relation of the dry mass. The data at hand suggests that the characteristics of the cell population depend on size in the sense that in larger worms there are more cells with an increased mass. Note, however, that a first approximate estimation of protein content in the worm does indicate that the protein content per cell roughly stays constant, see Appendix~\ref{app:growthallometric}. Thus, a possible explanation for the increase in cell mass could be the storage of lipids as energy resources in larger worms.

\noindent Having established size measurements, next, we aim to analyze growth dynamics.\\

\subsection{Characterizing the immediate growth response upon feeding}\label{sec:growthresponse}
\begin{figure}[tbp]
  \centering
\includegraphics[width=1\textwidth]{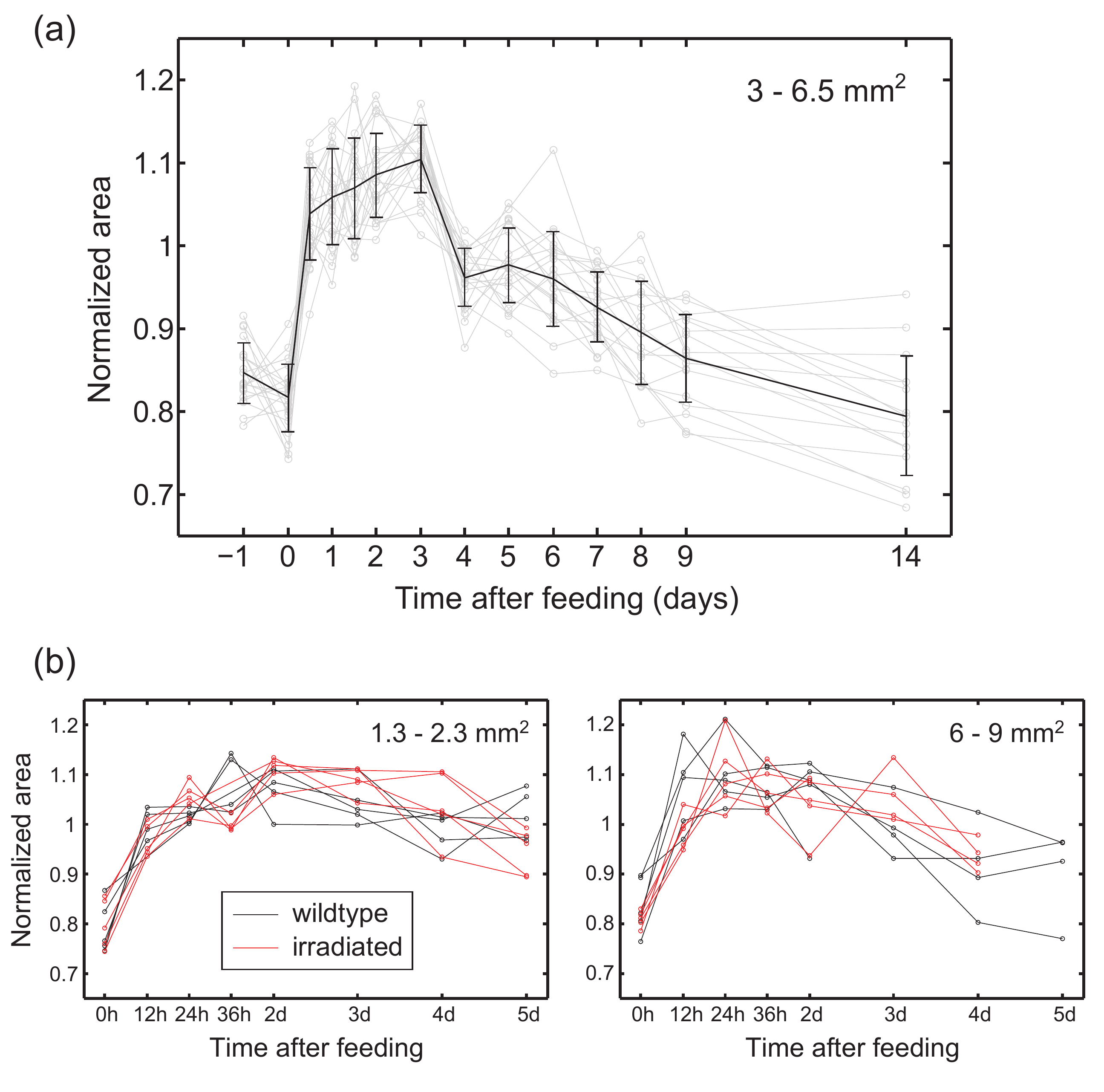}
  \caption[Growth response upon feeding]
  {Growth response upon feeding (after one week of starvation): (a) The worm area increases rapidly within the first day and returns again to its initial value after 2 weeks. The area is normalized by the average of day $0-7$. The individual worm tracks are shown in gray, the black line illustrates the average behavior across all worms (Imaging by Johanna Richter, Jordan Ferria and Nicole Alt under the supervision of the author, analysis by the author, 21 worms). (b) Irradiated worms (red) and non-irradiated worms show the same initial increase in area due to intake of food. The area is normalized by the average of day $0-4$ (Imaging by Ashutosh Mishra under the supervision of the author, analysis by the author, 5 irradiated and 5 non-irradiated worms each).}
  \label{fig:feedingpeak}
\end{figure}

Worms respond to feeding with a fast mitotic peak within the first 12 hours that lasts for $3-5$ days \cite{gonzalezestevez2012decreased, baguna1976mitosisI, baguna1974dramatic, baguna1981quantitative, newmark2000bromodeoxyuridine, nimeth2004stem}. We would like to relate this mitotic feeding response to the increase in worm size. Fig.~\ref{fig:feedingpeak}(a) shows the changes in worm area immediately after feeding. After an initial increase of the worm area, the worms shrink again and return to their initial sizes within approximately two weeks.

  \noindent
 Note that the worm area already approaches its maximum very rapidly within the first 12 hours after feeding, although the mitotic response lasts for following next days. Thus, we were wondering which part of the peak can be explained by pure stuffing of the gut and which part reflects the actual cell division response. To this end, we measured the growth response also for irradiated animals without stem cells, see red curves in Fig.~\ref{fig:feedingpeak}(b). Interestingly, the stuffing effect last for at least the first $4$ days until the irradiated worms dissociate. Thus, the initial increase in worm area after feeding actually reflects stuffing.
 
\noindent Nevertheless, these measurements provide us with the following information: First, the food fills the gut for at least $4$ days, most likely providing the dividing cells with nutrients during this time. Second, the growth response appears to be rather generic and food intake is approximately proportional to worm size. We also have not found an obvious dependence on feeding history. Growth peaks after longer starvation periods look similar to the results above, see Appendix~\ref{app:growthallometric}. Finally, the growth effect decays within two weeks. Thus, worms should approximately keep their size when being fed every two weeks. After one week, the worm size has been roughly increased by $2\%$. Further experiments will aim to extract changes in cell number upon feeding to reveal the actual growth response.

\subsection{Small worms grow and degrow faster than large worms}
\subsubsection{Measurement of growth rates}
\begin{figure}[tbp]
  \centering
\includegraphics[width=1\textwidth]{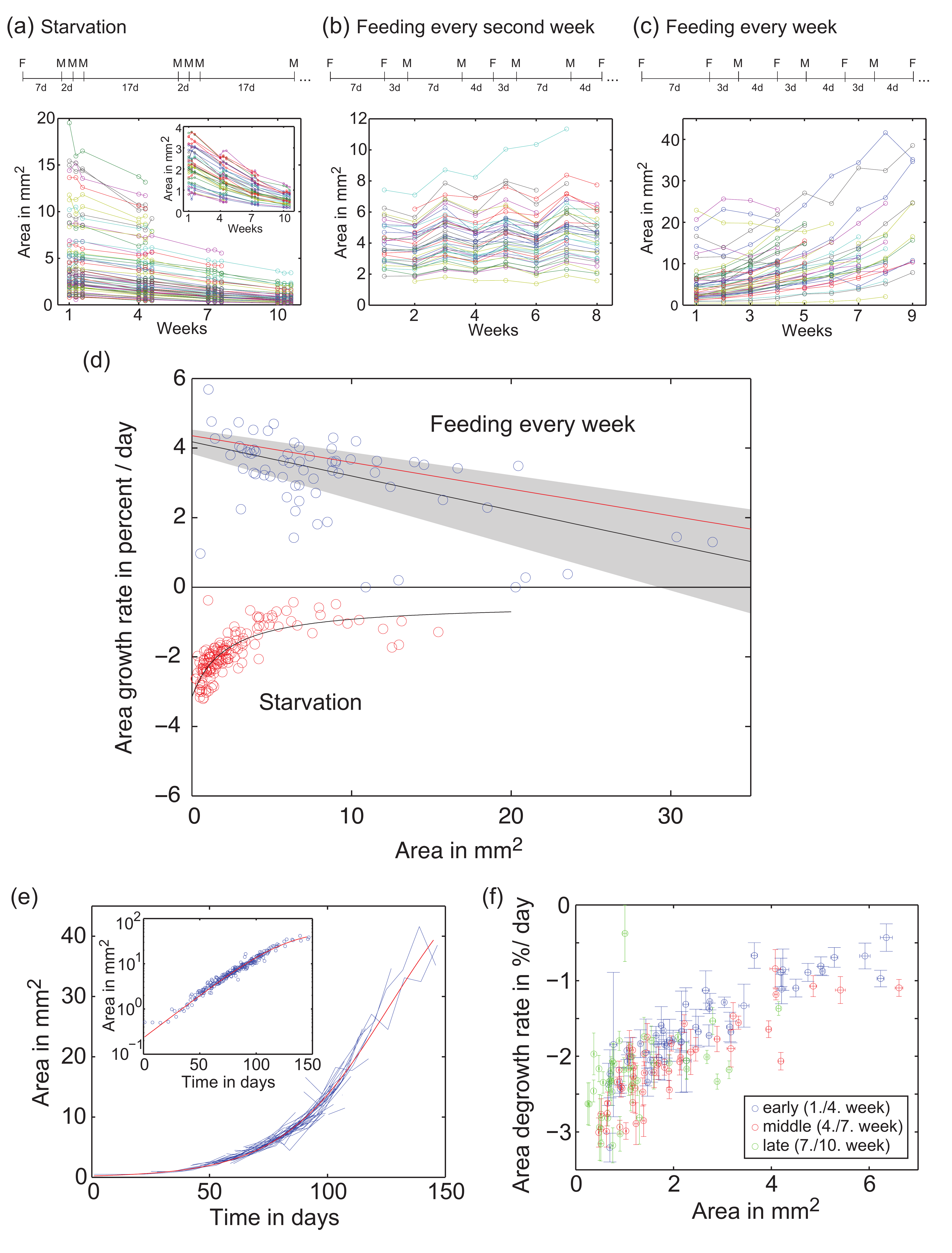}
  \captionof{figure}[Growth and degrowth rates]
  {(a)-(c) Measurements of worm area for different feeding conditions (Ima\-ging by Nicole Alt und Johanna Richter under the supervision of the author, analysis by the author includes 66, 41 and 48 worms, respectively). (d) Degrowth rates (red circles) have been computed from data in panel (a) by linear fits across three windows with 6 data points each using a robust algorithm. The size dependence is fitted by the function $c_1+c_2\,A/(c_3 + A)$ with $c_1=-3.17\pm0.38$ \%/day, \mbox{$c_2=2.72\pm0.33$ \%/day,} $c_3=2.11\pm 1.26$ mm$^2$ (black curve). {\bf(continued on next page)}}
  \label{fig:GrowthRates}  
\end{figure}

Even though we have learned that stuffing effects compromise tracking of the short term growth pulse, we can still quantify the averaged growth behavior on the time scale of weeks and in particular analyze how the growth and degrowth rates depend on feeding. Fig.~\ref{fig:GrowthRates}(a)-(c) shows the growth tracks for individual worms of different sizes for three feeding conditions: starvation, feeding every second week and feeding every week. In agreement with Fig.~\ref{fig:feedingpeak}, the worms that are fed every second week approximately maintain a constant size. One can clearly recognize the growth response upon feeding that decays until the next feeding event. The concatenation of short-term growth pulses results in a zig-zag line. For the other two cases, we compute average growth rates, see Fig.~\ref{fig:GrowthRates}(d). The data is very noisy for the well-fed worms because of the stuffing effect and the excretion of digested food. Still, we can obtain a clear linear trend using two independent fitting procedures. Besides the robust regression of the growth rates (black, with 95\% confidence intervals), we construct a master curve from the growth tracks assuming negligible effects of the feeding history, see Fig.~\ref{fig:GrowthRates}(e). The growth tracks of Fig.~\ref{fig:GrowthRates}(c) are shifted horizontally in an iterative procedure to minimize the variance from the average curve. Finally, we fit a Fermi function as the master curve, which corresponds to a growth rate that linearly depends on size. Both approaches agree well with each other.

\begin{figure} [t!]
  \caption*{{\bf Figure \ref{fig:GrowthRates}.(continued from previous page):} Growth rates (blue cicles) have been determined from the growth tracks in (c) by an exponential fit across two windows with 5 data points each. The trends in the growth rates are fitted by the robust regression $c_1-c_2\,A$ with $c_1=4.18\pm0.35$ \%/day, $c_2=0.10\pm 0.03$ \%/day/mm$^2$ (black line with 95\% confidence intervals in gray). A direct fit of the master curve in (e) agrees well (red). (e) Growth tracks in (c) can be shifted horizontally (blue) to collapse onto a master curve (red), which is fitted by the Fermi function $c_1/(c_2+c_3\, e^{-c_1 t})$ with \mbox{$c_1=4.4\pm0.3$ \%/day,} $c_2=0.077\pm0.012$ \%/day/mm$^2$, $c_3=19\pm 5$ \%/day/mm$^2$. Inset with log-linear plot illustrates the deviation from an exponential law. (f) Magnification of the degrowth rates for time intervals after the initial feeding shows no obvious dependence on feeding history. Error bars represent the standard error of the mean.}%missing
\end{figure}
\noindent Fig.~\ref{fig:GrowthRates}(d) shows a clear size-dependence of the growth and degrowth rates. Small worms appear to grow faster than large worms. Note that even the largest worms have a positive growth rate, meaning worms usually undergo fissioning before reaching the limits to growth. Surprisingly, small worms also degrow faster. This trend appears to be rather independent from the feeding history, see Fig.~\ref{fig:GrowthRates}(f).

\subsubsection{Discussion of growth and degrowth dynamics}
As for the allometric scaling laws, our results on the growth dynamics rather agree with very early measurements and less so with more recent publications.

\noindent At the beginning of the 20th century, Abeloos investigated growth dynamics in another flatworm species~\cite{abeloos1929recherches}. In analogy to our conclusion, he found that the increase in dry mass is stronger for smaller species.

\noindent Bagu\~n\`a \textit{et al.}~have analyzed growth and degrowth dynamics of \textit{Girardia tigrina} for various feeding conditions  \cite{baguna2012planarian,baguna1990growth}. We can extract from these data sets that their growth rates are also decreasing with size during feeding, see Appendix~\ref{app:growthallometric}. The growth rates are smaller than in our case, which could be a species-specific effect or \revi{due to gut stuffing. Furthermore, it might also be related} to the lower temperature of $12^{\circ}$C, while our experiments are conducted at $20^{\circ}$C. More importantly, the degrowth rates during starvation seem not to depend on the sizes of the worms in contrast to our data.

\noindent Other groups have also measured growth dynamics and typically fitted exponential functions with constant growth rates \cite{gonzalezestevez2012decreased,calow1977joint,thomas2012size}. Gonz\'alez-Est\'evez \textit{et al.}~pooled all worms together and therefore their data cannot discriminate between constant and weakly size-dependent rates \cite{gonzalezestevez2012decreased}. Nevertheless, the value of $1.8$ \%/day for the degrowth rate agrees well with our results in the considered size range of $1-5$ mm$^2$. 
Thomas \textit{et al.}~tracked the growth of individual worms but did not explicitly analyze a size-dependence of the rates \cite{thomas2012size}. They obtain values below $5$ \%/day and most of them lie in the range of $2-3$ \%/day, which is in good agreement with our data. 

\noindent Oviedo \textit{et al.} even claim to find linear functions for growth and degrowth with size \cite{oviedo2003allometric}. In consequence, the absolute values for growth and degrowth rates become smaller in larger worms like in Fig.~\ref{fig:GrowthRates}(d). Yet, otherwise the result does only roughly agree with our data, mainly for the case of starvation, see Appendix~\ref{app:growthallometric}. Although they even feed twice a week, they obtain very low growth rates, which are even lower than their degrowth rates.

\noindent Finally, it has previously been claimed that nutritional status and not size determines the growth dynamics \cite{gonzalezestevez2012decreased}. Yet, the respective measurements might not be able to resolve this issue because it does not compare size matched worms of different feeding histories. In fact, our data with a controlled feeding history supports the opposite hypothesis.

\noindent In summary, our measurements of growth dynamics agree in some cases quantitatively and in others qualitatively with studies published by other groups but also go significantly beyond. In particular, we were able to reveal the size dependence of the growth rates by establishing a strict protocol of feeding and imaging and a highly accurate \mbox{image} analysis. Furthermore, we were tracking individual worms and our measurements span the full range in size in contrast to the previous works discussed.

\begin{figure}[tbp]
  \centering
\includegraphics[width=1\textwidth]{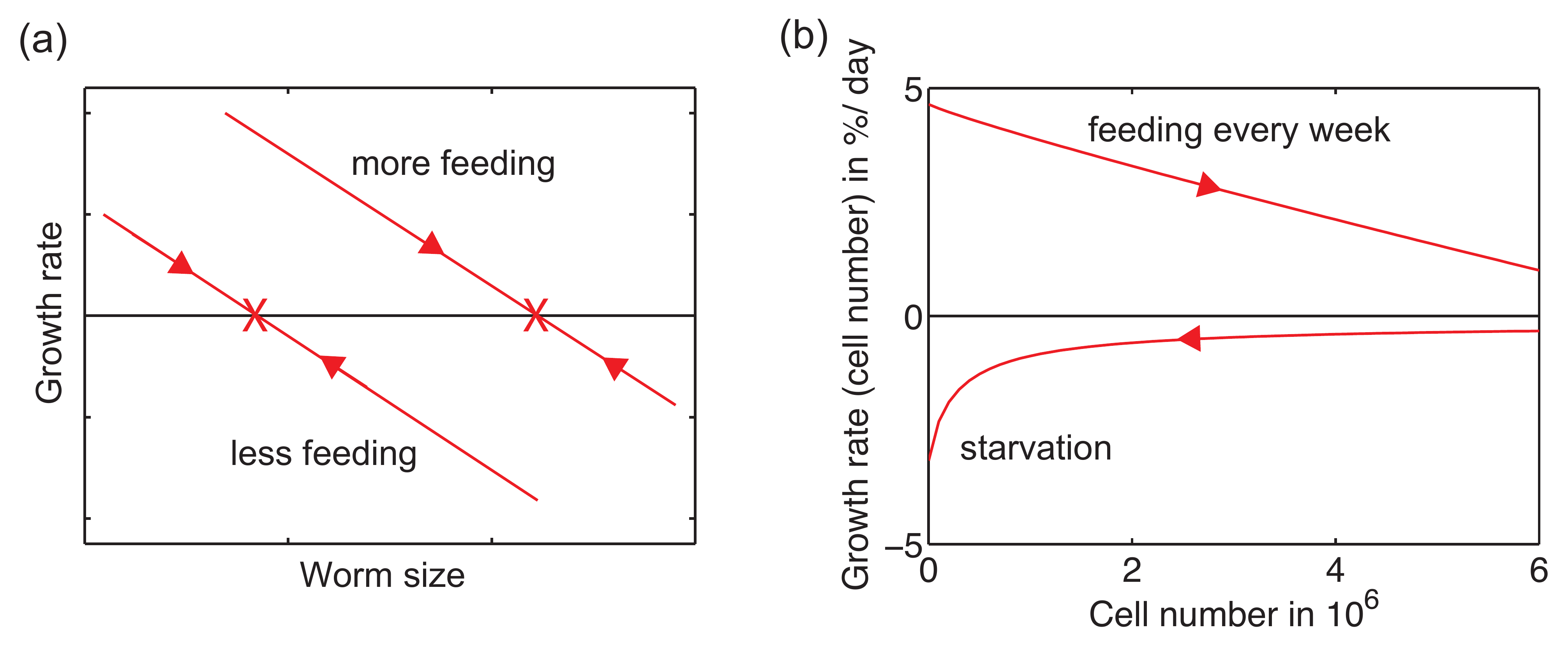}
  \captionof{figure}[Non-trivial growth dynamics in flatworms]
  {(a) Hypothetical case of growth and degrowth dynamics with steady state sizes (crosses). (b) Trends of the growth rates in terms of cell numbers for \smed{} obtained from fits to the growth rates of Fig.~\ref{fig:GrowthRates}(d) and the conversion curve to cell numbers of Fig.~\ref{fig:ScalingLaws}(b). \revi{Note that gut stuffing might slightly change the obtained values.}}
  \label{fig:GrowthRatesTrends}
\end{figure}

\noindent Our analysis clearly shows that small worms show faster growth and degrowth dynamics than large worms. Fig.~\ref{fig:GrowthRatesTrends} illustrates why this is a non-trivial observation. Naively one could have assumed that there is one limiting quantity that makes the use of food less efficient in larger worms, resulting in a decreasing growth rate with size, see  Fig.~\ref{fig:GrowthRatesTrends}(a). Different amounts of food would result in a shift of the curve. In consequence, worms would approach a steady state size, depending on the feeding frequency (crosses). In contrast, the trends of the curves for \smed{} change between feeding and starvation and the worms either grow or degrow. Fig.~\ref{fig:GrowthRatesTrends}(b) shows our result in terms of cell numbers. It suggests that there might be two different effects that depend in opposite ways on the size of the worm and each of which dominates for either starvation or maximum feeding. \revi{Stuffing effects as discussed in Sec.~\ref{sec:growthresponse} might introduce correction factors to the obtained growth rates. However, we expect no change in the qualitative trends.} In the next section, we explore three theoretical models that can account for the size-dependence of the growth rates. 
\clearpage

\section{Theoretical descriptions of cell turnover dynamics and energy flux}
Fig.~\ref{fig:GrowthRatesTrends}(b) shows nontrivial growth dynamics in flatworms depending on feeding conditions.
Growth corresponding to a change in cell number can be described by the two processes of cell division and cell loss with effective rates $k_{div}$ and $k_{loss}$, respectively:
\begin{equation} \dot{N}=k_{div} N - k_{loss} N\,.\end{equation}
With this, the growth rates plotted in Fig.~\ref{fig:GrowthRates}(d) are given by 
\begin{equation} \dot{N}/N=k_{div} - k_{loss}\,.\end{equation}
Thus, $K=k_{div} - k_{loss}$ has to depend on $N$ such that it decreases with $N$ for feeding and increases with $N$ for starvation.

\noindent One possible scenario to explain this behavior is a change in the fraction of stem cells. In fact, it has been reported that the neoblast fraction in \textit{Dugesia lugubris}  \cite{lange1967quantitative, lange1968possible} and \textit{Dugesia tigrina} \cite{baguna1981quantitative, baguna1976mitosisI,baguna1990growth} decreases as the worm grows. A higher stem cell fraction in smaller worms would lead to a larger growth rate during feeding. Additionally, it might be associated with a higher basal cell turnover, yielding a faster degrowth rate in the absence of food. However, the difference in the neoblast fraction was estimated to be less than $10\%$, which can not account by far for the observed size dependence of the growth rates. Thus, it is rather unlikely that the fraction of neoblasts in \smed{} varies by a factor of $2$ or $3$ like the growth rates. In fact, preliminary experimental data obtained by the group of Jochen Rink suggests that the stem cell fraction stays rather constant in animals of different sizes.

\noindent Feeding has a major effect on the growth dynamics. Therefore, we propose to consider the metabolic energy $E_t$ in the worm as a limiting factor that influences cell proliferation and cell death. At this point, we need to properly define this quantity, even though the conclusion will be largely independent of the exact definition. For the rest of the chapter, $E_t$ refers to the total amount of ATP and the ATP equivalent of any other molecule storing metabolic energy like glucose and lipids, whether it is found inside or outside the cells. Analogous descriptions can be found if only the freely available energy outside the cells is considered or if one takes into account the abundance of other limiting molecules like amino acids that need to be taken up by the food and that are necessary for the worm to stay healthy and alive.

\noindent In a minimal model, the total energy $E_t$ increases due to net influx $J_f$ by feeding and decreases due to consumption by metabolic housekeeping with rate $\mu$:
\begin{equation} \partial_t E_t= J_f-\mu N\,.\label{eq:Etdot}\end{equation}
A typical value for the metabolic consumption per cell $\mu$ would be of the order of \mbox{$10$ pW} as estimated for \textit{Schmidtea polychroa} \cite{mouton2011lack}. %1\mu W/mm^2
In humans with $4\cdot10^{13}$ cells and an energy consumption of $10^7$J/day \cite{milo2010bionumbers}, we estimate \mbox{$\mu\approx 1$ pW}. Further values obtained for 228 mammalian species from shrimp to elephant also show a decrease with size across the range of $1-10$ pW \cite{west2002allometric}.%ID 109708,100842

\noindent Eq.~\ref{eq:Etdot} assumes for simplicity that the energy stored in the dying cells can be completely taken up by the remaining cells.
We could consider an additional term of the form $-\rho_{loss}\,e_c\, k_{loss}N$, accounting for imperfect recycling upon cell death. Here, $e_c$ denotes the metabolic energy per cell and $\rho_{loss}$ is the fraction of it that is lost per dying cell. The main conclusions of this chapter remain unchanged even for $\rho_{loss}>0$.

\noindent We define the metabolic energy per cell as $e_c=E_t/N$ and find for its dynamics
\begin{equation} \partial_t e_c= j_f-\mu - K\, e_c\,.\label{eq:energypercell}\end{equation}
Here, $j_f=J_f/N$ is the net influx per cell and the last term describes a dilution effect. As the worm grows, the same amount of energy has to be shared among a larger number of cells.

\noindent In the following, we discuss three basic models on how the metabolic energy per cell might effect cell division and loss. Given the simplicity of the models, they all fit the noisy data reasonably well. Yet, they result in very different predictions about the respective variables and parameters such as metabolic consumption and storage of energy, which will be tested in future experiments.

\noindent The models describe three distinct scenarios of energy storage: (i) dynamic energy storage, for which feeding increases the fraction of energy stored in the worm, (ii) energy storage of fixed proportion, for which the worm is not able to store additional amount of energy upon feeding, and (iii) size-dependent energy storage, for which the fraction of energy stored depends on the size of the worm and not explicitly on feeding. In the first model, the metabolic energy per cell $e_c$ changes as a direct read-out of size and feeding conditions and directly regulates cell division and cell loss. In the second and third model, the metabolic energy is quickly regulated to a physiologically preferred target value $e_0$ by adjusting the cell turnover rates. Thereby, the mechanism implicitly draws on the idea of integral feedback control, which is commonly used in engineering applications to drive a dynamic variable to a pre-set value.

%\noindent In the following, we discuss three basic models on how the metabolic energy per cell might effect cell division and loss. In a first model, the metabolic energy per cell changes as a direct read-out of size and feeding conditions and directly regulates cell division and cell loss. In a second model, the metabolic energy is quickly regulated to a physiologically preferred target value by adjusting the cell turnover as a function of the deviation from this target value. In a third model, the target value representing the amount of metabolic energy stored in the system changes as a function of worm size.\\

\subsection{Model 1: Dynamic energy storage}
In a most simple model for metabolic growth control, the cells are assumed to directly respond to the amount of metabolic energy per cell $e_c$. If there is an excess in energy, cells divide more often, and if there is a lack of energy, the death rate increases. Let us consider a simple case, for which the growth rate depends linearly on $e_c$:
\begin{equation} K = K_0\,(e_c/e_{s}- 1)\label{eq:mod1K}\end{equation}
with the constant parameters $K_0$ and $e_{s}$, which determine the rate of growth and the energy, for which the worm switches between growth and degrowth, respectively.

\begin{figure}[tbp]
  \centering
\includegraphics[width=1\textwidth]{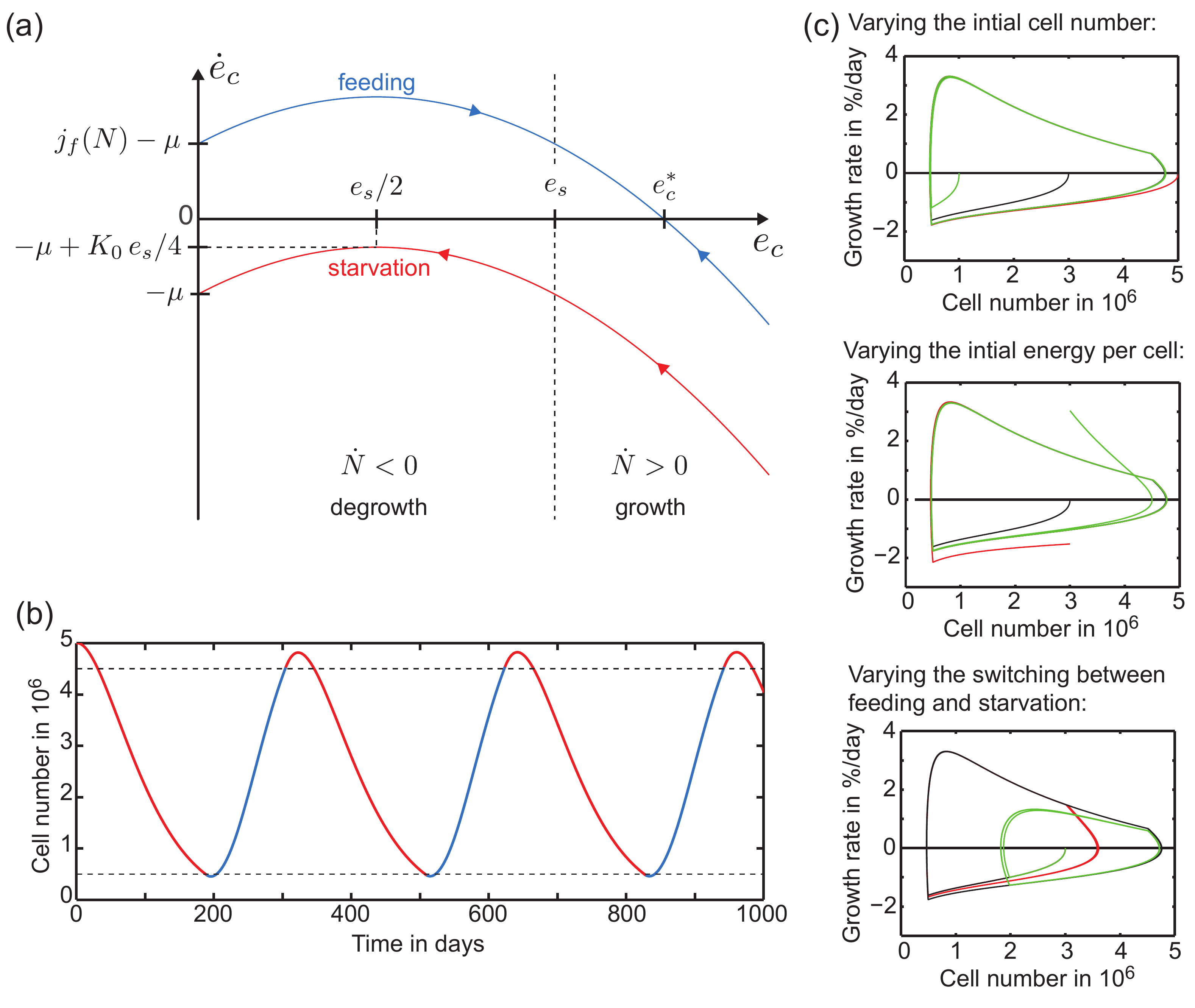}
  \caption[Dynamic energy storage]
  {Model 1 assumes that cell division and cell death directly depend on the metabolic energy per cell $e_c$, which represents a dynamic energy store. (a) During starvation, $e_c$ and thus the degrowth rate decreases. During feeding, $e_c$ approaches the growing regime. Note that $e_c^*>e_s$ is not a steady state of the total system because the influx $j_f(N)=j_0/(1+\sqrt{N})$ decreases with $N$. (b) Time course of the cell number when switching between feeding (blue) and starvation (red) for $N=0.5\cdot10^6$ and $N=4.5\cdot10^6$, respectively (dashed lines). (c) We observe a generic dynamics, irrespective of the initial cell number, energy or feeding scheme.}
  \label{fig:EnergyModel1}
\end{figure}

\noindent The change in energy $\dot{e}_c$ as a function of $e_c$ is plotted in Fig.~\ref{fig:EnergyModel1}(a). During starvation periods, when $j_f=0$, the growth rate is decreasing, which requires $\dot{e}_c<0$ (red curve). The maximum of $\dot{e}_c$ is at $e_c=e_s/2$ and from $\dot{e}_c<0$ follows that $\mu>e_sK_0/4$.
%In Fig.~\ref{}, we have seen that the degrowth rate hardly changes for large $N$. In Appendix~\ref{}, we show that this plateau is found at $e_c=\chi_0/(2\chi)$, yet only exists if $\mu_0\approx 1$. The degrowth rate at the plateau is $-\chi_0/2$ and the minimum degrowth rate at $e_c=0$ is $-\chi_0$, see Fig.~\ref{}. Therefore, it only increases by a factor of $2$, in contradiction to the data.\\

\noindent During feeding with $j_f>\mu$, the curve in Fig.~\ref{fig:EnergyModel1}(a) is shifted upwards and $e_c$ ends up in a regime, for which the organism grows (blue).
If $j_f$ was constant, there would be a stable steady state with $e^*_c=e_s/2+\sqrt{(e_s/2)^2+(j_f-\mu)e_s/K_0}$. We see that $e^*_c>e_s$ for $j_f>\mu$ and the system is in the growing regime.
In order for the growth rate to decrease with worm size as seen in Fig.~\ref{fig:GrowthRatesTrends}(b), the influx due to feeding $j_f(N)$ must not be constant but has to be a decreasing function of $N$. 
 \noindent

\noindent Fig.~\ref{fig:EnergyModel1}(b) illustrates a time course of $N$, when going through several rounds of feeding and starvation, always switching at a certain size (horizontal lines). Especially in the beginning of the starvation interval, we see an overshoot, for which the worm still grows although the feeding has stopped. Nevertheless, we want to stress that the growth and degrowth behavior is rather generic. The growth and degrowth rates in Fig.~\ref{fig:EnergyModel1}(c) collapse and show the same size-dependence, irrespectively of the initial values for energy and cell number. Any perturbation decays quickly and there is no strong dependence on feeding history.

\noindent We can fit this model to the experimental data, see Fig.~\ref{fig:EnergyModelsTogether}(a). The plots shows the size dependence of the influx $j_f$ and illustrate that the metabolic energy per cell $e_c$ acts as a read-out for both feeding conditions and size at the same time.
\clearpage

%The simple model might yield rates that change with size as observed in experiments, yet it might be questioned based on the timescales of the relaxation dynamics. After feeding has stopped, 
%%the $e_c$ values for organisms of different sizes quickly become similar due to the quadratic term in $\dot{e}_c$. However, 
%the switching from growth to degrowth happens at the same time scale as the very slow degrowth dynamics in contrast to the observation in the flatworm. Thus, next, we explore two further models, for which the level of metabolic energy per cell is quickly regulated to a target value.
%%This is only one of several problems with this approach. Besides the fact that the organism remains growing even for a long time though the food influx has stopped, we have also seen that the degrowth rate does only change by a factor of $2$ for $z=1$. For $z<1$, this range can be increased, see Appendix~\ref{}. Yet, this leads us to the third and and most important issue. Already for $z=1$, the energy levels have to change as much as the degrowth rates. For $z<1$, they even have to change more. It is

\subsection{Model 2: Fixed proportion energy storage}
We now assume that the energy per cell $e_c$ is quickly regulated to a physiologically preferred target value $e_0$. In this scenario, there might still be specialized energy stores like fat cells but they are only formed proportionally to the worm size and not in response to feeding. Any metabolic energy outside the cells is also quickly regulated, which is typical for the metabolism of animals. For example, sugar levels in the human blood are under tight regulation by the insulin pathway and insulin-like molecules are found even in the simplest unicellular eukaryotes \cite{leroith1985insulinrelated}. %LeRoith D, Shiloach J, Heffron R, Rubinovitz C, Tanenbaum R, Roth J (1985). "Insulin-related material in microbes: similarities and differences from mammalian insulins". Can. J. Biochem. Cell Biol. 63 (8): 839Ð49. doi:10.1139/o85-106. PMID 3933801
Yet again in contrast to humans, here we assumed that no additional long-term storage cells are formed upon feeding. Instead, the metabolic energy is regulated by adjustment of cell division and cell death. We can picture an integral feedback scheme, according to which $K$ changes depending on the difference between $e_c$ and $e_0$:
\begin{equation} \tau_K \dot{K}=K_0 (e_c/e_0-1)\end{equation}
 This control theoretic approach substantially differs from the case of model 1 in Eq.~\ref{eq:mod1K}.

\begin{figure}[tbp]
  \centering
\includegraphics[width=1\textwidth]{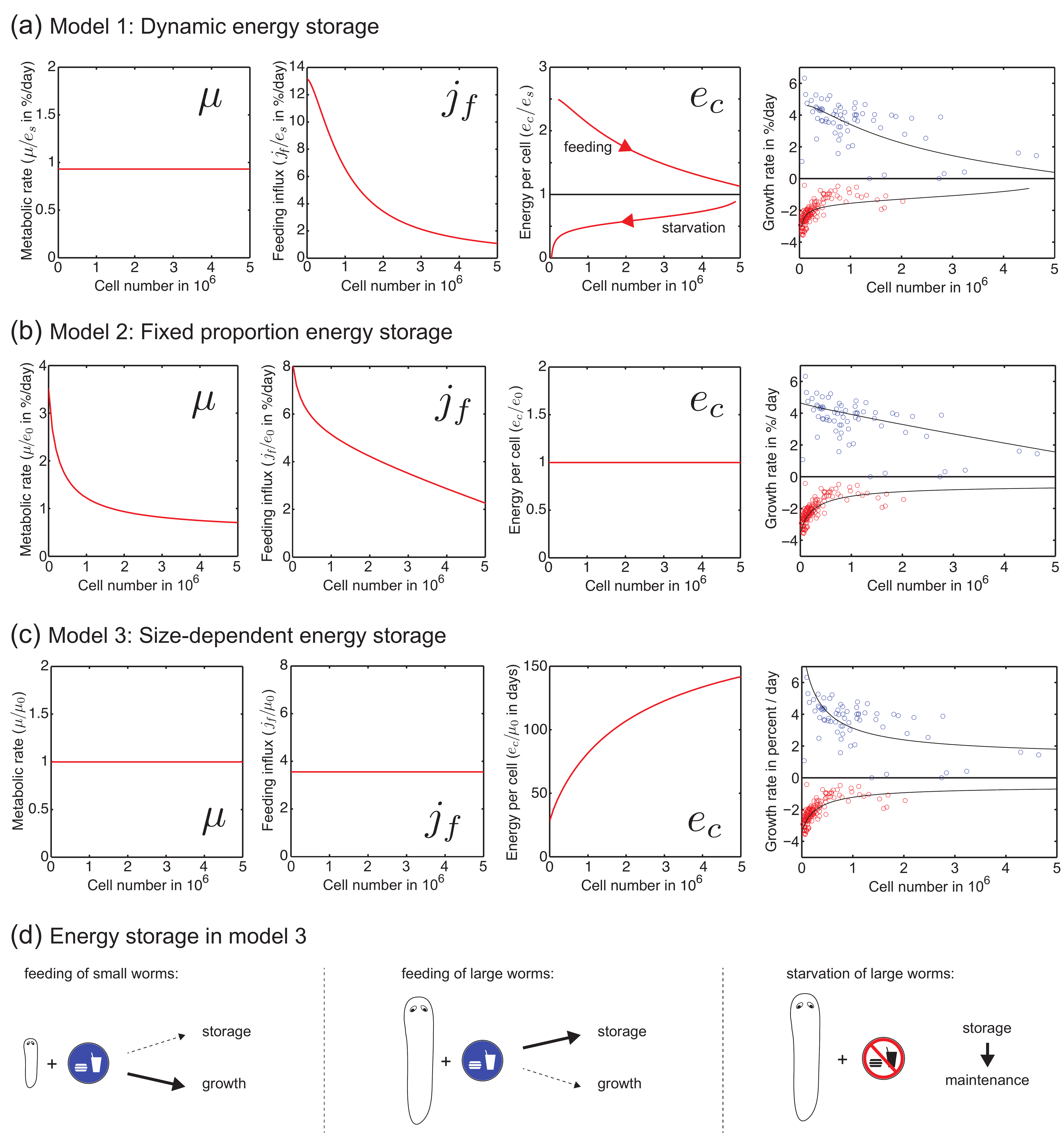}
  \caption[Growth models with energy storage]
  {(a) Model 1 is characterized by energy stores that are filled and depleted depending on feeding and worm size. The stored energy per cell $e_c$ as well as the metabolic rate $\mu$ and the feeding influx $j_f$ are obtained by minimizing the variance between  the experimental data and numerical solutions of our model. The values are given relative to the energy $e_s$, for which the worm switches between growth and degrowth.
(b) Model 2 draws on energy stores with fixed proportions. We compute $\mu$ and $j_f$ relative to the energy per cell from the fits to the growth and degrowth rates in Fig.~\ref{fig:GrowthRates}(d). The energy per cell $e_c=e_0$ is assumed to be constant.
(c) In model 3 with size-dependent energy stores, the energy per cell $e_c$ changes in a size-dependent manner, while $\mu=\mu_0$ and $j_f$ are constant. Degrowth and growth rates are fitted by the same function with a different amplitude corresponding to the feeding influx.
(d) Basic scheme of model 3: During feeding, small worms invest their metabolic energy in growth, yet large worms store an increasing fraction of the available energy. During starvation, large worms degrow more slowly as they can deplete their energy stores.}
  \label{fig:EnergyModelsTogether}
\end{figure}
\noindent Given a fast regulation to the target value $e_0$ ($\tau_K\rightarrow 0$), $K$ relaxes to
\begin{equation} K^*= \frac{j_f-\mu}{e_0}\,,\label{eq:kquickreg}\end{equation}
which defines the nullcline of $\dot{e}_c=0$ according to Eq.~\ref{eq:energypercell}.

\noindent  Fig.~\ref{fig:GrowthRatesTrends}(b) shows that the degrowth rate is decreasing during starvation, where \mbox{$j_f=0$}. Thus, $\mu$ has to increase for smaller worms. Fig.~\ref{fig:EnergyModelsTogether}(b) illustrates the size-dependence of $\mu$ based on our measurements. This can be interpreted in the sense that the metabolism is less efficient in smaller animals, which is in agreement with the typical metabolic sca\-ling laws found across the animal kingdom \cite{agutter2004metabolic, savage2007scaling} and which have also been observed in flatworms \cite{mouton2011lack, hyman1919physiological3, gonzalezestevez2012decreased}.

%\begin{figure}[tbp]
%  \centering
%\includegraphics[width=1\textwidth]{EnergyModel2.pdf}
%  \caption[Fixed proportion energy storage]
%  {Model 1 assuming energy stores with fixed proportions is fitted to the growth and degrowth rates in \smed{}. (a) The metabolic rate $\mu$ from the starvation curve, (b) the feeding influx $j_f$ is obtained from the difference between starvation and feeding and (c) the energy per cell is set to $e_0=1$.}
%  \label{fig:EnergyModel2}
%\end{figure}
\noindent During feeding the growth rate decreases with size. Thus, $j_f$ has to decrease with $N$ and even has to overcompensate for the opposite effect of $\mu$. This suggests that food uptake is less efficient in larger animals, see Fig.~\ref{fig:EnergyModelsTogether}(b).

\noindent Taken together, we assumed that the metabolic energy is the limiting factor determined by the corresponding nutrition influx and consumption, to which turnover rates are adjusted accordingly.
We obtained two functions, one for $\mu$ and one for $j_f$, which depend on worm size in opposite ways and therefore can account for the opposite trends in the growth and degrowth dynamics.

\noindent As a side note, one might question what is cause and what is consequence. While it is likely that the metabolic energy is the limiting factor during starvation periods, this is not necessarily the case for maximum feeding.
%The following scenario leads to analogous relations: during starvation, cell division might be set to a minimum and cell death has to increase the smaller the worm and the less efficient the metabolism, just like before. Yet, during feeding, there could be a well-controlled, size-dependent response of cell division (and maybe cell death) and in consequence the net influx $j_f$ is adjusted appropriately.
In the presence of an abundance of food in the gut, cell division response could rather be limited by other factors as the maximum rate of DNA replication and the available number of stem cells. This would invert the argument: there would be a generic, yet size-dependent response of cell division (and maybe cell death) and the size dependence of the net influx $j_f$ is merely a consequence of it.

\subsection{Model 3: Size-dependent energy storage}
In the scenario above, we introduced a constant target value $e_0$ for the metabolic energy that determined the turnover rates. In our third and final model, we again consider a quick relaxation to such a target energy, but now this energy $e_0(N)$ changes with worm size. For example, the fraction of energy-rich cells could increase in large worms. We again obtain Eq.~\ref{eq:kquickreg} but now $\mu$ and $j_f$ are assumed to be constant for simplicity, while $e_0$ depends on $N$. The three variables are plotted in Fig.~\ref{fig:EnergyModelsTogether}(b). Even with only one size-dependent quantity, we obtain a reasonable fit to the data. If additionally $\mu$ was allowed to change as suggested by the measurements of metabolic scaling laws \cite{agutter2004metabolic, savage2007scaling,mouton2011lack, hyman1919physiological3}, the fits would even improve.

% \begin{figure}[tbp]
%  \centering
%\includegraphics[width=1\textwidth]{EnergyModel3.pdf}
%  \caption[Size-dependent energy storage]
%  {Model 3 with size-dependent energy stores is fitted to the growth and degrowth rates in \smed{}. The metabolic rate $\mu$ (a) and the feeding influx $j_f$ (b) is assumed to be constant, while the energy per cell $e_c$ increases (c). (d) The curve obtained for the degrowth rate in Fig.~\ref{fig:GrowthRates}(d) is also fitted to the growth rates with a constant factor corresponding to the feeding influx. (e) During feeding, small worms invest their metabolic energy in growth, yet large worms store an increasing fraction of the available energy. During starvation, large worms degrow more slowly as they can deplete their energy stores.}
%  \label{fig:EnergyModel3}
%\end{figure}
\noindent The basic logic of such a mechanism based on energy storage cells is shown in Fig.~\ref{fig:EnergyModelsTogether}(d). While small worms invest all available energy into growth, large worms store part of their energy uptake and thus grow more slowly. In consequence, large worms
deplete these energy stores during starvation and initially degrow more slowly. Importantly, we assume that the size of the stores scales with worm size. Thus, only an effective size-dependence of growth and degrowth is observed, but no explicit dependence on feeding history.

% \begin{figure}[tbp]
%  \centering
%\includegraphics[width=1\textwidth]{FoodStorage.pdf}
%  \caption[Size-dependent energy storage]
%  {Size-dependent energy storage: During feeding, small worms invest their metabolic energy in growth, yet large worms store an increasing fraction of the available energy. During starvation, large worms degrow more slowly as they can deplete their energy stores.}
%  \label{fig:FoodStorage}
%\end{figure}\noindent

% Furthermore, a higher cell division rate might come at a cost of more energy consumption. Thus, we might have to modify Eq.~\ref{eq:energypercell} to account for these effects by introducing the recycled amount of energy per dying cell $e_r$ and an extra energy per cell division $e_d$:
%\begin{equation} \partial_t e_c= j_f-\mu - e_d k_{div} + e_r k_{loss} - (k_{div} - k_{loss}) e_c\,.\label{eq:energypercellextend}\end{equation}
%We obtain a similar relationship as before in Eq.~\ref{eq:kquickreg} with an additional factor depending on whether $k_{div}$ or $k_{loss}$ is adjusted:
%\begin{equation} \left(1+\frac{e_d}{e_0}\right)k_{div} - \left(1+\frac{e_r}{e_0}\right)k_{loss}= \frac{j_f-\mu}{e_0}\,.\end{equation}
%If these extra terms depend on worm size, they can change the scaling relationship obtained before. They might become size-dependent if the ratio of different cell types changes with worm size. For example, with increasing worm size there could be an increasing fraction of cells that require more energy when being made but also provide more energy when being recycled.

\subsection{Discussion of the turnover models}
We have explored three models that are based on very different assumptions and differ greatly in the microscopic details, yet they all can account for the observed size-dependent growth behavior. 
This is an interesting observation, which one should generally take into consideration when comparing microscopic models to more macroscopic data. A theoretical model is typically associated with a specific coarse-graining level. As a necessary condition, it needs to agree with observations on larger scales, but this is not sufficient for the model to be unique. % This means one can disprove a model by measurements on higher levels but one usually needs measurements on the same level to validate it.
In order to distinguish between several microscopic models, one usually needs measurements on the same level.

\noindent In our case, the theory makes specific testable predictions about the influx, consumption and storage of energy, as illustrated in Fig~\ref{fig:EnergyModelsTogether}. In close collaboration with our experimental colleagues, we have planned further experiments to measure these quantities and to parametrize the models. First, it needs to be verified that the fraction of stem cells is approximately constant in \smed{} and cannot explain the growth dynamics. Furthermore, our collaborators are conducting experiments at the time of writing to determine metabolic rates and feeding influx. Additionally, they aim to identify potential stores for the metabolic energy and probe the effects of inhibiting them. The existence of fat cells has already been reported for other flatworm species \cite{hyman1923physiological6}.

\noindent Fig.~\ref{fig:feedingpeak} already shows that the stuffing peak is approximately proportional to worm size. If therefore the average amount of food that becomes available per cell is independent of the size of the worm, it will support the third model.
Furthermore, we have discussed in Section~\ref{sec:AllScalLaw} that the scaling laws for worm mass and total protein mass might also be in agreement with an increased storage of lipids or glycogen in larger worms. It remains to be answered whether the scaling of metabolic rates and storage cells can explain the growth and degrowth behavior in \smed{}.

\noindent Even though the models are highly simplified \revi{and gut stuffing might introduce corrections}, we can extract \revi{order of magnitude estimates} from the fits to the data. For example, we can estimate from all three models that each cell stores on average as much energy as needed to maintain it for approximately $e_c/\mu = 50$ to $150$ days. In comparison, the metabolic rate in humans is $\mu\approx 1$ pW and the total energy stored in human fat cells (as the main energy storage) is \mbox{$4\cdot 10^{8}$ J} \cite{milo2010bionumbers}. Thus, $e_c=10^{-5}$ J, which amounts to a similar ratio of $e_c/\mu = 130$~days. %ID 109708,100842
%N=3.7 10^13, E_t=10^5 kcal, J=2.390 10^?4 kcal
%Et=10^9 /2.39 Ws approx 4 10^8 J
%ec=10^-4 /3.7/2.39 Ws
%ec/mu=10^8 /3.7/2.39 s=10^4 /3.7/2.39 /8.64 d
Furthermore, based on measurements in the flatworm \textit{Schmidtea polychroa} and on the typical metabolic scaling laws \cite{mouton2011lack,milo2010bionumbers,west2002allometric}, we expect the metabolic rate of \smed{} to be in the range of \mbox{$\mu\approx 10$ pW} per cell. Thus, according to our models, each cell stores as much as \mbox{$e_c\approx 10^{-5}$ to $10^{-4}$ J}. 
Note that this estimate is based on the assumption of perfect recycling of the energy upon cell death. For imperfect recycling, the value is reduced.

\noindent Metabolic energy can be stored in various ways: in lipids or glycogen, in specialized fat cells or distributed among many different cell types. An interesting observation in this respect is the presence of germ line precursor cells in asexual flatworms. Cells without a purpose should vanish during evolution. Thus, we might speculate that the germ line precursors in asexual worms might have an additional function for the metabolic housekeeping. In analogy to the third model, they are especially prominent in large worms and can potentially serve as energy stores during starvation.

\section{Control logic for cell turnover and growth}
\subsection{Measuring cell turnover on various scales}
So far, we have investigated how the cellular behavior might depend on feeding conditions and worm size but we have not explicitly distinguished between the adjustment of cell division on the one hand and cell loss or death on the other hand. In this section, we now discuss how these individual processes might be controlled. In order to unravel the control logic, we propose experiments on various scales, ranging from the level of individual cells to a whole tissue and to the averaged turnover dynamics across the entire worm. The main questions are:
\begin{itemize}
\item Does feeding and worm size affect rather cell proliferation or cell death?
\item How do both processes influence each other?
\item To what extent does aging play a role on the cellular scale?
\end{itemize}
The first question will be addressed mainly at the cellular level and the second and third question at the tissue level.
\vskip0.3cm

\begin{figure}[tbp]
  \centering
\includegraphics[width=0.55\textwidth]{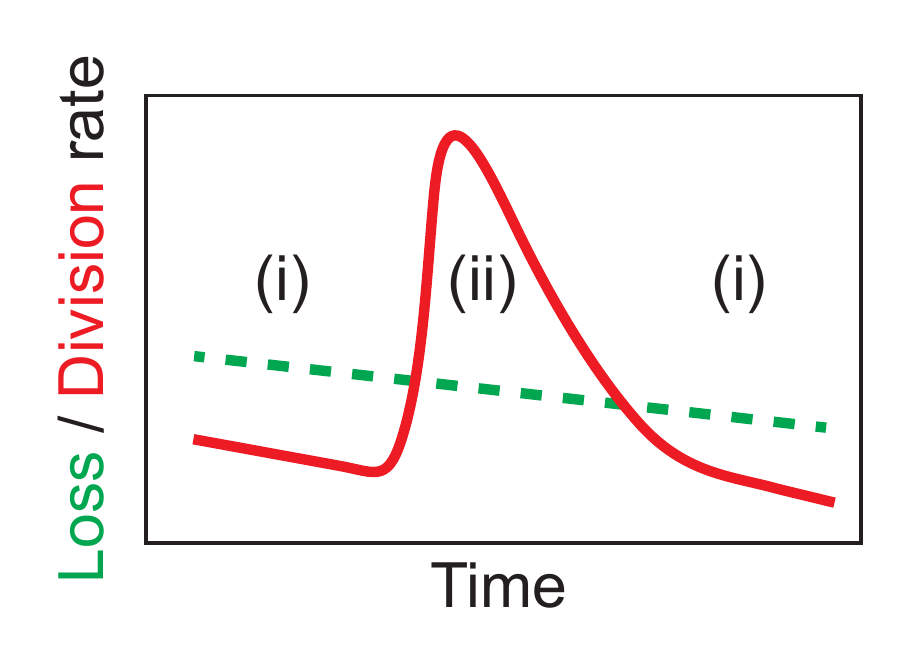}
  \caption[Catabolic default state and mitotic bursts upon feeding]
  {A hypothetical model that comprises (i) an catabolic default state, for which the cell loss rate (dashed green) is larger than the cell division rate (solid red), and (ii) short proliferation responses upon feeding.}
  \label{fig:anabolicdefault}
\end{figure}

\noindent {\it Turnover dynamics at the cellular level. --- }
%First, we aim to determine whether the feeding conditions and the worm size rather affect cell proliferation or cell death.
Previous measurements indicate that there is an catabolic default state during starvation periods, in which cell loss dominates over cell division. Upon feeding, there are short-term mitotic bursts leading to growth, as sketched in Fig.~\ref{fig:anabolicdefault} \cite{gonzalezestevez2012decreased, baguna1976mitosisI, baguna1974dramatic, baguna1981quantitative, newmark2000bromodeoxyuridine, nimeth2004stem, rink2013stem, kang2009flow}.
However, it is not clear whether the cell loss rate also changes as part of the feeding response and to what extent division and loss depends on feeding history. It has been suggested that apoptosis rates might in fact increase after the flatworms have been fed, especially after long starvation periods, as a means of cell renewal \cite{pellettieri2010cell, gonzalezestevez2012decreased}.

\noindent Assessing these questions is complicated by the fact that cell loss is poorly defined. Apart from the classical apoptosis pathway many other processes might contribute to cell loss such as other mechanisms of programmed cell death but also uncontrolled shedding of epidermis cells. In contrast, cell division can be well characterized. In Appendix~\ref{app:cellcycle} we discuss a measurement scheme to extract division rates. By comparing cell division to worm growth, we will also be able to determine to what extent cell death rates are regulated.

\vskip0.3cm
\noindent {\it Turnover dynamics at the tissue level. --- }
By monitoring the turnover dynamics in epidermis cells, we can exemplify how cell addition and removal work together to result in a well controlled size regulation. In the next section, we introduce a respective measurement protocol. We show preliminary data, which suggests that cell death depends on the age of the cells.

\noindent We also discuss further experiments to confirm the effect of aging and to investigate how the individual turnover processes influence each other.
On the one hand, turnover could be a rather stochastic process, in which cells are inserted in the tissue and deleted independently, yet on average at balanced rates. On the other hand, insertion and deletion could be tightly linked such that a cell is preferentially removed in the presence of a new cell that replaces it. We suggest a double labeling experiment which can potentially distinguish the two control paradigms.

\vskip0.3cm
\noindent {\it Turnover dynamics at the organismal level. --- }
 After constructing a comprehensive model on how turnover and growth is regulated based on the experiments on the cellular and the tissue scale, we can validate the model by measurements across the entire worm. To this end, we adapt a protocol that has originally been developed to analyze cell turnover in human brain cells \cite{spalding2013dynamics}. In Appendix~\ref{app:turnoverorganismlevel}, we provide the theoretical framework to set up the experiment.

%\subsection{Measuring cell turnover dynamics on the tissue level}
%Next, we provide a theoretical framework for measurements of cell turnover on the tissue level. We discuss the scope of the current experimental set-up, which takes the epidermis as an example tissue and we propose further experiments to infer more details on the turnover dynamics of epidermis cells. A pilot experiment has been performed by Sarah Mansour in the group of Jochen Rink.

\subsection{Analyzing turnover of the epidermis as an example tissue}\noindent
Epidermis cells as the outermost skin cells can be non-invasively labeled by soaking the worms in CFSE solution (carboxyfluorescein diacetate succinimidyl ester). It has been observed that the labeled cells disappear within days as a result of turnover of the epithelial tissue.
From this experiments, we can estimate turnover rates of the tissue and investigate whether older cells are more prone to die.  We will also propose a protocol with two consecutive labeling pulses, which might enable us to understand how insertion and deletion of cells is coordinated.
%\begin{figure}[tbp]
%  \centering
%\includegraphics[width=1\textwidth]{WildtypeLabeling.png}
%  \caption[CFSE labeling of epidermis cells]
%  {Labeled epidermis cells (CFSE staining - green) disappear within a few days due to turnover. Nuclei are labelled in blue (Hoechst staining). Experiment performed by Sarah Mansour in the group of Jochen Rink.}
%  \label{FigWTLabel}
%\end{figure}

\subsubsection{Measuring turnover rates by single pulse labeling}
\begin{figure}[tbp]
  \centering
\includegraphics[width=0.85\textwidth]{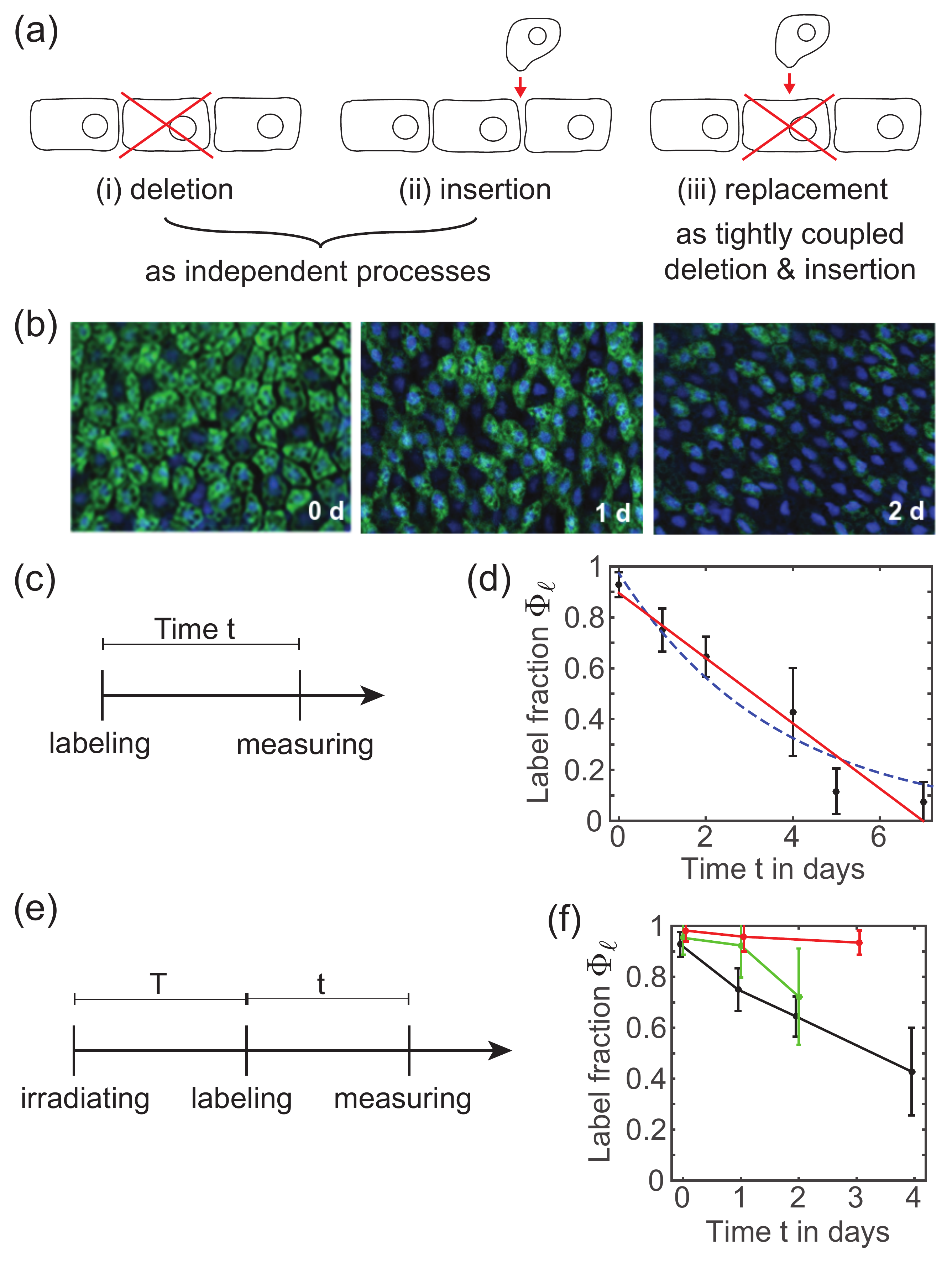}
  \caption[Turnover of epidermis cells]
  {(a) We distinguish three processes of cell turnover. (b) Labeled epidermis cells (CFSE staining - green) disappear within a few days due to turnover. Nuclei are labelled in blue (Hoechst staining).
(c) Measurement scheme of the CFSE labeling experiment. (d) Fraction of labeled cells at a certain time after the label is applied. We fit an exponential curve with a constant tissue turnover rate $k_{tis}=0.27/$day (dashed blue) as well as linear function corresponding to a fixed life time of the cells $a_{mx}=7.8$ days (red).
 (e)  Measurement scheme with irradiated (stem cell depleted) worms.
 (f) For $T=1$ day, the turnover  in irradiated worms (green) compares to non-irradiated worms (black). Yet, for $T=5$ days after irradiation, the labeled epidermis cells do not vanish from the tissue anymore (red). 
  All measurements were performed by Sarah Mansour in the group of Jochen Rink. The author analyzed the data and fitted the models.}
  \label{FigTissueTurnover}
\end{figure}

\noindent
{\it Cell turnover processes. --- }
We describe turnover in a cell population by three different processes: (i) deletion of cells, (ii)  insertion of cells and (iii) replacement of cells (i.e. coupled insertion and deletion), see Fig.~\ref{FigTissueTurnover}(a). Note that deletion potentially includes many different processes: the induced removal by long range signals or quorum sensing in the tissue, cell-autonomous decisions as well as uncontrolled shedding.

\noindent In order to measure turnover dynamics, we labeled all cells in the epidermis by soaking in CFSE and counted the fraction of labeled cells in the field of view after time $t$, see Fig.~\ref{FigTissueTurnover}(b)-(c). The experiments were performed by Sarah Mansour in the group of Jochen Rink. The author analyzed the data by automatically counting the cells with a custom MATLAB code.

\vskip0.3cm
\noindent{\it Dynamics of a labeled cell population. --- }
Let us consider the number density $n(a)$ of cells of a certain age $a$. The total number of cells $N$ in the tissue and the number of labeled cells $N_{\ell}$ are given by
\begin{equation} N=\int_0^{\infty}n(a)\,da\quad,\qquad N_{\ell }=\int_t^{\infty}n(a)\,da\,.\end{equation}
Due to cell turnover, these quantities change in time as follows
\begin{eqnarray} \dot{N}&=&(k_{i}-\bar{k}_d)\, N\\
\dot{N}_{\ell}&=&-(\bar{k}_{d\ell}+\bar{k}_{r\ell})\,N_{\ell}\,.
\end{eqnarray}
Here, we consider the rate for cell insertion $k_i$ and average rates for the replacement of labeled cells $\bar{k}_{r\ell}$ as well as for the deletion of labeled and unlabeled cells $\bar{k}_{d\ell}$ and $\bar{k}_d$:
\begin{equation} \bar{k}_{r\ell}=\int_t^{\infty} \frac{k_{r}(a)\,n(a)}{N_{\ell}}\,da\;,\quad  \bar{k}_{d\ell}=\int_t^{\infty} \frac{k_{d}(a)\,n(a)}{N_{\ell}}\,da\;,\quad  \bar{k}_{d}=\int_0^{\infty} \frac{k_{d}(a)\,n(a)}{N}\,da\,.\end{equation}
When taking snapshots of only part of the tissue, we do not obtain absolute numbers but the fraction of labeled cells in the field of view $\Phi_{\ell}=N_{\ell}/N$, which obeys
\begin{equation} \dot{\Phi}_{\ell}=-\big(\bar{k}_{r\ell}+k_{i}+(\bar{k}_{d\ell}-\bar{k}_d)\big)\,\Phi_{\ell}\,.\label{EqLabelFrac}\end{equation}
The fraction of labeled cells might not only decrease due to replacement (first term), but also due to a dilution effect as unlabeled cells get inserted into the tissue (second term). Furthermore, the third term arises if labeled cells and unlabeled cells are deleted at different rates (e.g.~in an age-dependent manner). In contrast, if cells are deleted stochastically (e.g.~according to a Poisson process), $k_d(a)$ is constant and the corresponding term vanishes: \mbox{$\bar{k}_{d\ell}-\bar{k}_d=0$}.
%\begin{figure}[tbp]
%%  \vskip1cm
%  \centering
%\includegraphics[width=0.8\textwidth]{CellProcesses03.pdf}
%  \caption[Turnover in epidermis cells of wildtype worms]
%  {We distinguish three processes of cell turnover: (a) cell deletion, (b) insertion of new cells and (c) cell replacement, which corresponds to a direct coupling between insertion and deletion. (d) Measurement scheme of the CFSE labeling experiment. (e) Fraction of labeled cells at a certain time after the label is applied. We fit an exponential curve $\Phi_{\ell}\propto e^{-t\,k_{tis}}$ with a constant turnover rate of the tissue $k_{tis}=0.39/$day (blue) as well as a model with a turnover rate that varies linearly in time $\Phi_{\ell}\propto \text{exp}(0.07\,t/\text{day}+0.12\,t^2/2/\text{day}^2)$ (red). Measurements are performed by Sarah Mansour, modeling and fitting is done by the author.}
%  \label{FigCellProcesses}
%\end{figure}

\noindent Feeding conditions and the sizes of the worms are approximately constant during the course of the experiment. Thus, if the turnover does not depend on the age of the cells (i.e.~$k_d$ and $k_r$ are constant), we can expect an exponential solution $\Phi_{\ell}= \Phi_{\ell,0}\, e^{-t\,k_{tis}}$ to Eq.~\ref{EqLabelFrac} with a constant rate $k_{tis}=k_{r}+k_{i}$. Yet, the preliminary data in Fig.~\ref{FigTissueTurnover}(d) does not strongly support an exponential law (dashed blue).

\noindent Another limiting case is that cells have a fixed live span and only die at age $a_{mx}$. Thus, the rates of  deletion and replacement are $k_d(a)=\delta(a-a_{mx})$ and $k_r(a)=\delta(a-a_{mx})$. Note that the worms only grow and degrow by a few percent per day, such that we can assume that the tissue size only changes very little in comparison to the turnover time scales of a few days. In consequence, the age distribution $n(a)$ will be approximately homogeneous and the solution to Eq.~\ref{EqLabelFrac} a linear function $\Phi_{\ell}=\Phi_{\ell,0}-t/a_{mx}$.
This simple linear model (solid red) appears to agree better with the data in Fig.~\ref{FigTissueTurnover}(d).
In fact, the truth might be in between the two limiting cases. This would resonate with an age-dependent turnover of the cells in the tissue such that old cells are more prone to deletion or replacement.

\noindent The time dependence enters because the age distribution of labeled cells is shifted over time. In order to extract in what way the cellular age influences deletion or replacement and which of the two processes is mostly affected, we need more sophisticated experiments as discussed below. Also note that the presented data is only very preliminary and has to be confirmed by further measurements.

%If tissue growth can be neglected in the time course of the experiment, $\dot{N}=0$, it follows $k_{i}=k_d$ and we can infer the averaged life time of cells in the tissue
%\begin{equation} \tau_{cell}=(k_{r\ell}+k_{d\ell})^{-1}\,.\end{equation}
%Time dependent rates indicate age-dependent deletion or replacement.
%\begin{figure}[tbp]
%  \includegraphics[width=0.65\textwidth]{IrradScheme02.pdf}\\
%\includegraphics[width=0.95\textwidth]{IrradiationLabeling.png}
%  \caption[Turnover in epidermis cells of irradiated worms]
%  {(a) Measurement scheme with irradiated (stem cell depleted) worms. (b) Labeled epidermis cells do not vanish from the tissue anymore $T=5$ days after irradiation (as shown for the time steps $t=$ 0 d, 1 d, 2 d). Experiment performed by Sarah Mansour in the group of Jochen Rink.}
%  \label{FigIRRLabel}
%\end{figure}

\vskip0.3cm
\noindent {\it Labeled cell fraction in irradiated worms. --- }
We briefly comment on a second pilot experiment by Sarah Mansour,  see Fig.~\ref{FigTissueTurnover} (e)-(f). Here, the measurement started at varying times $T$ after the worms were depleted of all stem cells by $\gamma$-irradiation. First, we observe that tissue turnover is not directly affected by the loss of stem cells: for small $T$ the label fraction (green) decreases in the same way as for non-irradiated worms (black). Labeled cells are replaced by progenitor cells, which were not deleted by irradiation. Thus, there seems not to be a long-range signaling effect from the stem cell pool.
In contrast, $T=5$ days after irradiation (red), the fraction of labeled cells does not decrease anymore. In conclusion, dividing cells (which are affected by irradiation) or corresponding signals need 5 days to reach the epidermal tissue.

\noindent\subsubsection{Monitoring cell turnover by double pulse labeling}
\begin{figure}[tbp]
  \centering
  \includegraphics[width=1\textwidth]{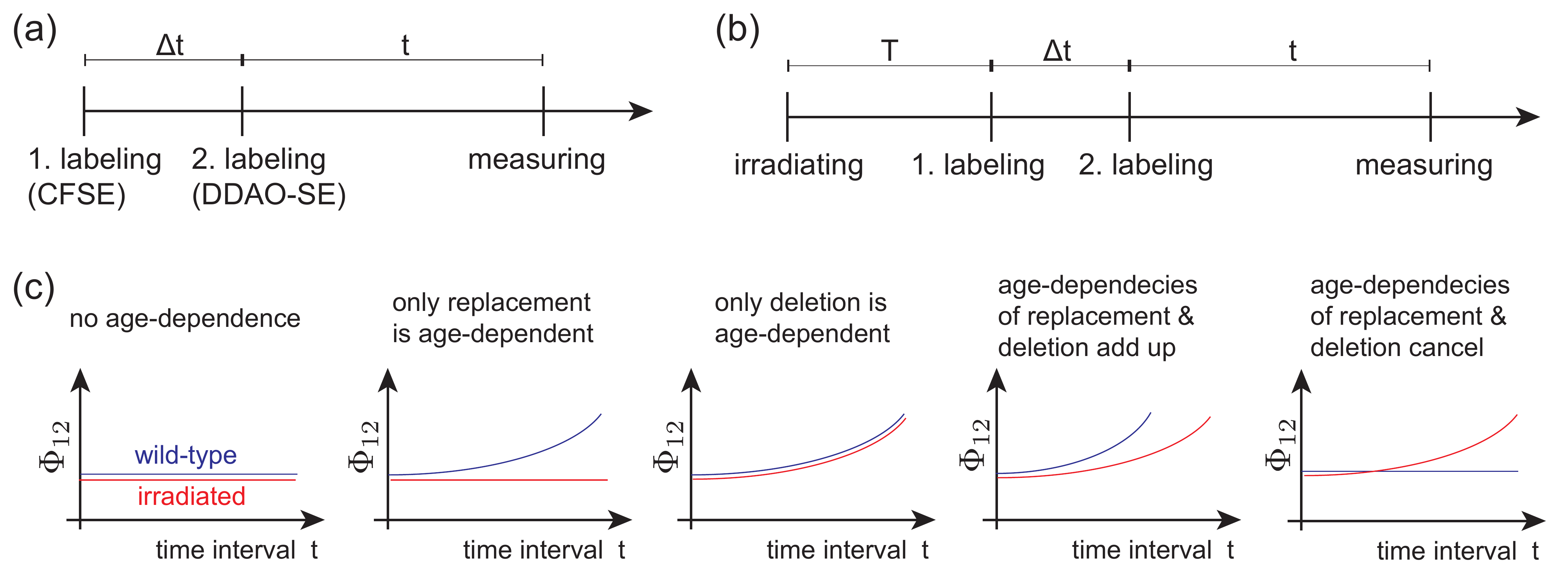}
  \caption[Measuring cell turnover dynamics with two labels]
  {Double pulse labeling scheme in (a) non-irradiated and (b) irradiated animals. (c) The double labeling experiment allows us to deduce whether deletion or replacement depend on the age of the cells.}
  \label{FigDoubleLabel}
\end{figure}

\noindent
We propose a double-labeling experiment using a second labeling pulse of a different marker (DDAO-SE) applied after a short time interval $\Delta t$, see Fig.~\ref{FigDoubleLabel}(a). As a result, there are two labeled cell populations -- cells with both labels $N_2$ and young cells with only the second label $N_1$:
\begin{equation} N_1=\int_t^{t+\Delta t} n(a)\,da\quad,\qquad N_{2}=\int_{t+\Delta t}^{\infty}n(a)\,da\,.\end{equation}
Similar to Eq.~\ref{EqLabelFrac}, the ratio $\Phi_{12}=N_1/N_2$ obeys
\begin{equation} \dot{\Phi}_{12}=-\big((\bar{k}_{d1}-\bar{k}_{d2})+(\bar{k}_{r1}-\bar{k}_{r2})\big)\,\Phi_{12}\,,\label{eq:f12nonirr}\end{equation}
which includes the deletion and replacement rates of both labeled cell populations.

\noindent Additionally, we propose an irradiation experiment as depicted in Fig.~\ref{FigDoubleLabel}(b). 
For this, we choose the time interval $T$ such that there are still new cells arriving in the epidermis during $\Delta t$ but not so after the second labeling pulse. According to our measurements with a single label in Fig.~\ref{FigTissueTurnover}(f), $T+\Delta t\approx 5$. In consequence, $(\bar{k}_{r1}-\bar{k}_{r2})=0$ during the time interval $t$:
\begin{equation} \dot{\Phi}_{12,irr}=-(\bar{k}_{d1}-\bar{k}_{d2})\,\Phi_{12}\,.\label{eq:f12irr}\end{equation}
This shows that $\Phi_{12,irr}$ only changes for age-dependent deletion rates. Thus, by comparing the dynamics of $\Phi_{12}$ in non-irradiated and $\Phi_{12,irr}$ in irradiated worms given by Eq.~\ref{eq:f12nonirr} and \ref{eq:f12irr}, respectively, we can deduce whether deletion or replacement depends on the age of the cell, as illustrated in Fig.~\ref{FigDoubleLabel}(c). If $\Phi_{12,irr}$ changes according to a time-dependent rate, deletion is related to the age of the cells. If instead only the dynamics of $\Phi_{12}$ are determined by a time-dependent rate, replacement relies on the age of the cells.

\noindent Furthermore, we can extract actual rates for the individual processes by comparing the results to the previous experiments. For example, $\bar{k}_{d2}+\bar{k}_{r2}$ in Eq.~\ref{eq:f12nonirr} is equivalent to $\bar{k}_{r\ell}+\bar{k}_{d\ell}$ of the experiment with a single label. By combining these measurements, we can find the functional relationship for how deletion and replacement changes with the age of the cells. 

%\noindent Together with the dynamics of the young cells that were added in the interval $\Delta t$ and only carry the second label, we can further discriminate between different contributions to tissue turnover. The label fraction $f_1=N_1/N$ of these young cells with only one label changes according to
%\begin{equation} \dot{f}_{1}=-\big(k_{r1}+k_i+(k_{d1}-k_{d})\big)\,f_{1}\,.\end{equation}
%The interpretation of measurements of $f_1$ strongly depends on the outcome of the previous experiments.
%As a specific example, let us assume, we have obtained from Eq.~\ref{eq:f12irr} that deletion does not depend on age, such that $k_{d1}-k_{d}=0$. Furthermore, let us assume we measure a curve like in \ref{FigDoubleLabel}(b). The initial dynamics for small $t$ should provide an upper bound for the insertion rate $k_i$, which probably does not depend on the age of the existing cells in the tissue. In contrast, for larger $t$, we can extract the age-dependent replacement rate, which should agree with the result we obtained by comparing Eq.~\ref{eq:f12nonirr} and Eq.~\ref{eq:f12irr}. Thus, we might be able to estimate whether replacement or insertion dominates. This provides a first evidence whether the tissue turnover arises due to a tight coupling between stem cells and differentiated cells or whether a stochastic turnover model applies where insertion and deletion are rather independent processes.

\noindent The same experiments can be performed in starving and well fed animals to reveal the connection between feeding status and turnover rates. One hypothesis would be that there is no insertion during starvation, yet replacement dominates and the tissue slowly shrinks due to a small deletion rate. After feeding, insertion leads to growth of the tissue.

\section{Summary}
Our aim is to study cell turnover and growth in a comprehensive way, bridging the scales from a single cell to the average behavior at the level of the organism. For this, we have carefully designed several experiments to determine the flatworm size and to characterize its growth dynamics. In particular, we developed a protocol for precise measurement of the worm area based on the analysis of movie sequences with a custom MATLAB routine.
The data was compared to previous studies, which in parts yielded controversial results. %Our carefully designed and performed experiments contribute to settle the questions about .
%put these earlier results into question

\noindent We observed that the growth and degrowth rates are dependent on worm size. Smaller worms grow and shrink faster. This resonates with the picture that small worms might be more juvenile with a potentially higher turnover rate and a different metabolism \cite{gonzalezestevez2012decreased,lange1968possible}. Based on the idea that the available metabolic energy is the main determinant for growth, we have developed three distinct classes of microscopic models, which can all be fitted to the macroscopic growth data. The models describe dif\-ferent paradigms of energy storage and make specific predictions that are tested by our collaborators in current experiments. In particular, they also differ in the way energy inflow (i.e.~digestion of food) and energy outflow (i.e.~metabolic consumption) varies with size, which directly relates to the question about the limits to growth.

\noindent Finally, we propose further experiments, which can unravel the control logic of cell and tissue turnover. We provide the theoretical framework and discuss first preliminary data suggesting an age-dependent death of cells.
%Such signatures of cellular aging would rise the question whether the worm as a whole ages when growing and whether it can rejuvenate during degrowth, as discussed previously \cite{pellettieri2010cell, gonzalezestevez2012decreased,lange1968possible}. 
The experiments will be performed in the laboratory of Jochen Rink by the author and other members of the group.

% ---------------------------------------------------------------------------
% ----------------------- end of thesis sub-document ------------------------
% ---------------------------------------------------------------------------			
% this file is called up by thesis.tex
% content in this file will be fed into the main document

\chapter{Summary and outlook} \label{summary}% top level followed by section, subsection

% the code below specifies where the figures are stored
%\ifpdf
%    \graphicspath{{5_conclusion/figures/PNG/}{5_conclusion/figures/PDF/}{5_conclusion/figures/}}
%\else
%    \graphicspath{{5_conclusion/figures/EPS/}{5_conclusion/figures/}}
%\fi

% ----------------------- contents from here ------------------------

In this thesis, we analyzed various aspects of growth and body plan scaling, drawing particular inspiration from the flatworm \textit{Schmidtea mediterranea} (\smed{}) as a model animal.  For this purpose, we combined theoretical descriptions and analysis of experimental data.

\noindent As a first approach, we analyzed previously proposed mechanisms for self-organized pattern formation as well as pattern scaling in Chapter~\ref{patterningA}. We systematically extracted the requirements for scaling and demonstrated that they are challenged by the fascinating regeneration and re-patterning capabilities of flatworms.
On this basis, we presented a minimal model for fully self-organized and self-scaling pattern formation in Chapter~\ref{patterningB}. This mechanism is capable of spontaneously generating concentration gradients and expression regions of involved molecules that robustly adjust to the size of the system.  

\noindent In the future, it will be interesting to further compare the theoretical framework and the derived predictions to the Wnt/$\beta$-catenin system, which is associated with head-tail (AP) polarity in \smed{} \cite{gurley2010expression,almuedocastillo2012wnt,gurley2008betacatenin,adell2010gradients}. Preliminary measurements by our collaborators indicate that the $\beta$-catenin gradient as well as expression profiles of various Wnts in fact scale with worm size. Our theory enables us to interpret the data and to suggest additional experiments to unravel the mechanism of scaling in flatworms. Further beyond, the results might also enrich the discussions on signaling gradients in other systems, such as the developing fly wing, for which it is still a matter of debate how scaling is established \cite{benzvi2011expansion,averbukh2014scaling,wartlick2011understanding,fried2014dynamic}.

%\noindent The concentration profiles of signaling molecules are also found to scale in other organisms. For example for the case of the developing fly wing, promising mechanisms and potential molecular candidates have been discussed, yet it is still a matter of debate how scaling is established \cite{}. Our theory might provide an additional viewpoint and help to understand scaling also in systems different from flatworms.

\noindent \smed{} only recently evolved to become a model organism, thus the basic mechanisms for axis specification are far from being well understood \cite{lobo2012modeling}. It is known that components of the Wnt/$\beta$-catenin system are especially present in the tail \cite{gurley2010expression,almuedocastillo2012wnt,gurley2008betacatenin,adell2010gradients}. However, there could be also an additional patterning system originating from the head or even a compartmentalization by several systems along the body axis. If there are several anterior-posterior (AP) polarity systems, how do they interact? And to what extent is the patterning self-organized or relies on pre-existing cues such as a polarized tissue or wound-specific signals after amputation? We are currently analyzing a large data set comprising the spatial changes in gene expression after RNAi treatment and expression time courses during regeneration in order to answer these questions.

\noindent Perpendicular to the AP axis, there are also two more body axes in dorsal-ventral and medial-lateral direction \cite{lobo2012modeling}. How do the patterning systems for the different body
axis influence each other? Is there a hierarchy of axis formation or do the axes emerge
simultaneously? One particular curious case is the formation of a straight midline as an important signaling center.  The experimental data suggests that a repulsive signal from the body margin positions the midline \cite{meinhardt2004different,lobo2012modeling,adell2009smedevi}. Yet, is it sufficient to account for this narrow, straight row of distinct cells, which tends to re-emerge from the anterior and posterior poles during regeneration? As a complementary approach, we have started to design a model explaining midline formation on the basis of cell proliferation and resulting cellular flows in the tissue. 

\noindent Besides these long-range patterning mechanisms that are classically assumed to rely on secreted, motile molecules, there are also other polarity cues. For example, the planar-cell-polarity (PCP) system is based on direct cell-cell interactions and results in cellular polarisation consistently across the tissue. In order to determine the coupling between both mechanisms, we have started analyzing the motility of flatworms, when various Wnt-pathway components have been knocked out by RNAi feeding, see Chapter~\ref{pca}. Resulting movement phenotypes report on a dysfunctional cilia carpet associated with an impaired PCP system.

\noindent  For a further, more detailed investigation of worm motility, we adapted principal component analysis to apply it to the highly deformable worm body. We could demonstrate
that during normal gliding motion worms steer their path by bending in the preferred direction of movement. Additionally, we provided the first quantitative account of an alternative motility mode, called inchworming. This appears to be a tightly controlled
behavioral response to impaired cilia functionality with a characteristic frequency of about \mbox{$1/4$ Hz}. Such  stereotypic behaviors have been previously described in the much simpler nematode \mbox{\textit{C. elegans}} \cite{stephens2011emergence, Stephens:2008,stephens2010modes}. It is a fascinating observation that more complex organisms such as flatworms still show very generic motility patterns, which might have emerged during evolution as an optimized strategy and are expected to relate to the structure of the muscular plexus and the nervous system.
Similarly, we were able to demonstrate the applicability of shape mode analysis to the variable head shapes between different species. We are planning to extend this analysis to many more species of the large flatworm collection in the laboratory of Jochen Rink. As both, specific morphologies and specific movement strategies, are expected to emerge due to the evolutionary pressure in the respective environmental niche, our analysis provides the basis for future research relating form and function.

\noindent In the last chapter, we quantified growth and turnover dynamics in flatworms. Interes\-tingly, we obtained size-dependent growth rates. Small worms appear to grow and
degrow faster than large worms. This particular growth and degrowth behavior of the worms is the coarse-grained result from the underlying processes of cell division and cell loss. Thus, we devised several microscopic models describing specific rules on the cellular scale to explain the observed growth dynamics on the macroscopic level of the organism. The models are based on the idea that there is a limiting quantity that 
\sidenote{0.45\textwidth}{``Nothing in biology makes sense except in the light of evolution.'' \;---\; ditto, Theodosius Dobzhansky, 1973 \cite{dobzhansky1973nothing}}
(i) is provided by feeding, (ii) is permanently consumed by the cells and (iii) affects the division and loss rates of cells. As an example, we expli\-cit\-ly consider the metabolic energy as the limiting quantity. All of our three models fit the growth data equally well. Importantly, each model is based on distinct assumptions on how the energy is stored and how the energy availability affects the division and loss rates. Furthermore, each model makes specific predictions about how the three considered processes of energy influx by feeding, energy storage and energy consumption have to depend on the size of the worm to account for the measured growth data.

\noindent With this result in mind, we contribute to the design of future experiments that will enable us to distinguish between the models. Thereby, we aim to bridge scales by providing an explanation for the organismal growth dynamics on the macroscopic level in terms of cellular behavior on the microscopic scale. The outcome might also hint at systemic limitations to growth. For example, if the food uptake is size-dependent and relatively decreases with worm size, there will be an upper limit at which the worm is not capable to sustain growth anymore. It then poses the question whether this is a physical limit set by constraints of the worm body or whether evolution has not selected for a more efficient uptake because worms anyways do not grow bigger for other reasons.
Similarly, energy stores that depend on either size or feeding will trigger further investigations on the specific cells that store the energy. Are there specialized fat cells or does a broad range of cells store the energy? An interesting idea is related to the formation of the germ line. For sexual flatworms, there might be a tradeoff between growth and the development of a reproductive system. Small worms would rather grow to survive, large worms would rather invest in reproduction. Yet, germ line precursor cells still exist in the asexual strain used in the experiments. From an evolutionary viewpoint, this might hint at a dual role of these cells, suggesting that they are involved in growth control or energy storage. 

\noindent The level of detail in the theory reflected the coarse-graining level of the experimental data. In a next step, we aim to go beyond and include further measurements to dissect the control logic of growth and cell turnover. Which of the two processes of cell division and loss is affected by feeding status and worm size in which way? How do the two processes influence each other? To what extent are these turnover processes related to aging of the cells as well as aging of the organism? There are many open questions and we started looking into some of them. For example, we have discussed a theoretical framework for measurements from the cellular to organismal scale and also analyzed preliminary data, which suggests an age-dependent replacement of cells. Eventually, we aim to develop a comprehensive picture of how cell turnover and organism growth is controlled.

\noindent This thesis addresses various aspects of growth, cell turnover and scalable body plan patterning in flatworms. It illustrates the sheer diversity of biological processes, which are fascinating for biologists and physicists alike. Within a collaboration between experimentalists and theoreticians, we applied physical concepts to quantitatively analyze and interpret experimental data. Importantly, the gained insights enabled us to ask new questions and to suggest future experiments, which will ultimately help to grow our general understanding of homeostasis and growth, development and regeneration in multicellular organisms.

% ÔNothing in biology makes sense except in the light of evolutionÕ Theodosius Dobzhansky (1973)

% ---------------------------------------------------------------------------
% ----------------------- end of thesis sub-document ------------------------
% ---------------------------------------------------------------------------

%\include{8/materials_methods}        % description of lab methods

% --------------------------------------------------------------
%:                  BACK MATTER: appendices, refs,..
% --------------------------------------------------------------

% the back matter: appendix and references close the thesis
%\backmatter

%: Appendix
%\renewcommand{\appendixname}{}
\begin{appendix}
% this file is called up by thesis.tex
% content in this file will be fed into the main document

\chapter[Reaction-diffusion systems: fixed points and scaling]{Reaction-diffusion systems:\\ fixed points and scaling} \label{appreactdiff}% top level followed by section, subsection
\chaptermark{Reaction-diffusion systems: fixed points \& scaling} %only change chapter heading

% the code below specifies where the figures are stored
%\ifpdf
%    \graphicspath{{6_appendix/figures/PNG/}{6_appendix/figures/PDF/}{6_appendix/figures/}}
%\else
%    \graphicspath{{6_appendix/figures/EPS/}{6_appendix/figures/}}
%\fi

% ----------------------- contents from here ------------------------

\section{Morphogen dynamics with linear degradation}\label{appreactdiff:lindeg}
\subsection{Reaction, diffusion, advection and dilution}\label{appreactdiff:generaldyn}
Here, we consider general morphogen dynamics in a growing tissue. We assume that the morphogens are produced by a source term $\nu=\nu(\vec{r})$ and spread in the system with an effective diffusion coefficient $D$ while being subject to linear degradation with rate $\beta$. The dynamics can be described by the following convection-diffusion equation for the morphogen concentration $C=C(t,\vec{r})$ \cite{averbukh2014scaling,bittig2008morphogenetic}
%%Fischer, H. B., List, E. G., Koh, R. C. Y., Imberger, J. & Brooks, N. H. (1979), Mixing in Inland and Coastal, Waters, Academic Press, New York, NY OR https://ceprofs.civil.tamu.edu/ssocolofsky/cven489/Book/Book.htm
\begin{equation}\label{eq:convectdiff}
\partial_t C=D\,\nabla^2\,C-\beta\,C+\nu-\nabla \big(\vec{u}\, C\big)\,.
\end{equation}
The last term corresponds to dilution and convection due to tissue growth with a velocity field $\vec{u}=\vec{u}(t,\vec{r})$.

\noindent Spreading of morphogens from a localized source results in graded concentration profiles.
Many systems can be considered to be homogeneous with respect to all but one direction (e.g. the $x$-direction). In this case, the graded profiles $C=C(t,x)$ only depend on the respective spatial coordinate and the differential equation simplifies to
\begin{eqnarray}\label{eq:convecdiffx}
\partial_t C &=& D\,\partial_x^2\,C-\beta\,C+\nu-u_x\partial_x C-C \nabla \vec{u}\nonumber\\
 &=& D\,\partial_x^2\,C-(\beta+\partial_y u_y+\partial_z u_z)\,C+\nu-\partial_x (u_x C)\,.
\end{eqnarray}
Note that dilution by growth in the directions perpendicular to the graded profile can effectively be described by a degradation term. For slow growth dynamics, we can neglect the convection-dilution terms and recover Eq.~\ref{eq:mdiff}.

\subsection{Steady state solution neglecting tissue growth}\label{appreactdiff:steadystatenogrowth}
The steady state solution of Eq.~\ref{eq:convecdiffx} can be found in \cite{bittig2008morphogenetic}. Yet, often the effect of tissue growth is considered to be negligible or a second order perturbative effect to the solution of Eq.~\ref{eq:mdiff}. Thus, let us consider this limit of slow tissue growth. Furthermore, $D$ and $\beta$ are assumed to be constant in space across the tissue and constant in time within the time scale of the morphogen dynamics. As a result, the morphogen concentration approaches a steady state profile of the form
\begin{equation}
C^*(x)=c_1\,\text{exp}(x/\lambda)+c_2\,\text{exp}(-x/\lambda)\,.
\end{equation}
The steady state solution for reflecting boundary conditions is given by
\begin{equation}\label{eq:mstcomplete}
C^*(x)=\frac{\alpha}{\beta}\,\begin{cases}
1-\frac{\sinh(L/\lambda-w/\lambda)}{\sinh(L/\lambda)} \cosh(x/\lambda) & \text{for }0\leq x< w\\[0.3cm]
\frac{\sinh(w/\lambda)}{\sinh(L/\lambda)} \cosh(L/\lambda-x/\lambda) & \text{for }w\leq x\leq L
\end{cases}
\end{equation}
In the limit of a small source and a large system ($w\ll\lambda$, $\lambda\ll L$), the steady state simplifies to
\begin{equation}\label{eq:mstexp}
C^*(x)\approx \frac{\alpha\,w}{\sqrt{D\,\beta}}\,\begin{cases}e^{-x/\lambda}& \text{for}\, x\ll L\\[0.1cm]
2\,e^{-L/\lambda}& \text{for}\, x=L\,.
\end{cases}
\end{equation}
The square-root of $D$ and $\beta$ in the amplitude also emerges if there is no degradation in the source region, see Appendix~\ref{appreactdiff:nodegsource}.

\subsection{Relaxation to the steady state}\label{appreactdiff:relaxtosteadystate}
The degradation rate defines the time scale of relaxation to the steady state of Eq.~\ref{eq:mdiff}. Let us consider a concentration $C(t,x)=C^*(x)+\epsilon_c(t,x)$ that deviates from the steady state by a perturbation $\epsilon_c(t,x)$. The dynamics of $\epsilon_c(t,x)$ are described by
\begin{equation}
\partial_t\,\epsilon_c=D\,\partial_x^2\,\epsilon_c-\beta\,\epsilon_c\,.
\end{equation}
The concentration in Fourier space $\tilde{\epsilon}_c=\tilde{\epsilon}_c(t,s)$ obeys
\begin{equation}
\partial_t\,\tilde{\epsilon}_c=-\big((2\pi s)^2 D+\beta\big)\,\tilde{\epsilon}_c\quad\Rightarrow\quad \tilde{\epsilon}_c=\tilde{\epsilon}_{c,0}\, e^{-\big((2\pi s)^2 D+\beta\big)t}
\end{equation}
and the back-transform yields (see Eq.~\ref{appgauss:FBT} for details)
\begin{equation}
\epsilon_c(t,x)=e^{-\beta t}\,\int_{-\infty}^{\infty} dx'\,\epsilon_c(0,x')\, \frac{ e^{\frac{-(x'-x)^2}{4 D t}} }{\sqrt{4\pi Dt}}\,.
\end{equation}
Besides the term for diffusive spreading that homogenizes the initial perturbation $\epsilon_c(0,x)$, there is an exponential damping term. Thus, the perturbation of the steady state $C^*$ decays on a time scale of $1/\beta$.

\subsection{Steady state without morphogen degradation in the source}\label{appreactdiff:nodegsource}
The reaction-diffusion equation without morphogen degradation in the source region is
\begin{equation}\label{eq:reactdiffnosourcedeg}
\partial_t C=D\,\partial_x^2\,C-\beta\,C\,\Theta(x-w)+\alpha\,\Theta(w-x)\,.
\end{equation}
With reflecting boundary conditions as in Eq.~\ref{eq:mdiffbound}, we obtain
\begin{equation}\label{eq:mstnosourcedeg}
C^*(x)=\begin{cases}
\frac{\alpha\,w}{\sqrt{D\beta}\,\tanh(L/\lambda-w/\lambda)}+\frac{\alpha}{2D}(w^2-x^2)  & \text{for }0\leq x< w\\[0.3cm]
\frac{\alpha\,w}{\sqrt{D\beta}\,\sinh(L/\lambda-w/\lambda)} \cosh(L/\lambda-x/\lambda) & \text{for }w\leq x\leq L
\end{cases}
\end{equation}
While we obtain a power law inside the source, the expression outside the source is identical to the limit $w\ll\lambda$ of Eq.~\ref{eq:mstcomplete}.

\section{Gradient scaling with expander}
\subsection{On the scaling with an autonomously controlled expander}\label{appreactdiff:autonexp}
We consider a three-dimensional domain of volume $V=AL$, in which the graded morphogen profile forms along the $x$-direction of length $L$, whereas $A$ denotes the cross-sectional area. Now, we aim to achieve scaling with respect to some part of the system $L_1$ with $L=L_1+L_2$. For example, $L_1$ could be the target tissue without the morphogen source (if $L_2=w$) or the whole system (if $L_2=0$).

\noindent The expander dynamics are described by
\begin{equation}\label{eq:genedyn}
\partial_t E=D_E\,\nabla^2\,E-\beta_E\,E+\nu_E
\end{equation}
with a constant diffusion term $D_E$ and the terms for degradation $\beta_E=\beta_E(x) $ and production $\nu_E=\nu_E(x)$ that depend on the position in the tissue. In the limit of a fast diffusing expander, the general steady state of Eq.~\ref{eq:genedyn} is given by
\begin{equation}\label{eq:genest}
E^*=\langle \nu_E\rangle\,\left/\,\langle \beta_E\rangle\right.\,,
\end{equation}
where
\begin{equation}\label{eq:spataverage}
\langle \beta_E\rangle=\frac{1}{V}\int \beta_E\,dV\quad\text{and}\quad\langle \nu_E\rangle=\frac{1}{V}\int s_E\,dV
\end{equation}
denote the spatial averages. In order to achieve scaling of the morphogen profile with respect to $L_1$, the expander level has to be a function of $L_1$ in the steady state. Hence, $\langle \nu_E\rangle$ and $\langle \beta_E\rangle$ must not depend on $L_1$ in the same way.

\noindent If the expander is degraded everywhere in the system part of length $L_1$ und produced in a source of constant width $w_E$ like in Eq.~\ref{eq:ediffED}, we recover the result from Eq.~\ref{eq:est_eg} by computing
\begin{equation}\label{eq:estopt1}
\langle \beta_E\rangle= \frac{\beta_E\,A\,L_1}{V}\;,\quad\;\langle \nu_E\rangle= \frac{\alpha_E\,A\,w_E}{V}.
\end{equation}
The steady state of the expander would be a function of $L_1$ and thus could couple to the morphogen dynamics to generate scaling. Analogously, degradation could happen only at the boundary in a stripe of width $w_E$ and production might be turned on everywhere along the length $L_1$, thus
\begin{equation}\label{eq:estopt2}
\langle \beta_E\rangle= \frac{\beta_E\,A\,w_E}{V}\;,\quad\;\langle \nu_E\rangle= \frac{\alpha_E\,A\,L_1}{V}
\end{equation}
would also make $E^*$ a function of $L_1$.

\subsection{Scaling by expander feedback with a switch-like production}\label{appreactdiff:erstepprod}
We assume that the expander is suppressed by the morphogen in a switch-like manner according to Eq.~\ref{eq:ediffED} and discuss implications for scaling and robustness of the feedback schemes.

\noindent By combining Eq.~\ref{eq:exprepstM} and the steady state solution for reflecting boundary conditions given by Eq.~\ref{eq:mst}, we obtain
\begin{equation}\label{eq:expreprefl}
\frac{\alpha}{\beta^*}\,\frac{\sinh(w/\lambda^*)}{\sinh(L/\lambda^*)}\,\cosh(w_E/\lambda^*)=C_{th}\,.
\end{equation}
We can derive a relationship between $\beta^*$ and $E^*$ by replacing $w^*_E$ using Eq.~\ref{eq:exprepstE}
\begin{equation}\label{eq:exprepk}
\beta^*=\frac{\alpha}{C_{th}}\,\frac{\sinh(w/\lambda^*)}{\sinh(L/\lambda^*)}\,\cosh\left(\frac{\beta_E\,L\,E^*}{\alpha_E\,\lambda^*}\right).
\end{equation}
In order to illustrate the behavior of the feedback system, we are considering the two limiting cases of (i) a scaling morphogen source with $w=\chi_w L$ and (ii) a constant morphogen source. For both, we assume that there exists a perfectly scaling steady state with $\lambda^*=\chi_{\raisemath{-2pt}{\lambda}} L$ and a constant scaling factor $\chi_{\raisemath{-2pt}{\lambda}}$. Next, we show what this implies for the coupling between expander and morphogen and discuss the consequences. 

\noindent (i) For a scaling morphogen source, the following relation between $\beta^*$ and $E^*$ has to hold in order for the morphogen profile to scale in the steady state:
\begin{equation}\label{eq:exprepke}
\beta^*=\frac{\alpha}{C_{th}}\,\frac{\sinh(\chi_w/\chi_{\raisemath{-2pt}{\lambda}})}{\sinh(1/\chi_{\raisemath{-2pt}{\lambda}})}\,\cosh\left(\frac{\beta_E}{\alpha_E\chi_{\raisemath{-2pt}{\lambda}}}\,E^*\right)\,.
\end{equation}
The coefficient $\chi_{\raisemath{-2pt}{\lambda}}$ is a free parameter encoding the effect of the expander on $\beta$. Note that it is not possible to achieve scaling by only adjusting the diffusion coefficient $D$ instead of $\beta$ because this equation does not depend on $D$. Eq.~\ref{eq:exprepke} challenges the stability of the feedback loop because the expander has a suppressing effect on the morphogen (at least in the steady state), while according to our Eq.~\ref{eq:ediffED} the morphogen also suppresses the expander. Thus, this feedback tends to be unstable.

\noindent (ii) For a constant morphogen source, we obtain
\begin{equation}\label{eq:exprepkDe}
\left(\frac{\beta}{\sinh(w\sqrt{\beta/D})}\right)^*=\frac{\alpha}{C_{th}\,\sinh(1/\chi_{\raisemath{-2pt}{\lambda}})}\,\cosh\left(\frac{\beta_E}{\alpha_E\chi_{\raisemath{-2pt}{\lambda}}}\,E^*\right)\,.
\end{equation}
Now, either of both, $\beta$ or $D$, can be adjusted for scaling. The question on the stability is a bit more complicated than in the case above, where an increase in $\beta$ results in a decrease of the morphogen level everywhere in the system. In contrast, for example, an increase in $D$ results in a decrease of the morphogen level close to the morphogen source and an increase at the other side of the system because the morphogen is distributed more homogeneously. In consequence, the feedback can be stable in a limited size range, as long as the resulting expander source is sufficiently far away from the morphogen source.~As the size of the expander source changes stronger than linearly with $L$ (compare to Eq.~\ref{eq:est_eg}), the system will eventually reach a size at which this is not fulfilled anymore. Analogously, if $\beta$ is adjusted, there also is a size limit. At this size, $\beta$ is too small such that the effect of $E$ on $\beta$ is positive according to Eq.~\ref{eq:exprepkDe}. The size limit corresponds to $\lambda$ larger than $w$, which is likely the case for most of the systems as $w$ is kept constant.

\subsection{Scaling by expander feedback with a graded production}\label{appreactdiff:ercontprod}
The step-like production term in Eq.~\ref{eq:ediffER} yields rather complicated relationships between expander and morphogen dynamics, see Eq.~\ref{eq:exprepke} and \ref{eq:exprepkDe}. Instead, when considering a non-linear production term
\begin{equation}\label{eq:se2}
\partial_t E(t,x)=D_E\,\partial_x^2\,E(t,x)-\beta_E\,E(t,x)+\alpha_E\,C(x)^h
\end{equation}
with an arbitrary real number $h$, one obtains analogous results in a more general and much clearer fashion. The exponent $h$ can be positive or negative, corresponding to an enhancing or a repressing effect on the source size, respectively. The term can be understood as an approximation to a Hill function for concentrations below the saturation.

\noindent The expander concentration in the steady state for the limit of fast diffusion is
\begin{equation}\label{eq:est2}
E^*=\frac{\alpha_E\,\lambda^*}{\beta_E\,L}\int_0^{L/\lambda^*}C^*(x')^h\,dx'\quad\text{with}\quad x'=x/\lambda^*\,.
\end{equation}
Again, we will distinguish between the limiting cases of (i) a scaling morphogen source and (ii) a morphogen source of constant size.

\noindent (i) For a scaling source, we obtain from Eq.~\ref{eq:mst}
\begin{equation}\label{eq:scalsource2}
E^*=\frac{\alpha_E}{\beta_E}\,\left(\frac{\alpha}{\beta^*}\right)^h\,\mathcal{I}_1(L/\lambda^*)\quad\Rightarrow\quad \beta^*\propto E^{*\,-1/h}\,,
\end{equation}
where $\mathcal{I}_1$ is a short-hand form of the integral term. In analogy to the switch-like source in Section~\ref{appreactdiff:erstepprod}, we cannot obtain scaling by solely adjusting the diffusion properties. If the degradation rate is adjusted, the steady state tends to be unstable, irrespective of the exponent $h$, which characterizes whether the morphogen has an enhancing or suppressing effect.

\noindent (ii) If we assume $w$ is constant, it follows
\begin{eqnarray}\label{eq:constsource2}
&&E^*=\frac{\alpha_E}{\beta_E}\,\left(\frac{\alpha\,\sinh(w\, \sqrt{\beta/D})}{\beta}\right)^{*\,h}\,\mathcal{I}_2(L/\lambda^*)+\mathcal{O}(w/\lambda^*)^2\\[0.2cm]
&&\Rightarrow\quad (\beta\,D)^*\propto E^{*\,-2/h}\quad \text{for } w\ll\lambda^*\,\label{AppGradProdFeedii}
\end{eqnarray}
with $\mathcal{I}_2$ denoting the integral. A corresponding expression like in Eq.~\ref{AppGradProdFeedii} can be derived without the limit $w\ll\lambda$ if there is no morphogen degradation in the source, see Appendix~\ref{appreactdiff:nodegsource}. As before, there can be a stable scaling fixed point if either $\beta$ or $D$ is controlled, yet only in a limited size range.

%\section{Gradient scaling without expander}
%nonlinear degradation, growth, non-steady state

\section{Linear stability analysis of a Turing system}\label{appreactdiff:linstabTuring}
\subsection{Eigenvalues of the linearized reaction-diffusion matrix}
The linearized reaction-diffusion system in Fourier space given by Eq.~\ref{eq:linTuringsys} is characterized by the $2\times 2$-matrix $M_s$ in Eq.~\ref{eq:linTuringmat}:
\begin{equation}\label{eq:linTuringmatGeneric}
\text{\huge M}_{\text{\large s}}=\begin{pmatrix} M^1_{s} & M^2_{s} \\[0.3cm] M^3_{s}  & M^4_{s} \end{pmatrix}
\end{equation}
 The two eigenvalues $q^{I}_s,\,q^{II}_s$ of $M_s$ determine the linear stability of the homogeneous steady state, about which the system is linearized. If the real parts of both eigenvalues are negative, the steady state is stable, otherwise the steady state is unstable.
 
\noindent The eigenvalues are given by
\begin{eqnarray}
&&\left(M^1_s-q_s^{I/II}\right)\left(M^4_s-q_s^{I/II}\right)-M^3_s M^2_s=0\nonumber\\[0.3cm]
&&\left(q_s^{I/II}\right)^2-\text{Tr}[M_s]\,q_s^{I/II}+\text{Det}[M_s]=0\nonumber\\[0.3cm]
&&q_s^{I/II}=\frac{\text{Tr}[M_s]}{2}\pm\sqrt{\left(\frac{\text{Tr}[M_s]}{2}\right)^2-\text{Det}[M_s]}\,,
\end{eqnarray}
\begin{figure}[tbp]
\centering
\includegraphics[width=1\textwidth]{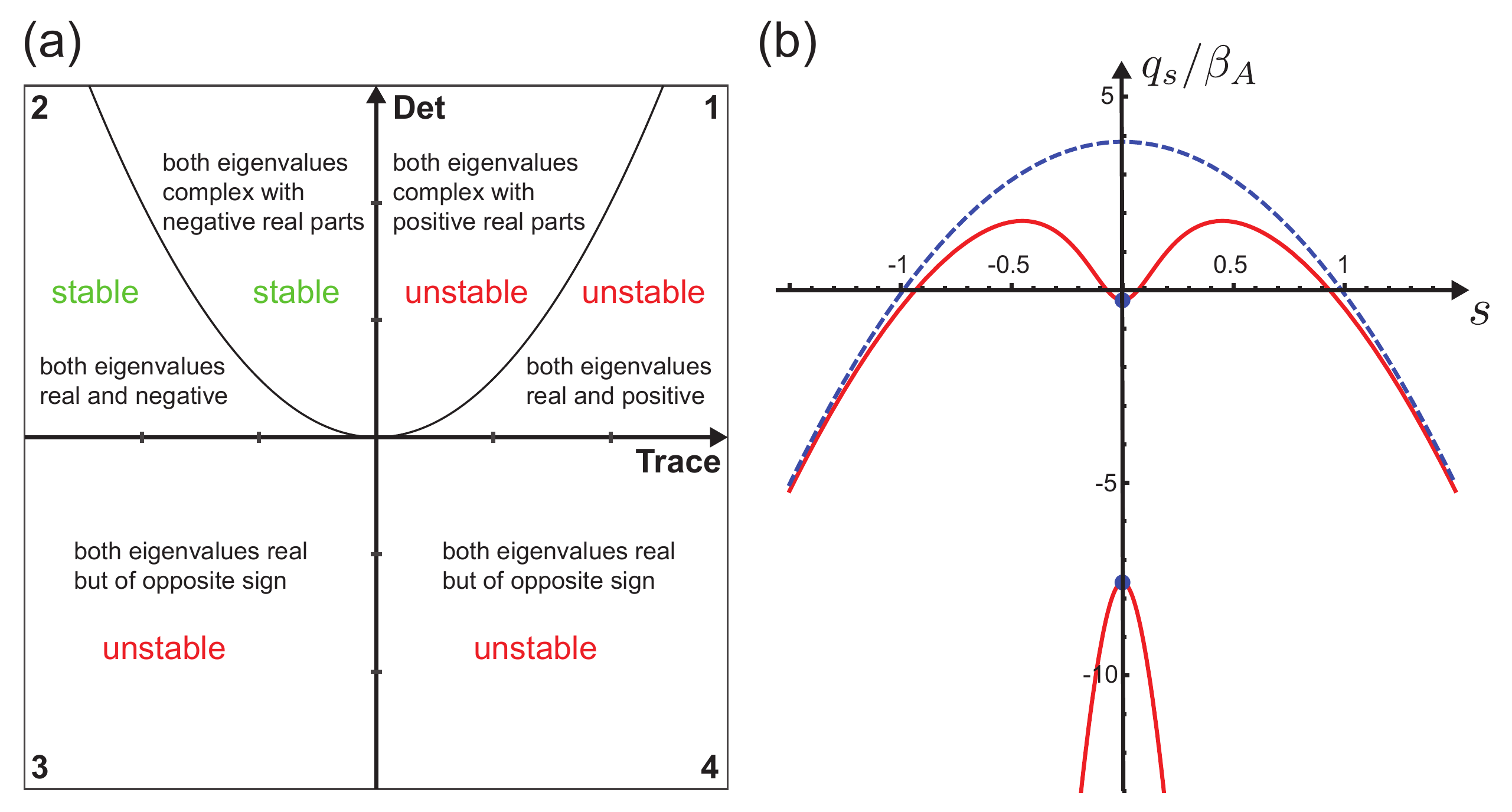}
\caption[Eigenvalues of the reaction-diffusion matrix]{\small 
(a) The trace-determinant plot summarizes the properties of the two eigenvalues of the $2\times 2$-matrix $M_s$. (b) The two eigenvalues of the matrix $M_s$ for the parameters used in Section~\ref{TuringNoScaling} with finite $D_B$ (red) and $D_B\rightarrow\infty$ (blue). For our choice of parameters and reflecting boundary conditions only the first mode ($s=0.5$) is linearly unstable, compare to Fig.~\ref{ModeStabPlot}.
Parameters: $D_B/D_A = 30$, $\alpha_B/\alpha_A = 4$, $\beta_B/\beta_A = 2$, $\lambda_A/L=\sqrt{0.1}\approx 0.3$, $\lambda_B/L=\sqrt{1.5}\approx 1.2$, $h=5$ (if not stated otherwise).
}
  \label{fig:TrDetEigval}
\end{figure}\noindent
where ``Tr" denotes the trace and ``Det" denotes the determinant. The properties of the eigenvalues can be summarized in the Tr-Det-diagram in Fig.~\ref{fig:TrDetEigval}(a). If $\text{Tr}[M_s]^2<4\,\text{Det}[M_s]$, both eigenvalues are complex and, in fact, are complex conjugates of each other. For $\text{Tr}[M_s]<0$ both real parts are negative, otherwise both real parts are positive.

\noindent Instead, if $\text{Tr}[M_s]^2<4\,\text{Det}[M_s]$, both eigenvalues are real. For $\text{Det}[M_s]>0$ both eigenvalues have the same sign: a positive sign for $\text{Tr}[M_s]>0$ and a negative sign for $\text{Tr}[M_s]<0$. This can be easily seen from 
\begin{equation}
q^I_s+q^{II}_s=\text{Tr}(M_s) \quad\text{and}\quad q^I_s\cdot q^{II}_s=\text{Det}(M_s)\,.
\end{equation}
Finally, for $\text{Det}[M_s]<0$ both eigenvalues are real but of opposite sign.

\noindent For the homogeneous steady state to be stable with respect to homogeneous perturbation ($s=0$), both real parts of the eigenvalues $q^{I}_0,\,q^{II}_0$ have to be negative. Thus, the matrix $M_0$ should be located in the second quadrant, where
\begin{equation}
\text{Tr}(M_0)<0 \quad\text{and}\quad \text{Det}(M_0)>0\,.
\label{eigcond1}\end{equation}
For the homogeneous steady state to be unstable with respect to inhomogeneous perturbation ($s\neq 0$), at least one of the eigenvalues $q^{I}_s,\,q^{II}_s$ has to have a positive real part.
It turns out that by changing $s$, the trace cannot become positive if $\text{Tr}(M_0)<0$:
\begin{equation} \text{Tr}(M_s)=M^1_s+ M^{4}_s= M^1_0+ M^{4}_0-(D_A+D_B)(2\pi s/L)^2< 0\,.
\end{equation}
Thus, the instability for $s\neq 0$ has to correspond to the third quadrant, for which both eigenvalues are real and one of them is positive:
\begin{equation}
\text{Det}(M_s)<0 \quad\text{with }s\neq 0\,.\label{eigcond2}
\end{equation}
Note that $\text{Tr}(M_s)$ decreases with increasing $s$, while $\text{Det}(M_s)$ first decreases and later increases again according to
\begin{equation} \text{Det}(M_s)=\text{Det}(M_0)-(D_A\,M^{4}_0+D_B\,M^{1}_0)(2\pi s/L)^2+D_A\,D_B\,(2\pi s/L)^4\,.
\end{equation}
As a consequence, the eigenvalues of a Turing system depend on the mode number $s$ as depicted in Fig.~\ref{fig:TrDetEigval}(b). For an intermediate range of $s$ one eigenvalue is positive and the system is linearly unstable with respect to the corresponding modes.

\subsection{The principle of local activation and lateral inhibition}
We derive necessary conditions for spontaneneous pattern formation in a two-component reaction diffusion system.
This section is based on analogous derivations in the literature \cite{turing1952chemical, segel1972dissipative, gierer1988biological, gierer1981generation, okubo2002diffusion}.
 For the following, we define
\begin{eqnarray} R_{AA}=\partial_A R_A\big\vert_{A^*_h,B^*_h}\,,\;R_{AB}=\partial_B R_A\big\vert_{A^*_h,B^*_h},\nonumber\\
R_{BA}=\partial_A R_B\big\vert_{A^*_h,B^*_h}\,,\;R_{BB}=\partial_B R_B\big\vert_{A^*_h,B^*_h}\,.\end{eqnarray}
The conditions in Eq.~\ref{eigcond1} and \ref{eigcond2} result in the relations
\begin{eqnarray}
R_{AA}+R_{BB}&<&0\qquad\label{cond1}\\[0.3cm]
R_{AA}\,R_{BB}-R_{AB}\,R_{BA}&>&0\label{cond2}\\[0.3cm]
\Big( R_{AA}-D_A(2\pi s/L)^2\Big )\Big( R_{BB}-D_B(2\pi s/L)^2\Big)&<& R_{BA}\,R_{AB}\,.\label{cond3}
\end{eqnarray}
The last two inequalities can be combined to
\begin{equation}R_{AA}\,D_B> -R_{BB}\,D_A+D_A\,D_B\,(2\pi s/L)^2\,.\label{cond23}
\end{equation}
Eq.~\ref{cond1} tells us that at least one of the two species has to have a self-inhibiting effect. Here, we choose species $B$: $R_{BB}<0$. From Eq.~\ref{cond23}, we see that the other one, i.e.~$A$, has to be self-enhancing: $R_{AA}>0$. From Eq.~\ref{cond2} follows that
$ R_{AB}\,R_{BA}<0$. The cross-reaction terms must be of opposing signs.\\
As $R_{BA}\,R_{AB}<0$ and $R_{BB}-D_B(2\pi s/L)^2<0$ in Eq.~\ref{cond3}, it follows that
\begin{equation} R_{AA}-D_A(2\pi s/L)^2>0\,.\end{equation}
The diffusion coefficient $D_A$ of the self-activator has to be sufficiently small for certain modes to become unstable (``local activation"). This becomes even more apparent if we consider the reaction term $R_A=\alpha_A P(A,B)-\beta_A\,A$ like in Eq.~\ref{EqReactDiffChoice}. Now, the characteristic wavelength of the reaction-diffusion equation has to be sufficiently small in comparison to the system size: $(\lambda_A/L)^2<(\alpha_A\partial_A P\big\vert_{A^*_h,B^*_h}/\beta_A -1)/(2\pi s)^2$. The weaker the self-activating effect and the larger the wave number $s$ of the perturbation, the smaller $\lambda_A$ has to be.\\
Finally, from Eq.~\ref{cond23}, we obtain 
$R_{AA}\,D_B> -R_{BB}\,D_A$. With Eq.~\ref{cond1}, hence \mbox{$R_{AA}<-R_{BB}$}, we see that the self-inhibitor has to diffuse faster than the self-activator (``lateral inhibition"):
\begin{equation} D_B>D_A\,.
\end{equation}

\section{Motivation for the Hill function}\label{appreactdiff:HillFunc}\noindent
The production function $P(A,B)$ is a switch-like element that takes two inputs and compares them. In a biological system the inputs are the concentrations of two molecules that bind to the receptors of a cell.  We discuss two possibilities to compare the concentrations of activator $A$ and the inhibitor $B$ by simple generic mechanisms that describe binding and unbinding to receptors. The number of unbound receptors on a cell is given by $Q$. The output could be the number of receptors $Q_A$ that are bound to the activator molecule.\\[0.3cm]
{\bf 1.~Case:} The activator $A$ binds to the receptor $Q$ while the inhibitor $B$ leads to unbinding of $A$. The dynamic equation for concentration of receptors bound to $A$ is
\begin{equation}\dot{Q}_A=\gamma_1\,A^{h_1}\,Q-\gamma_2\,B^{h_2}\,Q_A-\gamma_3\,Q_A\,,\end{equation}
where $h_1$ and $h_2$ describe cooperativity effects for binding and unbinding. For example, if the activator only binds in pairs, $h_1=2$. The coefficients $\gamma_1$, $\gamma_2$ and $\gamma_3$ are the rates of binding as well as induced and spontaneous unbinding. If the total concentration of receptors $Q_0=Q+Q_A$ is constant, the steady state reads
\begin{equation}Q_A=\frac{\gamma_1\,A^{h_1}\,Q_0}{\gamma_3+\gamma_2\,B^{h_2}+\gamma_1\,A^{h_1}}\quad\underset{\gamma_3=0}{\xrightarrow{h=h_1=h_2}}\quad\gamma_1\,Q_0\;\frac{(A/B)^h}{\gamma_2/\gamma_1+(A/B)^h}\,.\end{equation}
If the cooperativity exponents are the same and the activator rarely unbinds sponaneously, the standard Hill function is recovered. A finite $\gamma_3\neq 0$ allows to compute the limit $A=B=0$.\\[0.3cm]
{\bf 2.~Case:} We have seen that if the inhibitor facilitates the unbinding of the activator, we obtain a Hill function for the concentration of receptors bound to the activator. Another option for implementing the inhibiting effect is competitive binding. Here, the inhibitor binds to the same receptors as the activator and thus blocks activator binding. Now, the dynamic equations are
\begin{eqnarray}\dot{Q}_A&=&\gamma_1\,A^{h_1}\,Q-\gamma_2\,Q_A\label{dynQA}\\
\dot{Q}_B&=&\gamma_3\,B^{h_2}\,Q-\gamma_4\,Q_B\label{dynQB}\\
\dot{Q}&=&-(\gamma_1\,A^{h_1}+\gamma_3\,B^{h_2})\,Q+\gamma_2\,Q_A+\gamma_4\,Q_B\label{dynQ}\,.\end{eqnarray}
In the steady state Eq.~\ref{dynQA} results in
\begin{equation} Q=\frac{\gamma_2}{\gamma_1\,A^{h_1}}\,Q_A\end{equation}
and Eq.~\ref{dynQB} yields
\begin{equation} Q_B=\frac{\gamma_3\,B^{h_2}}{\gamma_4}\,Q=\frac{\gamma_3\,B^{h_2}(Q_0-Q_A)}{\gamma_4+\gamma_3\,B^{h_2}}\end{equation}
with $Q=Q_0-Q_A-Q_B$, again assuming a constant total receptor number $Q_0$. When inserting these two equations in the steady state of Eq.~\ref{dynQ}, we finally get
\begin{eqnarray}&&-(\gamma_1\,A^{h_1}+\gamma_3\,B^{h_2})\,\frac{\gamma_2}{\gamma_1\,A^{h_1}}\,Q_A+\gamma_2\,Q_A+\gamma_4\,\frac{\gamma_3\,B^{h_2}(Q_0-Q_A)}{\gamma_4+\gamma_3\,B^{h_2}}=0\nonumber\\[0.3cm]
&&\frac{\gamma_2\,\gamma_3\,B^{h_2}}{\gamma_1\,A^{h_1}}\,Q_A+\gamma_4\,\frac{\gamma_3\,B^{h_2}}{\gamma_4+\gamma_3\,B^{h_2}}\,Q_A=\gamma_4\,\frac{\gamma_3\,B^{h_2}}{\gamma_4+\gamma_3\,B^{h_2}}\,Q_0\nonumber\\[0.3cm]
&&\frac{\gamma_2\,\gamma_4+\gamma_2\,\gamma_3\,B^{h_2}}{\gamma_4\,\gamma_1\,A^{h_1}}\,Q_A+Q_A=Q_0\nonumber\\[0.3cm]
&&Q_A=\frac{\gamma_4\,\gamma_1\,A^{h_1}\,Q_0}{\gamma_2\,\gamma_4+\gamma_2\,\gamma_3\,B^{h_2}+\gamma_4\,\gamma_1\,A^{h_1}}\,.\end{eqnarray}
This is also a Hill function with exactly the same structure as the one above. It becomes apparent that this function stands for a wide class of realizations of a logical element that compares the magnitude of two inputs in a biological system.\\

\section{Homogeneous steady state of our Turing system}\label{appreactdiff:HomSteady}
\begin{figure}[tp]
\centering
\includegraphics[width=1\textwidth]{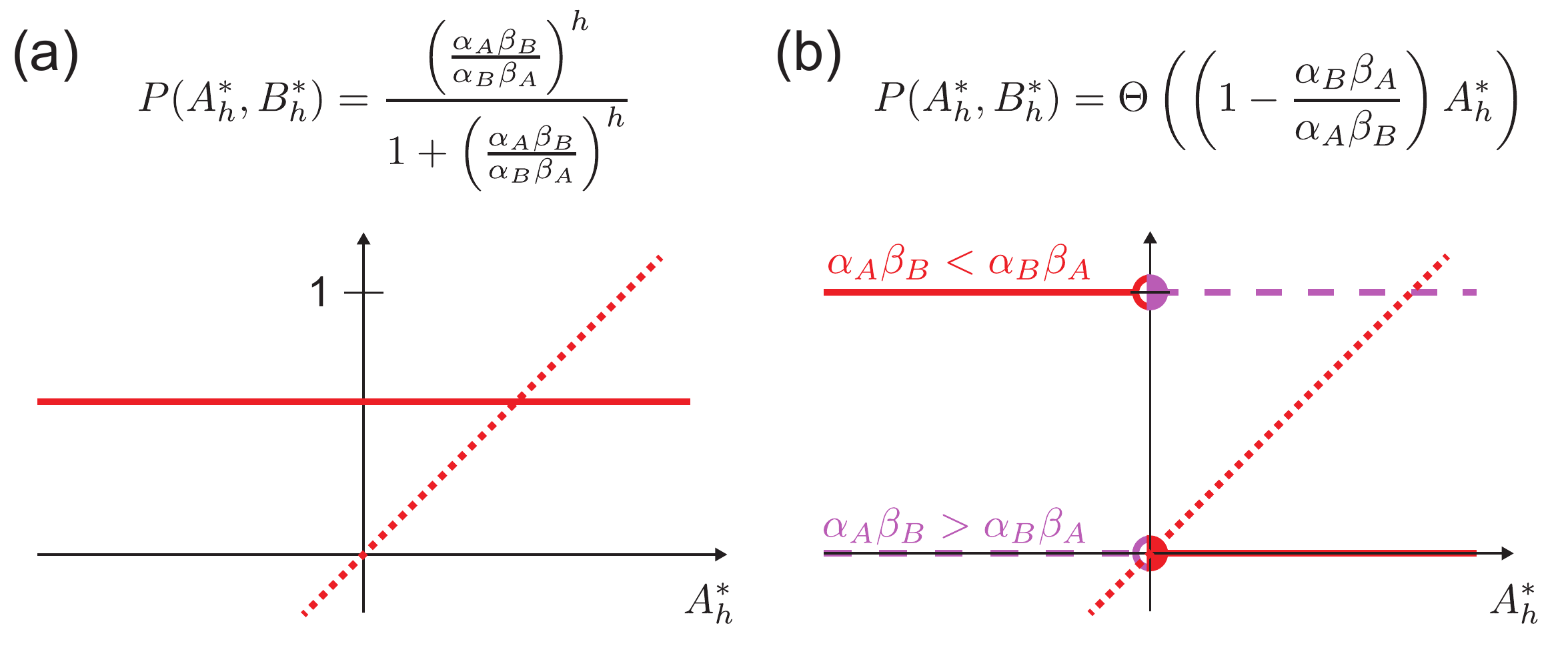}
\caption[Homogeneous steady state of our Turing system]{\small 
Homogeneous steady state of our Turing system for (a) a Hill function and (b) a theta function. Solid red (and dashed purple) lines correspond to the left side of Eq.~\ref{Eq:TuringHomStead2}, dotted red lines to the right side.
}
  \label{fig:HillThetaSS}
\end{figure}

\noindent
The homogeneous steady state of our choice of Turing system given by Eq.~\ref{EqReactDiffChoice} is characterized by
\begin{eqnarray} \label{Eq:TuringHomStead1}
&&A_h^*=\frac{\alpha_A\beta_B}{\alpha_B\beta_A}\,B_h^*\\[0.3cm]
&&P(A_h^*,B_h^*)=\frac{\beta_A}{\alpha_A}\,A_h^*\,. \label{Eq:TuringHomStead2}
\end{eqnarray}
Fig.~\ref{fig:HillThetaSS} illustrates the steady state in terms of $A_h^*$ for the Hill function of Eq.~\ref{EqHill} and the step function of Eq.~\ref{eq:prodtheta}. The solid red line corresponds to the left side and the dotted line to the right side of Eq.~\ref{Eq:TuringHomStead2}. The intersections mark the steady state value of $A^*_h$. There is a second case for the step function (purple dashed line).

\noindent We use a specific definition of the theta function at position zero
\begin{equation}
\Theta(0)\stackrel{!}{=}\begin{cases} 0& \text{if } \frac{\alpha_A\beta_B}{\alpha_B\beta_A}\leq 1\\[0.2cm]
1 & \text{if } \frac{\alpha_A\beta_B}{\alpha_B\beta_A}>1\,.
\end{cases}
\end{equation}
These two cases correspond to $A_h^*\leq B_h^*$ and $A_h^*>B_h^*$, respectively. This definition is chosen to ensure the existence of a steady state and to avoid artificial oscillations in the simulations. Moreover, it establishes the correspondence between the theta function and the Hill function in the limit of $h\rightarrow\infty$.

\noindent Next, we will show that only the case of $A_h^*\leq B_h^*$ ensures spontaneous pattern formation irrespectively of $h$.
For this, we explicitly perform a linear stability analysis of our specific Turing system. The  matrix of Eq.~\ref{eq:linTuringmatGeneric} for the linearized system reads
\begin{equation}\label{eq:linTuringmatExplicit}
\text{\huge M}_{\text{\large s}}=\begin{pmatrix} \alpha_A\,\partial_A P\big\vert_{A^*_h,B^*_h}-\beta_A-D_A(2\pi s/L)^2 & \alpha_A\partial_B P\big\vert_{A^*_h,B^*_h}\\[0.3cm] \alpha_B\,\partial_A P\big\vert_{A^*_h,B^*_h} & \alpha_B\partial_B P\big\vert_{A^*_h,B_h}-\beta_B-D_B(2\pi s/L)^2\end{pmatrix}\,.
\end{equation}
It includes the derivatives of the Hill functions with respect to $A$ and $B$ at steady state:
\begin{eqnarray}
\partial_A P\big\vert_{A^*_h,B^*_h}&=&\frac{h\,\beta_A/\alpha_A}{1+(A_h^*/B_h^*)^h}\\[0.2cm]
\partial_B P\big\vert_{A^*_h,B^*_h}&=&-\frac{A_h^*}{B_h^*}\;\partial_A P\big\vert_{A^*_h,B^*_h}\,.
\end{eqnarray}
The limit of $h\rightarrow\infty$ diverges for $\alpha_A\beta_B\leq\alpha_B\beta_A$:
\begin{equation}
\partial_A P\big\vert_{A^*_h,B^*_h}\rightarrow\begin{cases}\infty&\text{if } \frac{\alpha_A\beta_B}{\alpha_B\beta_A}\leq 1\\[0.2cm]
0 &\text{if }  \frac{\alpha_A\beta_B}{\alpha_B\beta_A}> 1\end{cases}\,.
\end{equation}
This is in in agreement with the theta function, for which the derivative diverges at position zero and is zero everywhere else.

\noindent The determinant of the matrix $M_s$ is
\begin{equation}
\text{Det}(M_s)=\Big(\beta_A+D_A(2\pi s/L)^2\Big)\Big(\beta_B+D_B(2\pi s/L)^2\Big)-\frac{\beta_A\,\beta_b\,(2\pi s/L)^2\, h}{1+(A_h^*/B_h^*)^h}\Big(\lambda_B^2-\lambda_A^2\Big)\,.
\end{equation}
Thus, the instability condition of Eq.~\ref{eigcond2} is given by
\begin{equation}
\Big(\beta_A+D_A(2\pi s/L)^2\Big)\Big(\beta_B+D_B(2\pi s/L)^2\Big)<\frac{\beta_A\,\beta_b\,(2\pi s/L)^2\, h}{1+(A_h^*/B_h^*)^h}\Big(\lambda_B^2-\lambda_A^2\Big)\,.
\end{equation}
First of all, this requires $\lambda_B>\lambda_A$, typical for Turing systems. Furthermore, in order to fulfill this inequality for arbitrary $h$ and in particular in the limit $h\rightarrow\infty$, it follows that
\begin{equation}  A^*_h\leq B_h^*\;, \qquad \frac{\alpha_A\beta_B}{\alpha_B\beta_A}\leq 1
\end{equation}
Our choice of parameters in Chapter~\ref{patterningA} amounts to $(\alpha_A\beta_B)/(\alpha_B\beta_A)=1/2$. The second set of conditions of Eq.~\ref{eigcond1} result in
\begin{eqnarray}\label{eigcond11}
&& \frac{h}{1+(A_h^*/B_h^*)^h}\,(\beta_A-\beta_B)<\beta_A+\beta_B\\[0.4cm]
&& \beta_A+\beta_B>0\label{eigcond12}
\end{eqnarray}
While Eq.~\ref{eigcond12} is trivially fulfilled, Eq.~\ref{eigcond11} requires $\beta_A<\beta_B$ in order to hold for arbitrarily large $h$. In our parameter sets of Chapter~\ref{patterningA}, we have chosen $\beta_B/\beta_A=2$.\\

\section{Inhomogenous steady states of our Turing system}\label{appreactdiff:InhomSteady}\noindent
\subsection{First order steady state solution}
\begin{figure}[tp]
\centering
\includegraphics[width=0.9\textwidth]{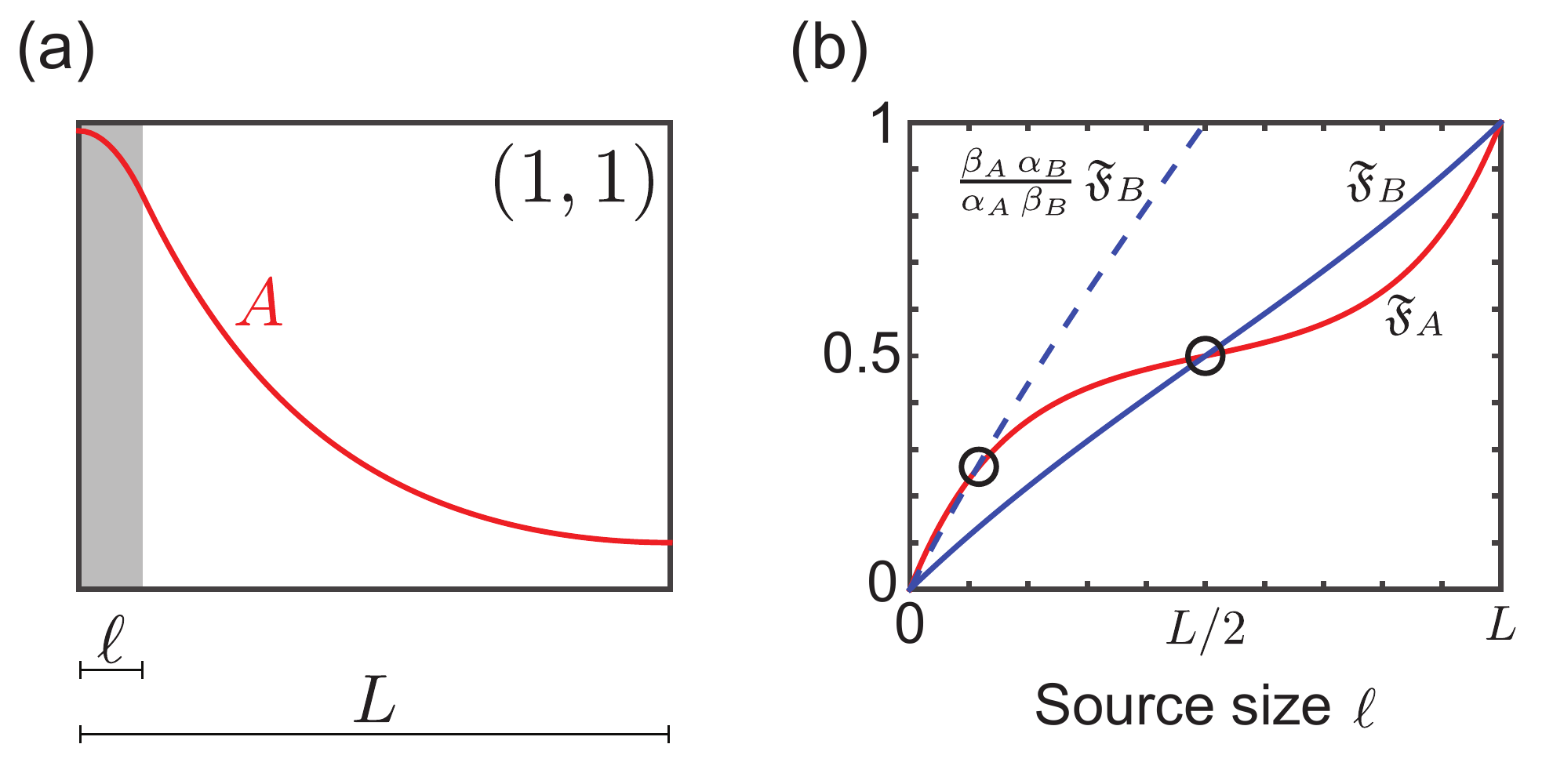}
\caption[$(1,1)$-pattern and its source size]{\small 
(a) The concentration profile of $A^*_{(1,1)}$ can be written as a piecewise function for the part inside and outside the source. (b) The intersections (circles) of the blue and the red curves mark the steady state source size according to Eq.~\ref{eq:Turingellsteadystate}. Parameters like in Section~\ref{TuringNoScaling}: $D_B/D_A = 30$, $\alpha_B/\alpha_A = 4$, $\beta_B/\beta_A = 2$, $\lambda_A/L=\sqrt{0.1}\approx 0.3$, $\lambda_B/L=\sqrt{1.5}\approx 1.2$.
}
  \label{fig:Turing11Source}
\end{figure}

First, we solve Eq.~\ref{EqReactDiffChoice} of our Turing system with a step-like production function by assuming a single source region which is touching the boundary as shown in Fig.~\ref{fig:Turing11Source}(a). We refer to this as the $(1,1)$-pattern. The general steady state solution for $A_{(1,1)}^*$ is a piecewise function
\begin{eqnarray}
&&A^*_{in}(x)=\mathfrak{A}_{in,1}\,e^{x/\lambda_A}+\mathfrak{A}_{in,2}\,e^{-x/\lambda_A}+\frac{\alpha_A}{\beta_A}\qquad \text{for } 0<x\leq \ell\\
&&A^*_{out}(x)=\mathfrak{A}_{out,1}\,e^{x/\lambda_A}+\mathfrak{A}_{out,2}\,e^{-x/\lambda_A}\qquad\quad\;\, \text{for } \ell<x<L
\end{eqnarray}
for the region inside and outside the source. In order to determine the four constant factors, we use the two reflecting boundary conditions and two continuity conditions at the source boundary:
\begin{eqnarray}
&&\partial_x A^*_{in}(x)\big\vert_{0}=0\label{eq:TuringSSBound1}\\
&&\partial_x A^*_{out}(x)\big\vert_{L}=0\label{eq:TuringSSBound2}\\
&&A^*_{in}(\ell)=A^*_{out}(\ell)\label{eq:TuringSSBound3}\\
&&\partial_x A^*_{in}(x)\big\vert_{\ell}=\partial_x A^*_{out}(x)\big\vert_{\ell}\,.\label{eq:TuringSSBound4}
\end{eqnarray}
From Eq.~\ref{eq:TuringSSBound1}-\ref{eq:TuringSSBound2}, we obtain $\mathfrak{A}_{in,1}=\mathfrak{A}_{in,2}$ and $\mathfrak{A}_{out,1}=\mathfrak{A}_{out,2}\,e^{-2L/\lambda_A}$, respectively. Thus, the solution becomes
\begin{eqnarray}
&&A^*_{in}(x)=\mathfrak{A}_{in,0}\,\cosh\left(\frac{x}{\lambda_A}\right)+\frac{\alpha_A}{\beta_A}\qquad \text{for } 0\leq x\leq \ell\\[0.2cm]
&&A^*_{out}(x)=\mathfrak{A}_{out,0}\,\cosh\left(\frac{x-L}{\lambda_A}\right)\qquad \text{for } \ell<x\leq L
\end{eqnarray}
Finally, from Eq.~\ref{eq:TuringSSBound3}-\ref{eq:TuringSSBound4}, we compute the $(1,1)$-solution
\begin{equation}A^*_{(1,1)}=\frac{\alpha_A}{\beta_A}\,\begin{cases}1-\frac{\sinh(L/\lambda_A-\ell/\lambda_A)}{\sinh(L/\lambda_A)}\,\cosh(x/\lambda_A)& \text{for }0\leq x\leq \ell\\[0.3cm]
\frac{\sinh(\ell/\lambda_A)}{\sinh(L/\lambda_A)}\,\cosh(x/\lambda_A-L/\lambda_A)& \text{for }\ell<x\leq L\,.
\end{cases}\label{eq:Asol}
\end{equation}
The solution for $B_{(1,1)}^*$ is determined  analogously, see Eq.~\ref{Bsol}.
%In complete analogy, the solution for $B_{(1,1)}^*$ is
%\begin{equation}B^*_{(1,1)}=\frac{\alpha_B}{\beta_B}\,\begin{cases}1-\frac{\sinh(L/\lambda_B-\ell/\lambda_B)}{\sinh(L/\lambda_B)}\,\cosh(x/\lambda_B)& \text{for }0\leq x\leq \ell\\[0.3cm]
%\frac{\sinh(\ell/\lambda_B)}{\sinh(L/\lambda_B)}\,\cosh(x/\lambda_B-L/\lambda_A)& \text{for }\ell<x\leq L\,.
%\end{cases}\label{eq:Bsol}
%\end{equation}

\subsection{Source size of the first order steady state}\label{appreactdiff:InhomSteadySource}
For the step-like production function, the source size is defined by
\begin{eqnarray} A^*_{(1,1)}(\ell)&=&B^*_{(1,1)}(\ell)\\
\mathfrak{F}_A(\ell)&=&\frac{\alpha_B\beta_A}{\alpha_A\beta_B}\,\mathfrak{F}_B(\ell)\label{eq:Turingellsteadystate}
\end{eqnarray}
with
\begin{equation} \mathfrak{F}(\ell)=\frac{\sinh(\ell/\lambda)}{\sinh(L/\lambda)}\,\cosh(\ell/\lambda-L/\lambda)=\frac{1}{2}\,\left(1-\frac{\sinh(L/\lambda-2\ell/\lambda)}{\sinh(L/\lambda)}\right)\,.\label{eq:frakF}
\end{equation}
Next, we show that the source size $\ell$ has only one non-trivial solution. First, we compute
\begin{equation}\mathfrak{F}(0)=0\;,\quad\mathfrak{F}(L/2)=1/2\;,\quad
\mathfrak{F}(L)=1\,,\end{equation}
and we see that $\ell=0$ is always a solution to Eq.~\ref{eq:Turingellsteadystate}. The curve $\mathfrak{F}(\ell)$ is essentially a shifted hyperbolic sine function und is monotonously increasing
\begin{equation}\partial_{\ell}\mathfrak{F}(\ell)=\frac{\cosh(2\ell/\lambda-L/\lambda)}{\lambda\,\sinh(L/\lambda)}> 0\,.\end{equation}
The largest slope is found at the interval bounds:
\begin{equation}\partial_{\ell}\mathfrak{F}(\ell)\big\vert_0=\partial_{\ell}\mathfrak{F}(\ell)\big\vert_L=\frac{1}{\lambda\,\tanh(L/\lambda)}\,.\end{equation}
In particular, we see that
\begin{equation} \partial_{\lambda}\left(\frac{1}{\lambda\,\tanh(L/\lambda)}\right)=\frac{2\,L/\lambda-\sinh(2L/\lambda)}{2\,\lambda^2\sinh(L/\lambda)^2}<0\quad\text{because }L/\lambda>0\,.
\end{equation}
Thus, the larger $\lambda$ the smaller the slope at $\ell=0$ and $\ell=L$. In consequence,  the curve for $\lambda_B$ lies below the curve for $\lambda_A$ if $\ell<L/2$ and above if $\ell>L/2$, as shown in Fig.~\ref{fig:Turing11Source}(b). We can demonstrate this fact explicitly by computing
\begin{equation}
\partial_{\lambda}\mathfrak{F}(\ell)=\frac{(L-\ell)\,\ell}{\lambda^3\sinh(L/\lambda)^2}\,\left(\frac{\sinh(2\ell/\lambda)}{2\ell/\lambda}-\frac{\sinh(2L/\lambda-2\ell/\lambda)}{2(L-\ell)/\lambda}\right)\lessgtr 0 \quad\text{if }\ell \lessgtr L/2\,.
\end{equation}
Therefore, $\mathfrak{F}_A(\ell)$ and $\mathfrak{F}_B(\ell)$ always have three intersection points at $0$, $L/2$ and $L$. If we now consider Eq.~\ref{eq:Turingellsteadystate} with $\alpha_A\beta_B<\alpha_B\beta_A$ and $\lambda_A<\lambda_B$ (as required for the Turing system to form patterns, see Section~\ref{appreactdiff:HomSteady}), $\ell=L$ is not a solution anymore and the second solution is shifted to smaller values of $\ell$ (intersection with dashed blue line in Fig.~\ref{fig:Turing11Source}(b)).

\noindent In summary, there is always a solution with $\ell=0$ (no source). Additionally, for our Turing condition $\alpha_A\beta_B\leq\alpha_B\beta_A$, there can be a second solution with a finite source size $\ell<L/2$ if
\begin{equation} \frac{\alpha_B\beta_A}{\alpha_A\beta_B}\,\lambda_A\,\tanh(L/\lambda_A)<  \lambda_B\,\tanh(L/\lambda_B)\,.\label{eq:bifurcpoint}
\end{equation}
For $\alpha_A\beta_B<\alpha_B\beta_A$, there are no further solutions. The condition of Eq.~\ref{eq:bifurcpoint} marks a bifurcation point of the system, beyond which the $(1,1)$-solution exists in addition to the $(0,0)$-solution.

\noindent From Eq.~\ref{eq:bifurcpoint}, we see that $\lambda_A$ is constrained by either $L$ or $\lambda_B$, depending on which one is smaller. If $L\ll\lambda_B$, the equation becomes
\begin{equation} \tanh(L/\lambda_A)< \frac{\alpha_A\beta_B}{\alpha_B\beta_A}\,L/\lambda_A\,.
\end{equation}
For $\alpha_A\beta_B<\alpha_B\beta_A$, this is only fulfilled if $L/\lambda_A$ is large enough.
In contrast for $L\gg\lambda_B$, we obtain
\begin{equation} \lambda_A< \frac{\alpha_A\beta_B}{\alpha_B\beta_A}\,\lambda_B\,.
\end{equation}
For $\alpha_A\beta_B<\alpha_B\beta_A$, this is only fulfilled if $\lambda_A$ is sufficiently smaller than $\lambda_B$.

\noindent Finally, we consider two limiting cases of Eq.~\ref{eq:frakF}:
\begin{equation}
\mathfrak{F}(\ell)=\frac{1}{2}\,\left(1-\frac{\sinh(L/\lambda\,(1-2\ell/L))}{\sinh(L/\lambda)}\right)\rightarrow\begin{cases}\ell/L& \text{for } L/\lambda\ll 1\\
1/2& \text{for } L/\lambda\gg 1\,.\end{cases}
\end{equation}
Note that the absolute value of $1-2\ell/L$ is smaller than $1$. Thus, for $L\ll\lambda_B$ and  $L\gg\lambda_A$, we obtain $\ell/L=(\alpha_A\beta_B)/(2\alpha_B\beta_A)$, see Fig.~\ref{fig:RelSource}.

\subsection{Hierarchy of higher order steady states}
We can successively derive the higher order steady states.
First, we consider the se\-cond order solution with two sources at the boundaries, denoted as $(2,2)$-pattern, see Fig.~\ref{appreactdiff:StabInhom}(a).
Thus, there are maxima of concentration $A^*_{(2,2)}$ and $B^*_{(2,2)}$ at each boundary. Furthermore, there is a minimum for each concentration in between the two source regions. We now prove that the system is completely symmetric as depicted in Fig.~\ref{appreactdiff:StabInhom}(b). For that it is sufficient to show that the minima for $A^*_{(2,2)}$ and $B^*_{(2,2)}$ are at the same position, i.e.~$L_A=L_B$.
Let us consider the solutions left of the respective minima in Fig.~\ref{appreactdiff:StabInhom}(a). These solutions correspond to the $(1,1)$-pattern, which we have characterized above. If the minima are at the same position ($L_A=L_B$), there is at most one $\ell$ with $0<\ell<L$ in agreement with $A^*_{(1,1)}(\ell)=B^*_{(1,1)}(\ell)$, see Section~\ref{appreactdiff:InhomSteadySource}. However, $B^*_{(1,1)}(\ell)$ is
a strictly monotonic function of $L_B$ if $\ell>0$:
\begin{equation}
\partial_{L_B}B^*_{(1,1)}(\ell)=-\frac{\alpha_B\,\sinh(\ell/\lambda_B)\cosh(\ell/\lambda_B)}{\sinh(L/\lambda_B)^2\lambda_B}<0\,.
\end{equation}
Therefore, there is no other choice of $L_B$ also solving the equation for $\ell$. In consequence we conclude that $L_A=L_B$ in the steady state.

\noindent For $L_A=L_B$, we can cut the (2,2)-pattern at this position, which results in two (1,1)-patterns. But since these two (1,1)-patterns have in fact the same concentration $A$ at their common boundary, their system size must be equal. 
An analogous reasoning can be applied to the $(1,0)$-pattern. Finally, Fig.~\ref{appreactdiff:StabInhom}(b) exemplifies that any higher order pattern can be considered as a concatenation of $(1,1)$-patterns as the basic building blocks. Each adjacent pair of these fundamental sections is either a $(1,0)$-pattern or a $(2,2)$-pattern and thus completely symmetric. In consequence, any higher order pattern can be constructed by a concatenation of $(1,1)$-patterns of identical size.

\begin{figure}[tp]
\centering
\includegraphics[width=0.9\textwidth]{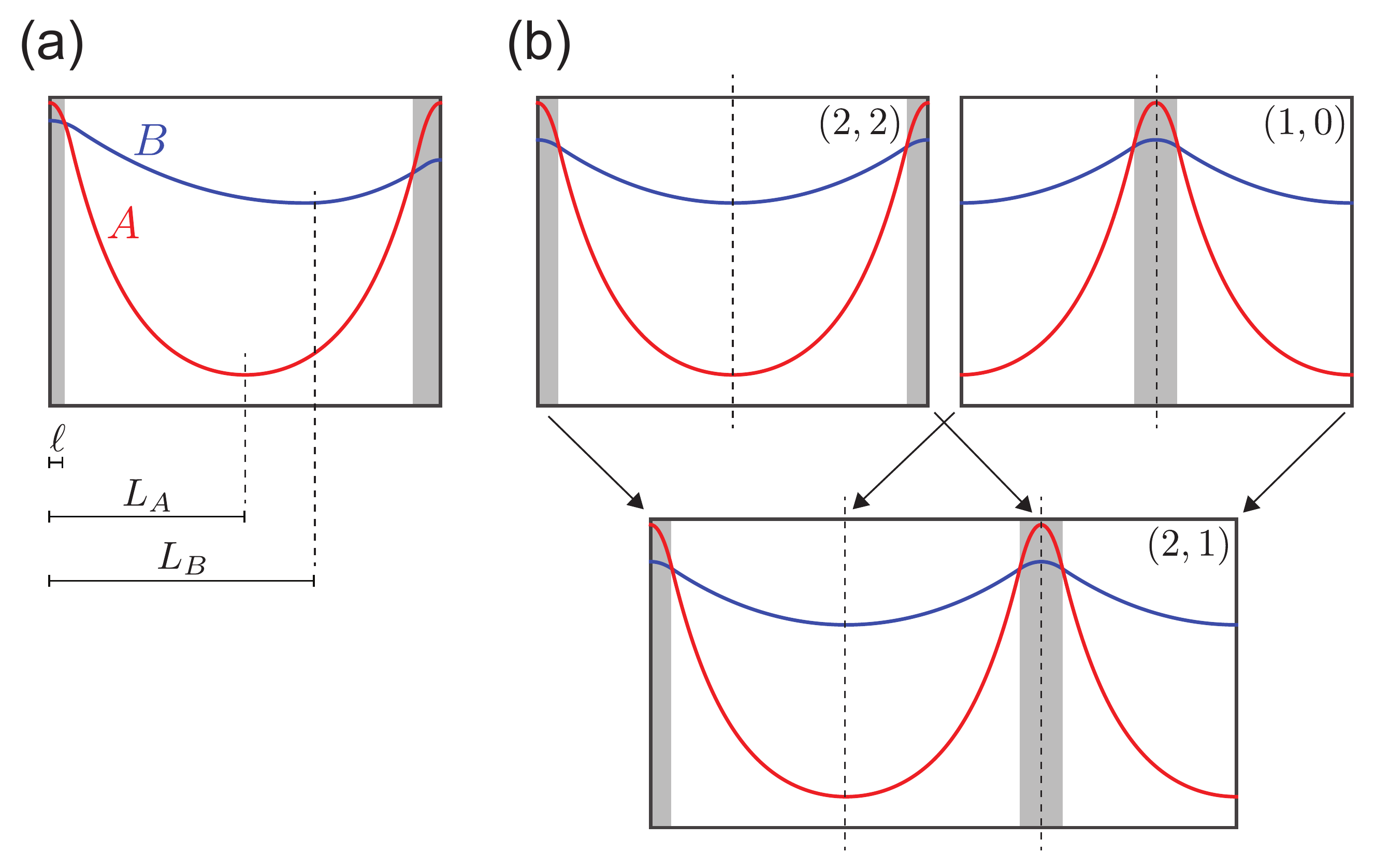}
\caption[Constructing a hierarchy of Turing patterns]{\small 
(a) The concentrations $A$ and $B$ of the ${(2,2)}$-pattern possess maxima at boundaries and one minimum each between the source regions. (b) All patterns can be constructed by a concatenation of $(1,1)$-patterns of identical size.
Parameters like in Section~\ref{TuringNoScaling}: $D_B/D_A = 30$, $\alpha_B/\alpha_A = 4$, $\beta_B/\beta_A = 2$, $\lambda_A/L=\sqrt{0.1}\approx 0.3$, $\lambda_B/L=\sqrt{1.5}\approx 1.2$.
}
  \label{fig:TuringHierarchy}
\end{figure}

\subsection{Stability  of the inhomogeneous steady state of our Turing system}\label{appreactdiff:StabInhom}\noindent
We are probing the stability of the inhomogeneous steady states in various ways. First, we can add a set of small perturbations to the steady state profiles and monitor the relaxation behavior in numerical simulations. Second, we can start with a linear combination of two steady state patterns with varying weights and observe to which steady state they converge, as shown in Fig.~\ref{ModeStabPlot}(b).
Finally, for the case of the Hill-type production function, we can numerically perform a linear stability analysis, as we will show next.

\noindent Let us consider small perturbations $a(t,x)$ and $b(t,x)$ about the inhomogeneous steady states:
\begin{equation}
A(t,x)=A^*_{(m,n)}(x)+a(t,x)\;,\quad B(t,x)=B^*_{(m,n)}(x)+b(t,x)\,.
\end{equation} 
The linearized dynamics of the perturbations is given by
\begin{equation}
\partial_t \begin{pmatrix} a\\[0.2cm] b\end{pmatrix}=
\begin{pmatrix} \mathfrak{P}_1 & \mathfrak{P}_2 \\[0.2cm] \mathfrak{P}_3 & \mathfrak{P}_4 \end{pmatrix}
\begin{pmatrix} a\\[0.2cm] b\end{pmatrix}
\end{equation}
with the operators
\begin{eqnarray}
\mathfrak{P}_1&=& \alpha_A\,\partial_A P\big\vert_{A^*_{(m,n)},B^*_{(m,n)}}-\beta_A+D_A\partial^2_x\\
\mathfrak{P}_2&=&\alpha_A\partial_B P\big\vert_{A^*_{(m,n)},B^*_{(m,n)}}\\
\mathfrak{P}_3&=& \alpha_B\,\partial_A P\big\vert_{A^*_{(m,n)},B^*_{(m,n)}}\\
\mathfrak{P}_4&=& \alpha_B\partial_B P\big\vert_{A^*_{(m,n)},B^*_{(m,n)}}-\beta_B+D_B\partial^2_x\,.
\end{eqnarray}
Now, we express the perturbations in terms of a set orthonormal modes that agree with the reflecting boundary conditions:
\begin{equation} a(t,x)=a_0+\sum_{j=1} a_j(t)\,\mathfrak{m}_j(x)
\quad\text{with}\quad\mathfrak{m}_j(x)=\sqrt{2}\,\cos(\pi\,j\,x/L)\,.\end{equation}
The coefficients $a_j(t)$ can be determined by
\begin{equation}  a_j(t)=\int_0^L a(t,x)\,\mathfrak{m}_j(x)\,dx\,.
\end{equation}
The linearized system in terms of these modes is
\begin{equation}
\partial_t \begin{pmatrix} a_1\\ a_2\\ ...\\ b_1\\b_2\\...\end{pmatrix}=
\begin{pmatrix} &\\
\int_0^L \mathfrak{m}_j\mathfrak{P}_1\mathfrak{m}_i\,dx \;&\; \int_0^L \mathfrak{m}_j\mathfrak{P}_2\mathfrak{m}_i\,dx \\[0.6cm]
 \int_0^L \mathfrak{m}_j\mathfrak{P}_3\mathfrak{m}_i\,dx \;&\;  \int_0^L \mathfrak{m}_j\mathfrak{P}_4\mathfrak{m}_i\,dx\\
 & \end{pmatrix}
\begin{pmatrix} a_1\\ a_2\\ ...\\ b_1\\b_2\\...\end{pmatrix}
\end{equation}
If the largest eigenvalue of the linear operator matrix is negative, the steady state is stable and this maximum eigenvalue provides the time scale of relaxation of the slowest decaying mode. If at least one eigenvalue is positive, the steady state is linearly unstable. 

\begin{figure}[tp]
\centering
\includegraphics[width=0.65\textwidth]{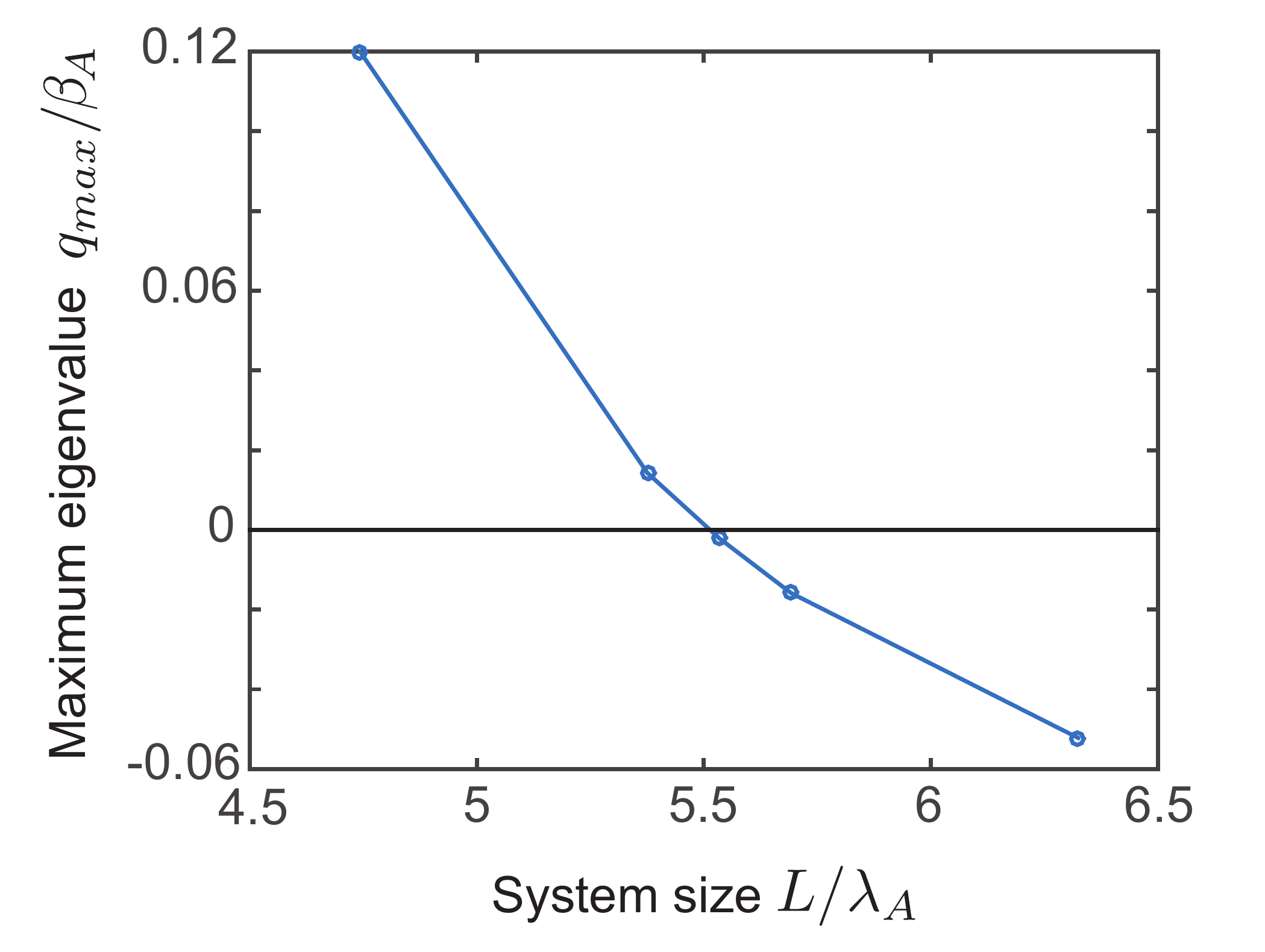}
\caption[Linear stability analysis of inhomogeneous Turing patterns]{\small 
Maximum eigenvalue of the linear operator matrix for the $(1,0)$-pattern as a function of system size. The pattern becomes stable at $L/\lambda_A\approx 5.5$, see also Fig.~\ref{ModeStabPlot}. Approximation with 20 modes and a spatial discretization of the profiles with 200 grid points. 
Parameters like in Fig.~\ref{ModeStabPlot}: $D_B/D_A = 30$, $\alpha_B/\alpha_A = 4$, $\beta_B/\beta_A = 2$, $h=5$.
}
  \label{fig:LinStabInhom}
\end{figure}

\noindent Fig.~\ref{fig:LinStabInhom} shows the maximum eigenvalue for the $(1,0)$-pattern as a function of the system size. We can determine the lower bound for which this pattern becomes stable, as illustrated in  Fig.~\ref{ModeStabPlot}. 
For none of the inhomogeneous patterns, we observe a maximum system size, at which these patterns would become unstable again.

\section{On our scalable Turing system}

\subsection{A homogeneous dynamic state for low expander levels}\label{appreactdiff:LowExpander}\noindent
For low expander values, the concentrations in the Turing system with expander feedback become homogenous. Here, we consider the adiabatic limit for which the expander relaxation is much faster than the dynamics of the Turing molecules. In consequence, the expander tightly follows the much slower dynamics of the inhibitor concentration $B$:
\begin{equation} E=\frac{\alpha_E}{\kappa_E\,B}\,.\label{eq:EBHomS}\end{equation}
It exists a dynamic state of the Turing system for which the concentrations $A$ and $B$ relax in synchrony such that the ratio $\chi=B/A$ stays constant. As a result, the Hill-type production function has a constant value $P(A,B)=g(\chi)$ given by
\begin{equation} g(\chi)=\frac{1}{1+\chi^h}\,.\end{equation}
The ratio $\chi$ is is obtained from
\begin{eqnarray}
&&\partial_t \chi= 0\nonumber \\[0.2cm]
&&\alpha_B g(\chi)-\kappa_BEB-\alpha_A g(\chi)\chi+\kappa_A EB=0\nonumber \\[0.2cm]
&&g(\chi)=\frac{\alpha_E}{\kappa_E}\,\frac{\kappa_A-\kappa_B}{\alpha_A\,\chi-\alpha_B}\,.
\end{eqnarray}
The system can leave the homogeneous regime if $E$ increases. For this, $B$ has to decrease, according to Eq.~\ref{eq:EBHomS}. We have discussed for Eq.~\ref{eq:BdynHomS} that this requires $g(\chi)<\ell^*/L$, where $\ell^*/L=(\alpha_E\kappa_B)/(\kappa_E\alpha_B)$. For the threshold where $g(\chi)=\ell^*/L$, we obtain 
\begin{eqnarray}
\frac{\alpha_E}{\kappa_E}\,\frac{\kappa_A-\kappa_B}{\alpha_A\,\chi-\alpha_B}=\frac{\alpha_E\kappa_B}{\kappa_E\alpha_B}\nonumber\\[0.3cm]
\chi=\frac{\alpha_B\kappa_A}{\alpha_B\kappa_B}\,.
\end{eqnarray}
Thus, the ratio $B/A$ equals the ratio of the homogeneous steady state concentrations $B_h^*/A_h^*$. Therefore, from $g(\chi)=\ell^*/L$ follows that $f_{(0,0)}=\ell^*/L$.

\subsection{Generalized scalable Turing system}\label{appreactdiff:GenScalTuring}\noindent
We can generalize our approach of scalable Turing patterning by considering the following generic equations for $A$, $B$ and $E$
\begin{eqnarray}
\partial_t A&=&\alpha_A\,P(A,B)+\mathfrak{R}_A(A,E)+D_A\,\partial_x^2\,A\label{eq:StrukRob1}\\
\partial_t B&=&\alpha_B\,P(A,B)+\mathfrak{R}_B(B,E)+D_B\,\partial_x^2\,B\label{eq:StrukRob2}\\
\partial_t E&=&\alpha_E+\mathfrak{L}_E\Big(\mathfrak{R}_A(A,E),\mathfrak{R}_B(B,E)\Big)+D_E\,\partial_x^2\,E\label{eq:StrukRob3}\,.
\end{eqnarray}
Here, $\mathfrak{R}_A$ and $\mathfrak{R}_B$ are generic functions fulfilling the Turing conditions and $\mathfrak{L}_E$ is a linear function of $\mathfrak{R}_A(A,E)$ and $\mathfrak{R}_B(B,E)$. By computing the spatial averages in the steady state, one obtains
\begin{eqnarray}
&&\alpha_E+\langle \mathfrak{L}_E\Big(\mathfrak{R}_A(A,E),\mathfrak{R}_B(B,E)\Big)\rangle=0\nonumber\\
&&\alpha_E+\mathfrak{L}_E\Big(\langle \mathfrak{R}_A(A,E)\rangle,\langle \mathfrak{R}_B(B,E)\rangle\Big)=0\nonumber\\
&&\alpha_E+\mathfrak{L}_E\Big(\alpha_A\langle P\rangle,\alpha_B\langle P\rangle\Big)=0\,.\label{eq:StrukRobSconst}
\end{eqnarray}
If $\mathfrak{R}_A$, $\mathfrak{R}_B$ and $\mathfrak{L}_E$ are chosen such that $\tilde{\mathfrak{L}}_E(\langle P\rangle)=\mathfrak{L}_E(\alpha_A\langle P\rangle,\alpha_B\langle P\rangle)$ is a monotonic function in $\langle P\rangle$, this uniquely determines the source size in the steady state. Thus, if a steady state exists, the size of the source will be independent of the system size.

\noindent Let us again consider a homogeneous expander distribution. Furthermore, we assume that the relaxation of the Turing system is much faster than the relaxation of the expander. Thus, Eq.~\ref{eq:StrukRob1}-\ref{eq:StrukRob2} define the size of the source as a function of slowly varying $E$. The intersection of this nullcline with the solution for $\langle P\rangle$ from Eq.~\ref{eq:StrukRobSconst} provides the steady state values of $E$. For the steady state to be linearly stable, it requires that $\partial_{\langle P\rangle} \tilde{\mathfrak{L}}_E(\langle P\rangle)\cdot \partial_{E} \langle P\rangle\vert_{E^*}<0$.

\subsection{Scaling of downstream targets with a constant amplitude}\label{appreactdiff:ScalConstAmpl}\noindent
One important feature of our scalable Turing system is that the amplitude of the morphogens increases quadratically with $L$, see Section~\ref{Sec:TuringPredictions}. This is a general property of such scaling mechanisms for which the degradation rate is adjusted, see Section~\ref{Sec:ScalingLesson}. As the degradation rate changes proportional to $L^{-2}$, the products $\beta_A A$ and $\beta_B B$ are characterized by a constant amplitude independent of system size. If the cell responds to this flux, the expression of down-stream targets scales with a constant amplitude. Yet, how might the cell read out the degradation flux? 

\noindent Signaling often requires binding of the molecules to receptors on the cell surface. The concentration of activator molecules bound to receptors is $A_b\propto \beta_{Ab}\, A$, where $\beta_{Ab}$ is the binding rate. If the molecules become internalized or degraded upon binding, this binding removes the molecules from the system and the binding rate $\beta_{Ab}$ in fact contributes to the degradation rate $\beta_A$.

\noindent As a simple example, let us consider the following dynamics of the bound molecules
\begin{equation}
\partial_t A_b=\beta_{Ab}\, A -\beta_{Ar}\,A_b\,.
\end{equation}
Free molecules bind with rate $\beta_{Ab}$ and bound molecules are removed with rate $\beta_{Ar}$. The steady state of this equation is
\begin{equation}
A^*_b=\frac{\beta_{Ab}}{\beta_{Ar}}\, A^*\,.
\end{equation}
If spontaneous degradation of the free molecules can be neglected in comparison to binding and internalization, it follows that $\beta_A=\beta_{Ab}$ and the cells read out the flux $\beta_{A}\, A^*$. In this scenario, the expander could be a co-receptor that controls the binding rate and thus the effective degradation rate in a size-dependent manner. The downstream targets would perfectly scale with a constant amplitude.

\subsection{On knockout experiments in scalable Turing systems}\label{appreactdiff:knockout}\noindent
In Chapter~\ref{patterningB}, we presented a simple mechanism that couples a classical Turing system including two molecules of concentration $A$ and $B$ to the dynamics of an additional expander molecule. Here, we will show that knockout of $A$ or $B$ (for example by RNAi feedings) can yield misleading results.

\begin{figure}[tp]
\centering
\includegraphics[width=1\textwidth]{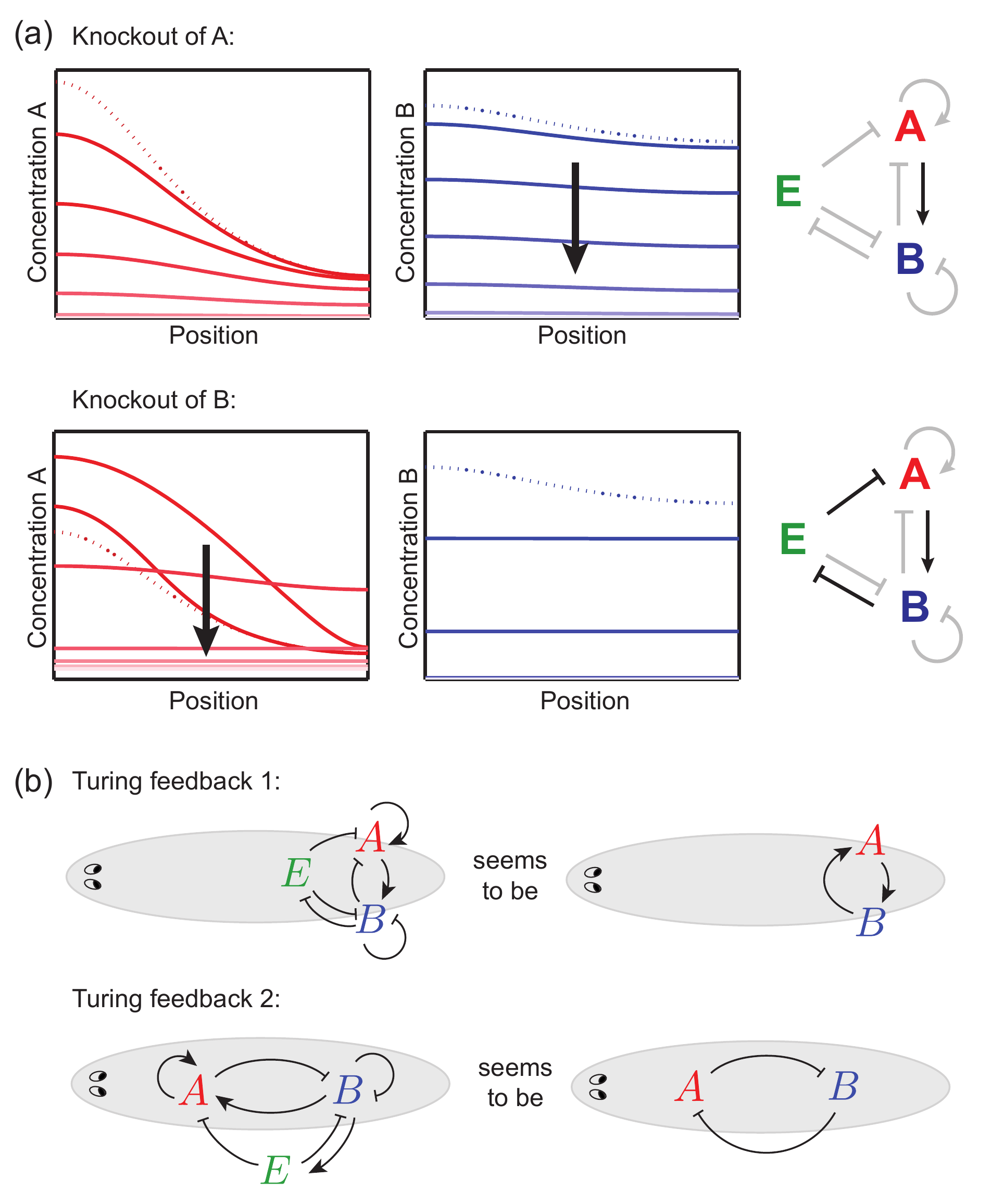}
\caption[RNAi experiments in Turing systems with expander]{\small 
(a) The scalable Turing system discussed in Chapter~\ref{patterningB} can appear like a mutually activating feedback loop in knockout experiments of $A$ and $B$ (dashed lines are the initial steady states, fading of colors denote the time evolution).
Parameters like in Chapter~\ref{patterningB}: $D_B/D_A = 30$, $\alpha_B/\alpha_A = 4$, $\beta_B/\beta_A = 2$, $\lambda_A/L=\sqrt{0.1}\approx 0.3$, $\lambda_B/L=\sqrt{1.5}\approx 1.2$, $h=5$.
(b) Examples of the two possible Turing topologies plus an appropriate expander feedback and the interpretation of knockouts of the two Turing molecules. The results compare to experimental observations with respect to the Wnt/$\beta$-catenin  system in flatworms.
}
  \label{fig:TuringRNAi}
\end{figure}

\noindent If the production of the activator is blocked, one observes that both concentrations fade away, see Fig.~\ref{fig:TuringRNAi}(a). Thus, one would correctly conclude that $A$ activates $B$. Yet, after removing the inhibitor from the system, both concentrations decrease as well. In particular, $A$ is vanishing after an initial increase because of the indirect effect via $E$. In consequence one would conclude that $B$ is activating $A$, despite its direct inhibiting effect within the Turing feedback loop.

\noindent In the case of the flatworms, the Wnt/$\beta$-catenin system has an instructive role for tail identity. RNAi experiments indicate that respective pathway components are expressed in the tail and have a positive effect on themselves and each other. In contrast, the inhibitors of this system reside in the head. This is not in agreement with any of the two most simple Turing systems based on two key players. Fig.~\ref{fig:turingtopologies} illustrates that we would expect either additional inhibiting effects of molecules that are expressed in the same region or additional activating effects of molecules that are expressed on opposite ends of worm. Including an expander feedback could solve this problem, as shown in Fig.~\ref{fig:TuringRNAi}(b). A Turing system of the first type, with both Turing molecules expressed in the same region, might appear like a mutually enhancing feedback during gene knockout. A Turing system of the second type, with both Turing molecules expressed on opposite ends, might appear like a mutually suppressing feedback.

\noindent Monitoring the dynamics of the concentrations upon RNAi feeding over time is more informative than endpoint assays, see Fig.~\ref{fig:TuringRNAi}(a). After knockout of the inhibitor, the activator first expands before the pattern vanishes, hinting at the suppressive effect of $B$.

% ---------------------------------------------------------------------------
% ----------------------- end of thesis sub-document ------------------------
% ---------------------------------------------------------------------------
% this file is called up by thesis.tex
% content in this file will be fed into the main document

\chapter[On the numerical solution of reaction-diffusion equations]{On the numerical solution\\ of reaction-diffusion equations} \label{appgauss}% top level followed by section, subsection
\chaptermark{On the numerical solution\\ of reaction-diffusion equations} %only change chapter heading

% the code below specifies where the figures are stored
%\ifpdf
%    \graphicspath{{6_appendix/figures/PNG/}{6_appendix/figures/PDF/}{6_appendix/figures/}}
%\else
%    \graphicspath{{6_appendix/figures/EPS/}{6_appendix/figures/}}
%\fi

% ----------------------- contents from here ------------------------

\section{Euler method and Courant criterion}\label{appgauss:EulerCourant}
A reaction-diffusion equation of type
\begin{equation} \partial_t C(t,x) = -\beta\, C(x,t)+D\partial_x^2 C(t,x)\end{equation}
describes diffusion and linear degradation of a chemical species of concentration $C$ in a one-dimensional domain. It can be solved numerically by discretizing time and space and using the Euler method as the most simple approach \cite{press2002numerical}. The concentration $C_i$ at position $i$ at time $t+\Delta t$ is then given by
\begin{equation} C_i(t+\dt) = C_i(t)-\beta\,C_i(t)\, \Delta t+\frac{D \dt}{\Delta x^2} \Big(C_{i+1}(t)-2C_i(t)+C_{i-1}(t)\Big)\,.\label{EulerDiff}\end{equation}
In order for the Euler update to be numerically stable and not to violate causality, the following Courant-L\'evy-Friedrich criterion has to be fulfilled \cite{press2002numerical}
\begin{equation} \dt < \frac{2 \dxs}{4 D + \beta \dxs}\,.\end{equation}
In this thesis, we often analyze systems of several coupled reaction-diffusion equations. At least one of which usually contains a very fast diffusing chemical species, like the inhibitor in the Turing system or the expander in Chapter~\ref{patterningA} and \ref{patterningB}. Usually $4D_{f}\gg \beta\dxs$ with $D_{f}$ of the fastest diffusing species and $\beta$ of any species. Thus, $\dt < \dxs/(2D_{f})$ has to be chosen to be very small. Note that $\dx$ is bounded from above especially by the non-homogeneity of the slowest diffusing species. Yet, the time scale of relaxation to the steady state is determined by the reaction rates $\beta$. The number of iterations to reach the steady state is of order $\mathcal{O}\big[1/(\beta\dt)\big]=\mathcal{O}\big[2D_f/(\beta \dxs)\big]$, which is very large as stated above. Next, we show how to avoid such long computing times.
%As a result, the Euler time step would have to be chosen very small. As the reaction rates are typically rather small and therefore the relaxation to the steady state rather slow, we are confronted with long simulation times.

\section{Algorithmic speed-up using a convolution with a Gauss kernel}\label{appgauss:gauss}
If diffusion of one or several species is fast in comparison to the other processes of the system, we can apply a separation of time scales to avoid long computing times. During one rather large time step of the Euler scheme that would violate the Courant criterion due to the fast diffusion but would be still in agreement with the other slow processes, we account for the fast diffusion by using the analytical solution of the pure diffusion equation
\begin{equation} \partial_t C(t,x) = D\partial_x^2 C(x,t)\,.\label{appgauss:diffeq}\end{equation}
A Fourier transform yields
\begin{equation} \partial_t \tilde{C}(t,s) = - (2\pi s)^2 D \tilde{C}(t,s)\end{equation}
with the solution
\begin{equation} \tilde{C}(t,s) = \tilde{C}_0(s)\, e^{- (2\pi s)^2 D t}\end{equation}
in Fourier space. The back-transform yields an integral with a Gauss kernel
\begin{eqnarray}\label{appgauss:FBT}
C(t,x)&=&\int_{-\infty}^{\infty}ds\,  \tilde{C}(t,s)\,e^{2 i\pi x s} \nonumber\\
&=&\int_{-\infty}^{\infty}ds\int_{-\infty}^{\infty}dy \,  C_0(y)\,e^{-2 i\pi y s}\, e^{- (2\pi s)^2 D t}\,e^{2 i\pi x s}\nonumber \\
&=&\int_{-\infty}^{\infty}dy \,  C_0(y)\int_{-\infty}^{\infty}ds\,e^{- 4\pi^2Dt\,\left(s+\frac{i(y-x)}{4\pi^2 Dt}\right)^2}\,e^{\frac{-(y-x)^2}{4Dt}}\nonumber\\
&=&\int_{-\infty}^{\infty}dy \,  C_0(y)\,e^{\frac{-(y-x)^2}{4Dt}}\,\frac{1}{\sqrt{4\pi Dt}}\,.
\end{eqnarray}
This solution can be used to describe the propagation within one time step of the Euler scheme.\\
The solution has to be computed in discrete spatial coordinates. It can be written as
\begin{equation}
C(t+\dt,x)=\sum_j \int_{y_j-\dy/2}^{y_j+\dy/2}dy \,  C(t,y)\,e^{\frac{-(y-x)^2}{4D\dt}}\,\frac{1}{\sqrt{4\pi D\dt}}\,.
\end{equation}
We define $C_j(t)$ as the average of $C(t,y)$ in the interval $[y_j-\dy/2,\,y_j+\dy/2]$.
If $\dy$ is sufficiently small, such that $C(t,y)\approx C_j(t)$ (i.e. approximately homogeneous) in the interval $[y_j-\dy/2,\,y_j+\dy/2]$, we find
\begin{equation}
C(t+\dt,x)\approx\sum_j \frac{C_j(t)}{2}\left(\text{Erf}\left[\frac{(y_j+\dy/2)-x}{\sqrt{4D\dt}}\right]-\text{Erf}\left[\frac{(y_j-\dy/2)-x}{\sqrt{4D\dt}}\right]\right)\,,
\end{equation}
including the error function $\text{Erf}[z]=\tfrac{2}{\sqrt{\pi}}\int_0^z e^{-\xi^2}d\xi$. In order to obtain the discretized concentration at the next time point, we compute the average
\begin{eqnarray}
C_i(t+\dt)&=&\frac{1}{\dx}\int_{x_i-\dx/2}^{x_i+\dx/2}dx \,  C(t+\dt,x)\nonumber\\[0.3cm]
&\approx&\sum_j \frac{C_j(t)}{2}\Bigg((i-j+1)\, \text{Erf}\left[\frac{(i-j+1)\dx}{\sqrt{4D\dt}}\right]\\[0.3cm]
&&-2(i-j)\, \text{Erf}\left[\frac{(i-j)\dx}{\sqrt{4D\dt}}\right]+(i-j-1)\, \text{Erf}\left[\frac{(i-j-1)\dx}{\sqrt{4D\dt}}\right]\nonumber\\[0.3cm]
&&+\frac{\sqrt{4D\dt}}{\dx\sqrt{\pi}}\left(e^{\frac{-(i-j+1)^2\dx^2}{4D\dt}}-2e^{\frac{-(i-j)^2\dx^2}{4D\dt}}+e^{\frac{-(i-j-1)^2\dx^2}{4D\dt}}\right) \Bigg)\,.\nonumber
\end{eqnarray}
Here, we used $\dx=\dy$ and identified $x_i-y_j=(i-j)\dx$. Note that this approximation is well justified for small $\dx$, in particular $\dx<\sqrt{4 D \dt}$. Next, we demonstrate that for $\dx\gg\sqrt{4 D \dt}$ the result is not in agreement with Eq.~\ref{appgauss:diffeq}. We can substitute $k=i-j$ and $\dr=\dx/\sqrt{4D\dt}$.
\begin{eqnarray}
C_i(t+\dt)&\approx&\sum_k \frac{C_{i-k}(t)}{2}\Bigg((k+1)\, \text{Erf}\Big[(k+1)\dr\Big]+(k-1)\, \text{Erf}\Big[(k-1)\dr\Big]\nonumber\\[0.3cm]
&&-2k\, \text{Erf}\Big[k\dr\Big]+\frac{e^{-(k+1)^2\dr^2}-2e^{-k^2\dr^2}+e^{-(k-1)^2\dr^2}}{\dr\sqrt{\pi}}\Bigg)\nonumber\\
&=&\frac{C_{i}(t)}{2}\Bigg(2\, \text{Erf}\Big[\dr\Big]+\frac{2\,e^{-\dr^2}-2}{\dr\sqrt{\pi}}\Bigg)\\
&&+\frac{C_{i-1}(t)}{2}\Bigg(2\, \text{Erf}\Big[2\dr\Big]-2\, \text{Erf}\Big[\dr\Big]+\frac{e^{-(2\dr)^2}-2\,e^{-\dr^2}+1}{\dr\sqrt{\pi}}\Bigg)\nonumber\\
&&+\frac{C_{i+1}(t)}{2}\Bigg(2\, \text{Erf}\Big[2\dr\Big]-2\, \text{Erf}\Big[\dr\Big]+\frac{e^{-(2\dr)^2}-2\,e^{-\dr^2}+1}{\dr\sqrt{\pi}}\Bigg)\nonumber\\
&&+\sum_{|k|>2} \frac{C_{i-k}(t)}{2}\Bigg((k+1)\, \text{Erf}\Big[(k+1)\dr\Big]-2k\, \text{Erf}\Big[k\dr\Big]+\nonumber\\[0.3cm]
&&(k-1)\,\text{Erf}\Big[(k-1)\dr\Big]+\frac{e^{-(k+1)^2\dr^2}-2e^{-k^2\dr^2}+e^{-(k-1)^2\dr^2}}{\dr\sqrt{\pi}}\Bigg)\nonumber
\end{eqnarray}
The Taylor expansion of the error function is given by
\begin{equation}
\text{Erf}[1/x]=\text{sign}[x]-e^{-x^2}\left(\frac{x}{\sqrt{\pi}}+\mathcal{O}[x^3]\right)\,.
\end{equation}
Thus, we obtain
\begin{eqnarray}
C_i(t+\dt)&\approx& C_i(t)+\frac{C_{i+1}(t)-2C_i(t)+C_{i-1}(t)}{2\sqrt{\pi}\dr}\nonumber\\
&=&C_i(t)+\frac{\sqrt{D\dt}}{\sqrt{\pi}\Delta x} \Big(C_{i+1}(t)-2C_i(t)+C_{i-1}(t)\Big)\nonumber\\
&=&C_i(t)+\frac{D_{eff} \dt}{\Delta x^2} \Big(C_{i+1}(t)-2C_i(t)+C_{i-1}(t)\Big)\,,
\end{eqnarray}
with $D_{eff}=\sqrt{D/(\pi\dt)}\dx$. Even though the result looks similar to Eq.~\ref{EulerDiff} of the Euler method, the diffusion appears to be effectively faster than $D$. As we consider the limit $\dx\gg\sqrt{4 D \dt}$, it follows that
\begin{eqnarray}
\left(\frac{\sqrt{4 D \dt}}{\dx}\right)^2&<&\frac{\sqrt{4 D \dt}}{\dx}\nonumber\\
\frac{D \dt}{\dxs}&<&\frac{\sqrt{D \dt}}{2\dx}<\frac{\sqrt{D \dt}}{\sqrt{\pi}\dx}\nonumber\\
D&<&\sqrt{\frac{D}{\pi\dt}}\dx=D_{Eff}\,.
\end{eqnarray}
The effectively faster diffusion comes from the fact that the concentration within each interval $[y_j-\dy/2,\,y_j+\dy/2]$ becomes equally distributed instantaneously by setting $C(t,y)=C_j(t)$.\\
In order to avoid this numerical effect, $\dx<\sqrt{4 D \dt}$ is required. In a system with a slow and a fast diffusion species ($D_s\ll D_f$), which combines the Euler method with a Gauss kernel, two conditions have to be fulfilled:
\begin{equation}\frac{\dxs}{4 D_f}<\dt<\frac{2\dxs}{4 D_s+\beta_s\dxs}\,.\end{equation}

% ---------------------------------------------------------------------------
% ----------------------- end of thesis sub-document ------------------------
% ---------------------------------------------------------------------------
% this file is called up by thesis.tex
% content in this file will be fed into the main document

\chapter[Worm handling and measurements of size and shape]{Worm handling and\\ measurements of size and shape} \label{appmeasurement}% top level followed by section, subsection

% the code below specifies where the figures are stored
%\ifpdf
%    \graphicspath{{6_appendix/figures/PNG/}{6_appendix/figures/PDF/}{6_appendix/figures/}}
%\else
%    \graphicspath{{6_appendix/figures/EPS/}{6_appendix/figures/}}
%\fi

% ----------------------- contents from here ------------------------

\section{Worm handling}
If not specified otherwise, we use a clonal line of an asexual strain of \textit{Schmidtea mediterranea} 
\cite{sanchezalvarado2002schmidtea, benazzi1972fissiparous}. 
Worms were maintained at $20^{\circ}$C as described in \cite{cebria2005planarian}. Other flatworm species were taken from the flatworm collection established by the lab of Jochen Rink (CBG, Dresden).

\noindent Large worms kept in isolation increasingly tend to fission, while social crowding is known to reduce fissioning rates \cite{pigon1974cephalic, dunkel2011memory}. In order to prevent fissioning during long term measurements but still track individual worms, we added small worm pieces in the dishes of large worms. In long term experiments, we occasionally also added antibiotics (ciprofloxacin, gentamicin) to ensure that the worms stay healthy during the entire time course. 

\section{Image acquisition}
Flatworms were placed one at a time into a plastic petri dish ($90$ mm) and filmed using a Nikon macroscope (AZ 100M, 0.5x objective) and a Nikon camera set-up (DS-Fi1, frame rate 3 Hz, exposure time 6 ms, total observation period 15 s, resolution 1280 x 960 pixel, conversion factor 44 pixel/mm). The movies were converted from AVI (provided by the Nikon imaging software) to MP4 using Handbreak, reducing the bit-size by a factor of 30 and facilitating further processing. For this purpose, we wrote a bash script which calls HandbrakeCLI (the command line interface version) and transcodes the AVI files using the H.264 standard for video compression and fixed optimized settings. We carefully tested the conversion procedure to exclude distorting effects on the data. It turns out that the worm tracking and shape analyses are even more robust if we first transcode the files in comparison to loading the AVI files directly into the MATLAB program.  

\noindent For the size measurements, we aimed to obtain movies in which the worms are moving rather straight and stretched out. This behavior corresponds to a flight response which can be seen after the worms have been exposed to light. After the worms are placed in the center of the petri dish, they try to escape the illumination of the macroscope by moving towards the slightly darker region near the rim of the dish. The stage is adjusted such that the worms traverse the field of view and are imaged just after an initial acceleration phase. Student helpers have been thoroughly trained at the macroscope to strictly adhere to the fixed imaging protocol. 

\section{Extracting size and shape from worm movies}
\begin{figure}[tp]
\centering
\includegraphics[width=0.96\textwidth]{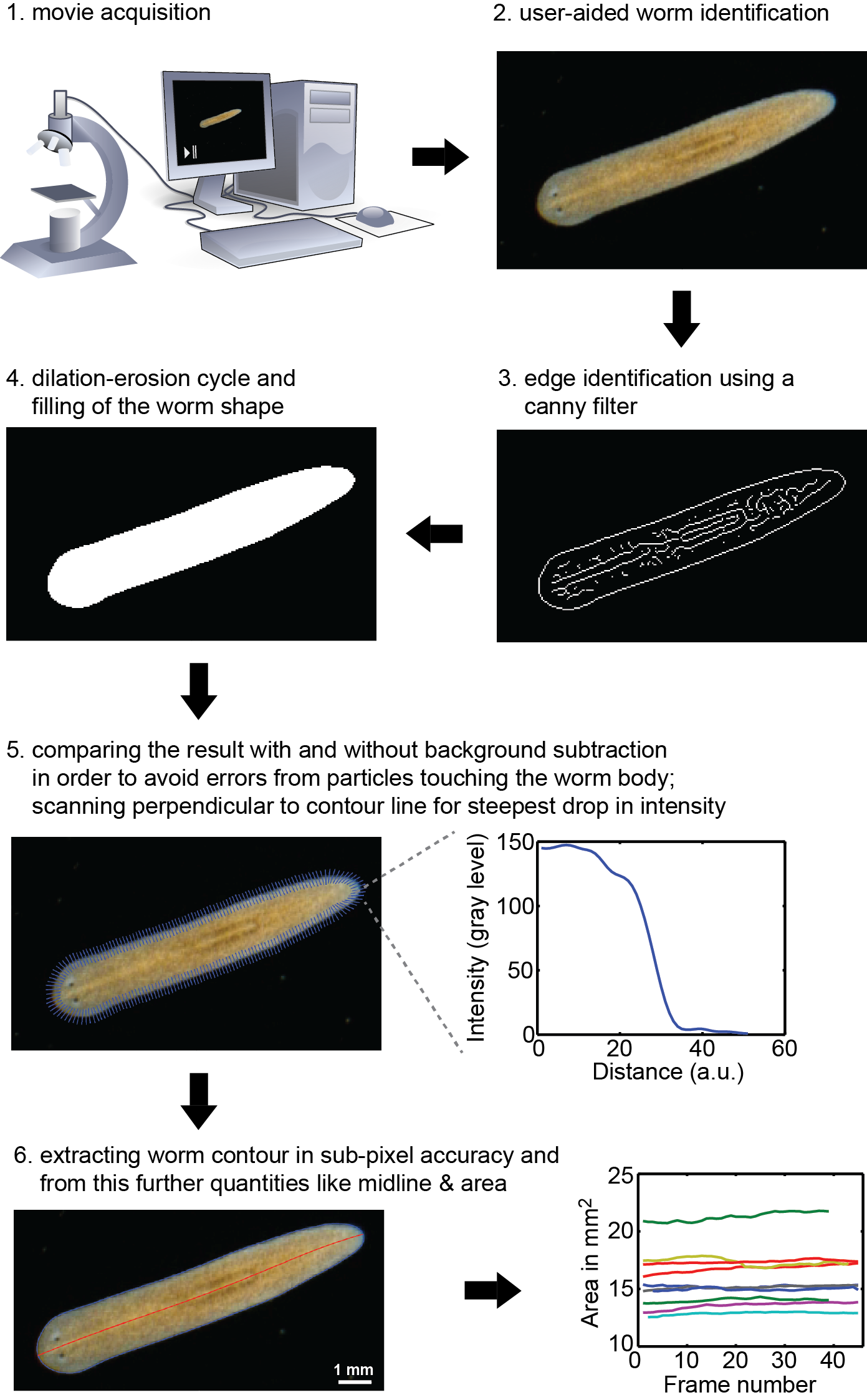}
\caption[Extracting size and shape from worm movies]{\small 
Extracting size and shape information from worm movies.
}
  \label{fig:ImageAnalysis}
\end{figure}

Movies were analyzed off-line using custom-made MATLAB software, see Fig.~\ref{fig:ImageAnalysis}.
A first shape proxy was determined from movie frames
via edge detection using a canny-filter, followed by a dilation-erosion cycle and a filling of the shape. 
If possible, the shape was extracted twice for each frame -- with and without background correction. By comparing the two results, the program can reduce errors from particles touching the worm boundary.
In a subsequent refinement step, the worm perimeter was adjusted by
finding the steepest drop in intensity along directions transverse to the perimeter proxy.
This procedure together with interpolation steps allowed us to extract the worm shape with sub-pixel accuracy,

\noindent After the perimeter was determined, we were able to compute further quantities such as the worm area. The midline was obtained by searching for points with equal distances to both sides of the worm boundary. Due to the large body plasticity, the measured quantities such as worm area or length can easily vary by up to $30\%$ during normal, rather stretched gliding motion. It requires an extensive imaging training and a large number of worms to extract reliable and reproducible data. Often, we analyzed several movies per worm and experimental condition. In contrast to the biological variability, the analysis software works very robustly and can cope with different noise levels and manyfold intensity variations.

% ---------------------------------------------------------------------------
% ----------------------- end of thesis sub-document ------------------------
% ---------------------------------------------------------------------------
% this file is called up by thesis.tex
% content in this file will be fed into the main document

\chapter[Shape reconstruction and worm bending]{Shape reconstruction\\ and worm bending} \label{apppca}% top level followed by section, subsection

% the code below specifies where the figures are stored
%\ifpdf
%    \graphicspath{{6_appendix/figures/PNG/}{6_appendix/figures/PDF/}{6_appendix/figures/}}
%\else
%    \graphicspath{{6_appendix/figures/EPS/}{6_appendix/figures/}}
%\fi

% ----------------------- contents from here ------------------------

\section{On the reconstruction of a closed worm outline}\label{apppca:reconstruction}
In order to reconstruct a worm shape from a given radial distance profile, we have to determine the constant increment $d\hat{s}$ along the worm perimeter such that the outline closes on itself. This corresponds to
\begin{equation} \sum_i \alpha_i=2\pi\label{closingcondition}\end{equation}
with the angle $\alpha_i$ between two consecutive radii $\hat{\rho}_i$ and $\hat{\rho}_{i+1}$. The angle $\alpha_i$ is monotonously increasing with $d\hat{s}$ for radially convex shapes. $\sum_i \alpha_i$ can even be monotonously increasing with $d\hat{s}$ if the shapes are not strictly radially convex but close to it. For a monotonously increasing $\sum_i \alpha_i$, there exists one choice for $d\hat{s}$ such that Eq.~\ref{closingcondition} is fulfilled. Small variations of the radial distance profile yield similar results for $d\hat{s}$.

%\section{Correlation between turning and the bending mode}\label{apppca:turningbending}
%Irrespective of worm size, we find a correlation between the rate of turning and the bending mode (first mode), see Fig.~\ref{}. Smaller worms turn quicker given the same degree of relative bending. Interestingly, larger worms cover a larger range of the first mode. This might be explained by a finite stiffness of the worm body.

\section{On the reconstruction of head shapes}\label{apppca:head}
The head perimeter does not completely close on its own. Instead of Eq.~\ref{closingcondition}, the following condition has to be fulfilled
\begin{equation} \sum_i \alpha_i=2\pi-\alpha_0\,,\end{equation}
including the opening angle $\alpha_0$ towards the trunk. The angle $\alpha_0$ can be determined under the assumption of an approximately symmetric head shape. For $N$ radial distances $\hat{\rho}_{i}$ describing the head, we obtain
\begin{equation}\alpha_0= \arccos(\hat{\rho}_{N/2}/\hat{\rho}_{1})+\arccos(\hat{\rho}_{N/2}/\hat{\rho}_{N})\,.
\end{equation}

% ---------------------------------------------------------------------------
% ----------------------- end of thesis sub-document ------------------------
% ---------------------------------------------------------------------------
% this file is called up by thesis.tex
% content in this file will be fed into the main document

\chapter{On growth and cell turnover} \label{appgrowth}% top level followed by section, subsection

% the code below specifies where the figures are stored
%\ifpdf
%    \graphicspath{{6_appendix/figures/PNG/}{6_appendix/figures/PDF/}{6_appendix/figures/}}
%\else
%    \graphicspath{{6_appendix/figures/EPS/}{6_appendix/figures/}}
%\fi

% ----------------------- contents from here ------------------------

\section{Signatures of aging in flatworms}\label{app:flatwormaging}
\noindent Large and small worms are not only different in size but also show size-dependent changes in system behavior, which have been previously related to characteristics of more adult or juvenile animals, respectively \cite{gonzalezestevez2012decreased,lange1968possible,baguna1990growth, baguna2012planarian, mouton2011lack}. 
For this, we need to distinguish between two different notions of ``aging". The first one refers to the progression of the physical time since birth. Secondly, there would be physiological aging that is determined by varying characteristics of the body such as sexual maturity and also symptoms of decline as well as an increased mortality. Aging in the second sense might potentially be independent of the life time of the organism and might instead rather depend on its size or environmental conditions.

\noindent For sexual flatworms, it has in fact been reported that growth, metabolic rate, regene\-ration abilities and reproduction change with age \cite{lange1968possible, hyman1919physiological3, haranghy1964ageing,abeloos1929recherches}. Additionally, aging is often attributed to a decreasing length of telomeres (i.e.~the end regions of the chromosomes) with each cell division. Indeed, this telomere erosion has been described in sexual \smed{} and it correlates with the total life time of the worm \cite{tan2012telomere}. In contrast, other features seem to rather depend on size instead of the life time. For example, \mbox{sexual} flatworms develop reproductive organs when growing but also reabsorb their germ line cells again when shrinking during starvation periods \cite{newmark2002not, bowen1976effects, schultz1904reduktionen,berninger1911uber}.

\noindent For asexual worms, as we consider in the experiments, aging effects are even more difficult to discuss. There is no well-defined birth and, at most, fissioning could be considered as death and birth at the same time. Nevertheless, asexual \smed{} still show size-dependent physiological changes. For example, the fissioning rate increases for large worms \cite{thomas2012size} and precursor cells of the germ line, which still exist in asexual flatworms, show a similar size-dependence as the germ line in sexual worms \cite{handbergthorsager2007planarian, wang2007nanos}. Telomere erosion has also been observed in asexual \smed{}, yet the length is restored during regeneration in contrast to the sexual strains \cite{tan2012telomere}. In this sense, physiological aging in asexual worms might be considered to be reversible and regeneration might be interpreted as a rejuvenation event, which also includes the renewal of many somatic cells during remodeling \cite{egger2006freeliving, haranghy1964ageing}. It is an interesting question whether the worms also show a physiologically rejuvenation while starving \cite{hyman1919physiological1, haranghy1964ageing, newmark2002not}.

\noindent In summary, there is some evidence that sexual worms show irreversible signatures of aging with progression of their life time while asexual strains might be able to return to a more juvenile state. As a sidenote, sexual worms survive longer than asexual worms after irradiation \cite{wagner2011clonogenetic}, which could be due to a reduced turnover rate. One might speculate that sexual worms rather invest in producing a germ line on the cost of body maintenance. Such a potentially insufficient replacement of damaged cells and the corresponding deterioration of function alongside with a higher mortality can be considered as a hallmark of aging \cite{pellettieri2007cell}.
However, most importantly, the discussion shows how difficult it is to unambiguously characterize aging, especially as it is in general lacking a clear definition \cite{flatt2012new}. It will be left for future works to investigate whether size-dependent changes in flatworms might be linked to aging and rejuvenation. %Below, we will discuss size-dependent changes in asexual \smed{}, yet it will be left for future works to investigate whether this might be linked to aging and potentially even rejuvenation.

\section{Additional size measurements and growth dynamics}\label{app:growthallometric}
In this section, we provide more details and additional measurements of worm sizes and growth dynamics to  back up our discussion in Section~\ref{sec:growthdegrowth}.

\begin{figure}[tbp]
  \centering
\includegraphics[width=1\textwidth]{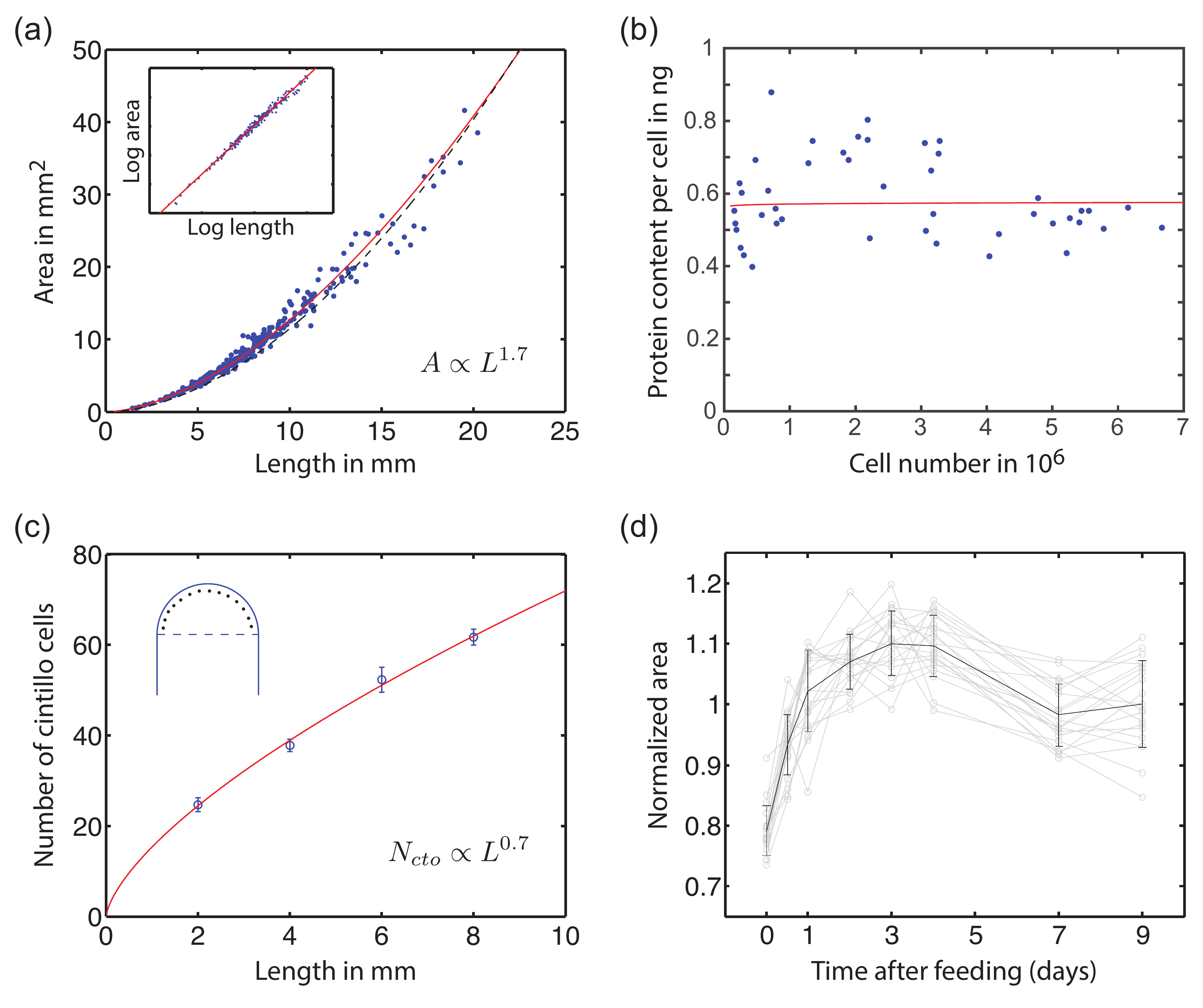}
  \caption[Additional scaling laws and feeding peaks]
  {(a) Relationship between area and length in well-fed worms. The red curve represents the result of a robust fit in the double-logarithmic plot (inset). In comparison, the dashed line denotes the scaling law of starved worms with exponent $1.8$ (Imaging by Nicole Alt under the supervision of the author, analysis by the author includes 281 measurements 3 days after feeding). (b) Protein content per cell does not depend on worm size (measurement by Albert Thommen, analysis by the author, 45 worms, absolute values are only approximate). (c) Scaling of the number of cintillo cells along the head margin with worm length. Data measured by Oviedo \textit{et al.}~\cite{oviedo2003allometric}. (d) Feeding response after two weeks of starvation (Imaging by Ian Smith under the supervision of the author, analysis by the author, 20 worms).}
  \label{fig:AreaLengthProtein}
\end{figure}

\noindent We define the worm area for a particular measurement as the average over the $10$ frames with the largest values. Scaling laws are fitted by a robust algorithm with bi-squared weights. Thus, the spread of the data is assumed to correspond to the measurement uncertainty. This circumvents the difficulties with estimating the error for each data point from the size measurements.
Fig.~\ref{fig:AreaLengthProtein}(a) shows the scaling relation between area and length for well-fed worms ($3$ days after feeding). The fit yields a scaling exponent of $1.69\pm0.01$, which is slightly lower than for the starving worms in Fig.~\ref{fig:ScalingLaws}(a).

 \noindent
As a byproduct of the cell number measurements using histones, Albert Thommen also determined the protein mass in the worms. Even though the measurements are not sufficiently calibrated to obtain absolute numbers, the respective scaling laws might still be meaningful. While the total mass per cell seems to change with worm size, the protein content per cell appears to stay rather constant, see Fig.~\ref{fig:AreaLengthProtein}(b). It remains to validate this result with additional experiments and to investigate whether it could be linked to potential lipid stores.

\noindent Oviedo \textit{et al.}~have counted the number of cintillo cells around the head margin \cite{oviedo2003allometric}. If we assume an equidistant positioning of these sensory cells and approximate the head by a semi-circle, the width of the worm head can be considered to be proportional to the number of cintillo cells, see Fig.~\ref{fig:AreaLengthProtein}(c). The resulting scaling law for the width with an exponent of $0.67\pm0.03$ is in agreement with our measurements of area versus length.

\noindent Upon feeding, worms show a characteristic growth response. The immediate increase in worm size is mainly due to stuffing as discussed in Section~\ref{sec:growthresponse}. When comparing the feeding response after two weeks of starvation with the peak after one week of starvation of Fig.~\ref{fig:feedingpeak}, we see no obvious dependence on feeding history.

\begin{figure}[tbp]
  \centering
\includegraphics[width=1\textwidth]{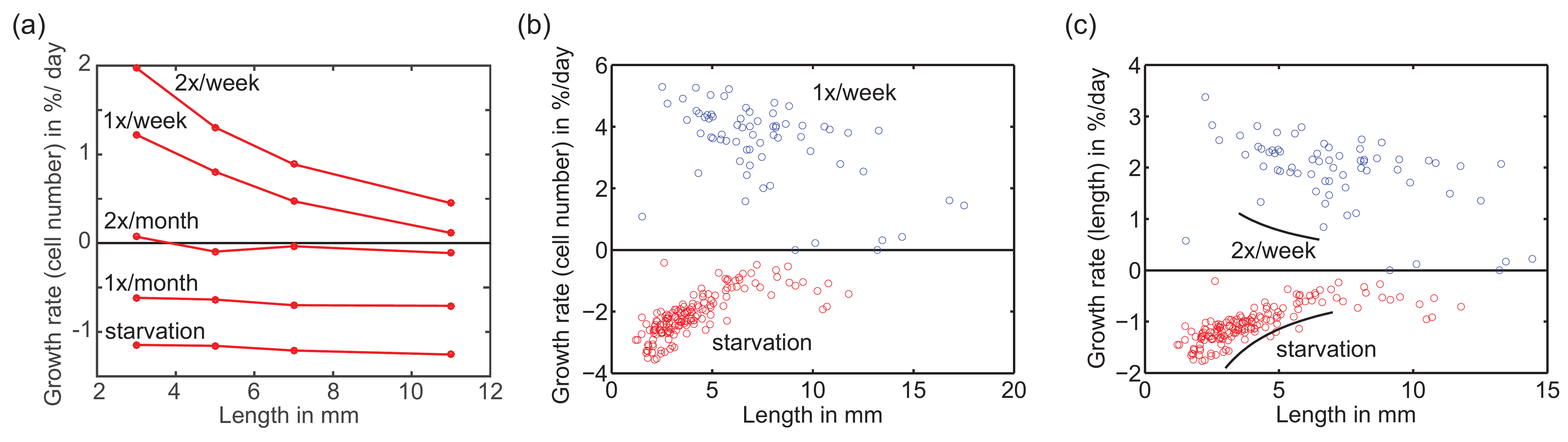}
  \caption[Previously measured growth dynamics in flatworms]
  {(a) Growth dynamics in \textit{Girardia tigrina} based on measurements by Bagu\~n\`a \textit{et al.}~\cite{baguna2012planarian,baguna1990growth}. (b) The plot of our data for \smed{} corresponding to the plot in (a) shows a similar trend for the growth rates (blue) but a different behavior for the degrowth rates (red). (c) Growth and degrowth rates extracted from the data of Oviedo \textit{et al.}~for \smed{} are shown as black curves \cite{oviedo2003allometric}. These curves (for starvation and feeding twice a week) show a similar trend to our data (for starvation and feeding once a week), yet they do not agree quantitatively.}
  \label{fig:OtherGrowthDynamics}
\end{figure}

\noindent When being fed regularly, worms grow and degrow depending on the feeding frequency. Bagu\~n\`a \textit{et al.}~have determined the rates of addition and removal of cells in \textit{Girardia tigrina} for various feeding conditions \cite{baguna2012planarian,baguna1990growth}. From this data, we can back-calculate the growth rates which they originally measured, see Fig.~\ref{fig:OtherGrowthDynamics}(a). While the growth trends are similar to our results, the degrowth rates seem not to vary with size in contrast to our data, see Fig.~\ref{fig:OtherGrowthDynamics}(b).

\noindent Oviedo \textit{et al.}~have obtained linearly increasing and decreasing functions of the worm length with time during feeding and starvation, respectively \cite{oviedo2003allometric}. We can compute growth and degrowth rates from their data sets and compare them to our measurements in terms of worm length, see Fig.~\ref{fig:OtherGrowthDynamics}(c). The linear growth behavior yields $\dot{L}/L\propto \pm 1/L$. Thus, the absolute rates are decreasing for large worms in agreement with our results. However, the curves do not fit our data points very well. They are shifted downwards even though the growing worms are even fed twice a week. In particular, the absolute values for the growth rate are smaller than for the degrowth rate, which contradicts our observation.

\section{Measuring cell cycle times}\label{app:cellcycle}
\begin{figure}[tbp]
  \centering
\includegraphics[width=0.5\textwidth]{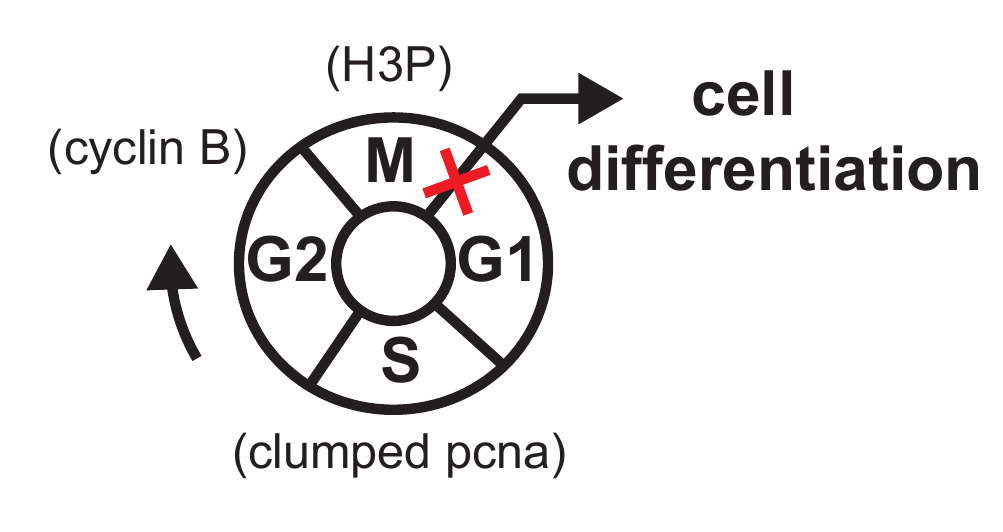}
  \caption[Cell division rates from cell cycle dynamics]
  {Dividing stem cells progress through four distinct cell cycle stages, which can be identified by respective markers. DNA replication happens in S-phase and cells divide in M-phase, while G1 and G2 are merely resting phases including check points for error control. We discuss an experiment to measure the cell division rates by blocking the transition from M-phase to G1 (red cross).}
  \label{fig:cellcycle}
\end{figure}

Stem cells move through four distinct cell cycle phases, which can be characterized by different biochemical markers: pcna (all stem cells), clumping pcna (S-phase), cycling B (G2-phase), H3P (M-phase), see Fig.~\ref{fig:cellcycle}.
An increase of H3P marker has been measured upon feeding \cite{gonzalezestevez2012decreased}. Under the assumption that the duration of the M-phase is relatively constant, this can be interpreted as an increase in cell division rate. However, one cannot extract actual rates from this measurement if one does not know the duration of the M-phase.

\noindent Bagu\~n\`a \textit{et al.}~have performed experiments from which they obtained cell division rates \cite{baguna1976mitosisI, baguna1990growth, baguna2012planarian, baguna1974dramatic, romero1988quantitative, romero1991intraspecific}. They blocked the progression through M-phase \mbox{using} colchicine as indicated by the red cross in Fig~\ref{fig:cellcycle}, which results in an increasing number of cells in M-phase. Under the assumption that the M-phase has a constant duration and cell death can be neglected, the time derivative of the number of accumulating cells in M is a direct read-out of cell division rate. We plan to perform similar experiments using the cell cycle markers discussed above, which have not been available for the previous works. In particular, we aim to reveal how the division rates depend on size and feeding.  By subtracting division rates from growth rates, we will also be able to extract the corresponding cell loss rates. Thus, we hope to reveal whether the control points of cell turnover are associated with the mechanisms for the addition or loss of cells.

\noindent Furthermore, the time scale of the first response upon feeding might also hint at how the cell division rate is controlled. It has been discussed that there might be a population of slowly cycling cells that mainly stay in G2 to allow a fast regeneration and feeding response  \cite{rink2013stem, baguna1976mitosisI, newmark2000bromodeoxyuridine,nimeth2004stem,salo1984regeneration, baguna1974dramatic}.
This represents a major investment of cellular resources as G2 cells already have a duplicated DNA. It would also be in contrast to the arrest in G1, typically observed for other organisms. If we were to measure a very fast division response, it would support the hypothesis of G2 arrested cells.

\section{Measuring cell turnover dynamics on the organism level}\label{app:turnoverorganismlevel}
Here, we discuss the theoretical framework for measuring turnover dynamics on the organismal level. It will enable us to test and compare various models with specific assumptions on how cell division and loss rates depend on feeding and worm size. The experimental approach relies on Histone labeling by heavy isotopes (SILAC). It is inspired by the measurements of neurogenesis dynamics in adult humans exploiting the known C$^{14}$ concentration in the atmosphere due to nuclear bomb tests \cite{spalding2013dynamics}. First, we present the main principles of this paper. Next,  we apply the idea to flatworms suggesting an adapted experimental protocol. Finally, we discuss details of histone labeling.

\subsection{Measurement of C$^{14}$ reveals dynamics of neurogenesis in humans}\noindent
First, we briefly review the main ideas of the work by Spalding \textit{et al.} \cite{spalding2013dynamics} and introduce our notation. 
The paper describes a measurement of the turnover of hippocampal neurons in humans. It is based on the known C$^{14}$ concentration in the atmosphere due to nuclear bomb tests, which becomes incorporated into the DNA as a label when a new cells is made. From this, we can compute the total C$^{14}$  concentration in the DNA of a person of a particular age if we assume a particular model for the turnover dynamics. This result can then be benchmarked using the measured C$^{14}$ in dead people.

\noindent The starting point is that we have a list of several possible scenarios for how cell birth and cell death depends on each other as well as on the age of the cells and the age of the person. For each of these scenarios, we can compute a distribution $n(A,a)$ that describes the number of cells of age $a$ in a person of age $A$. This distribution changes over time because of aging (left) and because of cell death (right)
\begin{equation} \partial_A\,n(A,a)+\partial_a\,n(A,a)=-k_{loss}(A,a)\,n(A,a)\,,\label{partialC14}\end{equation}
where $k_{loss}$ is the loss rate which might depend on the age of the person as well as the age of the cell. Cell birth is considered by the boundary condition
\begin{equation}n(A,0)=n_0(A)\,,\label{boundC14}\end{equation}
where $n_0(A)$ is the influx of cells due to cell division. Cell birth might depend on the age of the person explicitly but also implicitly by e.g.~depending on the number of dying cells at each time point.

\noindent Furthermore, as an initial condition it is assumed that a first set of cells is made in the embryo half a year before the person is born, resulting in an initial condition
\begin{equation}n(-0.5,a)=\delta(a)\,.\label{iniC14}\end{equation}
This includes a delta function $\delta$ (which is zero for $a\neq0$), describing the fact that cells are born with age $0$.

\noindent By solving the system of Eqs.~\ref{partialC14} - \ref{iniC14}, we determine the number of cells $n(A,a)$ of age $a$ for any age $A$ of the person, given a specific model for $k_{loss}(A,a)$ and $n_0(A)$. In particular, we can predict the age distribution $n(A_d,a)$ in a person who died at age $A_d$. For a person who died in year $t_d$, we can compute the total concentration of C$^{14}$ label by
\begin{equation}C(t_d,A_d)=\int_0^{A_d+0.5}\frac{C_{in}(t_d-a)\,n(A_d,a)}{N(A_d)}\,da\,,\label{totcC14}\end{equation}
if the concentration $C_{in}(t)$ of C$^{14}$ inside the body (which could thus be incorporated in new cells) is known at any time point. The integral sums over all cells of all ages, multiplied by the amount of label $C_{in}$ at the respective time of birth $t_d-a$.

\begin{figure}[tbp]
  \centering
\includegraphics[width=0.9\textwidth]{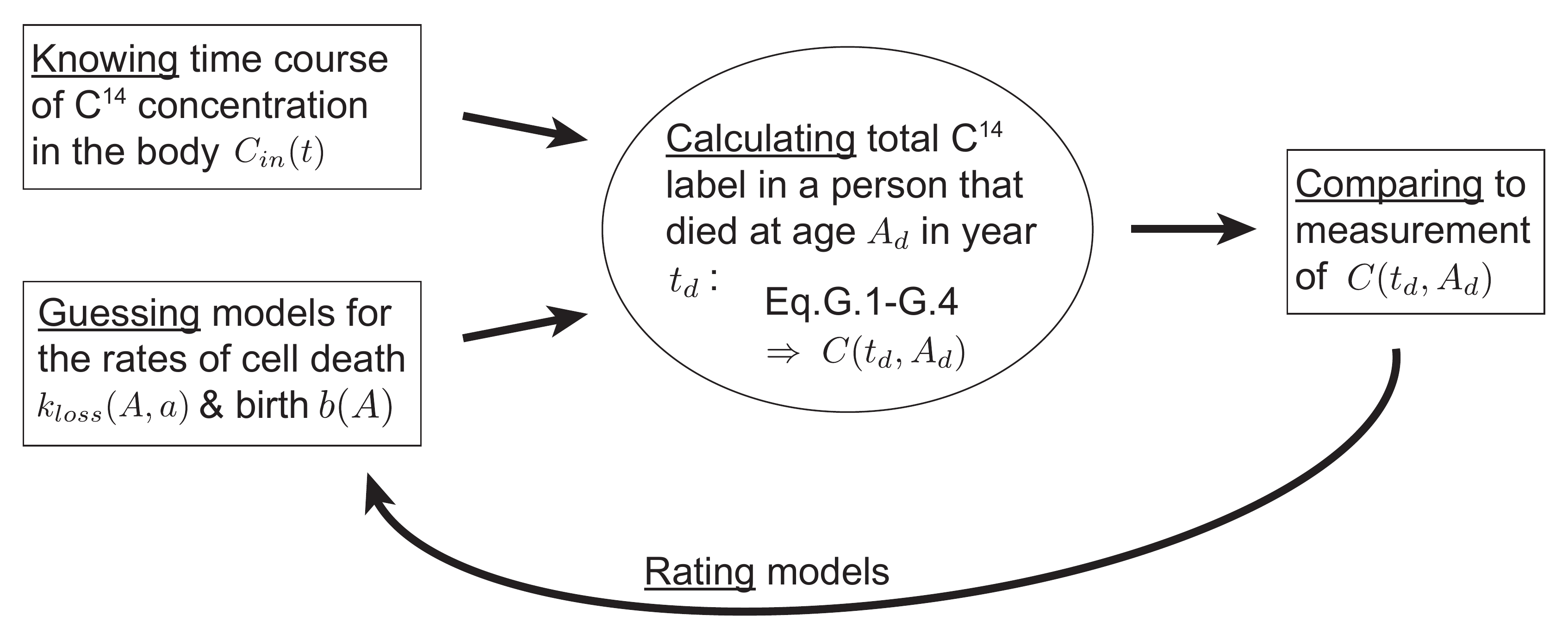}
  \caption[Measurement of turnover dynamics in humans]
  {\small Schematic representation of the C$^{14}$ experiment for the measurement of turnover dynamics in human neurons \cite{spalding2013dynamics}.}  \label{C14Scheme}
\end{figure}

%If the person has died in year $t_d$, this tells us how many of the present cells were born at the time point $t_d-a$ before.
\noindent The result is normalized by the total number of all considered cells in a dead person, which is given by $N(A_d)=\int_0^{A_d+0.5} n(A_d,a)\,da$. The upper bound of integration reflects the maximum age obtained by cells, which persist during the full life span of the person. Note that the summand of $0.5$ was missing in the original publication.\\
The concentration $C_{in}$ of C$^{14}$ inside the body is estimated from the known C$^{14}$ concentration in the atmosphere by introducing a heuristic time lag of one year. Variations in the time lag are claimed not to significantly change the result.

\noindent Different turnover models can be rated using the Akaike Information Criterion (AIC), which avoids over-fitting when comparing models with a different number of parameters. In brief, this technique calculates the likelihood that a certain model has generated the measured data. As a more complex model with more parameters tends to fit the data better, it introduces a simple but mathematically well-justified penalty term for the number of parameters based on information theory.

\subsection{Adapting the C$^{14}$ technique to cell turnover in flatworms}\noindent
In contrast to the neurogenesis paper, we need to consider a different label than the C$^{14}$ concentration in the atmosphere for flatworms. We choose to use the SILAC protocol (stable isotope labeling of amino acids) \cite{boeser2013silac}. %http://www.cell.com/cell-reports/fulltext/S2211-1247%2813%2900619-0
The worms are fed with mouse liver, in which more than 96\% of the amino acid lysine is replaced by the heavy isotope form $^{13}$C$_6$-lysine. In consequence, the labeled lysine gets incorporated into proteins and in particular into histones, which package the DNA in the nucleus. Histones are believed to be only synthesized when a cell is made and to remain stable throughout the lifetime of the cell. Thus, the fraction of labeled histones can serve as a read-out for cell turnover.

\vskip0.3cm
\noindent {\it The technique from the neurogenesis paper has to be modified due to the unknown initial age distribution. --- }
In flatworms, we encounter two main differences: On the positive side, we can directly measure (and to some extent even deliberately vary) the label content in the body $C_{in}(t)$. On the negative side, we do not know an initial age distribution analogous to Eq.~\ref{iniC14}. In the following, we suggest experimental protocols that resolve this problem.\\
Similar to Eq.~\ref{partialC14}-\ref{boundC14}, we can describe the time evolution by
\begin{eqnarray} &&\partial_t\,n(t,a)+\partial_a\,n(t,a)=-k_{loss}(t,a)\,n(t,a)\label{partialHist}\\[0.3cm]
&&n(t,0)=n_0(t)\,.\label{boundHist}
\end{eqnarray}
This set of equations can be solved analytically
\begin{equation} n(t,a)=n_0(t-a)\,\exp\left(-\int_0^ak_{loss}(t-a+a',a')\,da'\right)\,.\label{solHist}\end{equation}
The number of cells of age $a$ at time $t$ is the same number that was born at time $t-a$, but reduced by all the death events in between, described by the exponential term with the integral over the cell death time course. If we knew or were able to model all the cell birth and death processes for all times in the past, this solution would give us the full age distribution at time $t$. Alternatively, one could only consider birth and death rates starting from a particular time point, if one knew the initial distribution at this time, see Eq.~\ref{iniC14}.

\noindent Unfortunately, in flatworms we do neither know the age distribution at one initial time point nor the full feeding and growth history, which would enable us to model $n_0$ and $k_{loss}$. Luckily, there are two options to solve this problem, each relies on a particular assumption.

\vskip0.3cm
\noindent {\it First option: Assuming a stem cell control model. --- }
If there is no feedback from $k_{loss}(t,a)$ or $n(t,a)$ on cell birth $n_0(t)$ (i.e.~mainly negligible apoptosis-induced cell division), we can determine the age distribution $n(t,a)$ at least partially for all ages $a<t-t_0$, assuming the experiment (i.e.~the well-controlled feeding conditions) starts at $t_0$. The longer we perform the experiment, the more of the distribution we can compute. As cells do not live forever and the number of old cells decays exponentially (see Eq.~\ref{solHist}), we will approximately obtain the full age distribution with time.

\vskip0.3cm
\noindent {\it Second option: Assuming the system relaxes to a steady state for a reference feeding condition. --- }
If we provide a constant feeding condition, for which the worms neither grow nor degrow, we might assume that the age distribution eventually relaxes to a steady state $n^*(a)$, which is a solution to the equation
\begin{equation} \partial_a\,n^*(a)=-k_{loss}(a)\,n^*(a)\,.\end{equation}
This equation can be solved for a particular choice of $k_{loss}(a)$ and $n^*_0$:
\begin{equation} n^*(a)=n^*_0\,e^{-\int_0^a\,k_{loss}(a')\,da'}\,.\end{equation}

\noindent In order to obtain such a steady state, we would need to constantly provide the worms with a small and well-controlled amount of food. Even though, theoretically it would be the most clean approach, experimentally it is rather difficult to frequently pipette a well-defined amount of food to each worm.

\vskip0.3cm
\noindent {\it Modified second option: Assuming a quasi-steady state for a reference feeding condition. --- }
If worms are fed every second week, on average, they will also neither grow nor degrow but show size oscillations with a frequency of two weeks. Importantly, the steady state argument can be modified for such a periodically oscillating quasi-steady state $n^*(t,a)$.
This quasi-steady state can be used as an initial condition for other feeding schemes in analogy to Eq.~\ref{iniC14}.

\subsection{Label dynamics after a single feeding event}\noindent
Here, we illustrate the measurement scheme by discussing the effect of a single feeding pulse with labeled lysine in a simplified example.
The total influx of the amino acid lysine is given by
\begin{equation} J_{AF}=\frac{N_{AF}}{\tau_F} \,e^{-t/\tau_F}\,,\end{equation}
where $N_{AF}$ is the total amount of the amino acid lysine in the gut after feeding and $\tau_F$ is the digestion time of the food. A certain fraction $\Phi_{AF}$ of the lysine in the food is assumed to be labeled by heavy isotopes.

\noindent After feeding with labeled food, the fraction $\Phi_A$ of the labeled amino acid lysine increases in the animal. Lysine belongs to the building blocks of histones and other proteins, which also will become labeled by the incorporation the amino acid. As a simple example, let us assume there are only two proteins in the body that take up lysine: a fast turnover protein and histone. The result is qualitatively the same, independent of the number of proteins that incorporate lysine and might be turned over at very different rates. We define the label fraction $\Phi_H$ of lysine in histones as the ratio between all labeled lysine in histones and the total amount of lysine in histones. Analogously, we also define the label fraction of the lysine in the fast turnover protein $\Phi_P$.

\noindent  If we furthermore assume that cell death happens stochastically and does not depend on the age of the cell, the dynamics of $\Phi_A$ are captured by
\begin{equation}
\partial_t  \Phi_A\, =\frac{J_{AF}}{N_A}\,(\Phi_{AF}-\Phi_A)+\frac{(1-\rho_P)N_{AP}}{N_A}(\Phi_P-\Phi_A)J_P+\frac{(1-\rho_H)N_{AH}}{N_A}(\Phi_H-\Phi_A)\,k_{loss}\,N\,.
\end{equation}
Here, $N_A$ denotes the total amount of lysine in the worm, $(1-\rho_P)$ is the fraction of lysine that can be recycled from fast turnover proteins and $N_{AP}$ is the amount of lysine per protein. Analogously, $(1-\rho_H)$ is the fraction of lysine that can be recycled from histones and $N_{AH}$ is the amount of lysine in all the histones in a cell. The label fraction of the amino acid only changes via recycling or feeding. Each further protein that incorporates lysine would be represented by an analogous term.

\noindent Similarly, the dynamics of the label fractions of the fast turnover proteins and the histones are
\begin{eqnarray}
\partial_t  \Phi_P\, &=&\frac{N_{AP}}{N_P}(\Phi_A-\Phi_P)\,J_P\\
\partial_t  \Phi_H\, &=&\frac{N_{AH}}{N_H}(\Phi_A-\Phi_H)\,n_0\,.
\end{eqnarray}

\begin{figure}[tbp]
  \centering
\includegraphics[width=1\textwidth]{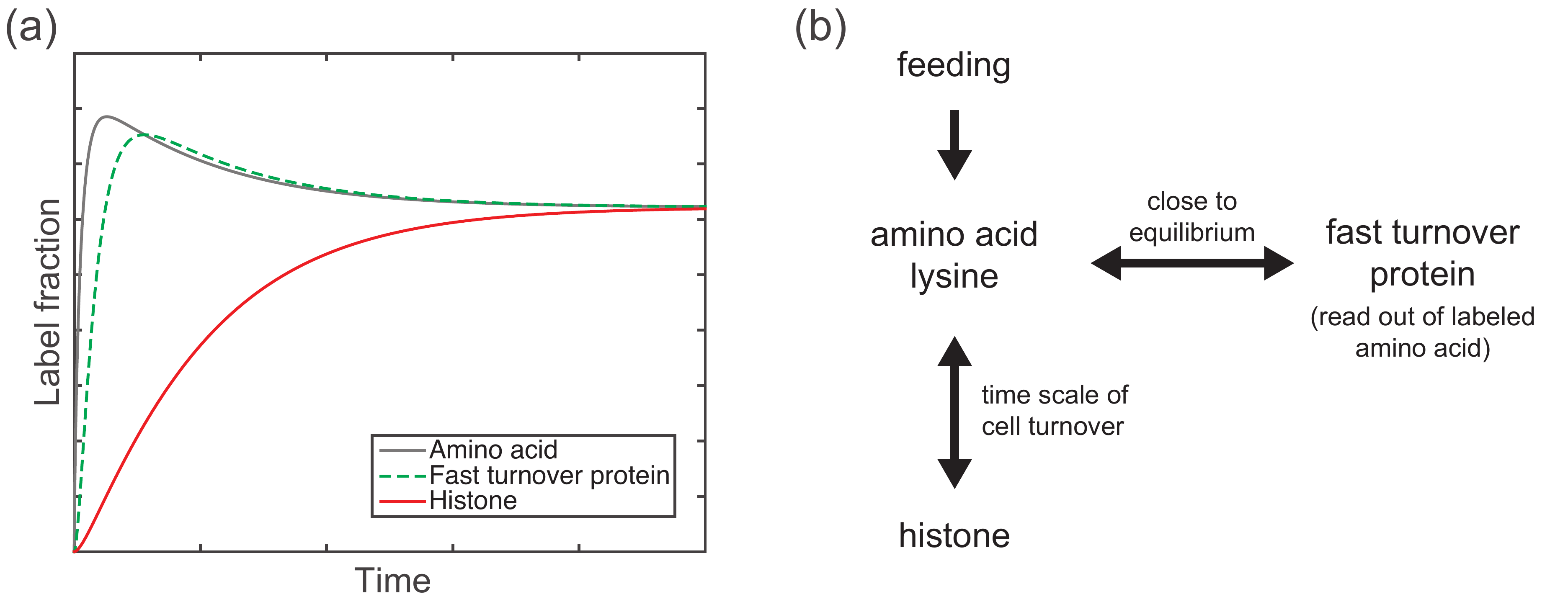}
  \caption[Measuring turnover dynamics on the organism scale in flatworms]{(a) Example dynamics for a single labeling pulse. The label fraction of a fast turnover protein (dashed green) acts as an approximate read-out of the available amount of labeled amino acid lysine (gray), which can be incorporated into other proteins such as histones (red). As an example, we have chosen the fast turnover half as fast as the label digestion and ten times as fast as the cell turnover. If digestion and the fast turnover happen at comparable time scales, the gray and the green curves collapse. (b) Labeling scheme: the label dynamics in the fast turnover protein corresponds to labeled lysine in the worm, while the label incorporation in histones reflects the cell turnover.}  \label{HistScheme}
\end{figure}

\noindent The label fractions of lysine $\Phi_A$, the fast turnover protein $\Phi_P$ and histone $\Phi_H$ are shown in Fig.~\ref{HistScheme}(a) for a single feeding pulse with labeled food. $\Phi_A$ (gray) cannot be measured directly, yet the fast turnover protein (dashed green) immediately follows and can be used as a read-out of labeled lysine. Finally, the histones (red) also incorporate the label on a much longer time scale given by the life times of the cells. Fig.~\ref{HistScheme}(b) sketches the main principle of the experiment.

\noindent By combining the measured fraction of labeled fast turnover proteins with a model on cell turnover, we can compute
\begin{equation} \Phi_H = \int_0^{\infty}\frac{\Phi_P(t-a)\,n(t,a)}{N(t)}\,da\end{equation}
and compare it to the measured value of $\Phi_H(t)$. Here, $\Phi_P(t)$ takes over the role of $C_{in}(t)$ in Eq.~\ref{totcC14}. Yet, in contrast to the neurogenesis paper, the concentration of label in the worm (which can be incorporated into histones) can be directly measured by monitoring the label in high turnover proteins.

\subsection{Challenges and advantages of the application to flatworms}
Lacking knowledge about an initial age distribution, we have to rely on experimental protocols and specific assumptions that eliminate a potential history-dependence. We can tackle this problem by assuming a periodically varying quasi-steady state when feeding every second week and maybe independently measure the respective cell birth rate. We will be able to estimate the time scale of the relaxation to this quasi-steady state based on the spreading of neoblast clones in irradiated worms. Once it has been characterized, the quasi-steady state can act as the initial configuration for other feeding schemes.

\noindent As an advantage in flatworms, we might be able to deliberately vary the time course of the internalized label by alternating between labeled and non-labeled food. This might help to distinguish different models for birth rate $n_0$ and death rate $k_{loss}$ to a higher resolution. Note that the more $C_{in}(t)$ or $\Phi_P(t)$ varies with time, the more details of the models can be resolved. If the internal label was kept constant, almost any model would fit the data equally well. As a second advantage, we can directly determine the label in the body using high turnover proteins.

% ---------------------------------------------------------------------------
% ----------------------- end of thesis sub-document ------------------------
% ---------------------------------------------------------------------------
\end{appendix}

%: ----------------------- list of figures/tables ------------------------

\listoffigures	% print list of figures

%\listoftables  % print list of tables

% List of variables -------------------------------------

\chapter*{List of Variables}
\addcontentsline{toc}{chapter}{List of Variables}

{\bf\large In Chapter~\ref{patterningA}. \large\nameref{patterningA},}\\
{\bf\large and Chapter~\ref{patterningB}. \large\nameref{patterningB}\large:}

%\begin{table}[h]
%\begin{center}
  \begin{longtable}[l]{ l  l }
    $C$ & concentration profile of morphogens \\
    $C_0$ & amplitude of morphogen profile\\ 
    $\mathcal{Z}$ & shape function of morphogen profile \\ 
    $\nu$ & production function of morphogen\\
    $w$ & source size of morphogen\\
    $\alpha$ & production rate of morphogen\\
    $\beta$ & effective degradation rate of morphogen\\
    $D$ & effective diffusion coefficient of morphogen\\
    $\lambda=\sqrt{D/\beta}$ & length scale of morphogen profile\\
    $E$ & concentration profile of expander \\
    $\nu_E$ & source function of expander\\
    $w_E$ & source size of expander\\
    $\alpha_E$ & production rate of expander\\
    $\beta_E$ & effective degradation rate of expander\\
    $D_E$ & effective diffusion coefficient of expander\\
    $\lambda_E$ & length scale of expander profile\\
    $C_{th}$ & threshold for expander production (expansion-repression model) \\
    $\beta_{0}$ & morphogen degradation in the absence of expander (exp.-rep.~model) \\
    $E_{th}$ & conc.~threshold for feedback of expander on morphogen degradation\\
    $\chi_{\raisemath{-2pt}{\lambda}}=\lambda/L$ & proportionality factor between $\lambda$ and $L$\\
    $\chi_w=w/L$ & proportionality factor between $w$ and $L$\\   
    $A$, $B$ & concentration profiles of Turing molecules (activator and inhibitor) \\
    $B_c$ & formal shift of inhibitor concentration between the two Turing topologies\\
    $a$, $b$ & small perturbations of steady state profiles\\ 
    $R_A$, $R_B$ & generic reaction functions of Turing system\\ 
    $P$ & switch-like source function of Turing system (Hill or Heaviside function)\\
    $h$ & exponent for Hill function\\
    $\ell=\langle P\rangle\,L$ & size of self-organized source region (includes spatial average of $P$)\\
    $\alpha_A$, $\alpha_B$  & production rates of Turing molecules\\
    $\beta_A$, $\beta_B$ & degradation rates of Turing molecules\\
    $D_A$, $D_B$ & diffusion coefficients of Turing molecules\\
    $\lambda_A$, $\lambda_B$ & characteristic length scales of Turing molecules\\
    $M$ & linearized reaction-diffusion matrix\\
    $s$ & wavenumber of Fourier modes\\
    $q$ & eigenvalues of reaction-diffusion matrix\\
    $m$ & number of contiguous source regions\\
    $n$ & number of source regions touching the system boundary\\
    $\sigma=2m-n$ & pattern number\\
    $L_{\sigma}$ & system size beyond which a Turing pattern of pattern number $\sigma$ exists\\
    $\kappa_{A}$, $\kappa_{B}$, $\kappa_{E}$ & coupling constant for degradation rates in scalable Turing system\\
    $E_{0}$ & hypothetically imposed expander level in scalable Turing system\\
    $f_{(m,n)}$ & source size on the nullcline of scalable Turing system for $\partial_t E=0$\\
    $g$ & source size of a homogeneous dynamic state for low expander levels\\
    $\chi=B/A$ & constant ratio in a homogeneous dynamic state for low expander levels
      \end{longtable}
%\end{center}
%\end{table}
\vskip0.4cm

\noindent {\bf\large In Chapter~\ref{pca}. \nameref{pca}:}

 \begin{longtable}[l]{ l  l }
$s$ & arc-length along worm circumference\\
$L$ & total length of circumference\\
$\hat{s}=s/L$ & normalized arc-length\\
$\r$ & radial position vector\\
$\r_0$ & center point of worm\\
$\rho=\vert\r-\r_0\vert$ & radial distance\\
$\overline{\rho}$ & mean radius\\
$\hat{\rho}=\rho(s)/\overline{\rho}$ & normalized radius profile defining the shape\\
$\hat{\rho}_0$ & average shape variations\\
$N_w$ & number of worm outlines\\
$N_r$ & number of data points along worm perimeter\\
$\mathcal{R}_{i,j}=\hat{\rho}_i(s_j)$ & data matrix\\
$\mathcal{C}$ & covariance matrix\\
$\v$ & eigenvectors or shape modes\\
$B$ & eigenvalues or shape scores
\end{longtable}

\vskip0.4cm
%\newpage
\noindent {\bf\large In Chapter~\ref{growth}. \nameref{growth}:}

 \begin{longtable}[l]{ l  l }
$N$ & total cell number\\
$k_{div}$ & average cell division rate\\
$k_{loss}$ & average cell loss rate\\
$K=k_{div}-k_{loss}$ & average growth rate\\
$E_t$ & total stored metabolic energy in the worm\\
$e_c=E/N$ & stored metabolic energy per cell\\
$J_f$ & influx of metabolic energy due to feeding\\
$j_f=j_f/N$ & influx of metabolic energy per cell\\
$\mu$ & metabolic consumption rate per cell\\
$K_0$ & degrowth rate in the absence of any stored energy (model 1)\\
$e_s$ & energy threshold for the switch between growth and degrowth (model 1)\\
$\tau_K$ & relaxation time of the growth rate to a quasi steady state (model 2,3)\\
$e_0$ & target value of the metabolic energy (model 2,3)\\
$a$ & age of cells\\
$n$ & number density of cells of age $a$\\
$N_{\ell}$ & number of labeled cells\\
$k_i$, $k_d$, $k_r$  & age-dependent rates of insertion, deletion and replacement of cells\\
$\bar{k}_{d}$ &  averaged rate of deletion across the tissue\\
$\bar{k}_{d\ell}$, $\bar{k}_{r\ell}$  & averaged rates of deletion and replacement of labeled cells\\
$\phi_{\ell}$ & fraction of labeled cells\\
$N_{1}$, $N_{2}$ & number of cells with one or two labels\\
$\bar{k}_{d1}$, $\bar{k}_{r1}$  & averaged rates of deletion and replacement of cells with one label\\
$\bar{k}_{d2}$, $\bar{k}_{r2}$  & averaged rates of deletion and replacement of cells with two labels\\
$\phi_{12}=N_{1}/N_{2}$ & label ratio
\end{longtable}

% ---------------------------------------------------------------------- %list of variables

%: ----------------------- bibliography ------------------------

% The section below defines how references are listed and formatted
% The default below is 2 columns, small font, complete author names.
% Entries are also linked back to the page number in the text and to external URL if provided in the BibTex file.

% PhDbiblio-url2 = names small caps, title bold & hyperlinked, link to page 
%\begin{multicols}{2} % \begin{multicols}{ # columns}[ header text][ space]
\begin{small} % tiny(5) < scriptsize(7) < footnotesize(8) < small (9)

\bibliographystyle{jmb} % Title is link if provided (Latex/Classes/PhDbiblio-url2, acm, siam)
\renewcommand{\bibname}{References} % changes the header; default: Bibliography
%\bibliography{/Users/admin/Documents/Publications/AllometricScaling,/Users/admin/Documents/Publications/CellKineticsTurnover,/Users/admin/Documents/Publications/DevBioHistory,/Users/admin/Documents/Publications/DrosophilaMorphogen,/Users/admin/Documents/Publications/GradientScaling,/Users/admin/Documents/Publications/InformationANDTaxis,/Users/admin/Documents/Publications/Midline,/Users/admin/Documents/Publications/MiscGeneral,/Users/admin/Documents/Publications/MiscOrganisms,/Users/admin/Documents/Publications/MiscPatternScaling,/Users/admin/Documents/Publications/PaperTuringScaling,/Users/admin/Documents/Publications/PatterningMisc,/Users/admin/Documents/Publications/PlanarianGeneral,/Users/admin/Documents/Publications/PlanarianGermline,/Users/admin/Documents/Publications/PlanarianMisc,/Users/admin/Documents/Publications/PlanarianMorphogenPatterning,/Users/admin/Documents/Publications/PlanarianNeoblastsTurnover,/Users/admin/Documents/Publications/PotentialGapjuncPlanarianOthers,/Users/admin/Documents/Publications/MorphogenTheory,/Users/admin/Documents/Publications/ShapeMotility,/Users/admin/Documents/Publications/TuringMeinhardt,/Users/admin/Documents/Publications/TuringScaling,/Users/admin/Documents/Publications/TuringSystems} %IMPORTANT "no spaces"
\bibliography{thesis}

\end{small}
%\end{multicols}

% --------------------------------------------------------------
% Various bibliography styles exit. Replace above style as desired.

% in-text refs: (1) (1; 2)
% ref list: alphabetical; author(s) in small caps; initials last name; page(s)
%\bibliographystyle{Latex/Classes/PhDbiblio-case} % title forced lower case
%\bibliographystyle{Latex/Classes/PhDbiblio-bold} % title as in bibtex but bold
%\bibliographystyle{Latex/Classes/PhDbiblio-url} % bold + www link if provided

%\bibliographystyle{Latex/Classes/jmb} % calls style file jmb.bst
% in-text refs: author (year) without brackets
% ref list: alphabetical; author(s) in normal font; last name, initials; page(s)

%\bibliographystyle{plainnat} % calls style file plainnat.bst
% in-text refs: author (year) without brackets
% (this works with package natbib)

% --------------------------------------------------------------

%: Declaration of originality

% Thesis statement of originality -------------------------------------

% Depending on the regulations of your faculty you may need a declaration like the one below. This specific one is from the medical faculty of the university of Dresden.

\begin{declaration}        %this creates the heading for the declaration page

%I herewith declare that I have produced this thesis without the prohibited assistance of third parties and without making use of aids other than those specified; notions taken over directly or indirectly from other sources have been identified as such. This thesis has not previously been presented in identical or similar form to any other German or foreign examination board.
%
%\noindent The thesis work was conducted from October 2011 to October 2015 under the supervision of Prof. Dr. Frank Jlicher at the Max-Planck-Institute for the Physics of Complex Systems in Dresden.
%
%\vspace{10mm}

%Hiermit versichere ich, dass ich die vorliegende Arbeit ohne unzul\"assige Hilfe Dritter und ohne Benutzung anderer als der angegebenen Quellen und Hilfsmittel angefertigt und die den benutzten Quellen w\"ortlich oder inhaltlich entnommenen Stellen als solche kenntlich gemacht habe. Diese Arbeit wurde nicht schon in dieser oder \"ahnlicher Form bei einem Pr\"ufungsamt in Deutschland oder einem anderen Land eingereicht.

Hiermit versichere ich, dass ich die vorliegende Arbeit ohne unzul\"assige Hilfe Dritter und ohne Benutzung anderer als der angegebenen Hilfsmittel angefertigt habe; die aus fremden Quellen direkt oder indirekt \"ubernommenen Gedanken sind als solche kenntlich gemacht. Die Arbeit wurde bisher weder im Inland noch im Ausland in gleicher oder \"ahnlicher Form einer anderen Pr\"ufungsbeh\"orde vorgelegt.

\vspace{5mm}

\noindent Die vorliegende Arbeit wurde im Zeitraum zwischen Oktober 2011 und Dezember 2015 unter der Betreuung durch Prof. Dr. Frank J\"ulicher und Dr. Benjamin M. Friedrich am Max-Planck-Institut f\"ur Physik komplexer Systeme in Dresden angefertigt.

\vspace{5mm}

\noindent Ich versichere, dass ich bisher keine erfolglosen Promotionsverfahren unternommen habe. Ich erkenne die Promotionsordnung der Fakult\"at der Mathematik und Naturwissenschaften der Technischen Universit\"at Dresden an.

\vspace{5mm}

\noindent Teile dieser Arbeit wurden schon in folgender Form ver\"offentlicht:\\
- S. Werner, J.C. Rink, I.-H. Riedel-Kruse, B.M. Friedrich: Shape mode analysis exposes movement patterns in biology, PLoS ONE, 9(11), e113083, 2014\\
- S. Werner, T. St\"uckemann, M. Beiran Amigo, J.C. Rink, F. J\"ulicher, B.M. Friedrich: Scaling and regeneration of self-organized patterns, Phys. Rev. Lett. 114, 138101, 2015 (Editors' suggestion)

\vspace{15mm}

Dresden,

\end{declaration}

% ----------------------------------------------------------------------

\end{document}